\documentclass[final,1p,11pt]{elsarticle}

\usepackage{epubtk}    
\usepackage{booktabs}  
\usepackage{graphicx}  
\usepackage{xcolor,color,graphics}
\usepackage{makeidx}
\usepackage{epsfig,subfigure}
\usepackage{epstopdf}
\usepackage{bm}
\usepackage{amsmath}
\usepackage{amssymb}
\usepackage{amsthm}
\usepackage{color}
\usepackage{ucs}
\usepackage[utf8]{inputenc}
\usepackage{fontenc}
\usepackage[numbers]{natbib}
\usepackage{bbm}
\usepackage{calligra}
\usepackage{soul}

\usepackage[makeroom]{cancel}

\usepackage{amssymb}

\journal{Physics Reports}

\newcommand{\be}{\begin{equation}}
\newcommand{\ee}{\end{equation}}
\newcommand{\ba}{\begin{eqnarray}}
\newcommand{\ea}{\end{eqnarray}}

\newcommand{\lsim}   {\mathrel{\mathop{\kern 0pt \rlap
  {\raise.2ex\hbox{$<$}}}
  \lower.9ex\hbox{\kern-.190em $\sim$}}}
\newcommand{\gsim}   {\mathrel{\mathop{\kern 0pt \rlap
  {\raise.2ex\hbox{$>$}}}
  \lower.9ex\hbox{\kern-.190em $\sim$}}}

\newcommand{\Tr}{{\rm Tr}}

\newcommand{\Id}{\mathbbm 1}

\newcommand{\mpl}{M_{\rm Pl}}

\def\mpl{M_{\rm Pl}}
\def\e{{\epsilon}}
\def\E{\mathcal{E}}

\def\K{{\cal K}}

\def\L*{{\cal L}_*}
\def\L{\mathcal{L}}
\def\({\left(}
\def\){\right)}

\def\nn{\nonumber}

\def\mn{_{\mu \nu}}

\def\<{\langle}
\def\>{\rangle}

\def\cs2{c_{s}^{2}}

\def\be{\begin{equation}}
\def\ee{\end{equation}}
\def\ba{\begin{eqnarray}}
\def\ea{\end{eqnarray}}
\def\beq{\begin{eqnarray}}
\def\eeq{\end{eqnarray}}

\def\mpl{M_{\rm Pl}}
\def\e{{\epsilon}}
\def\E{\mathcal{E}}

\def\K{{\cal K}}

\def\L*{{\cal L}_*}
\def\L{\mathcal{L}}
\def\({\left(}
\def\){\right)}

\def\nn{\nonumber}

\def\mn{_{\mu \nu}}

\def\<{\langle}
\def\>{\rangle}

\def\Tr{{\rm Tr}}

\begin{document}

\begin{frontmatter}



\title{\huge{A systematic approach to generalisations of General Relativity and their cosmological implications}}


\author{Lavinia Heisenberg}
\address{Institute for Theoretical Studies, ETH Zurich, Clausiusstrasse 47, 8092 Zurich, Switzerland.}
\ead{lavinia.heisenberg@eth-its.ethz.ch}
\cortext[cor1]{Corresponding author}

\begin{abstract}
A century ago, Einstein formulated his elegant and elaborate theory of General Relativity, which has so far withstood a multitude of empirical tests with remarkable success. Notwithstanding the triumphs of Einstein's theory, the tenacious challenges of modern cosmology and of particle physics have motivated the exploration of further generalised theories of spacetime. Even though Einstein's interpretation of gravity in terms of the curvature of spacetime is commonly adopted, the assignment of geometrical concepts to gravity is ambiguous because General Relativity allows three entirely different, but equivalent approaches of which Einstein's interpretation is only one. From a field-theoretical perspective, however, the construction of a consistent theory for a Lorentz-invariant massless spin-2 particle uniquely leads to General Relativity. Keeping Lorentz invariance then implies that any modification of General Relativity will inevitably introduce additional propagating degrees of freedom into the gravity sector. Adopting this perspective, we will review the recent progress in constructing consistent field theories of gravity based on additional scalar, vector and tensor fields. Within this conceptual framework, we will discuss theories with Galileons, with Lagrange densities as constructed by Horndeski and beyond, extended to DHOST interactions, or containing generalized Proca fields and extensions thereof, or several Proca fields, as well as bigravity theories and scalar-vector-tensor theories. We will review the motivation of their inception, different formulations, and essential results obtained within these classes of theories together with their empirical viability.

\end{abstract}


\end{frontmatter}



\newpage

\tableofcontents

\section{Introduction}

Physics stands for \textquotedblleft knowledge of Nature \textquotedblright and studies the matter of our world and its evolution determined by the laws of physics. The macroscopic and the microscopic world are described by two conceptually simple standard models: the Standard Model of Particle Physics and the Standard Model of Big Bang cosmology. They are based on the physical assumptions and techniques of Quantum Field Theory and General Relativity. The unification of these two worlds would mark a breakthrough step forward in our understanding of Nature and is still a monumental challenge for modern physics. 

Nature manifests itself through four fundamental forces: the electromagnetic, the strong, the weak and the gravitational force. Despite the fact, that the gravitational force was the first to be revealed, it still constitutes the most puzzling one, posing the most tenacious problems.
Gravitational physics recently witnessed an important centenary and a breathtaking event. The year 2015 saw the 100th anniversary of Albert Einstein's elegant and elaborate theory of General Relativity and the year 2016 marked a milestone in the detection of gravitational waves \cite{GW150914}. Einstein successfully constructed General Relativity based on a geometrical interpretation of gravity. He inferred from his principal of equivalence that the gravitational interactions are due to the curvature of spacetime. This intimate relation of General Relativity with the geometry of spacetime puts it into a very extraordinary and distinctive place compared to other forces in Nature, even if its geometrical interpretation is ambiguous \cite{Nester:1998mp,Aldrovandi:2013wha,BeltranJimenez:2017tkd}. 

More than 100 years after its inception General Relativity is still the best theory to describe the underlying gravitational physics on a vast range of scales. A multitude of laboratory, ground-based and space-borne experiments were applied to an intense scrutiny of General Relativity in the last decades. It was put on trial by laboratory measurements at sub-mm scales and constrained with an exquisite precision by the observations of the Solar System \cite{Will:2005va}. The predictions of General Relativity for the  emission of gravitational waves is compatible with the decrease of the orbital period of binary systems at the percent level. Furthermore, it is in perfect agreement with the observations of black hole and neutron star mergers \cite{TheLIGOScientific:2017qsa}. After a great deal of inquiry General Relativity has outlived most competitors among the alternative gravity theories. 

The underlying physics on cosmological scales is described by the Standard Model of Big Bang cosmology on the basis of two fundamental pillars: General Relativity and the cosmological principle. The latter relies on the assumption that the universe is spatially homogeneous and isotropic on cosmological scales. Different combinations of cosmological observations like those of the Cosmic Microwave Background (CMB), supernovae, lensing and baryon acoustic oscillations (BAO) have firmly established the standard $\Lambda$CDM model of cosmology, which requires an accelerated expansion of the universe at late times. This simple world model puts us in an excellent position to account for the observed phenomenology on cosmological scales. Nevertheless, this simple picture forces us to introduce three unknown ingredients: dark energy in form of a cosmological constant, dark matter and the inflaton field. Despite many years of research their origin has not yet been identified. 

Within the simple cosmological standard model, there thus remain severe theoretical challenges. The most persistent theoretical obstacle without a satisfactory foundation is the cosmological constant problem \cite{Weinberg:1988cp,Martin12}, representing the largest discrepancy between theoretical predictions and observations in all of science. From an effective field theory perspective, the cosmological constant has to be added to the Einstein-Hilbert action and there is no symmetry recovered in the limit of a vanishing cosmological constant. Since the de Sitter and Poincar\'e groups have the same number of generators, any fine-tuning of the cosmological constant is technically unnatural. In other words, the classical value of the cosmological constant receives large quantum corrections. Within the realm of General Relativity one cannot address its smallness. 

Another burdensome theoretical challenge is a successful description of gravity according to the principles of quantum physics. Unfortunately, the attempts to apply the usual prescriptions of quantum field theory to the force of gravity fail drastically, since the underlying theory is not renormalizable. One can only use the theory as an effective field theory valid up to the Planck scale and a more complete theory is needed for physics beyond that scale. It could be that the cosmological constant problem is not just an IR problem but reflects the missing piece of a consistent quantum gravity. These are the two biggest theoretical puzzles of the Standard Model of Big Bang cosmology, that concern physics both at the largest and at the smallest scales.

As we mentioned above the Standard Model of Big Bang cosmology is in excellent agreement with observations. However, apart from the aforementioned theoretical challenges, it faces some anomalies from the observational side. Just to mention a few: the tension in the value of the Hubble constant obtained from the CMB and local measurements, the hemispherical asymmetry in the CMB, the lack of power in the CMB on large scales, the possible existence of large scale bulk flows and unexpected large scale correlations inferred from studies of distant quasars among other things. In the context of dark matter, galaxy clusters also seem to allude to a slightly different cosmology than CMB measurements \cite{Ade:2015fva}. 

Several of these tensions and anomalies are still controversial and of an unclear statistical significance. However, jointly they might signal the failure of the cosmological principle. These observational anomalies together with the theoretical challenges have motivated the study of modifications of gravity in the far IR and UV. Assuming that General Relativity is still the right effective theory for the intermediate scales, one can tackle these problems by modifying the gravitational interactions in the IR and UV, respectively. 

With the increase of high precision of the cosmological measurements, cosmology is becoming a fertile testing ground for fundamental physics. It is the unique place to test gravitational interactions. Its multi-disciplinary nature merges different concepts from gravity, field theory, particle physics, quantum mechanics, astrophysics, fluid mechanics, statistics and mathematics, which facilitates the creation of synergies between all these different fields.

The other three forces of Nature are successfully described by the Standard Model of Particle Physics and unified with an exquisite experimental success. The electromagnetic, weak and strong interactions together with the elementary particles have masterly been put together piece by piece. The last missing piece, the Higgs boson, has also joined the crew and gave an additional reassuring support for the Standard Model \cite{Aad:2012tfa,Chatrchyan:2012xdj}. Even though the Standard Model of Particle Physics accounts in a consistent way for most of the particle physics phenomena, it suffers from puzzles similar to the cosmological constant problem, namely the Higgs hierarchy problem. This reflects the 
fact that the mass of the Higgs boson is so many orders of magnitude smaller than the Planck/unification scales. This hierarchy is puzzling as it does not seem to be protected without the help of new physics or symmetries. It originates from tunings that are not technically natural. As a counter example, the electron mass is also hierarchically smaller than the electroweak scale. However, in the limit of vanishing electron mass there is an enhancement of the underlying symmetry in form of an additional chiral symmetry, which protects the electron mass from receiving large quantum corrections in that limit. Hence, beyond that limit quantum corrections will only give rise to a renormalization of the electron mass proportional to itself, which renders the hierarchy between the electron mass and the electroweak scale technically natural. This is 't Hooft's naturalness argument \cite{tHooft:1979bh,Dimopoulos:1979es}. Since this argument can not be applied to the Higgs mass, it exigently enforces the necessity of new physics, such as supersymmetry \cite{Weinberg:2000cr}, extra dimensions \cite{ArkaniHamed:1998rs}, or D-brane configurations in string theory \cite{Antoniadis:1998ig,Antoniadis:2001cv}. Unfortunately, solutions proposed to solve the Higgs hierarchy problem are not applicable to tackle the cosmological constant problem since they rely on physics at different scales. Other challenges of the Standard Model of Particle Physics are the finite neutrino masses, their hierarchy and oscillations. It also fails to explain the baryon asymmetry, reflecting the imbalance between baryonic and antibaryonic matter.

Most of the problems of the two Standard Models motivate the exploration of new physics and modifications of gravity, both in the IR and the UV \cite{Copeland:2006wr,Sotiriou:2008rp,DeFelice:2010aj,Clifton:2011jh,Amendola:2012ys,Amendola:2016saw,Joyce:2014kja,Bull:2015stt}. 
Imposing the conditions of Lorentz symmetry, unitarity, locality and a (pseudo-)Riemannian spacetime, any attempt of modifying gravity inevitably introduces new dynamical degrees of freedom. They could be additional scalar, vector or tensor fields. Specially, if the modifications are such that gravity is weakened on cosmological scales, one could hope for not only tackling the cosmological constant problem, but also for a mechanism responsible for the late-time acceleration enigma. Promising approaches in this context arise in massive gravity or in higher-dimensional frameworks. 

In the latter case, the Dvali-Gabadadze-Porrati (DGP) model is an important large scale modification of gravity \cite{Dvali:2000hr}, which is based on a three-brane embedded in a five dimensional bulk. The brane curvature sources an intrinsic Einstein Hilbert term, that helps to recover the four dimensional gravity on small scales. On the opposite scales, gravity is systematically weakened since it acquires a soft mass. From the four dimensional point of view the effective graviton carries five degrees of freedom: two helicity-1 modes and one helicity-0 mode on top of the usual helicity-2 modes. The decoupling limit of DGP revealed interesting derivative interactions for the helicity-0 mode, that gives rise to second order equations of motion and invariance under Galileon symmetry. The reason why the DGP model received quite some attention in the literature is the existence of a self-accelerating solution, which is sourced by the helicity-0 mode of the graviton. However, it was quickly realised that this branch of solutions seems to be plagued by ghost-like instabilities \cite{Koyama:2005tx,Charmousis:2006pn}.

Another promising approach to the cosmological constant problem and the late-time acceleration enigma arises by replacing the soft mass by a hard mass of the graviton, as is the case in massive gravity. From a purely theoretical perspective, the existence of a graviton mass is an important fundamental question: Is the graviton really massless or does its mass just happen to be so small that it can be safely neglected on sufficiently small scales? Exactly this question was considered long ago. However, introducing a consistent mass term for the graviton is not as easy as it might look at first sight. Starting with the linear interactions one can construct a unique mass term which does not lead to the presence of ghostly degrees of freedom in the theory, namely the Fierz-Pauli action \cite{Fierz:1939ix}. 

Unfortunately, albeit theoretically consistent this linear theory of massive gravity suffers from the vDVZ discontinuity, i.e., in the limit of vanishing graviton mass the predictions of General Relativity are not recovered. The main reason for this is the existence of the helicity-0 degree of freedom, that couples to the trace of the energy momentum tensor. Clearly, the vDVZ discontinuity is just an artefact of the linear approximation and signals the exigent necessity of non-linear completion. Lamentably,  non-linear extensions to the theory usually introduce the Boulware-Deser ghost instability when non-trivial backgrounds are considered \cite{BoulwareD}. This was a challenging task for more than forty years and many authors devoted large efforts to overcoming this difficulty. The attempt by \cite{Creminelli:2005qk} for constructing non-linear interactions of the helicity-0 mode yielded a negative conclusion due to an erroneous decomposition of the massive graviton into its helicity modes. This was successfully corrected in \cite{deRham:2010ik}, which marked a milestone in the construction of ghost-free non-linear interactions for a massive spin-2 field. This was performed in the decoupling limit of massive gravity. The resummation of the non-linear interactions beyond that limit was then triumphantly performed in \cite{deRham:2010kj}, now known as the dRGT model. It constitutes the first consistent example of a ghost free non-linear covariant theory of massive gravity in four dimensions, that avoids the no-go result of Boulware and Deser \cite{Hassan:2011hr}. The construction of a Lorentz invariant massive gravity theory is based on a framework in which the massive graviton propagates on top of a fixed background reference metric. Ghost-free bimetric gravity theories can be constructed by adding an additional kinetic term for the reference metric and hence promoting the reference metric dynamical \cite{Hassan:2011zd}.

In both frameworks, DGP as well as massive gravity, the cosmological constant problem is addressed via degravitation \cite{ArkaniHamed:2002fu,Dvali:2007kt,deRham:2010tw}. Due to the soft or hard mass of the graviton, which plays the role of a high-pass filter, gravity is weakened in the IR and the vacuum energy has a weaker effect on the geometry than anticipated, and hence can reconcile a natural value for the vacuum energy as expected from particle physics with the observed late time acceleration. Both theories contain five propagating degrees of freedom, with the most important protagonist being the helicity-0 mode. Typically, the helicity-1 modes decouple from the conserved stress energy tensor, whereas the helicity-0 mode does not and therefore can mediate an extra fifth force. A successful recovery of General Relativity in the decoupling limit then relies on the Vainshtein mechanism \cite{Vainshtein:1972sx,Babichev:2009jt,Babichev:2013usa}. In this limit the usual helicity-2 modes are treated linearly while the helicity-0 mode is subject to non-linear interactions. The basic idea behind the Vainshtein mechanism is that the helicity-0 mode can be decoupled from the gravitational dynamics via its nonlinear interactions. In the vicinity of matter, these non-linear interactions of the helicity-0 mode become appreciable and after canonical normalisation its coupling to matter becomes suppressed.

Motivated by the non-linear interactions of the helicity-0 mode in the decoupling limit of DGP, more general Galilean invariant interactions have been proposed \cite{Nicolis:2008in}. From the DGP perspective, the invariance under internal Galilean and shift transformations is a relict of the five dimensional Poincar\'e invariance. From the point of view of a four dimensional effective field theory, one can construct the most general Lagrangian for this Galileon scalar field, that is invariant under these residual symmetries and gives rise to second order equations of motion. The latter condition is relevant for the absence of ghostly degrees of freedom behind the derivative interactions. In four dimensions one can construct five interactions of this type. 

There has been a flurry of investigations of these Galileon interactions. One worrying property is the superluminal propagation of spherically symmetric solutions around compact sources \cite{Hinterbichler:2009kq,deFromont:2013iwa} and the danger of constructing closed time-like curves \cite{Evslin:2011vh}. However, a similar Chronology Protection conjecture as in General Relativity is realized in Galileon models and any attempt to form a closed time-like curve faces the fact that Galileon inevitably becomes infinitely strongly coupled and the effective field theory breaks down \cite{Burrage:2011cr}. Even if the Galileon interactions are interesting in their own right as scalar field theories, an interesting question is how one can promote the Galileon to a non-flat background. The covariantization of the decoupling limit of the DGP model was performed in \cite{Chow:2009fm}. The naive covariantization of the entire Galileon interactions would yield higher order equations of motion unless appropriate non-minimal couplings to gravity are added \cite{Deffayet:2009wt}. This covariant Galileon led to the rediscovery of the Horndeski scalar-tensor theories \cite{Horndeski:1974wa}. From a five dimensional perspective these covariant Galileon interactions are consequences of Lovelock invariants in the bulk of generalized braneworld models \cite{deRham:2010eu}. In analogy to the covariantization of the DGP decoupling limit \cite{Chow:2009fm}, one can also directly covariantize the decoupling limit of massive gravity and construct consistent scalar-tensor theories, which are a subclass of the Horndeski interactions \cite{deRham:2011by,Heisenberg:2014kea}. In all these theories the covariantization procedure removes the Galileon symmetry. However, one can devote an additional effort to successfully generalising the Galileon interactions to the (Anti-) de Sitter background and ultimately to maximally symmetric backgrounds, where a generalized Galileon symmetry becomes apparent \cite{Goon:2011qf,Burrage:2011bt}. 

Among the modified gravity theories, those based on scalar fields are the most extensively explored models. In the cosmological context, one practical benefit is that scalar fields can give rise to accelerated expansion without breaking the isotropy of the universe. They do not need to be the scalar fields that we know from the Standard Model of Particle Physics, like for instance the Higgs boson, but may belong to the gravity sector.
The basic inflationary paradigm providing an initial phase of accelerated expansion of the universe is commonly ascribed to a scalar field with flat potential. On the other hand, if one wants a scalar field to act as dark energy, it has to be very light, which then results in long-range forces. Since such fifth forces have never been detected in local gravity tests, the scalar field has to be hidden on small scales via screening mechanisms whereas being unleashed on large scales to produce cosmological effects. The aforementioned Horndeski scalar-tensor theories with second order equations of motion have been extensively explored for this purpose \cite{Kimura:2011dc,Koyama:2013paa,Kase:2013uja}. Giving up the restriction about the second order nature of the equations of motion, one can construct beyond Horndeski theories with higher order equations of motion but still retaining the right number of propagating degrees of freedom \cite{Zumalacarregui:2013pma,Gleyzes:2014dya}. The scalar-tensor theories do not only have important cosmological, but also astrophysical implications \cite{DeFelice:2011hq,Babichev:2013cya,Sotiriou:2013qea}.

The Standard Model of Particle Physics represents the fundamental fields of the gauge interactions by abelian and non-abelian vector fields. 
After their experimental discovery, we at least know that vector fields exist in Nature. This motivates the exploration of the role of bosonic vector fields in the cosmological evolution of the universe apart from scalar fields. Furthermore, some of the aforementioned anomalies could be indicating a preferred direction effect, which could be naturally generated by a vector field. For an abelian vector field with $U(1)$ gauge symmetry it is not possible to obtain homogeneous and isotropic background solutions. Therefore one would need to promote it to the non-abelian case, where three orthogonal vector fields can realise the symmetry together via their spatial components. An alternative relatively unexplored route emerges if one explicitly breaks the underlying $U(1)$ symmetry of the vector field and constructs generalisations of the Proca vector field \cite{Heisenberg:2014rta,Tasinato:2014eka,Allys:2015sht,Jimenez:2016isa}. The generalised Proca theories are the most general vector-tensor theories with second order equations of motion for both the tensor and vector field. 

In analogy to the scalar-tensor theories one can construct more general interactions by abandoning the restriction of second order equations of motion. In this way one can construct beyond generalized Proca theories with higher order equations of motion but still maintaining the three propagating degrees of freedom without the Ostrogradski instability \cite{Heisenberg:2016eld,Kimura:2016rzw}. Instead of breaking the $U(1)$ symmetry of an abelian vector field, one can consider the breaking of the $SU(2)$ symmetry of a non-abelian gauge field and construct derivative self interactions for such a field. Performing an analogous construction scheme as in generalised Proca theories, one can construct the consistent interactions for a set of vector fields, the multi-Proca theories \cite{Jimenez:2016upj,Allys:2016kbq}. Last but not least, an important step towards unifying the general classes of Horndeski and generalized Proca theories was undertaken in \cite{Heisenberg:2018acv}, which gave rise to consistent ghost-free scalar-vector-tensor gravity theories with second order equations of motion with derivative interactions, for both the gauge invariant and the broken-gauge cases.

When Einstein constructed his theory of General Relativity, he pursued the purely geometrical interpretation of gravity. In his formulation the gravitational interactions are ascribed to the curvature of spacetime. In this review we will adopt the more modern particle physics perspective of gravity and represent General Relativity as the unique theory of a massless spin-2 particle. Imposing locality, unitarity, and Lorentz invariance as the fundamental principles, we will construct alternative consistent field theories of gravity based on additional scalar, vector and tensor fields. Since there exist already exhaustive reviews on massive gravity \cite{Hinterbichler:2011tt,deRham:2014zqa,Schmidt-May:2015vnx}, we only aim at discussing the very recent developments in this respect and the points relevant for the rest of this review. It is an important concern to us to fill the gap in the recent literature of alternative theories of gravity by gathering a comprehensible account of the many different constructions of field theories, including the recent developments of vector-tensor theories and beyond. We shall devote an additional effort to classify and unify the different theoretical assumptions of gravity theories and discuss their defining properties together with their theoretical and phenomenological results. It will be hopefully useful both for particle physicists and astrophysicists/cosmologists interested in submerging into the framework of consistent gravity theories and in search of alternatives to the standard $\Lambda$CDM and inflation model. The review addresses both the familiarised reader and the non-expert and contains parts with an exhaustive and rigorous exposition for the newcomer but also brief compendiums of useful concepts for experts. 

Before diving into the consistent construction of field theories, we will give a brief skim through the different geometrical interpretations of General Relativity in the next section and reinforce our statement that the assignment of geometrical concepts to gravity is ambiguous. We will see that General Relativity allows three entirely different, but equivalent approaches of which Einstein's interpretation is only one. We have grown accustomed to assigning gravity to the curvature of a given spacetime. However, this perception has masked the fact that differential geometry affords much wider classes of geometric objects to represent the geometrical properties of manifolds. Besides the curvature, these are torsion and non-metricity. In Einstein's formulation of General Relativity, both non-metricity and torsion vanish. An equivalent representation of General Relativity can be established based on a flat spacetime with a metric, but asymmetric connection. In this teleparallel description, gravity is entirely ascribed to torsion \cite{Aldrovandi:2013wha}. A third equivalent and simpler representation can be constructed on an equally flat spacetime without torsion, in which gravity is purely assigned to non-metricity. In terms of a suitable gauge choice, the connection vanishes completely, depriving gravity from any inertial character, and the resulting action is purged from the boundary term \cite{BeltranJimenez:2017tkd}.

\section{Different Interpretations of Gravity}\label{section_GeomGrav}
When Einstein built his fundamental theory of gravity, he did not follow a modern field theory approach, but rather chose a geometrical formulation. You might wonder why one should expect to be able to describe gravity by geometry. The key word is: equivalence principle. The possibility to ascribe gravity to geometry is granted by the equivalence principle. The origin of this conception dates back to the somewhat unsatisfactory role of inertial frames in Newtonian physics. According to Newton, the force acting on an object is equal to the inertial mass $m_I$ of that object multiplied by the acceleration $F=m_Ia$, assuming a constant mass. This force is an inertial force if it is an apparent force acting on a mass whose motion is described using a non-inertial frame of reference and is always proportional to the inertial mass $m_I$. This inertial force, sometimes referred to as fictitious force, can be gotten rid of in an inertial frame of reference. Examples are for instance the centrifugal force and the Coriolis force on earth. 

The extraordinary achievement of Einstein began with the innocent wonder, whether gravity could be seen as a fictitious force as well, since an observer in a freely falling reference frame is not exposed to the gravitational force. Hence, Einstein wondered whether freely falling reference frames could replace inertial frames. If the gravitational mass $m_g$ equaled the inertial mass $m_g=m_I$, then the gravitational force $F=m_g g$ would be of the same form as an inertial force. The equivalence principle relies strongly on the assumption $m_g=m_I$, which is observationally confirmed with a tiny little uncertainty. In a modern language, the equivalence principle dictates that all the matter fields couple with the same strength to gravity. Therefore, the geometrical nature of gravity emerges from the universality dictated by the equivalence principle and the motion of particles is solely determined by the geometrical properties of spacetime.

In Newtonian physics gravitation is due to a force instantly propagating between bodies. According to that, every point mass attracts every other point mass by a force pointing along the line connecting both points, which is directly proportional to the product of the two masses and inversely proportional to the square of the distance between these two masses $\frac{G_N Mm}{r^2}=mg$. This force is conservative, meaning that one can write the acceleration as a negative gradient of a potential $g=-\nabla\phi$. Newtonian gravity is invariant under Galilean transformations. Thus, the gravitational force is transmitted instantaneously and there is the notion of an absolute universal time. Observationally, it has been confirmed that Newtonian physics works very well as long as the gravitational potential is weak $\phi/c^2\sim 10^{-8}$ and the involved velocities are sufficiently small $v^2/c^2\sim 10^{-8}$. For large gravitational potentials and speeds, it has to be complemented by a relativistic theory. In this sense, General Relativity is modified gravity with respect to Newtonian gravity, extended to cover large gravitational potentials and speeds.

Einstein was following the conviction that the laws of physics should be identical in all inertial systems. For all observers, the speed of light in vacuum should have the same value, independently of the proper motion of the source. The invariant time interval between two events should be promoted to an invariant spacetime interval. In this way, the notion of time would become dependent on the reference frame and spatial position and similarly time and space would not be defined independently of each other. All this would boil down to the requirement that the Galilean transformations should be replaced by the Lorentz transformations. These concepts led Einstein to develop his theory of Special Relativity. It is special in the sense that it corresponds to spacetime with Minkowski metric. The generalisation to a curved spacetime took him almost ten years since he had to familiarise himself with differential geometry. However, the necessary approach was clear to him. 

Einstein was convinced that gravitation should be regarded as an attribute of curved spacetime instead of a force that propagates between two bodies. The presence of energy and momentum would distort spacetime and particles in the vicinity would move along trajectories determined by the geometry of spacetime. In this way, the resulting gravitational force can be viewed as a fictitious force due to the curvature of spacetime. He could substantiate his ideas by borrowing concepts from differential geometry. In order to set up the geometrical framework, he assumed that spacetime is represented by a four dimensional manifold $\mathcal{M}$. 

A manifold $\mathcal{M}$ is a topological space, that resembles a Euclidean space around each point, meaning that it is homeomorphic to Euclidean space locally (preserving the topological properties). The manifolds interesting for us are differentiable. At each point $p\in\mathcal{M}$ on the manifold we can define a tangent space $T_p\mathcal{M}$, which represents a vector space containing all the possible directions tangential to curves containing $p$. In other words, at each point we can define directional derivatives $v^\mu\frac{\partial}{\partial x^\mu}$ that pass through $p$ and are tangent, where $e_\mu=\frac{\partial}{\partial x^\mu}$ is the underlying basis for $T_p\mathcal{M}$. Similarly, we can define a cotangent space $T^*_p\mathcal{M}$ as the dual vector space to the tangent space with the basis $\tilde{e}^\mu=dx^\mu$. Now, when we take a vector field $\bold{v}=v^\mu \frac{\partial}{\partial x^\mu}$ and a covector field $\bold{u}=u_\mu dx^\mu$, then the inner product $\langle\bold{u},\bold{v}\rangle$ of these two vector fields is invariant under general coordinate transformations, since $\langle\frac{\partial}{\partial x^\mu},dx^\nu\rangle=\delta_\mu^\nu$. With the basis elements of $T_p\mathcal{M}$ and $T^*_p\mathcal{M}$ we can define any tensor field with arbitrary covariant and contravariant indices. According to the transformation laws under change of coordinates (going from coordinates $x^\mu$ to coordinates $\tilde{x}^\mu$), we can define the objects of interest. For instance a vector density with weight $w$ transforms as $\tilde A^\mu=J^w\frac{\partial \tilde{x}^\mu}{\partial x^\nu}A^\nu$, with the Jacobian $J=\det\frac{\partial \tilde{x}^\rho}{\partial x^\sigma}$. It means that some objects might have a scaling power $w$ of the Jacobian under general coordinate transformations. 

Next, we can define tensorial derivatives. Let us consider an infinitesimal change of coordinates $\tilde{x}^\mu=x^\mu+\epsilon^\mu$, such that the functional variation of a scalar field would be
$\delta \phi=\tilde{\phi}(\bold{x})-\phi(\bold{x})=\tilde{\phi}(\tilde{\bold{x}}-\bold{\epsilon})-\phi(\bold{x})$, 
which can be approximated to first order as $\delta \phi\approx \tilde{\phi}(\tilde{\bold{x}})-\epsilon^\mu \partial_\mu\phi-\phi(\bold{x})$. On the other hand, we know that a scalar density changes under infinitesimal coordinate transformations as $\tilde{\phi}(\tilde{\bold{x}})=J^w\phi(\bold{x})$, where the Jacobian for the infinitesimal transformation, that we are considering, simply reads $J^w=\left(\det\frac{\partial (x^\rho+\epsilon^\rho)}{\partial x^\sigma}\right)^w=\left(\det(\delta^\rho_\sigma+\partial_\sigma\epsilon^\rho)\right)^w$. In fact, from linear algebra we know that for an arbitrary matrix $M$ we can write $\det(\Id+M)$ in terms of the elementary polynomials of $M$. Thus, to first order we can approximate $\det(\Id+M)\approx 1+\Tr M+\cdots$. For the Jacobian this simply means that we can approximate $J^w=1+w\partial_\mu \epsilon^\mu$, such that the scalar density transforms as $\tilde{\phi}(\tilde{\bold{x}})=J^w\phi(\bold{x})\approx (1+w\partial_\mu \epsilon^\mu)\phi(\bold{x})$. If we now return to the functional variation of the scalar field, we see immediately that the functional variation becomes $\delta \phi\approx (1+w\partial_\mu \epsilon^\mu)\phi(\bold{x})-\epsilon^\mu \partial_\mu\phi-\phi(\bold{x})$, which is the Lie derivative of the scalar field $\delta \phi\approx w\partial_\mu \epsilon^\mu \phi(\bold{x})-\epsilon^\mu \partial_\mu\phi=-\mathcal{L}_\epsilon\phi$. This quantifies the change of the scalar field along the vector $\epsilon^\mu$. In a similar way we can compute the functional variation of any tensor field with an arbitrary rank. In general, the Lie derivative is the change of a tensor field along the flow of a vector field. At this stage, let us emphasise that we have not yet introduced neither the connection nor the metric, nevertheless we could define tensorial derivatives and the Lie derivative.

\begin{center}
\begin{figure}[h]
\begin{center}
 \includegraphics[width=8.5cm]{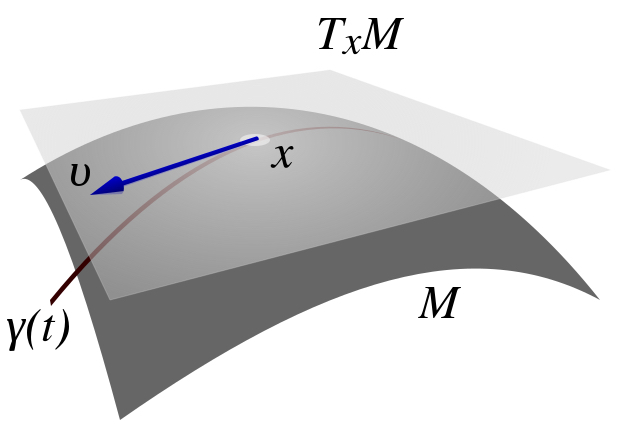}
\end{center}
  \caption{A pictorial representation of a tangent space $T_x\mathcal{M}$ at point $x\in\mathcal{M}$ on the manifold together with a differentiable curve $\gamma(t)$ on the manifold with its tangent vector $v=\dot{\gamma}(t)$. It is taken from \url{https://en.wikipedia.org/wiki/Tangent_space}.}
 \label{tangentspace}
\end{figure}
\end{center}

Apart from the Lie derivative, we cannot do much on our manifold if we do not introduce further structure. Next, we define the covariant derivate on our manifold. This stands for the derivative along tangent vectors of the manifold. Let us consider a differentiable curve $\gamma(t)$ on the manifold with its tangent vector $v=\dot{\gamma}(t)$ at a point $x\in\mathcal{M}$ (see figure \ref{tangentspace}). We can express the tangent vector in terms of the basis as $v^\mu e_\mu$. A variation of an arbitrary vector $u^\nu e_\nu$ along the flow of the tangent vector of the curve would be in terms of the basis vectors simply 
\begin{equation}
D_\bold{v}\bold{u}=v^\mu \partial_\mu (u^\nu e_\nu)=v^\mu(e_\nu\partial_\mu u^\nu+u^\nu \partial_\mu e_\nu)\,.
\end{equation}
The derivative of a basis vector along another basis vector is exactly the connection $\partial_\mu e_\nu=\Gamma_{\mu\nu}^\rho e_\rho$, such that the variation can be expressed as $D_\bold{v}\bold{u}=v^\mu(\partial_\mu u^\rho+\Gamma_{\mu\nu}^\rho u^\nu )e_\rho$. In other words, the covariant derivative applied to the vector field $u^\mu$ will be then given by $\nabla_\mu u^\rho=\partial_\mu u^\rho+\Gamma_{\mu\nu}^\rho u^\nu$. This can be generalised for tensors of arbitrary rank and defined in a very similar way. If we have a vector density with a given power of the Jacobian under coordinate transformations, then the weight of the vector density will also contribute to the covariant derivative 
\begin{equation}\label{covderwithWeight}
\nabla_\mu u^\rho=\partial_\mu u^\rho+\Gamma_{\mu\nu}^\rho u^\nu+w\Gamma_{\mu\alpha}^\alpha u^\rho\,. 
\end{equation}

A mathematically more robust definition of the connection arises after introducing a smooth fiber bundle. The fiber bundle $\mathcal{T}_p:\{e^a\}$ at point $p$ has the Lie group $GL(4,\mathbb{R})$ as its natural structure group, where the latin indices indicate here the gauge indices of the $GL(4,\mathbb{R})$. This can be easily understood recalling that at each point $p\in\mathcal{M}$ we can introduce an orthogonal frame, that is invariant under general linear transformations $GL(4,\mathbb{R})$. The fiber bundle and the tangent space have the same dimensions, which allows to go from the fiber space $\{e^a\}$ to the tangent space $\{\frac{\partial}{\partial x^\mu}\}$. If this map is invertible, we have an isomorphism between $\mathcal{T}_p:\{e^a\}$ and $T_p\mathcal{M}:\{\frac{\partial}{\partial x^\mu}\}$. Such a map is provided by the soldering of the fibre bundle. The solder form $e$ realises a linear isomorphism $e: T_p\mathcal{M} \to \mathcal{T}_p$ from the tangent space $T_p\mathcal{M}$ at $p$ to the vertical tangent space of the fibre $\mathcal{T}_p$. In other words, any vector field on the tangent space is mapped as $e: v^\mu\to e^\mu_a v^a$. The Lie group $GL(4,\mathbb{R})$ has its natural connection $w_{\mu\;\; b}^{\;\;a}$ defined such that $D_\mu v^a=\partial_\mu v^a+w_{\mu\;\; b}^{\;\;a}v^b$ transforms covariantly under the group $GL(4,\mathbb{R})$. We can then use the inverse solder form to map this natural connection on the fiber bundle to the tangent space
\begin{eqnarray}
\nabla_\mu v^\alpha&=& e^{-1}\left[ D_\mu \left[e(v) \right]\right]=e^\alpha{}_a D_\mu \left[ e^a{}_\beta v^\beta\right]\nonumber\\
&=& e^\alpha{}_a\left[ \partial_\mu e^a{}_\beta v^\beta+e^a{}_\beta \partial_\mu v^\beta +w_{\mu\;\; b}^{\;\;a}e^b{}_\beta v^\beta\right]\nonumber\\
&=& \partial_\mu v^\alpha+e^\alpha{}_a\left[  \partial_\mu e^a{}_\beta+w_{\mu\;\; b}^{\;\;a}e^b{}_\beta \right]v^\beta\nonumber\\
&=& \partial_\mu v^\alpha+\Gamma_{\mu\beta}^\alpha v^\beta \,.
\end{eqnarray}
Hence, the connection on the tangent space is given by $\Gamma_{\mu\beta}^\alpha=e^\alpha{}_a\left[  \partial_\mu e^a{}_\beta+w_{\mu\;\; b}^{\;\;a}e^b{}_\beta \right]$. As we have seen, we can use the solder form to map directly the natural connection $w_{\mu\;\; b}^{\;\;a}$ of the Lie group $GL(4,\mathbb{R})$ to the tangent space $\Gamma_{\mu\beta}^\alpha$. The soldering enables us to convert the gauge indices of the connection to the spacetime indices.

\begin{center}
\begin{figure}[h]
\begin{center}
 \includegraphics[width=5cm]{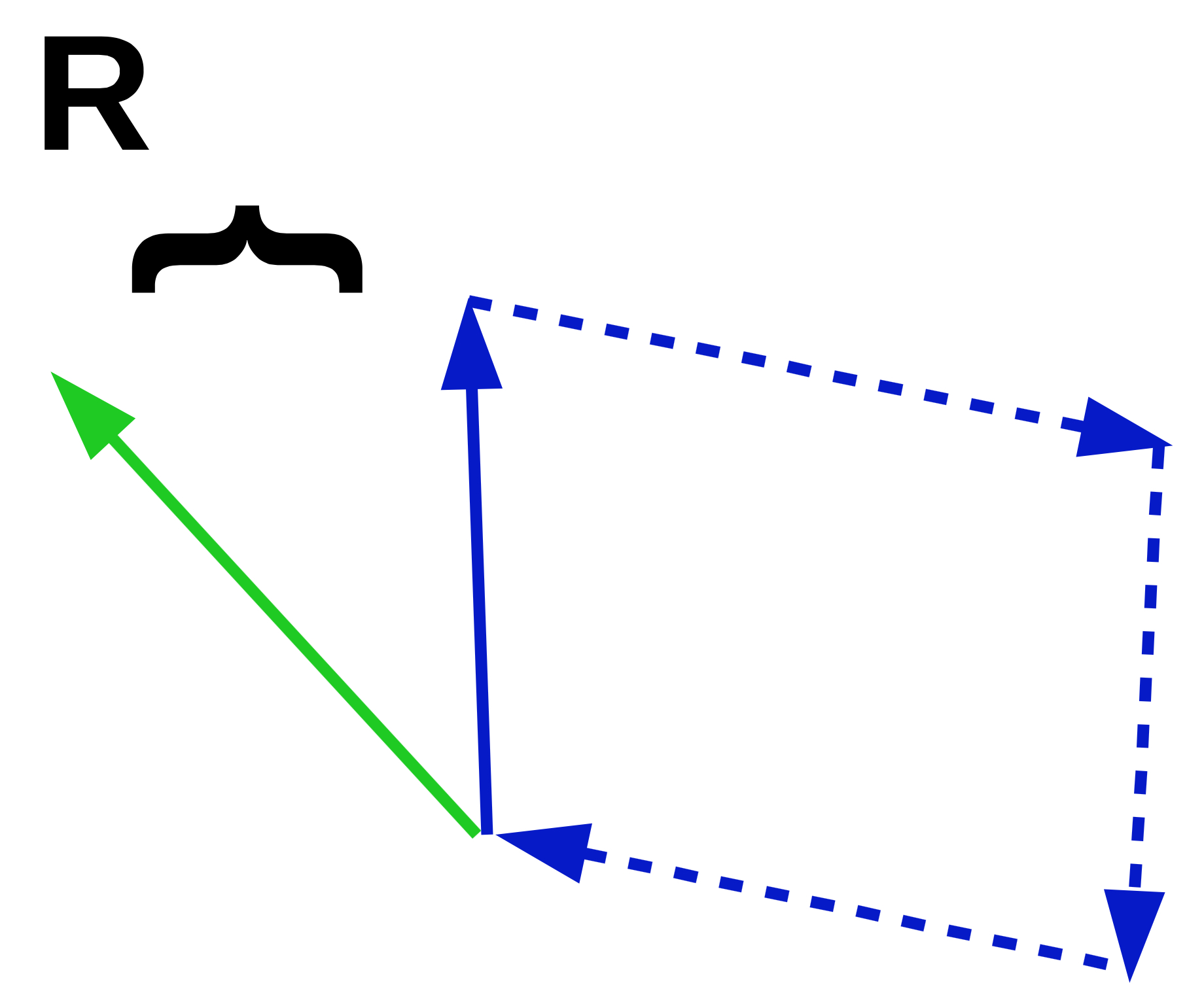}\hspace{1.5cm}\vspace{1.5cm}
  \includegraphics[width=6cm]{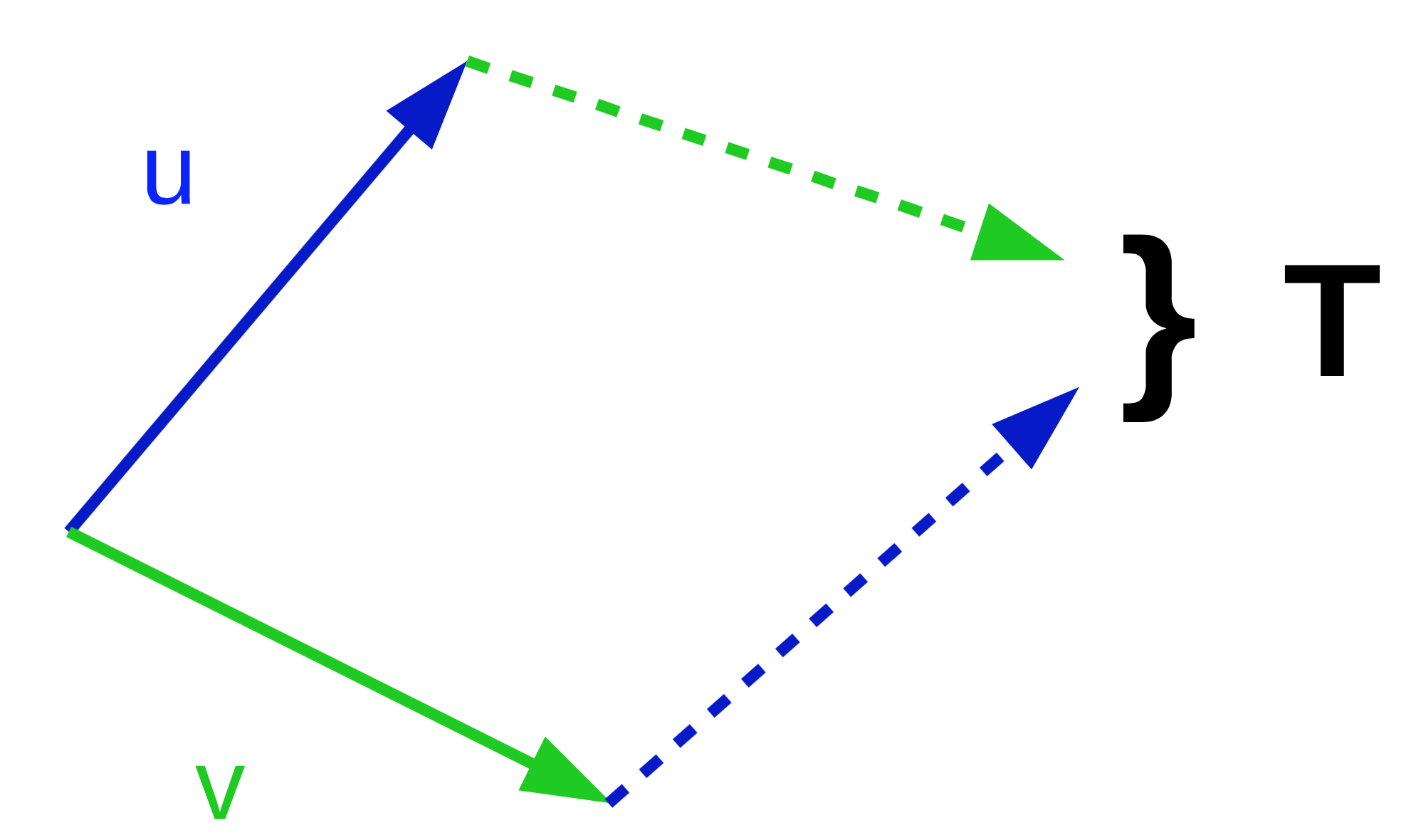} 
   \includegraphics[width=7cm]{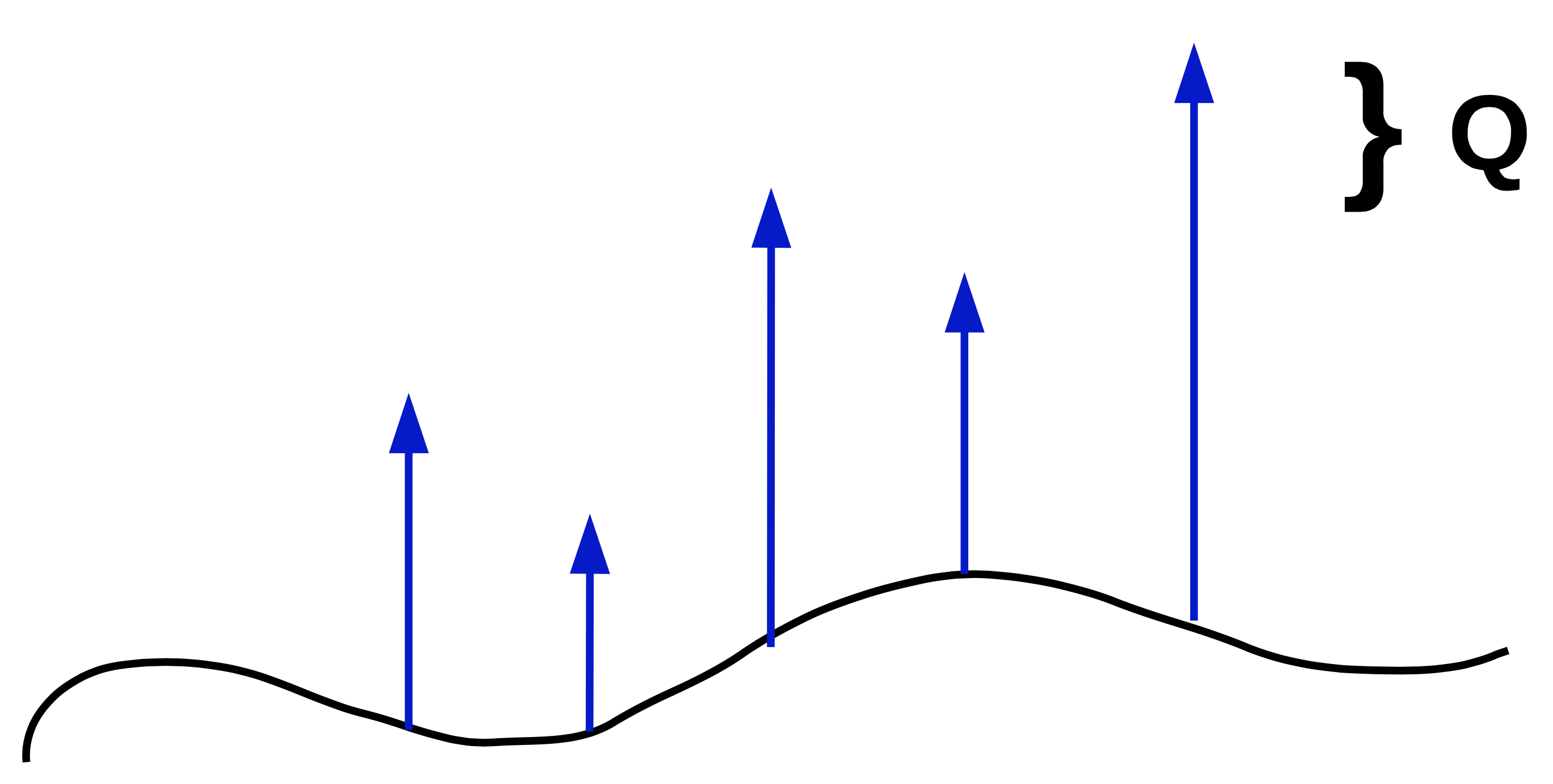}
\end{center}
  \caption{A schematic illustration of the roles of curvature $R^\alpha_{\beta\mu\nu}$, torsion $\mathcal{T}^\alpha_{\mu\nu}$ and non-metricity $\mathcal{Q}_{\alpha\mu\nu}$. We have omitted their indices in the figures for clarity of the presentation. If the spacetime contains curvature, then the direction of a vector field changes when we move it along a closed circle as shown in the upper left figure. In the presence of torsion, parallelograms do not close as shown in the upper right panel. Finally, non-metricity changes the norm of vector fields transported along a curve, as illustrated in the lower panel.}
 \label{geometricalObjects}
\end{figure}
\end{center}

With the covariant derivative at hand, we can define different geometrical objects by applying the commutator of covariant derivatives on different tensor fields. For instance, if we apply the commutator on a scalar field, we obtain 
\begin{equation}
[\nabla_\mu, \nabla_\nu]\phi=-(\Gamma_{\mu\nu}^\alpha-\Gamma_{\nu\mu}^\alpha)\partial_\alpha\phi. 
\end{equation}
We immediately observe, that the commutator vanishes if the connection is symmetric. The term in parentheses is defined as the torsion $\mathcal{T}_{\mu\nu}^\alpha=\Gamma_{\mu\nu}^\alpha-\Gamma_{\nu\mu}^\alpha$. Torsion represents one of the fundamental invariants of the connection and characterises the way how the tangent spaces twist about a curve after a parallel transport. One can visualise the presence of torsion by the non-closedness of parallelograms as shown in figure \ref{geometricalObjects}. 

Another important geometrical object can be constructed by applying the commutator of the covariant derivatives to a vector field       
\begin{equation}
[\nabla_\mu, \nabla_\nu]A^\rho=R^\rho{}_{\alpha\mu\nu}A^\alpha-\mathcal{T}_{\mu\nu}^\alpha\nabla_\alpha A^\rho,
\end{equation}
where the Riemann tensor is defined as $R^\rho{}_{\alpha\mu\nu}=\partial_\mu\Gamma^\rho_{\nu\alpha}-\partial_\nu\Gamma^\rho_{\mu\alpha}+\Gamma^\rho_{\mu\beta}\Gamma^\beta_{\nu\alpha}-\Gamma^\rho_{\nu\beta}\Gamma^\beta_{\mu\alpha}$. The Riemann tensor represents the other fundamental invariant of the connection and dictates the way how the tangent spaces roll along a curve, as shown in figure \ref{geometricalObjects}. On a general manifold with an arbitrary connection we can define different independent traces of the Riemann tensor, since the only symmetry property of the Riemann tensor is the antisymmetry in the last two indices. One such contraction corresponds to the Ricci tensor $R_{\mu\nu}=R^\rho{}_{\mu\rho\nu}$, contracting the first and third indices. There is another independent trace, which is called the homothetic tensor $Q_{\mu\nu}=R^\rho{}_{\rho\mu\nu}$, which is antisymmetric and corresponds to the contraction of the first two indices. It is often said in the literature that the Ricci tensor is symmetric if the torsion vanishes. That this is not the case can be easily verified by writing Bianchi's first identity $R^\mu{}_{[\alpha\nu\beta]}=0$ with a vanishing torsion. The right hand side of the Bianchi identity is exactly zero in absence of torsion. When we take the corresponding trace of the Ricci tensor, we have $R^\mu{}_{[\alpha\mu\beta]}=\frac{1}{3!}\left(R^\mu{}_{\alpha\mu\beta}+R^\mu{}_{\mu\beta\alpha}+R^\mu{}_{\beta\alpha\mu}-R^\mu{}_{\alpha\beta\mu}-R^\mu{}_{\mu\alpha\beta}-R^\mu{}_{\beta\mu\alpha}\right)$. In terms of the Ricci and the homothetic tensors, the Bianchi identity can be also written as $R^\mu{}_{[\alpha\mu\beta]}=\frac{1}{3!}\left(2R_{[\alpha\beta]}-2Q_{\alpha\beta}\right)=0$. This requires $R_{[\alpha\beta]}=Q_{\alpha\beta}$. So even if the torsion vanishes, the Ricci tensor is not symmetric and its antisymmetric part is given by the homothetic tensor $Q_{\alpha\beta}$.

Independently of the connection we can also introduce a metric $g_{\mu\nu}$ on our manifold in order to define the length of and angle between tangent vectors. A metric maps a pair of tangent vectors bilinearly into the real numbers. It is symmetric and non-degenerate, meaning $\det g\ne0$. Under coordinate transformations it transforms as $\tilde{g}_{\mu\nu}=\frac{\partial x^\alpha}{\partial \tilde{x}^\mu}\frac{\partial x^\beta}{\partial \tilde{x}^\nu}g_{\alpha\beta}$, therefore its determinant transforms as a scalar density of weight $w=-2$, i. e. $\det\tilde{g}_{\mu\nu}=J^{-2}\det g_{\mu\nu}$. Using its square root $\sqrt{-g}$ one can then write any tensorial density as an ordinary tensor with weight zero $T^{\mu\cdots}_{\nu\cdots}=(\sqrt{-g})^w \mathcal{T}^{\mu\cdots}_{\nu\cdots}$. The square root of the determinant of the metric is also crucial for constructing invariant volume elements $\sqrt{-g}dV$ and a covariant derivative applied to it gives $\nabla_\mu\sqrt{-g}=\partial_\mu\sqrt{-g}-\Gamma^\rho_{\mu\rho}\sqrt{-g}$. 

We can define the metric in a mathematically more robust way as we did with the connection. Using the inverse solder form we can map directly the natural metric of the fiber bundle to the tangent space. As before, we introduce a frame bundle $\mathcal{T}:\{e^a\}$ with the Lie group $GL(4,\mathbb{R})$. Now, we consider a frame which is orthonormal (orthogonal and unit). This reference frame introduces a natural metric. Restricting the symmetry group to the Lorentz group $SO(3,1)$ allows to define a reference frame orthonormal with respect to the Minkowski metric. The Lie algebra of the associated group defines the Killing metric as the natural metric on the frame bundle. The Killing metric is a symmetric bilinear map defined as $g(x,y)={\rm trace}(ad(x)ad(y))$ with $x$ and $y$ being elements of the Lie algebra. The adjoint endomorphism $ad(x)(y)$ (a specific linear group representation) is defined with the help of the Lie bracket $ad(x)(y)=[x,y]$. Thus, we can define in a natural way the Killing metric of the Lie group of the associated frame bundle and use the inverse soldering to induce a metric on the tangent space. In this way, we can attribute the metric to $g_{\mu\nu}=e^a_\mu e^b_\nu \eta_{ab}$.

We can use the metric to define yet a third rank-2 tensor from the Riemann tensor, which is the co-Ricci tensor $P^\mu{}_\nu=g^{\alpha\beta}R^\mu{}_{\alpha\nu\beta}$, representing the remaining independent contraction. Thus, in total we can have three independent contractions of the Riemann tensor, i.e. the Ricci tensor $R_{\mu\nu}=R^\rho{}_{\mu\rho\nu}$, the homothetic tensor $Q_{\mu\nu}=R^\rho{}_{\rho\mu\nu}$ and the co-Ricci tensor $P^\mu{}_\nu=g^{\alpha\beta}R^\mu{}_{\alpha\nu\beta}$.
They are all independent of each other. However, if one performs a further contraction, they all yield contributions given by the Ricci scalar.

Another important geometrical object arises when one applies the covariant derivative directly to the metric. In standard General Relativity this vanishes, but in a more general geometrical framework it will not
\begin{equation}\label{def_nonMetr}
\nabla_\rho g_{\mu\nu}=\mathcal{Q}_{\rho\mu\nu}\,.
\end{equation}
This non-metricity tensor represents the change of the norm of vector fields along a curve. It is symmetric in the last two indices.
In figure \ref{geometricalObjects} we illustrate schematically the effects of curvature $R^\alpha_{\beta\mu\nu}$, torsion $\mathcal{T}^\alpha_{\mu\nu}$ and non-metricity $\mathcal{Q}_{\alpha\mu\nu}$ on vector fields in a given spacetime. If one transports a vector field along a closed path and its direction at the end point differs from its initial direction, then this indicates a non-vanishing curvature. If one has two vectors $u$ and $v$ and transports each of them along the other and they do not form a closed parallelogram, this indicates the presence of a non-vanishing torsion. Similarly, if there is non-metricity, the transport of a vector field along a curve changes its norm. Note that the non-metricity satisfies $\nabla_{[\mu}\mathcal{Q}_{\nu]\alpha\beta} = R_{(\alpha\beta)\nu\mu}-\frac12T^\lambda{}_{\mu\nu} \mathcal{Q}_{\lambda\alpha\beta}$.

We re-write equation (\ref{def_nonMetr}) with cyclic permutations of the indices,
\begin{eqnarray}
&&\partial_\rho g_{\mu\nu}-\Gamma^\alpha_{\rho\mu}g_{\nu\alpha}-\Gamma^\alpha_{\rho\nu}g_{\mu\alpha}=\mathcal{Q}_{\rho\mu\nu} \label{nonmetricityEq1}\\
&&\partial_\mu g_{\nu\rho}-\Gamma^\alpha_{\mu\nu}g_{\alpha\rho}-\Gamma^\alpha_{\nu\mu}g_{\alpha\rho}=\mathcal{Q}_{\mu\nu\rho}  \label{nonmetricityEq2}\\
&&\partial_\nu g_{\mu\rho}-\Gamma^\alpha_{\nu\rho}g_{\alpha\mu}-\Gamma^\alpha_{\mu\rho}g_{\alpha\nu}=\mathcal{Q}_{\nu\rho\mu} \label{nonmetricityEq3} \,.
\end{eqnarray}
We now sum the first two equations, subtract the last and solve the result for the connection. This gives the general connection
\begin{eqnarray}
\Gamma^\rho_{\mu\nu}&=&\{^\rho_{\mu\nu}\}+L^\rho_{\mu\nu}(\mathcal{Q})+K^\rho_{\mu\nu}(\mathcal{T}) \\
&=&\frac12 g^{\rho\alpha}\Big( \partial_\nu g_{\mu\alpha}+\partial_\mu g_{\alpha\nu}-\partial_\alpha g_{\mu\nu}\Big)+L^\rho_{\mu\nu}(\mathcal{Q})+K^\rho_{\mu\nu}(\mathcal{T}) \nonumber\,.
\end{eqnarray}
We recognise immediately the standard Levi-Civita part of the connection 
\begin{equation}\label{connection_levicivita}
\{^\rho_{\mu\nu}\}=\frac12 g^{\rho\alpha}\Big( \partial_\nu g_{\mu\alpha}+\partial_\mu g_{\alpha\nu}-\partial_\alpha g_{\mu\nu}\Big)\,,
\end{equation}
but in general there are also contributions coming from the non-metricity, that we denoted in the disformation tensor $L^\rho_{\mu\nu}=\frac12\mathcal{Q}^\rho_{\mu\nu}-\mathcal{Q}_{(\mu}{}^\alpha{}_{\nu)}$, and also from torsion, which we represented by the contorsion tensor $K^\rho_{\mu\nu}=\frac12\mathcal{T}^\rho_{\mu\nu}+\mathcal{T}_{(\mu}{}^\alpha{}_{\nu)}$. Both contributions together are sometimes refereed as distorsion. Hence, General Relativity builds upon a manifold without distortion, with $\mathcal{Q}_{\rho\mu\nu}=0$ and $\mathcal{T}_{\mu\nu}^\alpha=0$. On a more general manifold the torsion contributes $d\frac{d(d-1)}{2}$ (24 in four dimensions) components and the non-metricity $d\frac{d(d+1)}{2}$ (40 in four dimensions) components on top of the standard contribution of the metric with $\frac{d(d+1)}{2}$ (10 in four dimensions) components. \\
Thus, a general manifold $(\mathcal{M}, g, \Gamma)$ can be equipped with
\begin{itemize}
\item Curvature $R^\alpha{}_{\mu\nu\beta}(\Gamma)\ne0$, 
\item Torsion $\mathcal{T}^\alpha{}_{\mu\nu}(\Gamma)\ne0$,
\item Non-Metricity $\mathcal{Q}^\alpha{}_{\mu\nu}(\Gamma,g)\ne0$.
\end{itemize}
In figure \ref{geometries} taken from \cite{Conroy:2017yln} we show all the permutations of the Riemann curvature $R^\alpha{}_{\mu\nu\beta}$, the torsion $\mathcal{T}^\alpha{}_{\mu\nu}$ and non-metricity $\mathcal{Q}^\alpha{}_{\mu\nu}$ in general spacetimes. The spacetimes denoted as {\it Riemann} and {\it Symmetric Teleparallel} in figure \ref{geometries} are characterised as torsionless, since the connection is symmetric $\mathcal{T}^\alpha{}_{\mu\nu}(\Gamma)=0$. The spacetime denoted as {\it Weitzenb\"ock} is characterised as metric compatible and flat, since in this case $\mathcal{Q}^\alpha{}_{\mu\nu}(\Gamma,g)=0$ and $R^\alpha{}_{\mu\nu\beta}(\Gamma)=0$.
Important relations between the curvature and the torsion tensors arise when one applies the Jacobi identity for the covariant derivative, 
\be
[\nabla_\alpha,[\nabla_\beta,\nabla_\gamma]] + [\nabla_\beta,[\nabla_\gamma,\nabla_\alpha]] + [\nabla_\gamma,[\nabla_\alpha,\nabla_\beta]]=0\,.
\ee
In terms of the torsion and curvature tensor the identity result in the following Bianchi identities: 
\ba
R^\alpha_{\phantom{\alpha}\beta(\mu\nu)}&= & 0\,, \\
R^\mu_{\phantom{\mu}[\alpha\beta\gamma]} - \nabla_{[\alpha}T^\mu_{\phantom{\mu}\beta\gamma]} + T^\nu_{\phantom{\nu}[\alpha\beta}T^\mu_{\phantom{\mu}\gamma]\nu} & = & 0\,, \label{Bianchi2}\\
\nabla_{[\alpha}R^{\mu}_{\phantom{\mu}\lvert\nu\rvert\beta\gamma]} - T^\lambda_{\phantom{\lambda}[\alpha\beta}R^{\mu}_{\phantom{\mu}\vert\nu\rvert\gamma]\lambda} & = & 0\,,
\ea
where the vertical lines exclude the indices of antisymmetrisation.
The first identity just reflects the defining property of the Riemann tensor, while the latter two are the outcome of the Jacobi identity applied to a scalar and a vector respectively.
\begin{center}
\begin{figure}[h]
\begin{center}
 \includegraphics[width=16cm]{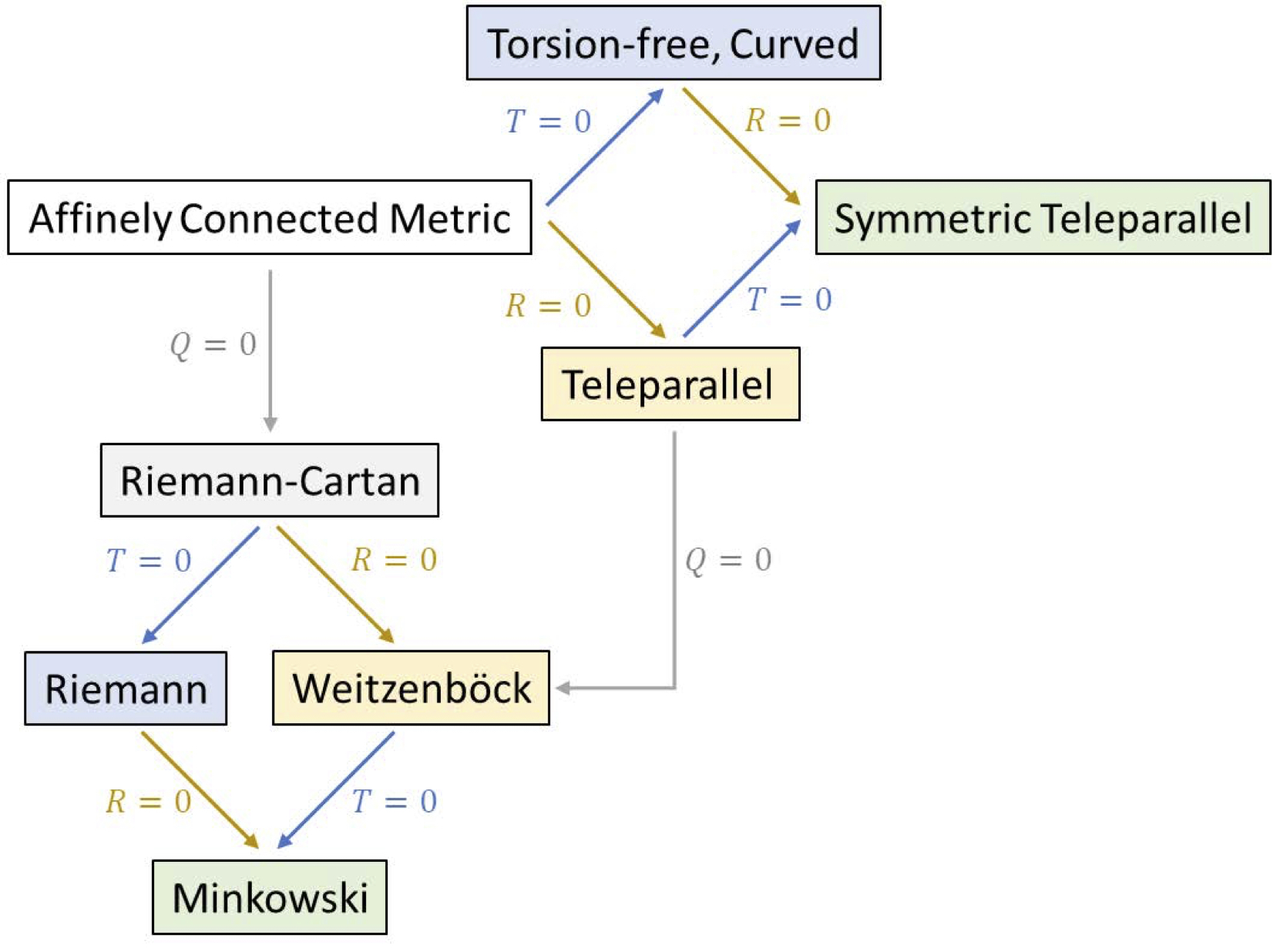}
\end{center}
  \caption{A diagram of all possible geometries based on the Riemann curvature $R^\alpha{}_{\mu\nu\beta}$, the torsion $T^\alpha{}_{\mu\nu}$ and the non-metricity $Q^\alpha{}_{\mu\nu}$. This is borrowed from \cite{Conroy:2017yln}.}
   \label{geometries}
\end{figure}
\end{center}

Concerning the matter fields living on this general manifold, the majority will be immune to the presence of the distortion. For instance, a point particle with its action of the form $\mathcal{S}=-mc^2\int d\tau$ will only see the Levi-Civita part of the connection. The same is true for all the bosonic particles that are minimally coupled to gravity. The matter fields will follow the geodesic equation
\begin{equation}\label{geodesich_pointpart}
\frac{d^2x^\mu}{d\tau^2}+\Gamma^\mu_{\nu\alpha} \frac{dx^\nu}{d\tau}\frac{dx^\alpha}{d\tau}=0\,,
\end{equation}
where the minimally coupled bosonic fields will have $\Gamma^\mu_{\nu\alpha}=\{^\mu_{\nu\alpha}\}$. In General Relativity starting from the action $\mathcal{S}=-mc^2\int d\tau$ or the geodesic equation (\ref{geodesich_pointpart}) gives rise to the same result. However, the coupling of matter can be somewhat ambiguous in generalised geometries if we do not follow the minimal coupling prescription.  
One immediate observation is that the geodesic equation in (\ref{geodesich_pointpart}) is symmetric under the exchange of $\nu \leftrightarrow \alpha$, therefore the torsion does not contribute. On the contrary, fermions will be very sensitive to any distortion of the connection. In spacetimes with torsion teleparallelism (with the connection of the parallelisation being the Weitzenb\"ock connection with vanishing curvature) the minimal coupling procedure fails for fermions \cite{Obukhov:2002tm}. Furthermore, considering the Weitzenb\"ock connection in the Dirac Lagrangian results in inconsistent field equations since the right hand side of the field equations receives antisymmetric contributions \cite{Obukhov:2004hv}. This could be made consistent by enforcing the connection in the Dirac Lagrangian to be the standard Riemannian Levi-Civita connection, introduced arguably ad hoc and less natural. On the other hand, in spacetimes with non-metricity, the minimal coupling prescription of the fermions works consistently without any adjustments, since the contribution of the disformation tensor $L^\alpha_{\mu\nu}$ drops out of the Dirac Lagrangian. Hence, the Dirac fields are insensitive to any disformation of geometry based on non-metricity \cite{BeltranJimenez:2018vdo}. Needless to say, that the dynamics of the matter fields will highly depend on the assumed matter action and whether the minimal coupling prescription is followed.\\


\textbf{Standard interpretation of General Relativity \`a la Einstein}: \\
In standard textbooks on gravity General Relativity is represented as describing the geometric property of spacetime, where the fundamental geometrical object is the curvature. Einstein's interpretation of gravity by spacetime curvature has been implanted in our minds whenever we think of gravitational interactions. After 100 years of this standard view it is challenging to unbind oneself from this firmly deep-set interpretation. However, as we will see in the following, there are two other justified and equivalent geometrical formulations of General Relativity that are not based on curvature, but rather on torsion or on non-metricity. The standard formulation of General Relativity is based on a manifold with non-vanishing curvature $R^\alpha_{\beta\mu\nu}\ne0$ but with zero torsion $\mathcal{T}^\alpha_{\mu\nu}=0$ and zero non-metricity $\mathcal{Q}^\alpha_{\mu\nu}=0$. The connection is simply given by the Christoffel symbols $\Gamma^\rho_{\mu\nu}=\{^\rho_{\mu\nu}\}=\frac12 g^{\rho\alpha}\Big( \partial_\nu g_{\mu\alpha}+\partial_\mu g_{\alpha\nu}-\partial_\alpha g_{\mu\nu}\Big)$ and the matter fields can be coupled minimally to the volume element. 

Now, we can build the underlying Einstein-Hilbert Lagrangian of the standard formulation and the field dynamics. Demanding locality, unitarity and causality together with covariance, Lorentz invariance, a pseudo-Riemannian manifold and second order equations of motion for the metric uniquely determines the action to be 
\begin{equation}\label{GR_pureMetric}
\mathcal{S}=\int d^4x \sqrt{-g} \left( \frac{\mpl^2}{2} \mathcal{R}-2\lambda\right)+\mathcal{S}_{\rm matter}\,,
\end{equation}
with a cosmological constant $\lambda$ allowed by the symmetries. The connection is simply the Levi-Civita connection $\Gamma^\rho_{\mu\nu}=\{^\rho_{\mu\nu}\}$ given by (\ref{connection_levicivita}), where $L^\rho_{\mu\nu}=0=K^\rho_{\mu\nu}$. Since the connection is uniquely determined in terms of the metric in this case, we only need to perform the variation with respect to the metric. This yields Einstein's field equations $G_{\mu\nu}=\frac{T_{\mu\nu}}{\mpl^2}$, where the Einstein tensor $G_{\mu\nu}=\mathcal{R}_{\mu\nu}-\frac12 \mathcal{R} g_{\mu\nu}$ is divergence-free and the stress energy tensor arises from the matter action $T_{\mu\nu}=\frac{-2}{\sqrt{-g}}\frac{\delta \mathcal{S}_{\rm matter}}{\delta g^{\mu\nu}}$. Actually, the fact that the connection is simply given by the Levi-Civita part is an intrinsic property of the action and does not need to be put by hand. In order to appreciate the uniqueness of General Relativity in terms of the Einstein-Hilbert action we could have written the action (\ref{GR_pureMetric}) alternatively as
\begin{equation}\label{GR_alaPalatini}
\mathcal{S}=\int d^4x \sqrt{-g} \left( \frac{\mpl^2}{2} g^{\mu\nu}R_{\mu\nu}(\Gamma)-2\lambda\right)+\mathcal{S}_{\rm matter}
\end{equation}
with an independent more general connection. This leads to the Palatini formulation of General Relativity. We can first perform the variation with respect to the metric at constant connection
\begin{equation}\label{GR_alaPalatini2}
\delta_g\mathcal{S}=\int d^4x \frac{\mpl^2}{2} \left[(\delta\sqrt{-g}) g^{\mu\nu}R_{(\mu\nu)}(\Gamma)+\sqrt{-g} \delta g^{\mu\nu}R_{\mu\nu}(\Gamma)\right]+\delta_g\mathcal{S}_{\rm matter}\,,
\end{equation}
where we have put $\lambda=0$ since it is irrelevant for the present discussion. The Ricci scalar is not varied with respect to the metric, since it is assumed to depend on an independent connection. Noting that $(\delta\sqrt{-g})=-\frac12\sqrt{-g}g_{\mu\nu}\delta g^{\mu\nu}$, the variation with respect to the metric simply yields $G_{\mu\nu}(\Gamma)=\frac{T_{\mu\nu}}{\mpl^2}$ as before, but with the Einstein tensor of the general connection. Since the connection is independent, we can also perform the variation of the action with respect to the connection at constant metric
\begin{equation}\label{GR_alaPalatini3}
\delta_\Gamma\mathcal{S}=\int d^4x \frac{\mpl^2}{2} \sqrt{-g} g^{\mu\nu} \delta R_{(\mu\nu)}(\Gamma)+\delta_\Gamma\mathcal{S}_{\rm matter}\,.
\end{equation}
The variation of the Ricci tensor is given by $\delta R_{\mu\nu}(\Gamma)=\nabla_\alpha \delta \Gamma^\alpha_{\nu\mu}-\nabla_\nu \delta \Gamma^\alpha_{\alpha\mu}-T^\alpha_{\beta\nu}\delta \Gamma^\beta_{\alpha\mu}$. We can integrate by parts in order to extract the $\delta \Gamma$ terms. Then, the variation becomes
\begin{eqnarray}\label{EOM_connection1}
\delta_\Gamma\mathcal{S}&=&-\int d^4x \frac{\mpl^2}{2}\left[ \nabla_\alpha(\sqrt{-g} g^{\mu\nu}) \delta \Gamma^\alpha_{\nu\mu} -\nabla_\nu(\sqrt{-g} g^{\mu\nu})\delta \Gamma^\alpha_{\alpha\mu}-\sqrt{-g} g^{\mu\nu}T^\alpha_{\beta\nu}\delta \Gamma^\beta_{\alpha\mu} \right] \nonumber\\
&&+\int d^4x \frac{\mpl^2}{2}\left[ \nabla_\alpha(\sqrt{-g} g^{\mu\nu} \delta \Gamma^\alpha_{\nu\mu}) -\nabla_\nu(\sqrt{-g} g^{\mu\nu}\delta \Gamma^\alpha_{\alpha\mu})\right]+\delta_\Gamma\mathcal{S}_{\rm matter}\,.
\end{eqnarray}
Now, some caution is needed here. In standard Riemannian manifolds with vanishing torsion, the terms in the last line simply correspond to total derivatives and can be discarded. As we mentioned above in equation (\ref{covderwithWeight}), if one has a vector density that generates a given power of the Jacobian under coordinate transformations, the weight of the vector density will inevitably contribute to the covariant derivative. If the weight of the vector density is $w=-2$, the covariant derivative acquires a contribution in form of the torsion $\nabla_\mu u^\mu=\partial_\mu u^\mu+\mathcal{T}_{\nu\mu}^\nu  u^\mu$. Exactly at this place this property will play a crucial role in equation (\ref{EOM_connection1}). These boundary terms will contribute whenever there is torsion. The variation with respect to the connection then results in
\begin{eqnarray}\label{EOM_connection2}
\delta_\Gamma\mathcal{S}&=&-\int d^4x \frac{\mpl^2}{2}\left[ \nabla_\alpha(\sqrt{-g} g^{\mu\nu}) \delta \Gamma^\alpha_{\nu\mu} -\nabla_\nu(\sqrt{-g} g^{\mu\nu})\delta \Gamma^\alpha_{\alpha\mu}-\sqrt{-g} g^{\mu\nu}T^\alpha_{\beta\nu}\delta \Gamma^\beta_{\alpha\mu} \right] \nonumber\\
&&+\int d^4x \frac{\mpl^2}{2}\left[\sqrt{-g} g^{\mu\nu} \delta \Gamma^\beta_{\nu\mu} -\sqrt{-g} g^{\beta\mu}\delta \Gamma^\rho_{\rho\mu})\right]\mathcal{T}_{\alpha\beta}^\alpha+\delta_\Gamma\mathcal{S}_{\rm matter}\,.
\end{eqnarray}
We can reshuffle the indices in order to factor out the common $\delta\Gamma$ parts. In this way the connection field equations can be equivalently written as
\begin{eqnarray}\label{EOM_connectionFinal}
\nabla_\rho\left[\sqrt{-g}g^{\mu\nu}\right]-\delta^\mu_\rho\nabla_\alpha\left[\sqrt{-g}g^{\alpha\nu}\right]=\sqrt{-g}\left[ g^{\mu\nu}\mathcal{T}^\alpha_{\alpha\rho}+g^{\alpha\nu}\mathcal{T}^\mu_{\rho\alpha}-\delta^\mu_\rho g^{\beta\nu}\mathcal{T}^\alpha_{\alpha\beta} \right]+\Delta^{\mu\nu}_\rho\,,
\end{eqnarray}
where we introduced the hypermomentum of the matter fields defined as $\Delta^{\mu\nu}_\rho=\frac{2}{\mpl^2}\frac{\delta \mathcal{S}_{\rm matter} }{\Gamma_{\mu\nu}^\rho}$. As we mentioned above the bosonic matter fields will not contribute to the hypermomentum. In the case of vanishing torsion and hypermomentum, only the left hand side of equation (\ref{EOM_connectionFinal}) survives $\nabla_\rho\left[\sqrt{-g}g^{\mu\nu}\right]-\delta^\mu_\rho\nabla_\alpha\left[\sqrt{-g}g^{\alpha\nu}\right]=0$. We can take the trace of these equations, i.e. set $\mu=\rho$. In this case the equations simplify to $\nabla_\alpha\left[\sqrt{-g}g^{\alpha\nu}\right]=0$, which can be plugged back into (\ref{EOM_connectionFinal}) giving rise to $\nabla_\alpha\left[\sqrt{-g}g^{\mu\nu}\right]=0$. Even in the presence of the hypermomentum, these equations can be solved for the connection algebraically and one finds that the connection has the Levi-Civita form up to a projective transformation $\Gamma_{\mu\nu}^\alpha\to \Gamma_{\mu\nu}^\alpha+\delta^\alpha_\nu\xi_\mu$ (see the detailed derivation in \cite{BeltranJimenez:2017doy}). Thus, even if one starts with a more general approach like the Palatini formulation, the connection ends up to be of Levi-Civita form up to a projective symmetry and General Relativity arises naturally. This is due to the fact that the action was chosen to be of Einstein-Hilbert form. In fact, if one wants to make use of the much richer geometrical structure, one has to modify the underlying action. One could for instance consider a general function of the form $\mathcal{S}=\int d^4x \sqrt{-g}F(g^{\mu\nu},R_{(\mu\nu)}(\Gamma))$, hence a general function of the inverse metric and the symmetric Ricci tensor depending on a general connection. There have also been purely affine suggestions such as Eddington gravity $\mathcal{S}=m_\lambda^4\int d^4x \sqrt{\det R_{(\mu\nu)}(\Gamma) }$. Similar considerations have been also studied in the context of the Born-Infeld inspired gravity theories, where the previous determinant is replaced by $\mathcal{S}=m_\lambda^4\int d^4x \sqrt{-\det (g_{\mu\nu}+m_\lambda^{-2}R_{(\mu\nu)}(\Gamma) )}$. For more details on this type of alternative gravity theories see \cite{BeltranJimenez:2017doy}. As we have seen in this section, if one is willing to modify General Relativity in the geometrical framework, then one has to consider manifolds that go beyond the restricted Riemannian setup and consider more general Lagrangians.
General Relativity arises uniquely if the dynamics is given by the Einstein-Hilbert action, even if it is formulated \`a la Palatini.\\

\textbf{Teleparallel Equivalent of General Relativity (TEGR)}: \\
There is an equivalent formulation of General Relativity based on torsion, which is also known as teleparallelism. A relevant object in teleparallelism is the conjugate torsion defined as
\be
{S}{}_\alpha^{\phantom{\alpha}\mu\nu}   =  a \mathcal{T}_\alpha^{\phantom{\alpha}\mu\nu} + b \mathcal{T}^{[\mu\phantom{\alpha}\nu]}_{\phantom{,\mu}\alpha} + c\delta^{[\mu}_\alpha \mathcal{T}^{\nu]}\,, \label{ConjugateTorsion}
\ee
where $a$, $b$ and $c$ are arbitrary constants. Using the conjugate torsion, we can define the torsion scalar as
\ba \label{mathbbT}
\mathbb{T} &= & \frac{1}{2}S_\alpha^{\phantom{\alpha}\mu\nu}\mathcal{T}^\alpha_{\phantom{\alpha}\mu\nu} \nonumber \\ 
& = &  \frac{1}{2}\left( a\mathcal{T}_{\alpha\mu\nu}+b\mathcal{T}_{\mu\alpha\nu}+cg_{\alpha\mu}\mathcal{T}_\nu\right) \mathcal{T}^{\alpha\mu\nu}\,.
\ea 
Note, that since the torsion is antisymmetric in the $\mu-$ and $\nu-$indices, there are only three possible independent contractions in the quadratic order of torsion and there is only one independent trace $\mathcal{T}_\nu=\mathcal{T}^\alpha_{\phantom{\alpha}\nu\alpha}$. 
The teleparallel formulation of the general quadratic torsion action is based on the constraints imposed by the teleparallelity and metricity in terms of appropriate Lagrange multipliers
\be \label{quadraticTorsionLag}
\mathcal{L}_G   =  \frac12\sqrt{-g}\mathbb{T}  
+ 
\lambda_\alpha^{\phantom{\alpha}\beta\mu\nu} R^\alpha_{\phantom{\alpha}\beta\mu\nu} + \lambda^\alpha_{\phantom{\alpha}\mu\nu}\nabla_\alpha g^{\mu\nu}\,,
\ee
where the two Lagrange multipliers are given by a rank-4 tensor density with the symmetry $\lambda_{\alpha}^{\phantom{\alpha}\mu\beta\nu}  = \lambda_{\alpha}^{\phantom{\alpha}\mu[\beta\nu]}$,
and a rank-3 tensor density with the symmetry $\lambda^\alpha_{\phantom{\alpha}\mu\nu}=\lambda^\alpha_{\phantom{\alpha}(\mu\nu)}$, both having weight $-1$.
The teleparallelism condition $R^\alpha_{\beta\mu\nu}=0$ is enforced by the variation of the action with respect to the 4-index Lagrange multiplier. This condition imposes the general teleparallel Palatini connection to be of the form ${\Gamma}^\alpha_{\phantom{\alpha}\mu\beta} = (\Lambda^{-1})^\alpha_{\phantom{\alpha}\nu}\partial_\mu \Lambda^\nu_{\phantom{\nu}\beta}$, with $\Lambda^\alpha{}_\beta$ being any element of the general linear transformations $GL(4,\mathbb{R})$. This implies the teleparallel torsion
to be constrained as $\mathcal{T}^\alpha_{\phantom{\alpha}\mu\nu} = 2(\Lambda^{-1})^\alpha_{\phantom{\alpha}\beta}\partial_{[\mu} \Lambda^\beta_{\phantom{\beta}\nu]}$. Similarly, the metricity condition $\nabla_\alpha g^{\mu\nu} =0$ is enforced after the variation of the action with respect to the 3-index Lagrange multiplier, which on the other hand relates directly the derivatives of the metric and of the inertial connection field to each other in the following form $g^{\lambda(\mu}\partial_\alpha\Lambda^{\nu)}{}_\rho(\Lambda^{-1})^\rho{}_\lambda =\frac{1}{2}\partial_\alpha g^{\mu\nu}{}$. Hence, we can integrate out the metric in terms of the inertial connection. The metric plays only the role of an auxiliary field. For the parameters $a=1/4$, $b=1/2$ and $c=-1$ in (\ref{quadraticTorsionLag}) we have $\mathbb{T} \to \mathring{\mathbb{T}} $ where $\mathring{\mathbb{T}} = \frac{1}{2}\mathring{S}_\alpha^{\phantom{\alpha}\mu\nu}\mathcal{T}^\alpha_{\phantom{\alpha}\mu\nu}$ and $\mathring{S}_\alpha^{\phantom{\alpha}\mu\nu}$ is the conjugate torsion given by equation (\ref{ConjugateTorsion}) with these parameters. Restricting the parameters to those results in the teleparallel equivalent of General Relativity (TEGR). This can be easily comprehended recalling the relations between the Levi-Civita connection and the general one. The Ricci tensor of the general connection $R_{\mu\nu}$ can be decomposed in terms of the Ricci curvature of the Levi-Civita connection $\mathcal{R}_{\mu\nu}$ as
\be
R_{\mu\nu} = \mathcal{R}_{\mu\nu} + \mathcal{D}_\alpha\left( \mathring{S}^{\phantom{\nu}\alpha}_{\nu\phantom{\alpha}\mu} + g_{\mu\nu} \mathcal{T}^\alpha\right) - \mathcal{T}_\alpha K^\alpha_{\phantom{\alpha}\nu\mu} - K_{\alpha\nu\beta}K^{\beta\alpha}_{\phantom{\beta\alpha}\mu}\,,
\ee
with the covariant Levi-Civita derivative $\mathcal{D}_\alpha$. The contraction of this relation gives the known relation of the curvatures
\be\label{relationRT}
R = \mathcal{R} + \mathring{\mathbb{T}} + 2\mathcal{D}_\alpha \mathcal{T}^\alpha\,.
\ee 
The flatness condition $R=0$ imposed on the left hand side of this relation then tells us that the Ricci scalar $\mathcal{R}$ of the Levi-Civita connection differs from $\mathring{\mathbb{T}}$ by a total derivative $2\mathcal{D}_\alpha \mathcal{T}^\alpha$. Hence, the dynamics of General Relativity are identically recovered. The standard formulation of TEGR as the gauge theory of translations in the literature relies on the tetrad fields, with the introduction of the frame bundle and the corresponding soldering form \cite{Aldrovandi:2013wha}. As we have seen, we can construct exactly the same theory in a manifestly covariant manner using the Lagrange multipliers. The restriction of the parameters, that recovers General Relativity, introduces an additional local Lorentz symmetry. Thus, out of the 16 components of $\Lambda^\alpha{}_\mu$, one can remove 8 due to diffeomorphisms and 6 more due to Lorentz transformations, leaving 2 propagating degrees of freedom as in General Relativity \cite{BeltranJimenez:2018vdo}.\\


\textbf{Coincident General Relativity (CGR)}: \\
As we have appreciated above, General Relativity \`a la Einstein corresponds to a torsion-free metric compatible curved spacetime. However, the geometrical richness allows an alternative description of the equivalent dynamics by a mere change of the geometrical stage. Alternatively, General Relativity can be described as a flat contorted spacetime based on the action $\mathcal{S}_{\rm TEGR}=\int d^4x \sqrt{-g}\mathring{\mathbb{T}}$ as we just saw above. As we mentioned, one can also ascribe gravity to the non-metricity $\mathcal{Q}_{\alpha\mu\nu}$ \cite{Nester:1998mp}. 
In \cite{BeltranJimenez:2017tkd}, yet another geometrical manifestation of General Relativity was considered, where gravity is attributed to a flat torsion-free spacetime with non-metricity. The findings of \cite{BeltranJimenez:2017tkd} can be summarised as:
\begin{itemize}
\item a simpler geometrical formulation of General Relativity is proposed without the concept of curved spacetime
\item the connection vanishes $\Gamma^\alpha_{\mu\nu}=0$, hence no inertial effects
\item the resulting theory is General Relativity without the boundary term.
\end{itemize}
This new formulation is based on a flat and torsion free geometry, which we can again impose with the help of appropriate Lagrange multipliers. Since the non-metricity tensor $\mathcal{Q}_{\alpha\mu\nu}$ is symmetric in the last two indices, we can construct five independent contractions at the quadratic order in non-metricity. Therefore, the most general quadratic action takes the form
\ba \label{quadraticNonMetricity}
\mathcal{L}_G  & = & \frac{1}{2}\sqrt{-g} \mathcal{Q}^\alpha_{\phantom{\alpha}\mu\nu}( c_1 \mathcal{Q}_\alpha^{\phantom{\alpha}\mu\nu} 
 +  c_2 \mathcal{Q}^{\mu\phantom{\alpha}\nu}_{\phantom{\mu}\alpha} 
+ c_3 g^{\mu\nu}\mathcal{Q}_\alpha + c_4\delta^\mu_\alpha\tilde{\mathcal{Q}}^\nu  + c_5\delta^\mu_\alpha {\mathcal{Q}}^\nu)     \\ 
& + & \lambda_\alpha^{\phantom{\alpha}\beta\mu\nu} R^\alpha_{\phantom{\alpha}\beta\mu\nu} + \lambda_\alpha^{\phantom{\alpha}\mu\nu}\mathcal{T}^\alpha_{\phantom{\alpha}\mu\nu}\,,\,\,\, \nn
\ea
where $\tilde{\mathcal{Q}}^\nu = \mathcal{Q}_\alpha^{\phantom{\alpha}\alpha\nu}$ and $\mathcal{Q}^\mu = \mathcal{Q}^{\mu\alpha}_{\phantom{\mu\alpha}\alpha}$, respectively. The 
Lagrange multipliers correspond to a rank-4 tensor density with the symmetry $\lambda_{\alpha}^{\phantom{\alpha}\mu\beta\nu}  = \lambda_{\alpha}^{\phantom{\alpha}\mu[\beta\nu]}$,
and a rank-3 tensor density this time with the symmetry $\lambda^\alpha_{\phantom{\alpha}\mu\nu}=\lambda^\alpha_{\phantom{\alpha}[\mu\nu]}$.
The teleparallelism condition $R^\alpha_{\beta\mu\nu}=0$ restricts again the connection to be only a pure $GL(4,\mathbb{R})$ transformation parametrised by $\Lambda^\alpha_{\phantom{\alpha}\mu}$, as in the TEGR case. The variation with respect to the 3-index Lagrange multiplier imposes this time the vanishing of the torsion tensor $\mathcal{T}^\alpha_{\phantom{\alpha}\mu\nu}=0$, which translates into a condition for the transformation matrix to fulfil $(\Lambda^{-1})^\alpha_{\phantom{\alpha}\nu}\partial_{[\mu} \Lambda^\nu_{\phantom{\nu}\beta]}=0$. In other words, we can parametrise the transformation matrix
as $\Lambda^\alpha_{\phantom{\alpha}\mu}=\partial_\mu\xi^{\alpha}$ with infinitesimal $\xi^\alpha\in GL(4,\mathbb{R})$. In these symmetric teleparallel theories (symmetric referring to the fact that $\mathcal{Q}$ is symmetric), the connection can then be exactly cancelled by means of a diffeomorphism. We will call the gauge in which the connection trivialises the {\it coincident gauge}. Hence, the connection is simply parametrised as
\begin{equation}
\Gamma^\alpha{}_{\mu\nu}=\frac{\partial x^\alpha}{\partial\xi^\lambda}\partial_\mu\partial_\nu\xi^\lambda\,.
\end{equation}
Let us emphasise again, that this seemingly innocent form of the connection implies an incredible property of the non-metricity representation, namely the connection can be put to zero by a coordinate transformation:
 The gauge choice $\xi^\alpha=x^\alpha$ makes the connection vanish. This can be interpreted as the gauge shifting the spacetime origin into the point parameterised by $\xi^\alpha$. Since then $\xi^\alpha$ coincides with the coordinate origin, this choice can be dubbed the coincident gauge.
 
 If the free coefficients $c_i$ in equation (\ref{quadraticNonMetricity}) are chosen in the following specific way
 \be \label{ci_qgr}
c_1 = -c_3= -\frac{1}{4}\,, \quad 
c_2 = -c_5= \frac{1}{2}\,, \quad
c_4 = 0\,,
\ee
one recovers General Relativity. The quadratic dependence on the non-metricity becomes for this choice
\be \label{Qgr}
\mathring{\mathcal{Q}} =   \frac{1}{4}\mathcal{Q}_{\alpha\beta\gamma}\mathcal{Q}^{\alpha\beta\gamma} -  \frac{1}{2}\mathcal{Q}_{\alpha\beta\gamma}\mathcal{Q}^{\beta\gamma\alpha} 
  -   \frac{1}{4}\mathcal{Q}_\alpha \mathcal{Q}^\alpha  
  + \frac{1}{2}\mathcal{Q}_\alpha\tilde{\mathcal{Q}}^\alpha\,.
\ee
In the coincident gauge $\Gamma^\alpha{}_{\mu\nu}$, the triviality of the connection directly imposes the relation $\left\{^{\phantom{i} \alpha}_{\mu\nu}\right\}=-L^\alpha_{\phantom{\alpha}\mu\nu}$. In this gauge, the action simplifies to
\be
\mathcal{S}_{\rm CGR}[\Gamma=0]=\frac{1}{16\pi G}\int d^4x\sqrt{-g}g^{\mu\nu}\Big(\left\{^{\phantom{i} \alpha}_{\beta\mu}\right\} \left\{^{\phantom{i} \beta}_{\nu\alpha}\right\} -\left\{^{\phantom{i} \alpha}_{\beta\alpha}\right\}\left\{^{\phantom{i} \beta}_{\mu\nu}\right\} \Big)\,.
\ee
This is the action of Coincident General Relativity, which corresponds exactly to the Hilbert action, but devoid of boundary terms.
The equivalence to General Relativity becomes apparent after decomposing the connection into the standard Christoffel symbols and the disformation tensor. The general curvature can then be expressed as
\ba \label{riccitensorq}
R_{\mu\nu}  =  \mathcal{R}_{\mu\nu} - L^\alpha_{\phantom{\alpha}\beta\mu}L^\beta_{\phantom{\alpha}\alpha\nu}  - \frac{1}{2}\mathcal{Q}_\alpha L^\alpha_{\phantom{\alpha}\mu\nu}
 +    \mathcal{D}_\alpha L^\alpha_{\phantom{\alpha}\mu\nu} +   \frac{1}{2}\mathcal{D}_\nu \mathcal{Q}_\mu \,.
\ea
Thus, the curvature scalar satisfies the relation
\be \label{ricciscalarq}
R   =  \mathcal{R}  + \mathring{\mathcal{Q}}+   \mathcal{D}_\alpha ( \mathcal{Q}^\alpha - \tilde{\mathcal{Q}}^\alpha )\,,
\ee
where $\mathring{\mathcal{Q}}$ is the quadratic scalar (\ref{Qgr}) after setting the parameters to the values given in (\ref{ci_qgr}).
One immediate distinctive property of CGR is that it only involves first derivatives of the metric. In the standard formulation of General Relativity \`a la Einstein the boundary term introduces second derivatives of the metric and as we will see in section \ref{subsection_GHY} this jeopardises the well-posed variational principle and makes the
introduction of the Gibbons-Hawking-York boundary term inevitable. This can be straightforwardly avoided in CGR. This formulation of General Relativity has the following advantages:
\begin{itemize}
\item there is no need for the Gibbons-Hawking-York boundary term for a well-defined variational principle
\item it has more direct contact with (the most fundamental) field theory description (Deser's resummation approach), which we will see in section \ref{subsec:masslessSpin2Non-Linear}
\item it is oblivious to the affine spacetime structure, thus fundamentally depriving gravity from any inertial character
\item the computation of the entropy of black holes based on Euclidean action is improved and unambigous (see section \ref{subsection_GHY} for more detail)
\item it represents a new tool to explore the holographic nature of General Relativity.
\end{itemize}

Summarising, we have seen that the geometrical interpretation of gravity introduces ambiguity in the formulation of General Relativity: the trinity of gravity.
\begin{center}
\begin{figure}[h]
\begin{center}
 \includegraphics[width=10.5cm]{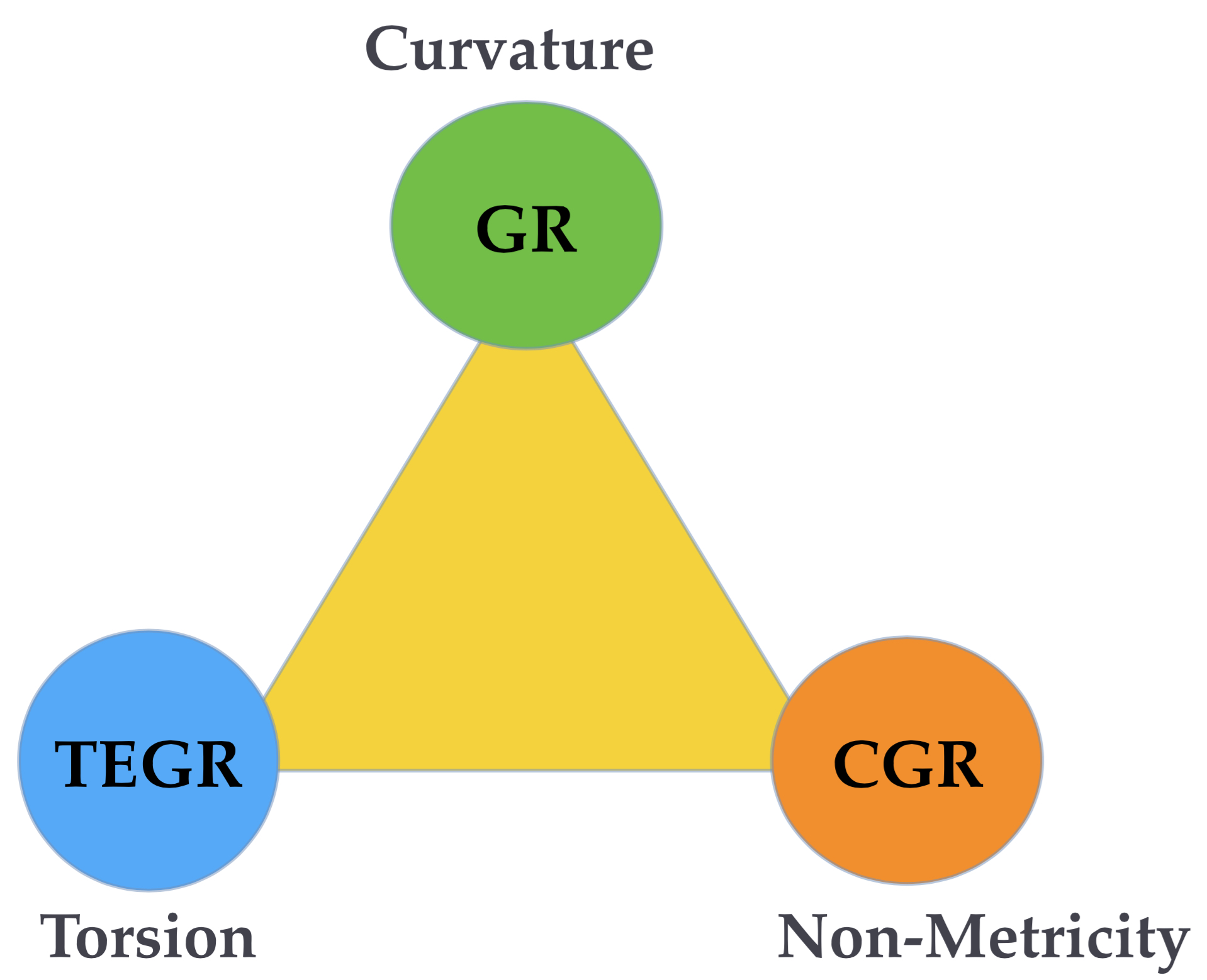}
\end{center}
  \caption{The triangle of General Relativity: the underlying physics behind General Relativity can be described in three equivalent manners, where the fundamental geometrical object is either the curvature (standard GR), the torsion (TEGR) or the non-metricity (CGR). They differ only by means of a boundary term and give rise to the same field equations.}
   \label{geometries}
\end{figure}
\end{center}
The standard formulation introduced by Einstein is based on curvature.
This perception brings along all the difficulties associated to a curved spacetime with inertial effects and the presence of a boundary term.
However, as we have seen the differential geometry provides a much wider class of geometrical objects to represent the geometrical properties of a given manifold, namely the torsion and the non-metricity. An equivalent representation of General Relativity can be achieved on a flat spacetime with an asymmetric connection, where the gravity is entirely assigned to torsion (TEGR). A third equivalent and simpler formulation of General Relativity arises on an equally flat spacetime without torsion, in which gravity is entirely ascribed to non-metricity (CGR). By means of a gauge choice, the connection can be made to vanish altogether in this representation. \\

\begin{center}
\begin{tabular}{|c|c|c| }
  \hline
  \multicolumn{3}{|c|}{\color{black}{different interpretations of gravity}} \\
  \hline
  \color{green}{GR} & \color{blue}{TEGR} & \color{orange}{CGR} \\
   \hline
  Curvature & Torsion  & Non-Metricity  \\
    \hline
  $ \mathcal{R}\ne0$, $\mathcal{T}=0$, $\mathcal{Q}=0$ & $ \mathcal{R}=0$, $\mathcal{T}\ne0$, $\mathcal{Q}=0$ & $ \mathcal{R}=0$, $\mathcal{T}=0$, $\mathcal{Q}\ne0$ \\
    \hline
     $\mathcal{S}=\int\sqrt{-g}\mathcal{R}$ &  $\mathcal{S}=\int\sqrt{-g}\mathring{\mathbb{T}}$  &  $\mathcal{S}=\int\sqrt{-g}\mathring{\mathcal{Q}}$ \\
    \hline
      $\Gamma^\rho_{\mu\nu}=\{^\rho_{\mu\nu}\}$ &  ${\Gamma}^\alpha_{\phantom{\alpha}\mu\beta} = (\Lambda^{-1})^\alpha_{\phantom{\alpha}\nu}\partial_\mu \Lambda^\nu_{\phantom{\nu}\beta}$  &   ${\Gamma}^\alpha_{\phantom{\alpha}\mu\beta} =0$ \\
    \hline
       $g_{\mu\nu}$(10)-$2\times4$ (Diffs)=2 &  $\Lambda^\alpha_{\phantom{\nu}\beta}$(16)-$2\times4$ (Transl.)-6(Lor.)=2  &  the same as in GR \\
    \hline
        $R=\mathcal{R}$ &  $R = \mathcal{R} + \mathring{\mathbb{T}} + 2\mathcal{D}_\alpha \mathcal{T}^\alpha$  &  $R   =  \mathcal{R}  + \mathring{\mathcal{Q}}+   \mathcal{D}_\alpha ( \mathcal{Q}^\alpha - \tilde{\mathcal{Q}}^\alpha )$ \\
    \hline
  \multicolumn{3}{|c|}{equivalent theories up to boundary terms} \\
  \hline
\end{tabular}
\end{center}

Even if the three actions based on the Levi-Civita curvature $\mathcal{R}$, the torsion scalar $ \mathring{\mathbb{T}}$ and the non-metricity scalar $\mathring{\mathcal{Q}}$ give rise to the same underlying physical theory (General Relativity), promoting these scalar quantities to arbitrary functions thereof yields distinctive modified gravity theories. 
\begin{eqnarray}
&&\int \sqrt{-q} \mathcal{R} \qquad \to \qquad \int \sqrt{-q} f(\mathcal{R}) \nonumber\\
&&\int \sqrt{-q} \mathring{\mathbb{T}} \qquad \to \qquad \int \sqrt{-q} f(\mathring{\mathbb{T}})\nonumber\\
&&\int \sqrt{-q} \mathring{\mathcal{Q}} \qquad \to \qquad \int \sqrt{-q} f(\mathring{\mathcal{Q}})\,.
\end{eqnarray}
Modifications based on $f(\mathcal{R})$ and $f(\mathring{\mathbb{T}})$ have been extensively studied in the literature. However, the simple geometrical formulation of General Relativity purified from any inertial effects introduces a promising alternative starting point for modified gravity
theories based on $f(Q)$, which is less explored. It would be interesting to study the distinctive features of $f(Q)$ modifications with respect to the other two already existing modifications. Specially, the cosmological implications deserve a detailed analysis.\\

In the remainder of this review we will abandon the view of geometrical formulation of gravity and adapt to the more modern perspective of field theoretical formulation of gravity. From a field theoretical perspective, the construction of a consistent theory for a Lorentz-invariant massless spin-2 particle uniquely leads to General Relativity.
Under the assumptions of unitarity, locality, Lorentz symmetry and a (pseudo-)Riemannian manifold, any attempt at generalising the theory of gravity inevitably leads to new propagating degrees of freedom, which can be scalar, vector, or tensor fields. In this review we will give a comprehensive overview over a conceptually complete landscape of theories and their consequences together with their already existing tight empirical constraints.

\section{Field theories in cosmology and particle physics}\label{sec_fieldsincosmology}

In cosmology as well as in particle physics the mostly studied fields are those with massless or massive particles of spin-0, 1/2, 1 and 2. Higher spin particles are considered only in theories beyond the Standard Model of Particle Physics. We can describe particles with zero mass by their helicities, representing the projection of their angular momentum onto the direction of motion. In the Standard Model of Particle Physics massless bosons play the role of long range forces since otherwise forces carried by massive particles would be Yukawa suppressed due to their masses. We shall discuss the protagonists of the particles present in the Standard Model of Particle Physics and the Standard Model of Big Bang cosmology and their main properties. The study of the Lorentz group plays a crucial role in these research fields. Many fundamental laws of Nature like special relativity, the theory of electromagnetism and the theory of fermions etc. are invariant under Lorentz transformations. 

The Lorentz group is a subgroup of the Poincar\'e group, which denotes all the isometries of the Minkowski spacetime. The Poincar\'e symmetry contains the translations on spacetime, the rotations in space and the Lorentz boosts. The latter two are the transformations in the Lorentz group. Hence, while the Poincar\'e group comprises ten degrees of freedom of the isometries, the Lorentz group has only six, namely rotations about the three orthogonal directions in space and boosts along these directions. The elements $\Lambda_{\alpha\mu}$ of the Poincar\'e group satisfy $\eta_{\mu\nu}=\Lambda_{\alpha\mu}\Lambda_{\beta\nu}\eta_{\alpha\beta}$. The determinant and the $00$ component of this relation give the conditions $\det(\Lambda_{\alpha\mu})=\pm1$ and $\Lambda_{00}\ge1$ or $\Lambda_{00}\le-1$, respectively, dividing the Poincar\'e group into subsets. The case with $\det(\Lambda_{\alpha\mu})=1$ and $\Lambda_{00}\ge1$ is the $SO(3,1)$ Lorentz group. It is classified as a Lie group since it describes the set of continuous infinitesimal physical transformations of rotations and boosts with the generators $J_{\mu\nu}=\partial_\mu x_\nu-\partial_\nu x_\mu$, that satisfy the commutation relation 
\begin{equation}
[J^{\mu\nu},J^{\alpha\beta}]=i(\eta^{\nu\alpha}J^{\mu\beta}-\eta^{\mu\alpha}J^{\nu\beta}-\eta^{\nu\beta}J^{\mu\alpha}+\eta^{\mu\beta}J^{\nu\alpha})\,.
\end{equation} 
The boost operations $K_i$ and the rotation operations $L_i$ of the Lorentz group in terms of these generators $L_i=-\frac12 \epsilon_{ijk}J_{jk}$ and $K_i=J_{0i}$ give the commutation relations of the Lorentz group as $[L_i,L_j]=i \epsilon_{ijk}L_k$, $[K_i,L_j]=i \epsilon_{ijk}K_k$ and $[K_i,K_j]=- \epsilon_{ijk}L_k$. The latter represents the interesting fact, that commuting two boosts results in a rotation, hence boosts do not form a subgroup. 

An element of the Lorentz group is represented by the exponent of the generators and the six parameters of the transformation $\Lambda=\exp(-\frac{i}{2}\omega_{\mu\nu}J^{\mu\nu})$. The form of Lorentz transformations can be given by finite or infinite dimensional representations. Since it is a non-compact group, the finite dimensional representations are not unitary, in other words the generators are not Hermitian and act on finite dimensional vector spaces. First of all, there is the one dimensional representation with $J^{\mu\nu}=0$ and hence $\Lambda=1$, acting on a one dimensional vector space, which is nothing else but a Lorentz scalar. This trivial representation of a Lorentz transformation acts on a scalar $\phi$ by 
\begin{equation}
\phi\to \Lambda(=1)\phi=\phi\,. 
\end{equation}
The vector representation on the other hand acts on a four dimensional vector space with the generators represented by a $4\times4$ matrix $(J^{\mu\nu})^\alpha{}_\beta=i(\eta^{\mu\alpha}\delta^\nu{}_\beta-\eta^{\nu\alpha}\delta^\mu{}_\beta)$. The elements of this four dimensional vector space are the Lorentz vectors. The Lorentz transformation acts on a vector $A^\mu$ as 
\begin{equation}
A^\mu\to (e^{-i\omega_{\alpha\beta}J^{\alpha\beta}})^\mu{}_\nu A^\nu\,,
\end{equation}
and the matrices of the vector representation are exactly the $SO(3,1)$ matrices themselves, hence they form the fundamental representation of the Lorentz group. Tensor representations are nothing else but direct tensor products of vector representations acting on tensors of a given rank, the Lorentz tensors. Hence, a $(2,0)$ rank tensor $g^{\mu\nu}$ with two contravariant indices transforms as 
\begin{equation}
g^{\mu\nu}\to \Lambda^\mu{}_\alpha \Lambda^\nu{}_\beta g^{\alpha\beta}\,. 
\end{equation}
Similarly, there are spinorial representations acting on the set of two component objects, the Lorentz spinors. 

These finite representations so far were constants. Quantum field theory on the other hand deals with fields, which depend explicitly on spacetime. Since the fields are functions of coordinates that are also subject to Lorentz transformations one needs to use infinite dimensional field representations. In this case a generic field $\phi_a(x)$ will transform as $\phi_a(x)\to \Lambda_{ab}\phi_b(\Lambda^{-1}x)$.\\

Particles are degrees of freedom in a spacetime with a given number of dimensions classified by their spin and masses, which are carried by fields. A spin-0 particle is simply a scalar field. In fact, we know that fundamental spin-0 particles do exist in nature. The missing piece of the Standard Model of Particle Physics, the Higgs boson, is a spin-0 particle, which has been found after adamant efforts. A scalar degree of freedom might also play a fundamental role in cosmology, as it is a natural candidate for inflation and dark energy providing accelerated expansion without breaking isotropy. Modified gravity theories incorporate scalar fields in a very natural manner. A free massless scalar field is simply described by the Lagrangian $\mathcal L_\phi=-\frac12(\partial\phi)^2$ consisting of the kinetic term. Adding a mass term for a massive scalar field does not require any caution in order not to alter the number of propagating physical degrees of freedom, since the massless limit does not contain any gauge symmetry. Hence, one can simply promote the Lagrangian to 
\begin{equation}
\mathcal L_\phi=-\frac12(\partial\phi)^2+\frac12m_\phi^2 \phi^2\,,
\end{equation}
 or even add a general potential interaction $V(\phi)$. Imposing Lorentz invariance, one can construct in a similar manner consistent self-interactions. The inclusion of derivative self-interactions requires peculiar caution in order to maintain the right number of propagating physical degrees of freedom, as is the case for the Galileon \cite{Nicolis:2008in} (which will be extensively studied in section \ref{sec_scalarFields}).

 A spin-1 particle on the other hand is simply carried by a vector field. We already observed the presence of spin-1 particles in Nature, represented by the abelian and non-abelian vector fields like the photon and the carriers of the weak and strong interactions in the Standard Model of Particle Physics. The latter two are massive vector fields while the photon is a massless vector field. Vector fields might also play a crucial role in cosmology. The only difficulty there is the generation of large scale anisotropies, which can be avoided by considering a massive vector field \cite{Tasinato:2014eka,DeFelice:2016yws,DeFelice:2016uil,Heisenberg:2016wtr}. A controlled way of their presence could however facilitate explaining some of the reported anomalies at large scales in the CMB.
In order to have a consistent theory for a massless spin-1 particle with manifest Lorentz symmetry and locality, the massless vector field has to carry a gauge symmetry $A_\mu \to A_\mu + \partial_\mu \theta$, which guarantees the propagation of only two physical degrees of freedom. This uniquely results in the Maxwell theory 
\begin{equation}
\mathcal L_{A_\mu}=-\frac14F_{\mu\nu}^2-J_\mu A^\mu\,,
\end{equation}
with the field strength $F_{\mu\nu}=\partial_\mu A_\nu -\partial_\nu A_\mu$ and an external source $J_\mu$. In a similar way, one can add consistent self interactions of the massless vector field, which on the other hand yields the non-abelian gauge Yang-Mills theory. Adding a mass term needs a little bit of caution since it alters the number of propagating degrees of freedom due to the broken gauge symmetry.  A massive spin one field carries three propagating degrees of freedom instead of two as the massless case. The consistent theory for a massive vector field is given by the Proca theory 
\begin{equation}
\mathcal L_{A_\mu}=-\frac14F_{\mu\nu}^2-\frac12m_A^2A_\mu^2-J_\mu A^\mu\,. 
\end{equation}
In the same way as in the massless spin-1 case, the consistent theory for a massless spin-2 field with manifest Lorentz invariance requires a gauge symmetry, in form of a general coordinate invariance. Demanding Lorentz invariance and locality uniquely gives General Relativity 
\begin{equation}
\mathcal L_{g_{\mu\nu}}=\sqrt{-g} R + \mathcal{L}_{\rm matter}\,,
\end{equation}
with the Ricci scalar $R$. This theory propagates only two physical degrees of freedom. Adding a mass term would instead give five propagating degrees of freedom. At linear order such a mass term corresponds to the Fierz-Pauli term  $\frac{m^2}{2}\left(h_{\mu\nu}h^{\mu\nu}-h_{\mu}^{\mu}h_{\nu}^{\nu}\right)$. Adding further interactions is not trivial and rather challenging. Arbitrary potential interactions introduce ghostly degrees of freedom and one has to construct the interactions in a very specific way in order to avoid this problem. We will discuss in section \ref{sec_MassiveSpin2} how this is achieved in the dRGT theory.

In the standard model of particle physics the spin-1/2 particles play also very important role, describing the elementary fermions, i.e. the leptons and quarks. They enter in form of the massless Weyl fermions or the massive Dirac or Majorana fermions. Among them the Dirac fermions deserve special attention since most of the Standard Model fermions are Dirac fermions described by 
\begin{equation}
\mathcal L_\psi=\bar\psi(i\gamma^\mu\partial_\mu-m)\psi\,,
\end{equation}
 with the Dirac spinor $\psi$ and $\bar\psi\equiv \psi^\dagger \gamma^0$. The $\gamma$ matrices form the Clifford algebra $\{\gamma^\mu,\gamma^\nu\}=2\eta^{\mu\nu}$ and are given in terms of the Pauli matrices $\sigma^i$ in the Dirac representation. In contrast to the Standard Model of Particle Physics, the standard spinorial fields have not been much explored in cosmology. On the one hand this is due to the fact that it is difficult to interpret classical fermionic fields in terms of their underlying quantum particles. They cannot produce classical fermionic fields as they cannot condensate in coherent states. On the other hand fermions are inefficiently produced during preheating and decay fast in an expanding universe. The role that fermions play in the cosmological evolution is in form of a thermal distribution, as it is the case for instance for neutrinos. 
Moreover, it is plausible to think that some composite systems might be well-described by an effective classical spinorial field. This has been explored in the so-called  ELKO spinors with the unusual field property of $(CPT)^2=-1$ \cite{Ahluwalia:2004ab, Boehmer:2010ma}. For instance, this framework has been used to drive an inflationary phase \cite{Boehmer:2008rz}. In order to describe spinors in a curved spacetime, using the tetrad formalism becomes essential, since it allows to generalize the gamma matrices to the case of a non trivial background.

In the following sections we will review the individual representations of the Lorentz group and construct consistent field theories together with their relevance for cosmology. We will start with the simplest scalar representation and discuss in this context the important classes of theories based on a scalar field, like k-essence and Galileons. We will not only pay special attention to their classical behaviour but also embrace their quantum nature. We will then formulate General Relativity as the unique fundamental theory for a massless spin-2 particle and emphasize the role of general coordinate invariance. On that occasion we will explicitly break this symmetry and discuss how a consistent theory of a massive spin-2 particle can be established. This will be the playground of steady tensor-tensor interactions. 

After appreciating the general properties of a massive spin-2 particle and its consistent couplings to matter fields, we will examine its stability under quantum corrections and prove its technical naturalness. Next, we will promote the scalar field theories discussed on flat space-time to the general curved space-time and construct viable scalar-tensor theories, known as covariant Galileon and Horndeski theories. We will accentuate the role of non-minimal couplings to gravity in order to have second order equations of motion. We will also draw some parallels between scalar-tensor theories and massive gravity by covariantizing the decoupling limit of massive gravity. Furthermore, we will mention how the technical naturalness argument is altered on general curved space-time. More general interactions can be constructed by giving up the restriction of second order equations of motion. In this context, we will discuss the beyond Horndeski and DHOST theories. 

After finalising the general properties attributed to scalar-tensor theories, we will concentrate on the spin-1 representation of the Lorentz group. We will see how the requirements of Lorentz symmetry and masslessness inevitably imposes a gauge symmetry on the vector field. The construction of the corresponding Lagrangian uniquely leads to the Maxwell theory of electromagnetism. We will then investigate how one can generate more general interactions for the vector field, depending on whether the vector field carries a mass or not. We meet with resistance and encounter a no-go result for constructing self derivative interactions for a massless vector field. This result can be bypassed in the case of massive vector fields, which enables the construction of generalized Proca theories. We will describe their properties using different formulations and derive them also from the decoupling limit in a bottom-up approach. New genuine vector interactions will arise, which do not have any analog in the scalar theories. We will quickly explore their behaviour under quantum corrections and comment on the differences to the scalar Galileon theories. We will then promote these vector interactions constructed on flat spacetime to the case of curved backgrounds. This will constitute the consistent vector-tensor theories. Some of the interactions will again require the introduction of non-minimal couplings between the vector field and the gravity sector. Among the two new genuine vector interactions, only one requires the presence of non-minimal coupling to the double dual Riemann tensor. The two important classes of field theories, Horndeski and generalized Proca, can be unified into consistent scalar-vector-tensor theories, for both the gauge invariant and the gauge broken cases. Moreover, we will promote the $U(1)$ symmetry of a massless vector field to a $SU(2)$ symmetry of a non-abelian vector field and discuss how the breaking of the $SU(2)$ symmetry yields multi-Proca theories with interactions among three vector fields. Finally, we will investigate in detail the cosmological implications of all these field theories, together with their common and distinctive features on cosmological scales.

\section{Scalar fields}\label{sec_scalarFields}
As we mentioned in the introduction, the spin-0 particles are associated with scalar fields in quantum field theory, either real or complex valued, where the latter serve as charged particles. The Higgs boson in the Standard Model of Particle Physics is such a charged scalar field. Extensions of the Standard Model of Particle Physics and modifications of gravity naturally encompass scalar degrees of freedom. It is a crucial question what kind of consistent interactions a scalar field can have and how one can construct them using a few standard techniques and concepts. 

First of all, in order to have a manifestly Lorentz invariant action, the terms to be considered have to be a scalar quantity in terms of the scalar field and its derivative. The simplest Lagrangian for a massive scalar field $\pi$ is given by
\begin{equation}
 \mathcal L_\pi=-\frac12\partial_\mu\pi\partial^\mu\pi-\frac12m_\pi^2 \pi^2\,.
 \end{equation}
As we mentioned above, the inclusion of a mass term (or in general any potential term) does not alter the number of propagating degrees of freedom since taking the limit $m_\pi\to0$ does not introduce a gauge symmetry. The variation with respect to the scalar field gives the well known free Klein-Gordon equation
 \begin{equation}
(\Box-m_\pi^2) \pi=0\,,
 \end{equation}
 with $\Box=\partial_\mu \partial^\mu$. 
 
 This equation is solved by a superposition of plane waves $e^{\pm i k_\mu x^\mu}$ with $k_\mu k^\mu =E^2-\vec{k}^2=m^2$
  \begin{equation}
 \pi(x)=\int \frac{d^3k}{(2\pi)^3\sqrt{2E}}\left(a_{\vec{k}}e^{-ik_\mu x^\mu}+a_{\vec{k}}^*e^{ik_\mu x^\mu} \right)\,,
 \end{equation}
 where $a_{\vec{k}}^*$ is the adjoint operator with the property $a_{\vec{k}}^{**}=a_{\vec{k}}$.
 The total energy is given by the Hamiltonian, which has to be positive definite or in other words bounded from below. It represents the conserved charge related to time translation invariance. In terms of the momentum $\Pi_\pi=\partial  \mathcal L_\pi/\partial \dot{\pi}=\dot{\pi}$ conjugate to $\pi$, the Hamiltonian is given by\footnote{At this stage it is useful to mention that in the presence of a ghost instability, the momentum conjugate would enter with a minus sign $ \mathcal H_\pi=\Pi_\pi \dot{\pi}- \mathcal L_\pi=\frac12\left(-\Pi_\pi^2+(\vec{\nabla}\pi)^2+m_\pi^2 \pi^2\right)$, which would render the Hamiltonian unbounded from below. See \ref{AppendixInstabilities} for different types of instabilities that one has to be careful about when constructing field theories.}
 \begin{equation}\label{Hamiltonian_freeScalar}
 \mathcal H_\pi=\Pi_\pi \dot{\pi}- \mathcal L_\pi=\frac12\left(\Pi_\pi^2+(\vec{\nabla}\pi)^2+m_\pi^2 \pi^2\right)\,.
 \end{equation}
 This Hamiltonian density corresponds to the 00-component of the energy momentum tensor $T^{00}= \mathcal H_\pi$. The remaining coefficients of the latter are given by
  \begin{equation}\label{Tuv_freeScalar}
T^{\mu\nu}=\partial^\mu\pi\partial^\nu\pi+\eta^{\mu\nu} \mathcal L_\pi \,.
 \end{equation}
 
Now, consider the following split into a background field and a small perturbation $\pi=\bar{\pi}+\delta\pi$. In this case, the Lagrangian of second order in perturbations reads
  \begin{equation}
 \mathcal L_\pi^{(2)}=\frac12 \delta \pi \left(\Box-m_\pi^2\right)\delta\pi\,.
 \end{equation}
 The corresponding propagator of the scalar field is simply the inverse of the expression in the brackets $D_\pi=\frac{1}{\left(\Box-m_\pi^2\right)}$. If we have two sources $J(x)$ and $J(y)$ at positions $x$ and $y$, respectively, the scalar exchange amplitude would yield
 \begin{eqnarray}\label{exchange_ampl_scalar}
 W(J)&=&-\frac12\int d^4x d^4y J(x) D_\pi(x-y) J(y) \nonumber\\
 &=&-\frac12\int \frac{d^4k}{(2\pi)^4}J^*\frac{1}{k^2-m^2}J\,.
  \end{eqnarray}
  Hence, the scalar field as an exchange particle would give rise to an attractive force between the two sources. If we did not know more about the gravitational force, in principle it could be attributed to a scalar field. 
 
 \subsection{K-essence}\label{subsec_kessence}
 An interesting and crucial question is whether one can construct other non-trivial interactions for the scalar field without spoiling the consistency of the theory and the number of propagating degrees of freedom. We already know, that we can promote the mass term to a general potential term without altering the consistency of the theory, yielding a Quintessence field \cite{Wetterich:1987fm,Ratra:1987rm}
 \begin{equation}
 \mathcal L_\pi=-\frac12\partial_\mu\pi\partial^\mu\pi-V(\phi).
 \end{equation}
In fact this type of scalar field theories has been for instance considered in the early universe cosmology with a particular flat potential playing the role of the inflaton field \cite{Liddle:2000cg}. A natural generalisation of the above Lagrangian would be to consider a general function of the field and its kinetic term as in K-essence \cite{Chiba:1999ka,ArmendarizPicon:1999rj,ArmendarizPicon:2000ah,ArmendarizPicon:2000dh}
 \begin{equation}
P(X,\pi) \qquad \text{with} \qquad X=-\frac12\partial_\mu\pi\partial^\mu\pi. 
 \end{equation}
The fact, that it contains only one derivative per field ensures that the equations of motion will be at most second order in the derivatives and hence the number of propagating degrees of freedom will be unchanged, namely just one scalar degree of freedom will propagate. One particular restriction widely considered in the literature is the factorised dependence $P(X,\pi)=f(\pi)\tilde{P}(X)$. This type of interactions naturally arises in the low energy limit of string theory of the dilaton field. Another special class of the $P(X,\pi)$ theories is the Dirac-Born-Infeld model, which has a higher-dimensional geometrical probe-brane origin \cite{Alishahiha:2004eh}. It has the restriction of the shift symmetry. It is also invariant under a global symmetry $\pi \to \pi+c+v_\mu x^\mu+\pi v^\mu \partial_\mu\pi$ with constants $c$ and $v_\mu$. It can be interpreted as the remnant of the invariance under higher-dimensional rotations and boosts. The effective Lagrangian is a functional in form of a square root $P(X)\sim \sqrt{1+X}$ in this case. These theories have been widely used in the literature and we refer to the review article \cite{Copeland:2006wr} for more detail and references therein.

 \subsection{Galileon}\label{subsec_Galileon}
Theories based on an arbitrary kinetic term of the scalar field do not alter the number of propagating degrees of freedom, since there is still only one derivative per field. How about other derivative self-interactions with two derivatives per field? Naively, one would expect to obtain higher order equations of motion. In fact, in the context of the decoupling limit interactions of the five dimensional DGP brane-world scenario \cite{Dvali:2000hr}, a self-derivative interaction for the helicity-0 mode of the form $\mathcal L_3=\Box\pi(\partial\pi)^2$ arises (where the subscript 3 in $\mathcal L_3$ refers to the number of appearance of the field $\pi$). Even if this interaction carries two derivatives at the level of the Lagrangian, the equation of motion is still second order in the field derivatives $\mathcal E_3=(\Box\pi)^2-(\partial_\mu\partial_\nu\pi)^2$. Furthermore, the interaction is invariant under internal Galilean and shift transformations $\pi \to \pi+c+x_\mu b^\mu$. This is also a remnant of the five dimensional symmetries. Within the effective field theory perspective, one could wonder whether further higher order derivative interactions could be constructed in the same spirit. The previous cubic interaction contained three fields and four derivatives. Next, one could try to construct interactions  with four fields and six derivatives and write down a general Ansatz with all possible contractions
\begin{equation}\label{Gal_symmetry}
\mathcal{L}_4= \alpha_1(\partial\pi)^2(\Box\pi)^2+\alpha_2 (\partial\pi)^2(\partial_\mu\partial_\nu\pi)^2\,,
\end{equation}
with arbitrary coefficients $\alpha_1$ and $\alpha_2$. Other contractions are equivalent to those after integration by parts. First of all, the equations of motion are not second order in the derivatives. We have a contribution of the form $(\alpha_1+\alpha_2)\dddot\pi$. In the presence of these higher order terms, an additional ghostly degree of freedom propagates, which hides itself behind $\dddot\pi$. In order to get rid of these constributions, one has to impose $\alpha_2=-\alpha_1$. Thus, the consistent interactions at this order have to be tuned in the following way
\begin{equation}\label{Gal_symmetry}
\mathcal{L}_4= (\partial\pi)^2\left((\Box\pi)^2-(\partial_\mu\partial_\nu\pi)^2\right)\,.
\end{equation}
One can construct the subsequent interactions following the same procedure and making sure that the equations of motion of the scalar field do not contain higher than second order time derivatives.

With the additional restrictions of invariance under internal Galilean and shift transformations, the requirement of second order equations of motion is fulfilled by only five interactions in four dimensions $\mathcal{L}_{\rm Gal}=\sum_{i=1}^5c_i\mathcal L_i$ \cite{Nicolis:2008in}
\begin{eqnarray}\label{Gal_interactions}
\mathcal L_1 & = & \pi\nonumber\\
\mathcal L_2 & = & (\partial\pi)^2\nonumber\\
\mathcal L_3 & = &(\partial\pi)^2\Box\pi\nonumber\\
\mathcal L_4 & = & (\partial\pi)^2\left[(\Box\pi)^2-(\partial_\mu\partial_\nu\pi)^2\right]\nonumber\\
\mathcal L_5 & = & (\partial\pi)^2\left[(\Box\pi)^3-3\Box\pi(\partial_\mu\partial_\nu\pi)^2+2(\partial_\mu\partial_\nu\pi)^3\right]\,,
\end{eqnarray}
with arbitrary coefficients $c_i$. These interactions are local and include terms higher order in the derivatives. Despite the presence of the higher order derivative operators the equations of motion are second order. The relative tunings of the coefficients guarantee that. Since the interactions contain only derivatives, they are invariant under the transformations
\begin{equation}\label{Gal_symmetry}
\pi \to \pi+c+x_\mu b^\mu\,,
\end{equation}
with constant $c$ and $b^\mu$. This constitutes the Galileon theories. A crucial property is that these symmetries are realised only up to total derivatives at the level of the Lagrangians. Thus, they are exact symmetries of the theory only at the level of the equations of motion. These interactions can also be written in a more convenient and compact form using the iterative relation
\begin{equation}\label{Gal_iterative}
\mathcal L_{n+1}=-(\partial\pi)^2 E_{n}
\end{equation}
where $n\ge 1$ and $ E_{n}=\frac{\delta \mathcal L_{n}}{\delta \pi}$ are the equations of motion
\begin{eqnarray}\label{EOMgalileons}
E_1 & = & 1\nonumber\\
E_2 & = & \Box\pi\nonumber\\
E_3 & = &(\Box\pi)^2-(\partial_\mu\partial_\nu\pi)^2\nonumber\\
E_4 & = & (\Box\pi)^3-3\Box\pi(\partial_\mu\partial_\nu\pi)^2+2(\partial_\mu\partial_\nu\pi)^3\nonumber\\
E_5 & = &(\Box\pi)^4-6(\Box\pi)^2(\partial_\mu\partial_\nu\pi)^2+8\Box\pi(\partial_\mu\partial_\nu\pi)^3+3((\partial_\mu\partial_\nu\pi)^2)^2
-6(\partial_\mu\partial_\nu\pi)^4\,.
\end{eqnarray}
As can clearly be seen, the equations of motion are at most second order in the derivatives, avoiding the Ostrogradski instability \cite{Ostrogradski1850,Woodard:2015zca}. The Galileon interactions can also be equally written in terms of the antisymmetric Levi-Civita tensors ${\mathcal{E}}^{\mu\alpha\rho\sigma}$ as $\mathcal{L}_{\rm Gal}^n=\sum_{i=1}^5\tilde{c}_i\mathcal L_i$
\begin{eqnarray}\label{eqsGal_LeviC}
\mathcal{L}_1&=&\pi{\mathcal{E}}^{\mu\alpha\rho\sigma}
{{\mathcal{E}}}_{\mu\alpha\rho\sigma}, \quad  \nonumber \\
\mathcal{L}_2&=&\pi{\mathcal{E}}^{\mu\alpha\rho\sigma}
{{\mathcal{E}}^{\nu}}_{\alpha\rho\sigma}\Pi_{\mu\nu}, \quad  \nonumber \\
\mathcal{L}_3&=&\pi{\mathcal{E}}^{\mu\alpha\rho\sigma}
{{\mathcal{E}}^{\nu\beta}}_{\rho\sigma}\Pi_{\mu\nu}\Pi_{\alpha\beta}, \nonumber \\
\mathcal{L}_4&=&\pi{\mathcal{E}}^{\mu\alpha\rho\sigma}
{{\mathcal{E}}^{\nu\beta\gamma}}_{\sigma}\Pi_{\mu\nu}\Pi_{\alpha\beta}\Pi_{\rho\gamma}\nonumber \\
\mathcal{L}_5&=&\pi{\mathcal{E}}^{\mu\alpha\rho\sigma}
{{\mathcal{E}}^{\nu\beta\gamma\delta}}\Pi_{\mu\nu}\Pi_{\alpha\beta}\Pi_{\rho\gamma}\Pi_{\sigma\delta}\, ,
\label{Gals}
\end{eqnarray}
with $\Pi_{\mu\nu}=\partial_\mu \partial_\nu \pi$. This fundamental object of the theory is symmetric under the exchange of the indices $\mu \leftrightarrow \nu$. Therefore, there is only one way of contracting it with the totally antisymmetric Levi-Civita tensors. Its two indices have to be contracted separately with the respective index of the two Levi-Civita tensors. In this way, one can construct the interactions order by order by augmenting in $\Pi_{\mu\nu}$. In this formulation it becomes apparent why the series stops at $\mathcal{L}_5$ in four dimensions, since all the indices of the two Levi-Civita tensors are then fully contracted. In five dimensions, one could for instance generate $\pi{\mathcal{E}}^{\mu\alpha\rho\sigma\kappa}
{{\mathcal{E}}^{\nu\beta\gamma\delta\xi}}\Pi_{\mu\nu}\Pi_{\alpha\beta}\Pi_{\rho\gamma}\Pi_{\sigma\delta}\Pi_{\kappa\xi}$ for $\mathcal{L}_6$, etc. The action of the Galileon interactions can be also rewritten as
\begin{equation}\label{GalinterpiPi}
\mathcal{S}_{\rm Gal}=\int d^4x\sum_{n=2}^5\frac{\tilde{c}_n}{n(n-2)!(5-n)!}\frac{1}{\Lambda^{6(n-2)/2}}\mathcal{L}_{\rm Gal}^n\,,
\end{equation}
where the interactions (\ref{eqsGal_LeviC}) in terms of the Levi-Civita tensors can be written in the compact form
\begin{equation}
\mathcal{L}_{\rm Gal}^n=\pi \mathcal{E}^{\alpha_1\cdots\alpha_4}\mathcal{E}^{\beta_1\cdots\beta_4}\prod_{i=1}^{n-1}\partial_{\alpha_i}\partial_{\beta_i}\pi\prod_{j=n}^{4}\eta_{\alpha_j \beta_j}\,.
\end{equation}
Due to this antisymmetric structure, the Galileon interactions can be also expressed in terms of a deformed determinant
\begin{eqnarray}\label{deformedDet_Gal}
\pi{\rm det}(\delta^\mu_\nu+\tilde{c}_i\partial^\mu\partial_\nu\pi)=\pi\sum_{i=0}^4\frac{-\tilde{c}_i}{i!(4-i)!}\mathcal{E}_{\mu_1\cdots \mu_i \alpha_{i+1}\cdots \alpha_4}\mathcal{E}^{\nu_1\cdots \nu_i \alpha_{i+1}\cdots \alpha_4}
\partial^{\mu_1}\partial_{\nu_1}\pi\cdots \partial^{\mu_i}\partial_{\nu_i}\pi\,.
\end{eqnarray}
Expanding the determinant gives exactly the elementary polynomials. 

The Hamiltonian of the Galileon can be straightforwardly computed and the simple expression of the Hamiltonian for a free scalar field in (\ref{Hamiltonian_freeScalar}) gets replaced by
 \begin{equation}
H_\pi=\int d^4x\mathcal H_\pi=\int d^4x \frac12\sum_n\left(\mathcal{K}\dot{\pi}^2+\mathcal{V}(\partial_i\pi)^2\right)\,.
 \end{equation}
where the kinetic and potential terms are respectively given by
\begin{eqnarray}
&&\mathcal{K}=\sum_{n=2}^5\frac{(n-1)\tilde{c}_n}{(n-2)!(5-n)!}E_{n-1}^{d-1} \,, \nonumber\\
&&\mathcal{V}=\sum_{n=2}^5\frac{\tilde{c}_n}{(n-2)!(5-n)!}E_{n-1}^{d-1}\,,
\end{eqnarray}
where the operator of the equation of motion is obtained by the variation $E_{n}^{d}=\frac{1}{n}\frac{\delta\mathcal{L}_n^d}{\delta\pi}$
\begin{equation}\label{EOMoperators}
E_n^d= \mathcal{E}^{\alpha_1\cdots\alpha_d}\mathcal{E}^{\beta_1\cdots\beta_d}\prod_{i=1}^{n-1}\partial_{\alpha_i}\partial_{\beta_i}\pi\prod_{j=n}^{d}\eta_{\alpha_j \beta_j}\,.
\end{equation}
Most non-trivial Galileon Hamiltonians on specific backgrounds are unbounded from below \cite{Zhou:2010di,Sivanesan:2011kw}. On the spherically symmetric backgrounds the Hamiltonian becomes unbounded from below when the kinetic term is negative. Let us consider a background field configuration $\pi=\pi(r,t)$, then for (\ref{EOMoperators}) we have $E_2=2\frac{d}{r^2dr}\left( r^2\pi^\prime\right)$, $E_3=2\frac{d}{r^2dr}\left( r(\pi^\prime)^2\right)$ and $E_4=2\frac{d}{r^2dr}\left( r(\pi^\prime)^2\right)$, respectively. The Hamiltonian in this case becomes
\begin{eqnarray}\label{HamiltonianSSB}
H&=&4\pi\int dr\dot\pi^2\frac{d}{dr}\left(\frac{1}{6}r^3+\tilde{c}_3r^2\pi^\prime+\frac32\tilde{c}_4r(\pi^\prime)^2 +\frac23\tilde{c}_5 (\pi^\prime)^3 \right) \nonumber\\
&+&(\pi^\prime)^2\frac12\frac{d}{dr}\left(\frac{1}{3}r^3+\tilde{c}_3r^2\pi^\prime+\tilde{c}_4r(\pi^\prime)^2  \right) \,.
\end{eqnarray}
Defining the new variable $\tilde{\pi}=\pi^\prime/r$, we can rewrite the Hamiltonian as $H=H_0+H_1$ after some integrations by parts, where we defined
\begin{eqnarray}\label{HamiltonianGalVar}
H_0&=&\frac{4\pi}{3!}\int dr\frac{d}{dr}\left(F_0(\tilde{\pi})\right)(\dot\pi^2+\kappa (\pi^\prime)^2)\nonumber\\
H_1&=&\frac{4\pi}{3!}\int dr r^4 \tilde{\pi}^2F_1(\tilde{\pi})\,,
\end{eqnarray}
with the functions $F_0$ and $F_1$ introduced for brevity
\begin{eqnarray}\label{HamiltonianFunctsFi}
F_0&=&1+6\tilde{c}_3\tilde{\pi}+9\tilde{c}_4\tilde{\pi}^2+4\tilde{c}_5\tilde{\pi}^3\nonumber\\
F_1&=&\frac{3}{2} (1-3\kappa)\tilde{c}_4\tilde{\pi}^2+4\tilde{c}_3\tilde{\pi}(1-2\kappa)+3(1-\kappa)\,.
\end{eqnarray}
There exists a relation between the $\tilde{c}_3$ and $\tilde{c}_4$ parameters, which allows to bring the Hamiltonian in a form, that is manifestly negative if only the kinetic term is negative. If these two parameters satisfy $\tilde{c}_4=p\tilde{c}_3^2$, where $p=\frac{8(1-2\kappa)^2}{9(1-\kappa)(1-3\kappa)}$, then the function $F_1$ becomes a perfect square, hence the part $H_1$ of the Hamiltonian positive definite. 
For $\kappa\ge0$ and $\tilde{c}_4=p\tilde{c}_3^2$ with $p\le0$ or $p\ge8/9$, the Hamiltonian of the Galileon becomes negative if only the kinetic term is negative. This causes the unboundedness in the case of spherical symmetry if the kinetic term becomes negative.

 \subsection{Stability under quantum corrections}\label{subsec_GalQuantum}
 One important theoretical question is whether the introduced tuning of the relative and overall coefficients of the scalar field interactions are subject to strong renormalisation under quantum corrections. A detuning of the relative coefficients within the scale of validity of the effective action would render the theories sick. The fact that the Galileon realises the internal Galileon symmetry up to total derivatives is a crucial property, that protects the interactions from large quantum corrections. 
 
Needless to say, the shift and Galileon symmetry will prevent the generation of local operators by loop corrections which would break explicitly these symmetries. However, one could still worry if quantum corrections might yield terms invariant under these symmetries and hence renormalise the Galileon interactions themselves. It is an appealing property of the Galileon that this actually does not happen and the tunings of the classical interactions are protected from the quantum corrections by the virtue of the non-renormalisation theorem. To appreciate the latter, the formulation of the interactions in terms of the two Levi-Civita tensors in (\ref{eqsGal_LeviC}) is very useful. 

Let us consider an arbitrary Feynman diagram with a vertex given by $\mathcal{L}_3=\pi{\mathcal{E}}^{\mu\alpha\rho\sigma}
{{\mathcal{E}}^{\nu\beta}}_{\rho\sigma}\Pi_{\mu\nu}\Pi_{\alpha\beta}$ and contract an external leg of momentum $q_\mu$ with the $\pi$ particle in the vertex without derivative and let the other two $\pi$-particles run in the loop with momenta $k_\mu$ and $(q+k)_\mu$. This vertex will then contribute to the scattering amplitude in the following form
 \begin{eqnarray}
\mathcal A
 \propto  \int \frac{\mathrm{d}^4 k}{(2\pi)^4} D_k \,
 D_{k+q}\  \, {\mathcal{E}}^{\mu\alpha\rho\sigma}
{{\mathcal{E}}^{\nu\beta}}_{\rho\sigma}
\, \, k_\mu \, k_\nu \, (q+k)_\alpha \, (q+k)_\beta \cdots \,,
\end{eqnarray}
with $D_k= k^{-2}$ representing the Feynman massless propagator for the Galileon field. Note, that the terms linear in the external momentum ${\mathcal{E}}^{\mu\alpha\rho\sigma}
{{\mathcal{E}}^{\nu\beta}}_{\rho\sigma}k_\mu k_\nu k_\alpha q_\beta$ or independent of it ${\mathcal{E}}^{\mu\alpha\rho\sigma}
{{\mathcal{E}}^{\nu\beta}}_{\rho\sigma}k_\alpha k_\beta k_\mu k_\nu$ will vanish exactly due to the contraction with the antisymmetric Levi-Civita tensors. The only surviving term will come in with at least two powers of the external momentum $q_\alpha q_\beta$. Thus, one generates only terms with higher order derivatives per field, which do not belong to the Galileon interactions and are suppressed in the effective action. 

The crucial point is that the Galileon symmetry is only realised up to total derivatives whereas the Feynman diagrams generate operators, which satisfy this symmetry exactly and therefore do not renormalise the classical Galileon interactions. This is the essence of the non-renormalization theorem of the Galileon (see \cite{Luty:2003vm,Nicolis:2004qq,Nicolis:2008in,Hinterbichler:2010xn,deRham:2012ew,dePaulaNetto:2012hm,Brouzakis:2013lla,Heisenberg:2014raa} for more detail on the quantum corrections of the Galileon). This means that the Galileon coupling constants are technically natural and remain radiatively stable. 
The non-renormalization theorem is not a unique attribute of the Galileon but also applies to the derivative theories $P(X)$, specially to the sub-class of the DBI scalar field models. Surprisingly, even if the DBI models are equipped with an additional symmetry, they are equally protected from the quantum corrections as the $P(X)$ theories. Independently of the symmetry difference they have the same regime of validity\cite{deRham:2014wfa}.

At this stage, it is important to emphasise that the non-renormalization theorem does not survive when one includes explicit couplings to matter. The scalar field can couple either via a conformal coupling $\pi T$, disformal coupling $\partial_\mu\pi\partial_\nu\pi T^{\mu\nu}$ or a longitudinal coupling $\partial_\mu \partial_\nu\pi T^{\mu\nu}$. The disformal coupling arises naturally in the context of massive gravity \cite{deRham:2010ik,deRham:2010tw}. In \cite{Heisenberg:2014raa} the contributions coming from the one-loop quantum corrections through the different couplings to matter were studied and it was shown that the terms generated by one-loop matter corrections not only renormalise the Galileon interactions themselves but also give rise to ghostly higher order derivative interactions. These contributions are suppressed by the coupling scale and hence harmless within the regime of validity of the effective field theory.\\

\begin{center}
\begin{figure}[h]
\begin{center}
 \includegraphics[width=14.5cm]{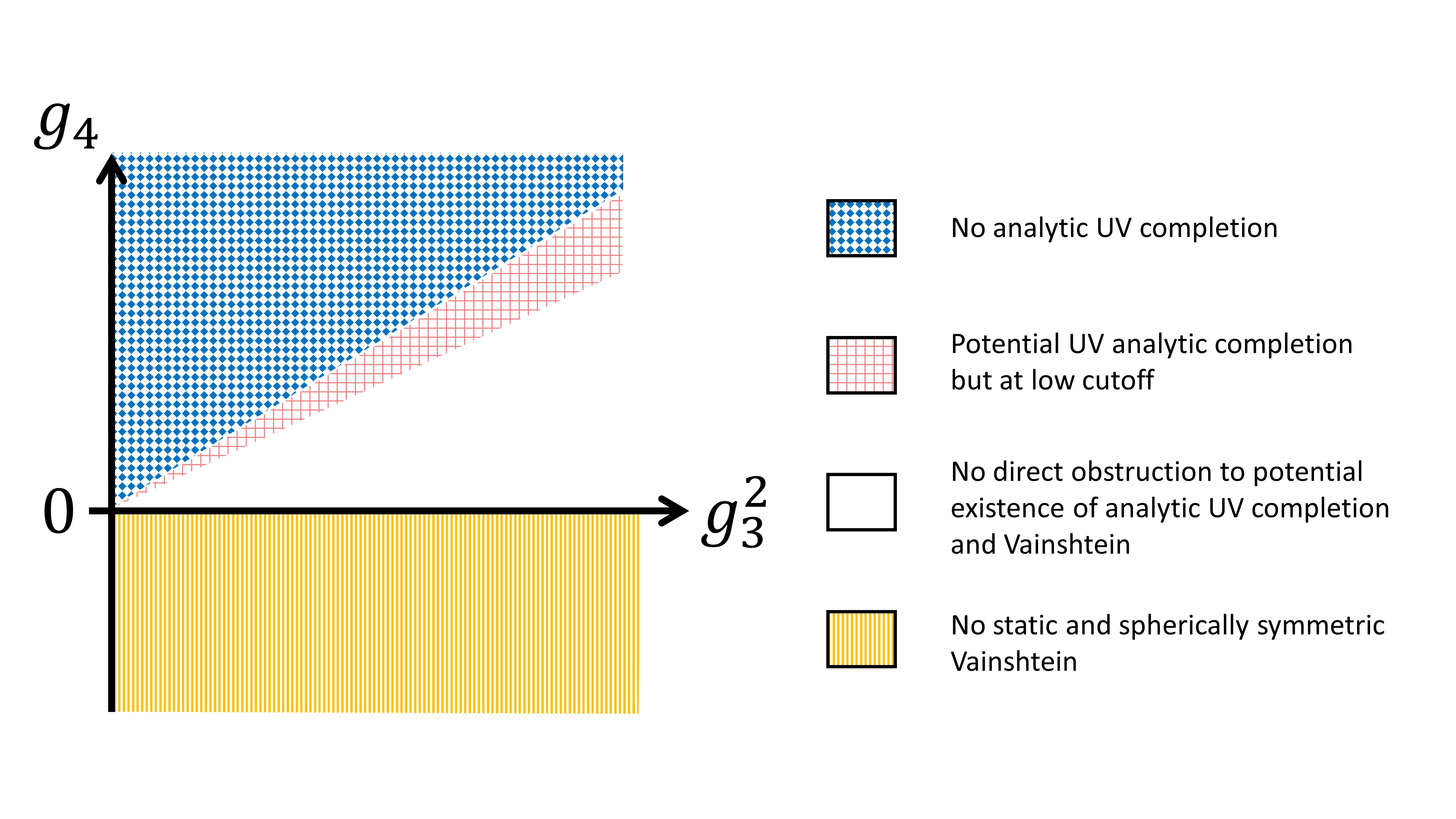}
\end{center}
  \caption{This figure is taken from \cite{deRham:2017imi}, where the positivity bounds on the coefficients of the cubic $g_3$ and quartic $g_4$ Galileon interactions are summarised. These two parameters correspond to $\tilde{c}_3$ and $\tilde{c}_4$ in our notation in (\ref{GalinterpiPi}). This allowed region of parameter space is the outcome of the imposed conditions for a local, analytic UV completion, for a cutoff above the Galileon mass, and for a static and spherically symmetric Vainshtein mechanism. For $\tilde{c}_4<3/4\tilde{c}_3^2$ there is no obstruction for a local and analytic UV completion.}
   \label{boundsg3g4}
\end{figure}
\end{center}

\textbf{Implications of the UV completion}: \\
Even if the UV completion of a theory is not known, one can infer important properties of the low energy effective field theory based on the unitarity and analyticity requirements of the scattering amplitudes \cite{Adams:2006sv,Nicolis:2009qm,Bonifacio:2016wcb,deRham:2017avq,deRham:2017imi}. They are known as positivity bounds and constrain the sign of the coefficients in the effective action. The 2-to-2 scattering amplitudes are the best apparatus for this purpose \cite{Mandelstam:1958xc}. One important precondition is that the probabilities should add up to 1 (conservation of probability), to wit the Hamiltonian should be hermetian $H^\dagger=H$. This on the other hand imposes that the S-matrix should be unitary $S^\dagger S=1$, with $S=e^{iHt}$. One immediate implication of unitarity is the optical theorem, which states that the imaginary part of the forward scattering amplitude $A$ is proportional to the total scattering cross section $\sigma$
\begin{equation}
{\rm Im}[A(s,t=0)]=\sqrt{s(s-4m^2)}\sigma(s)\,,
\end{equation}
with the centre of mass energy $s$ and the momentum transfer $t=0$. Unitarity translates into the positivity requirement of the scattering amplitude as \cite{deRham:2017avq}
\begin{equation}
\frac{\partial^n}{\partial t^n}{\rm Im}[A(s,t)]\Big|_{t=0}>0 \qquad \text{for} \qquad n\ge0 \qquad \text{and} \qquad s\ge 4m^2\,.
\end{equation}
Another important requirement is the analyticity of the scattering amplitude. This allows to imply positive bounds on the low energy scattering away from the forward limit
\begin{equation}
\frac{\partial^n}{\partial t^n}{\rm Im}[A(s,t)]>0 \qquad \text{for} \qquad s\ge 4m^2 \qquad \text{and} \qquad 0\le t < 4m^2 \,.
\end{equation}
Combining the unitarity and analyticity conditions enables to derive the Froissart-Martin bound on the scattering amplitude at fixed t.
\begin{equation}
{\rm lim}_{s\to \infty }|A(s,t)|<Cs^{1+\epsilon(t)} \qquad \text{for}\quad 0\le t < 4m^2 \,,
\end{equation}
with $C$ constant and $\epsilon(t)<0$, which facilitates that the fixed t amplitude can be dealt with as in the $t=0$ case \cite{Jin:1964zza}.
In \cite{deRham:2017avq} the reader can find the detail application of these bounds not only on the tree-level scattering amplitudes but actually on the coefficients in the tree-level Lagrangians of the low energy effective field theory. 
Since the quantic Galileon interactions do not contribute to the $2\to2$ tree-level amplitude, one only obtains bounds on the ratio of the cubic and quartic interactions. There is no obstruction for the potential existence of an analytic UV completion if the parameters of the Galileon satisfy the bound $\tilde{c}_4<3/4\tilde{c}_3^2$.
Violation of these bounds has tremendous implications: no local, unitary and Lorentz invariant UV completion of the underlying scalar theory. In this way one can put theoretical bounds on the coefficients of the effective action based on unitarity and analyticity, where the latter allows to constrain the scale of new physics beyond the perturbative unitarity argument. These positivity bounds are only applicable in the presence of a mass gap. Therefore, the forward limit positivity bounds show an obstruction to a local UV completion of the massless Galileon \cite{Nicolis:2008in,Adams:2006sv}. This problem was resolved in \cite{deRham:2017imi} by considering a massive Galileon theory. In this case the coefficients of the theory can follow the positivity bounds required for an unitary analytic Lorentz invariant UV completion. The bounds on the cubic and quartic interactions at the tree-level scattering amplitudes are summarised in figure \ref{boundsg3g4}.

  \subsection{Superluminal propagation}\label{subsec_GalSuperLum}
A potentially worrying phenomenon of the Galileon is the fact that the fluctuations of the Galileon field can propagate superluminally in the regime of validity of effective field theory \cite{Nicolis:2008in,Goon:2010xh}. The risk of superluminality is the construction of closed time-like curves and hence violation of causality. The superluminal propagation is unavoidable if one imposes the stability of the perturbations. Let us consider a localised point source $\rho=M\delta^3(r)$ and a spherically symmetric background for the Galileon field $\pi(r)$. The equation of motion of the Galileon reads for this background configuration \cite{Nicolis:2008in}
\begin{eqnarray}\label{Gal_sphericalBG}
\frac{1}{r^2}\frac{d}{dr}\left\{ r^3\left(c_2\frac{\pi^\prime}{r}+2c_3\left(\frac{\pi^\prime}{r}\right)^2+2c_4\left(\frac{\pi\prime}{r}\right)^3\right) \right\}=\frac{M}{M_{\rm Pl}}\delta^3(\vec{r})\,,
\end{eqnarray}
which can be integrated to write it as
\begin{eqnarray}
c_2\frac{\pi^\prime}{r}+2c_3\left(\frac{\pi^\prime}{r}\right)^2+2c_4\left(\frac{\pi^\prime}{r}\right)^3=\frac{M}{4\pi r^3M_{\rm Pl}}\,.
\end{eqnarray}
Now, a crucial condition is the stability of this background field configuration\footnote{Concentrating for instance up to the cubic Galileon, the background solution would be given by $\pi(r)=a_1\sqrt{r}+a_2r^2+\mathcal{O}(r)^{7/2}$, where the parameters $a_1$ and $a_2$ are given in terms of the parameters $M$, $M_{\rm Pl}$ and $c_3$. Introducing the strong coupling scale in the cubic interactions as $c_3/\Lambda^3$, the redressed strong coupling scale would become $\Lambda_{\rm red}=\sqrt{\Box\pi}\Lambda/\sqrt{\Lambda^3}$. } under small perturbations of the field $\pi=\pi(r)+\delta\pi(t, r,  \theta, \varphi)$. In this case the quadratic Lagrangian becomes
\begin{eqnarray}
\mathcal L= \frac12\partial_t\delta\pi \mathcal K^{tt}\partial_t\delta\pi -\frac12\partial_r\delta\pi  \mathcal U^{rr}\partial_r\delta\pi -\frac12\partial_\Omega\delta\pi  \mathcal V^{\Omega\Omega}\partial_\Omega\delta\pi 
\end{eqnarray}
with the solid angle $\Omega$ the following kinetic and potential terms
\begin{eqnarray}
\mathcal K_{tt}&=&\frac{c_2^2 + 4 c_2c_3 \frac{\pi^\prime}{r} +12(c_3^2-c_2c_4) (\frac{\pi^\prime}{r})^2 +24(c_3c_4-2c_2c_5)(\frac{\pi^\prime}{r})^3+12(3c_4^2-4c_3c_5)(\frac{\pi^\prime}{r})^4}{c_2+4c_3\frac{\pi^\prime}{r}+4c_4 (\frac{\pi^\prime}{r})^2}\nonumber\\
\mathcal U_{rr}&=&c_2 + 4 c_3 \left(\frac{\pi^\prime}{r}\right) +6c_4 \left(\frac{\pi^\prime}{r}\right)^2 \nonumber\\
\mathcal V_{\Omega\Omega}&=&\frac{c_2^2+2c_2c_3\frac{\pi^\prime}{r}+(4c_3^2-6c_2c_4)(\frac{\pi^\prime}{r})^2 }{c_2+4c_3\frac{\pi^\prime}{r}+6c_4(\frac{\pi^\prime}{r})^2}
\end{eqnarray}
In order to avoid any ghost instability $\mathcal K_{tt}>0$ (see \ref{AppendixInstabilities}), we have to impose the following conditions:
 \begin{eqnarray}
c_2> 0, \qquad
c_4\ge 0, \qquad
c_3 \ge \sqrt{\frac32 c_2 c_4} \qquad \text{and} \qquad
c_5 \le \frac34 \frac{c_4^2}{c_3}\,.
\end{eqnarray}
Together with these constraints we also need to ensure the absence of gradient instabilities related to the propagation speed of fluctuations. 
First of all, the radial speed of fluctuations is given by
 \begin{eqnarray}
c_r^2=\frac{\mathcal U_{rr}}{\mathcal K_{tt}}=1+4\frac{c_3}{c_2}\frac{\pi_0^\prime}{r}+\cdots>1\,.
\end{eqnarray}
One immediately observes that the radial propagation speed is always superluminal if one enforces the stability conditions.
We will also have a similar stability condition for the angular speed of fluctuations $c^2_{\Omega}=\mathcal V_{\Omega\Omega}/\mathcal K_{tt}<1$, which is forced to be subluminal after imposing the absence of ghost instabilities. As we have seen from this short analysis, the single Galileon field suffers always from the superluminal propagation as a consequence of the stability condition of the fluctuations. The same burden is shared by theories of multi-Galileon or massive gravity \cite{deFromont:2013iwa}.

A serious concern in the context of superluminal propagation comes from the possibility of acausality and construction of  closed time-like curves (CTCs). On a closer inspection there arise many facets of the configuration. There are studied cases in which the superluminal fluctuations come with their own metric and causal structure being completely different from those felt by photons. In some cases the causal cones of these superluminal fluctuations lie even outside the causal cones of the photons. As it was shown in \cite{Babichev:2007dw}, the causal structure of the spacetime can be protected if there exists one foliation of spacetime into surfaces which can be considered as Cauchy surfaces for both metrics. 

In the context of Galileon theories one can construct CTCs within the naive regime of validity of the effective field theory. Actually, already in GR one can construct such CTS. It is about curves which form closed loops in space and time, and which can be traversed in a time-like manner. Even if these solutions are theoretically possible, they actually never arise since the scalar Galileon inevitably becomes infinitely strongly coupled giving rise to an infinite amount of backreaction \cite{Burrage:2011cr}. Exactly this strong backreaction effect makes the effective field theory break down and so forbids the actual formation of CTCs. In other words, the background solutions with CTCs become unstable with an arbitrarily fast decay time. This reflects the fact, that theories based on Galileon interactions satisfy a direct analogue of Hawking's chronology protection conjecture \cite{Burrage:2011cr}. 

It is worthwhile to point out that the propagation of superluminal fluctuations in Galileon theories signals the failure of having a Wilsonian UV-completion \cite{Adams:2006sv}. However, the presence of the Vainshtein mechanism and the lack of a Wilsonian UV-completion could just mean the urgency of an alternative UV completion such as classicalization, \cite{Dvali:2010jz,Dvali:2012zc,Vikman:2012bx}.
A curios observation is that the Vainshtein mechanism and the superluminal propagation seem to appear hand in hand if one imposes trivial asymptotic conditions at infinity and wants to avoid quantum strong-coupling issues due to vanishing kinetic terms. Since the Vainshtein mechanism is inherently nonlinear, the perturbations depend strongly on the source distribution present. Hence, one would expect to always be able to find backgrounds around which there are superluminal propagation. 

One way of avoiding superluminal propagation but still having a fully functional Vainshtein mechanism seems to be only possible if the Galileon is not treated as an independent field in its own right, but rather as a fundamental component of a fully fledge theory, like it is the case for instance in massive gravity or generalized Proca theories. The crucial difference comes from the fact that in such a case one is not forced to impose trivial asymptotic conditions for the Galileon at infinity. 

By its nature, in massive gravity the helicity-0 mode does not need to vanish at infinity, as long as the metric itself is well defined at infinity. An example of this realisation was pushed forward in \cite{Berezhiani:2013dw} with a non-trivial asymptotic of the Galileon in massive gravity where the metric approaches to a cosmological one at large distances. An important ingredient in this setup is the presence 
of a disformal coupling to matter fields $\partial_\mu \pi \partial_\nu \pi T^{\mu\nu}$ which naturally arises in massive gravity in any case. In this way one could find configurations which explicitly exhibit the Vainshtein mechanism without necessarily giving rise to superluminal propagation on top of these configurations. Along these lines, it is essential to explore the deep physical relations between all these phenomena which are up to date still mysterious.

\subsection{Galileon dualities}\label{subsec_GalDual}

Another interesting property of the Galileon which is worth exploring is the existence of dual descriptions \cite{Creminelli:2013ygt,Fasiello:2013woa,deRham:2013hsa}. It has been shown that a specific subclass of Galileon interactions are dual to a free massless scalar field, which would suggest that the naive existence of superluminal propagation within this class can still give rise to causal theory with analytic and unitary S-matrix \cite{deRham:2013hsa}, since it corresponds to just a free luminal theory. Likewise, the idea of a UV completion via classicalization finds a promising realisation, since the duality maps a strongly coupled state to a weakly coupled state. Following this duality the original Galileon theory can be mapped to another Galileon theory by a non-trivial field redefinition $\tilde x^\mu=x^\mu+\partial^\mu\pi(x)/\Lambda^3$. Applying the inverse transformation gives the dual Galileon $x^\mu=\tilde x^\mu +\tilde \partial^\mu \rho(\tilde x)/\Lambda^3$. This corresponds to a Legendre transformation where the respective Hessian matrices should satisfy $|\eta_{\mu\nu}-\Pi_{\mu\nu}|=|\eta_{\mu\nu}+\rho_{\mu\nu}|^{-1}$. Hence, a crucial requirement is that the map is invertible. 
Starting with the original Galileon interactions $\mathcal{L}_{\rm Gal}^n$ with the coefficients $c_n$, we can map them to a set of Galileon interactions with different coefficients $p_n$, namely $\mathcal{L}_{\rm Gal}^{\rm dual}=\sum \frac{p_n}{\Lambda^{6(n-2)/2}}\mathcal{L}_n(\rho)$, where the relations between the coefficients are given by \cite{deRham:2013hsa}
\begin{equation}
p_n=\frac1n\sum_{k=2}^5(-1)^kc_k\frac{k(4-k+1)!}{(n-k)!(4-n+1)!}\,.
\end{equation}
For the specific coefficients $p_2=-1/12$, $p_3=-1/6$, $p_4=-1/8$ and $p_5=-1/30$, the duality maps a quintic Galileon to a free scalar field.
This on the other hand would strongly suggest, that the low energy classical superluminal propagation of the Galileon in a chosen field representation does not immediately mean acuasality. Even if group velocities could appear superluminal, the front velocities should be luminal and the same in both dual representations. This would also mean, that the S-matrix for this specific quintic Galileon is analytic. This duality was also applied to the generalised Galileon \cite{deRham:2014lqa}, which is Horndeski in flat space-time (we will discuss them in detail in section \ref{sec_scalar-tensor}). In specific classes of generalised Galileons the duality turns into a symmetry of the action. Following the coset construction of the Galileon theory, the dualities correspond to coordinate transformations on the coset space as a result of the spontaneously broken Galileon group \cite{Goon:2012dy,Kampf:2014rka}.

One immediate question arises concerning the fact that most non-tirival Hamiltonians of the Galileon is unbounded from below. Using the duality relation we might hope to shed light on the special class of Galileons dual to a free theory, which of course has manifestly bounded Hamiltonian. As an illustrative example, consider the Hamiltonian of a dual free scalar field in a $1+1$ dimensional toy model $H=\frac12\int d\tilde{x}(\dot{\rho}^2+(\rho^\prime)^2)$, which after applying the duality corresponds to a manifestly positive Hamiltonian $H=\frac12\int dx|1+\pi^{\prime\prime}|(\dot{\pi}^2+(\pi^\prime)^2)$. However, if we had started with the original formulation, this would have never been the corresponding Hamiltonian of the Galileon. The fundamental object $\Pi^\mu_\nu=\partial\tilde{x}^\mu/\partial x^\nu=\delta^\mu_\nu+\partial_\nu\partial^\mu\pi$ takes in this toy model the simple form $ \Pi^0_0=1-\ddot\pi$,  $ \Pi^0_1=\dot\pi^\prime$ and  $ \Pi^1_1=1+\pi^{\prime\prime}$, where we have omitted the scaling of $\Lambda$ for brevity. Now, some conditions need to be imposed on this fundamental object in order to guarantee the viability of the duality. First of all, we need to impose that $\det \Pi>0$ for the invertibility of the duality. Then, all the eigenvalues of $\Pi$ should be real (and even positive if we want to ensure that the Galileon metric has the same eigenvalues as the Minkowski metric). In order for the surface $\tilde{t}$ to be space-like eveywhere (or similarly $\Pi^0_\mu$ time-like) we need to have $\Pi^0_\mu\Pi^0_\nu\eta^{\mu\nu}<0$ and for the time-like unit vector to be forward directed we need $\Pi^0_0>0$. Furthermore, we need to enforce $\Pi^\mu_1\Pi^\nu_1\eta^{\mu\nu}>0$ ...etc. It could be that imposing all these conditions for the viability of the duality together with the equations of motion might make the obtained Hamiltonian bounded below. Could there exist Lorentz invariant conditions we can apply directly to the Galileon, so that the Hamiltonian of the Galileon would become bounded below and could the duality be a promising road to find these conditions? All these questions are worthwhile to explore and are still not-well understood mysteries.

\section{Massless Spin-2 Field}\label{sec:masslessSpin2field}

General Relativity is presented in textbooks of gravity as the geometric property of spacetime with the fundamental geometrical object being the curvature of spacetime, as formulated by Einstein and this interpretation has been incepted in our consciousness since then. 
After 100 years ingrained view it is an uncomfortable feeling to detach oneself from this interpretation. As we saw in section \ref{section_GeomGrav}, there exist two other legitimate alternative geometrical formulations of General Relativity, which are based on either torsion or non-metricity with zero curvature \cite{Hayashi:1979qx,BeltranJimenez:2017tkd}. Thus, one can geometrically interpret General Relativity as curvature, or torsion, or non-metricity. Curvature dictates how vectors change their directions along closed paths, torsion how a parallelogram does not close and non-metricity how the length of vectors change. So gravity can be attributed to these three completely exclusive views. Hence, the triangle of General Relativity shows that the geometrical interpretation of General Relativity is ambiguous. 

Let us carry out a gedankenexperiment and imagine that Einstein never existed. Admittedly, this would be a very sad thought. However, imagine a modern version of Einstein, who would have learned all the standard techniques of field theory description. This modern Einstein would have eventually constructed General Relativity from field theory perspective (even years later) based on the interactions of a massless spin-2 field. In this field theory perspective General Relativity emerges as a unique unambiguous theory of a spin-2 particle. Why should the gravitational force be carried by a spin-2 particle? First of all, we know that the gravitational force is long-range, which implies that the force carrier should be massless or with a very small mass (this possibility we will discuss in next section). We further know that gravity exists as a classical theory therefore the particle can be only bosonic with spin $s=0,1,2,\dots$. 

As we mentioned in section \ref{sec_fieldsincosmology} we do not know how to interpret classical fermionic fields in terms of their underlying quantum particles. The reason for why they cannot produce classical fermionic fields is that they cannot condensate in coherent states. Hence, the carrier of the gravitional force can not be a fermion. Among the bosonic candidates with spin $s=0,1,2,\dots$, it cannot be a spin-1 particle since we know that the gravitational force is purely attractive. There are theoretical challenges to construct theories based on higher spin $s>2$. Thus, only the spin-0 and the spin-2 particle remain as a possible candidate. A particle of spin-0 would quite naturally couple to the matter fields via $\pi T$, to wit to the trace of the stress energy momentum tensor. First and foremost it would not interact with light, which would evidently be in contraction with observations. Therefore, the spin-2 particle remains as the most promising survivor among the bosonic fields.

\subsection{Linear theory a la Fierz-Pauli and Weyl transverse diffeomorphism}\label{subsec:masslessSpin2Linear}
Following the introduction in section \ref{sec_fieldsincosmology}, a spin-2 particle is carried by a tensor field and in order to construct a manifestly Lorentz invariant theory, the interactions have to be scalar quantities built out of the tensor field and its derivatives. Locality and Lorentz invariance uniquely fixes the Lagrangian of the massless spin-2 particle to be given by General Relativity. We will see that manifest Lorentz invariance introduces diffeomorphism invariance.
Incorporating all the propagating physical degrees of freedom of the spin-2 field in $h_{\mu\nu}$, we write down the most general local and Lorentz invariant Lagrangian for a symmetric tensor as
\begin{eqnarray}\label{genAnsatzforh}
\mathcal{L}_{h}=\frac{\alpha_1}{2} \partial_\rho h_{\mu\nu}\partial^\rho h^{\mu\nu}-\alpha_2\partial_\mu h^\mu{}_\rho \partial_\nu h^{\nu\rho}+\alpha_3\partial_\mu h \partial_\nu h^{\mu\nu}-\frac{\alpha_4}{2} \partial_\mu h \partial^\mu h\,,
\end{eqnarray}
with arbitrary parameters $\alpha_{1,2,3,4}$. These are the most general Lorentz scalar quantities built out of the tensor field and one derivative per field. Since the spin-2 field is carried by a symmetric tensor field $h_{\mu\nu}$, it naively contains ten degrees of freedom. However, we know that a consistent theory for a massless spin-2 field should have only two propagating degrees of freedom. This means that we need additional constraint equations to eliminate the unphysical degrees of freedom. We can vary the action with respect to $h_{\mu\nu}$ in order to obtain the equations of motion. We will be aiming at rendering the $(0i)$ and $(00)$ components of the symmetric tensor non-dynamical. The equations of motion for the $(0i)$ components contain second time derivatives applied on $h_{0i}$ in form of 
\begin{equation}\label{EOMh0i_cond}
\mathcal{E}_{(0i)}=(-\alpha_1+\alpha_2)\ddot{h}_{0i}+\cdots
\end{equation}
Hence, in order to get rid of the dynamics for the $(0i)$ components we have to impose $\alpha_2=\alpha_1$. In this way the equations of motion for the $(0i)$ components generate three constraints associated to $h_{0i}$. Imposing this condition the above Lagrangian in equation (\ref{genAnsatzforh}) becomes invariant under transverse diffeomorphisms $h_{\mu\nu}\to h_{\mu\nu}+2\partial_{(\mu}\xi^T_{\nu)}$ with $\partial_{\mu}\xi^T_{\mu}=0$. As next, we shall construct one additional constraint that would render the $(00)$ component non-dynamical. Since the equations of motion for the $(00)$ and $(ii)$ components mix the second time derivatives applied on $h_{00}$ and $h_{ii}$, we have to first diagonalise the kinetic matrix. The obtained eigenvalue relevant for the dynamics of the $(00)$ component takes the form (with the canonically normalised field with $\alpha_1=1$)
\begin{eqnarray}\label{EOMh00_cond}
(-\alpha_3+2\alpha_4-\sqrt{1-2(\alpha_3+\alpha_4)+4(\alpha_3^2-\alpha_3\alpha_4+\alpha_4^2)}) \ddot{h}_{00}\,.
\end{eqnarray}
We impose the vanishing of this eigenvalue, that renders $h_{00}$ non-dynamical. This gives the relation
\begin{eqnarray}\label{condition1}
\alpha_4=\frac{1}{2}(1-2\alpha_3+3\alpha_3^2)\,.
\end{eqnarray}
The initial Lagrangian with the absence of the dynamics for the $(00)$ and $(0i)$ components becomes
\begin{eqnarray}\label{genAnsatzforhV2}
\mathcal{L}_{h}=\frac{1}{2} \partial_\rho h_{\mu\nu}\partial^\rho h^{\mu\nu}-\partial_\mu h^\mu{}_\rho \partial_\nu h^{\nu\rho}+\alpha_3\partial_\mu h \partial_\nu h^{\mu\nu}-\frac{1}{4}(1-2\alpha_3+3\alpha_3^2) \partial_\mu h \partial^\mu h.
\end{eqnarray}
with the remaining free parameter $\alpha_3$. This one parameter family of Lagrangians can be understood by considering via a non-derivative local field redefinition $h_{\mu\nu}\to h_{\mu\nu}+ \lambda h\eta_{\mu\nu}$. With this redefinition the original $\alpha_3$ and $\alpha_4$ parameters become $\tilde{\alpha}_3=\alpha_3-2\lambda(1-2\alpha_3)$ and $\tilde{\alpha}_4=\alpha_4-4(1+\alpha_3-4\alpha_4)\lambda+(1+4\alpha_3-8\alpha_4)\lambda^2$. For $\alpha_3=1=\alpha_4$, the new parameters are given by $\tilde{\alpha}_3=1+2\lambda$ and $\tilde{\alpha}_4=(1+2\lambda)(1+6\lambda)$. For $\lambda=\frac{\alpha_3-\tilde{\alpha}_3}{2(1-2\alpha_3)}$ the new parameters $\tilde{\alpha}_3$ and $\tilde{\alpha}_4$ still satisfy the relation in equation (\ref{condition1}). By means of this local field redefinition we can freely choose the value for $\alpha_3$. With the convenient choice $\alpha_3=1$, the Lagrangian in equation (\ref{genAnsatzforhV2}) then becomes
\begin{eqnarray}\label{LagFPm0}
\mathcal{L}_{h}=\frac{1}{2} \partial_\rho h_{\mu\nu}\partial^\rho h^{\mu\nu}-\partial_\mu h^\mu{}_\rho \partial_\nu h^{\nu\rho}+\partial_\mu h \partial_\nu h^{\mu\nu}-\frac{1}{2} \partial_\mu h \partial^\mu h.
\end{eqnarray}
Summarising, the conditions $\alpha_2=\alpha_1$ and $\alpha_4=\alpha_3$ with the canonical normalisation $\alpha_1=1$ and the choice $\alpha_3=1$ corresponds to the Fierz-Pauli case and has the $(00)$ and $(0i)$ components non-dynamical. This was already achieved for the one parameter family in equation (\ref{genAnsatzforhV2}). 

A special case is $\tilde{\alpha}_3=1/2$, since it can not be reached from $\alpha_3=1$ by means of the above mentioned local field redefinition with $\lambda\ne-1/4$. For this particular choice the redefinition becomes singular. Apart from this case, these one parameter family Lagrangians are equivalent to Fierz-Pauli. Another special case is $\lambda=-1/4$, when the local field redefinition is not invertible anymore for $\alpha_3=1/2$. In this special case the Lagrangian has an additional Weyl symmetry and is independent of the trace. In other words the equations of motion of the theory with Weyl transverse diffeomorphisms correspond to just the traceless part of the equations of motion of the Fierz-Pauli theory. Using the linearized Bianchi identity one observes that the trace of the latter ones can be recovered from the previous ones up to an integration constant, which stands for the cosmological constant. While the linear theory of Fierz Pauli satisfies the equations of motion $\Box h_{\mu\nu}=0$ with the spin-2 field having a vanishing trace $h=0$ and being transverse $\partial h_{\mu\nu}=0$ realising the linearised diffeomorphism $h_{\mu\nu}\to h_{\mu\nu}+\partial_\mu \xi_\nu+\partial_\nu \xi_\mu$, the linear theory of Weyl transverse diffeomorphism satisfies the equations of motion $\partial^\mu\partial^\nu h_{\mu\nu}=c$ (with constant $c$) with the field now being invariant under only transverse diffeomorphism $h_{\mu\nu}\to h_{\mu\nu}+2\partial_{(\mu}\xi^T_{\nu)}$ (with $\partial_{\mu}\xi_{\mu}=0$) but having the Weyl symmetry $h_{\mu\nu}\to h_{\mu\nu}+\phi \eta_{\mu\nu}$ and being also traceless $h=0$. Integrating the equations of motion of the latter gives $\partial^\nu h_{\mu\nu}=\frac14 \Lambda x_\mu+ t_\mu^T$ with an arbitrary transverse vector $t_\mu^T$, which can be get rid of by means of the transverse diffeomorphism. See \cite{Alvarez:2006uu} for useful discussions on this.

Thus, at the linear order, one can construct a massless spin-2 theory in form of
\begin{itemize}
\item Fierz-Pauli with diffeomorphism invariance $h_{\mu\nu}\to h_{\mu\nu}+\partial_\mu \xi_\nu+\partial_\nu \xi_\mu$ and equations of motion $\Box h_{\mu\nu}=0$ 
\item Weyl transverse diffeomorphism invariant theory with the equations of motion $\Box \bar{h}_{\mu\nu}=-\Lambda\eta_{\mu\nu}$ with $ \bar{h}_{\mu\nu}=h_{\mu\nu}-\frac{c}{8}x^2 \eta_{\mu\nu}$
\end{itemize}
Both theories contain the transverse diffeomorphism invariance, which can be either enlarged to the full diffeomorphism as in Fierz Pauli or to the linear theory with an additional Weyl symmetry.
As one can see, the difference between these two linear theories boils down to an integration constant $c$ that acts as a constant source.
The first one gives rise to General Relativity at the non-linear order and the latter one results in Unimodular gravity \cite{Unruh:1988in,Shaposhnikov:2008xb} (see also \cite{Blas:2011ac} for scale-invariant alternatives). It is equivalent to General Relativity and the cosmological constant appears as an integration constant. We will only focus on the linear theory \`a la Fierz Pauli. The linearized version of General Relativity given by the Fierz-Pauli action (\ref{LagFPm0}) can also be written compactly as
\begin{equation}\label{LagFPcompact}
\mathcal{L}_h=-h^{\mu\nu}\hat{\mathcal{E}}_{\mu\nu}^{\alpha\beta}h_{\alpha\beta}\,,
\end{equation}
with the Lichnerowicz operator $\hat{\mathcal{E}}$ standing for
\begin{equation}
\hat{\mathcal{E}}^{\alpha\beta}_{\mu\nu} h_{\alpha\beta}=-\frac 12 \left(\Box h_{\mu\nu}-2\partial_\alpha \partial_{(\mu}h^\alpha_{\nu)}+\partial_\mu\partial_\nu h-\eta_{\mu\nu} (\Box h-\partial_\alpha\partial_\beta h^{\alpha\beta})\right)\,.
\label{eq:epsilondef}
\end{equation}
Note, that the Lichnerowicz operator satisfies the identity $\partial^\mu\hat{\mathcal{E}}^{\alpha\beta}_{\mu\nu} h_{\alpha\beta}=0$, which will become important in the next subsection.

\subsection{Matter coupling}\label{subsec:masslessSpin2Matter}
One prompt question is how we can couple the matter fields to this linear theory of massless spin-2 field. The matter field could be for instance a scalar field with the stress energy tensor given by equation (\ref{Tuv_freeScalar}) or any other form of matter fields with their corresponding stress energy tensors. One immediate naive guess would be to couple the massless spin-2 field to the stress energy tensor of the matter fields
\begin{equation}
\mathcal{L}_h=-h^{\mu\nu}\hat{\mathcal{E}}_{\mu\nu}^{\alpha\beta}h_{\alpha\beta}
+\frac{1}{M_{\rm Pl}}h_{\mu\nu} T^{\mu\nu}\,,
\end{equation}
where the mass scale $M_{\rm Pl}$ enters in the coupling since the stress energy tensor has already dimension four and $M_{\rm Pl}$ should carry the same dimension of $h_{\mu\nu}$. We have seen in the previous subsection that for the right number of propagating degrees of freedom in the linear theory, it was crucial to have the diffeomorphism invariance $h_{\mu\nu}\to h_{\mu\nu}+\partial_\mu \xi_\nu+\partial_\nu \xi_\mu$. The coupling to the matter field should still satisfy this requirement in order not to spoil the right propagating degrees of freedom. This would mean that our naive Ansatz for the matter coupling $\frac{1}{M_{\rm Pl}} \partial_{\mu} \xi_{\nu}T^{\mu\nu}=-\frac{1}{M_{\rm Pl}}\xi_{\nu} \partial_{\mu}T^{\mu\nu}$ should satisfy $\partial_{\mu}T^{\mu\nu}=0$. Therefore the stress energy tensor would be the most natural candidate to achieve this. Since $T^{\mu\nu}$ is the Noether current of time translations, strictly speaking we should couple it to the gauge field of time translations which is the vielbein. Performing the analysis then in the vielbein formulation would yield the teleparallel formulation of gravity but we should not enter into this here. We definitely know that the Ansatz $h_{\mu\nu} T^{\mu\nu}$ is the simplest and most promising candidate we can try out. We already know the stress energy tensor of the free action, therefore let us denote it by $T^{(0)}_{\mu\nu}$ (and the free action by $\mathcal{S}^{(0)}_{\rm matter}$). The equations of motion of the linear action with this matter coupling become 
\begin{equation}\label{lineom_masslessSpin2}
\hat{\mathcal{E}}_{\mu\nu}^{\alpha\beta}h_{\alpha\beta}=\frac{1}{2M_{\rm Pl}} T^{\mu\nu}_{(0)} \,.
\end{equation} 
The stress energy tensor can be computed by either performing the variational principle with respect to an auxiliary field $\mathcal{K}_{\mu\nu}$, that replaces $\eta_{\mu\nu}$ in $\mathcal{S}^{(0)}_{\rm matter}$ (with the advantage of maintaining a possible gauge invariance of the matter fields) $T^{(0)}_{\mu\nu}=\frac{-2}{\sqrt{-\mathcal{K}}}\frac{\delta \mathcal{S}^{(0)}_{\rm matter}}{\delta \mathcal{K}^{\mu\nu}}\Big|_{\mathcal{K}=\eta}$ or via the canonical energy momentum tensor. In either way an inconsistency appears at the level of the equations of motion. A crucial reminder at this stage is in demand, namely the identity relation that the Lichnerowicz operator satisfies $\partial^\mu\hat{\mathcal{E}}^{\alpha\beta}_{\mu\nu} h_{\alpha\beta}=0$. This is nothing else but the linearized Bianchi identity. Lset us take the divergence of the equations of motion in (\ref{lineom_masslessSpin2})
\begin{equation}
0=\partial^\mu\hat{\mathcal{E}}_{\mu\nu}^{\alpha\beta}h_{\alpha\beta}=\frac{1}{2M_{\rm Pl}} \partial^\mu T_{\mu\nu}^{(0)}\ne0\,.
\end{equation}
The left hand side vanishes identically, but not the right hand side since $T_{\mu\nu}^{(0)}$ was computed from $\mathcal{S}^{(0)}_{\rm matter}$ without the new interaction term $h_{\mu\nu} T^{\mu\nu}_{(0)}$. The stress energy tensor $T_{\mu\nu}^{(0)}$ is conserved on-shell, with respect to the equations of motion arising from $\mathcal{S}^{(0)}_{\rm matter}$ and not from $\mathcal{S}^{(0)}_{\rm matter}+h_{\mu\nu} T^{\mu\nu}_{(0)}$. Thus, we need to compute the next order energy momentum tensor including this new interaction
\begin{equation}
\mathcal{L}_h=-h^{\mu\nu}\hat{\mathcal{E}}_{\mu\nu}^{\alpha\beta}h_{\alpha\beta}+\frac{1}{M_{\rm Pl}}h_{\mu\nu} T^{\mu\nu}_{(0)}+\frac{1}{2M_{\rm Pl}^2}h_{\mu\nu} h_{\alpha\beta}T^{\mu\nu\alpha\beta}_{(1)}\,,
\end{equation}
with the next leading operator $T_{\mu\nu\alpha\beta}^{(1)}=-\frac{2}{\sqrt{-\mathcal{K}}}\frac{\delta \sqrt{-\mathcal{K}}T^{(0)}_{\mu\nu}}{\delta \mathcal{K}^{\alpha\beta}}\Big|_{\mathcal{K}=\eta}$. Again, at the level of the equations of motion this gives the same inconsistency as in the previous order and one is forced to add the next leading operator etc. One immediately faces the fact that the series never stops and one has to some over an infinite number of interactions
\begin{equation}
\mathcal{S}^{\rm matter}=\sum_{n=0}^\infty\frac{1}{n!}\left( \frac{-2}{M_{\rm Pl}}\right)^n\int d^4x h_{\mu_1\nu_1}\cdots h_{\mu_n\nu_n} \frac{\delta^{(n)}\mathcal{S}_{(0)}^{\rm matter}}{\delta \mathcal{K}_{\mu_1\nu_1}\cdots  \mathcal{K}_{\mu_n\nu_n}}\Big|_{\mathcal{K}=\eta}\,.
\end{equation}
On closer inspection one quickly realises that this is nothing else but the Taylor expansion of the original matter field action, where $\eta_{\mu\nu}$ is replaced by $\eta_{\mu\nu}-\frac{2}{M_{\rm Pl}}h_{\mu\nu}$
\begin{equation}
\mathcal{S}_{(0)}^{\rm matter}[\eta_{\mu\nu}-\frac{2}{M_{\rm Pl}}h_{\mu\nu}], \qquad \text{where} \qquad \eta_{\mu\nu}\to\eta_{\mu\nu}-\frac{2}{M_{\rm Pl}}h_{\mu\nu} \,.
\end{equation}
At this stage, it is worthwhile to comment on the following: the stress energy tensor obtained by means of the variational principle and the canonical energy momentum tensor typically differ up to a divergence term $\partial_\mu \mathcal{D}^{[\mu\alpha]\beta}$. Adding such boundary terms in the above resummation procedure will yield non-minimal couplings to matter fields. We will discuss non-minimal couplings including scalars, vectors and tensors in later sections. Another useful comment concerns the arising of the infinite series of the matter couplings due to $\partial^\mu T_{\mu\nu}^{(0)}\ne0$. This could have been achieved as an identity if one was willing to introduce non-local operators. One could for instance introduce a divergenceless projector operator $P_{\mu\nu}=\eta_{\mu\nu}-\frac{\partial_\mu\partial_\nu}{\Box}$ and couple $h_{\mu\nu}$ with the matter coupling via this operator, i.e. $h_{\alpha\beta}\left(P^{\alpha\mu}P^{\beta\nu}+\cdots\right)T_{\mu\nu}$. In this way the linearised diffeomorphism could have been realised by identity relations at the prize of non-local operators. Throughout this review we will insist on local operators when constructing our consistent field theories. 

Now, if one works in the weak field limit with the first dominant contribution being $\frac{1}{M_{\rm Pl}}h_{\mu\nu} T^{\mu\nu}_{(0)}$, then 
the field equations at leading order in perturbation, as we saw, satisfy simply (ignoring for a moment the above mentioned inconsistency of the conservation of the equations of motion)
\begin{eqnarray}
\hat{\mathcal{E}}_{\mu\nu}^{\;\;\;\alpha\beta}h_{\alpha\beta}=\frac{1}{M_{\rm Pl}}T^{(0)}_{\mu\nu}.
\end{eqnarray}
These equations have the invariance under the linearized gauge symmetry transformations $h_{\mu\nu}\to h_{\mu\nu}+\partial_\mu \xi_\nu+\partial_\nu \xi_\mu$, hence we can choose a gauge. For instance, very often it is convenient to choose the Lorenz gauge $\partial^\mu \bar h_{\mu\nu}=0$. In this case, the equations of motion simply become
\begin{eqnarray}
-\frac12\Box\bar h_{\mu\nu}=\frac{1}{2M_{\rm Pl}}T^{(0)}_{\mu\nu}\,,
\end{eqnarray}
with $\bar h_{\mu\nu}= h_{\mu\nu}-\frac12\eta_{\mu\nu}h$. This choice of the gauge together with the remaining residual gauge symmetry $\Box\xi_\alpha=0$ eliminates eight out of ten degrees of freedom, as we saw above. Brimming over with enthusiasm for having found the linear ghost free theory for a spin-2 field one could take a step forward and compare this linear theory with solar system observations like for instance the bending of light. Then, unfortunately, one obtains for the bending angle $\theta_{\rm theory}=0.75\theta_{\rm obs}$. This means that the linear theory does not account correctly for the right observations and can not be the true theory behind gravity based on observational grounds. This signals the observational inconsistency of the linear theory.

\subsection{Non-linear interactions}\label{subsec:masslessSpin2Non-Linear}
In the previous subsections we have seen how one can successfully construct the linear theory of a massless spin-2 field with linearised diffeomorphism. Imposing this symmetry also on the matter field couplings resulted in an infinite series that had to be resummed. Nevertheless, the resulting leading order linear theory was in disagreement with observations. The reason for that was that there was still an important piece missing. Namely, the self coupling of the massless spin-2 field itself. In order words, we need to perform a non-linear completion of the interactions by adding the self energy of the massless spin-2 field to the stress energy tensor. We could again start with the linear theory as our zeroth order interaction $\mathcal{L}_h^{(0)}=-h^{\mu\nu}\hat{\mathcal{E}}_{\mu\nu}^{\alpha\beta}h_{\alpha\beta}$ and compute the associated self energy in terms of the stress energy tensor $T^{\mu\nu}_{(0)}(\mathcal{L}_h^{(0)})$ and add this contribution as a self coupling $\mathcal{L}_h^{(1)}=\mathcal{L}_h^{(0)}+\frac{1}{M_{\rm Pl}}h_{\mu\nu} T^{\mu\nu}_{(0)}(\mathcal{L}_h^{(0)})$ to the linear theory. This would again give rise to an inconsistency of the equations of motion since $T^{\mu\nu}_{(0)}$ is conserved on-shell with respect to $\mathcal{L}_h^{(0)}$ but not to $\mathcal{L}_h^{(1)}$. Thus, one would need to compute the next order stress energy tensor, and so on. This again gives rise to an infinite series in the same way as the matter coupling. One would need to resum an infinite series of self interactions in order to maintain the gauge symmetry at the non-linear level. There is a clever way of overcoming the resummation of this infinite series by introducing auxiliary fields. This is Deser's approach of the bootstrapping procedure \cite{Deser:1969wk}, that we shall quickly review. Starting with the lowest order Lagrangian $\mathcal{L}_h^{(0)}$ of equation (\ref{LagFPcompact}), we can rewrite it as
\begin{equation}
\mathcal{L}_h^{(0)}=-\bar{h}^{\mu\nu}\left( \partial_\mu \mathcal{G}^\rho_{\nu\rho}- \partial_\rho \mathcal{G}^\rho_{\mu\nu} \right)+\eta^{\mu\nu}\left(  \mathcal{G}^\sigma_{\mu\rho}\mathcal{G}^\rho_{\nu\sigma}- \mathcal{G}^\sigma_{\mu\nu}  \mathcal{G}^\rho_{\sigma\rho}\right)\,,
\end{equation}
where $\mathcal{G}^\rho_{\mu\nu} $ is an auxiliary field and $\bar{h}_{\mu\nu}=h_{\mu\nu}-\frac12h\eta_{\mu\nu}$. The Lagrangian (\ref{LagFPcompact}) rewritten in this way is now invariant under the transformations
\begin{equation}
\delta \bar{h}_{\mu\nu}=2\partial_{(\mu}\xi_{\nu)} \qquad \text{and} \qquad \delta \mathcal{G}^\rho_{\mu\nu}=\partial_{\mu}\partial_{\nu}\xi^\rho\,.
\end{equation}
The crucial simplification in writing the original Lagrangian in terms of the auxiliary field makes itself manifest by computing the associated self energy in terms of the stress energy tensor $T^{\mu\nu}_{(0)}$
\begin{equation}
T_{\mu\nu}^{(0)}=-\left(  \mathcal{G}^\sigma_{\mu\rho}\mathcal{G}^\rho_{\nu\sigma}- \mathcal{G}^\sigma_{\mu\nu}  \mathcal{G}^\rho_{\sigma\rho}\right)\,,
\end{equation}
which one can add to the zeroth order Lagrangian as self coupling
\begin{eqnarray}
\mathcal{L}_h^{(1)}&=&\mathcal{L}_h^{(0)}+\bar{h}_{\mu\nu} T^{\mu\nu}_{(0)}(\mathcal{L}_h^{(0)}) \nonumber\\
&=& -\bar{h}^{\mu\nu}\left( \partial_\mu \mathcal{G}^\rho_{\nu\rho}- \partial_\rho \mathcal{G}^\rho_{\mu\nu} \right)+\left(\eta^{\mu\nu}-\bar{h}_{\mu\nu} \right)\left(  \mathcal{G}^\sigma_{\mu\rho}\mathcal{G}^\rho_{\nu\sigma}- \mathcal{G}^\sigma_{\mu\nu}  \mathcal{G}^\rho_{\sigma\rho}\right)\,.
\end{eqnarray}
One can also add an appropriate boundary term to this Lagrangian 
\begin{equation}
\mathcal{L}_{\rm BT}=\eta^{\mu\nu}\left( \partial_\mu \mathcal{G}^\rho_{\nu\rho}- \partial_\rho \mathcal{G}^\rho_{\mu\nu} \right)\,,
\end{equation}
such that the Lagrangian now becomes
\begin{eqnarray}
\mathcal{L}_h^{(1)}&=&\mathcal{L}_h^{(0)}+\bar{h}_{\mu\nu} T^{\mu\nu}_{(0)}(\mathcal{L}_h^{(0)})+\mathcal{L}_{\rm BT} \nonumber\\
&=& \left(\eta^{\mu\nu}-\bar{h}^{\mu\nu}\right)\left( \partial_\mu \mathcal{G}^\rho_{\nu\rho}- \partial_\rho \mathcal{G}^\rho_{\mu\nu} \right)+\left(\eta^{\mu\nu}-\bar{h}_{\mu\nu} \right)\left(  \mathcal{G}^\sigma_{\mu\rho}\mathcal{G}^\rho_{\nu\sigma}- \mathcal{G}^\sigma_{\mu\nu}  \mathcal{G}^\rho_{\sigma\rho}\right)\,.
\end{eqnarray}
From this expression one already sees where one is heading to. By defining the variable $\hat{g}^{\mu\nu}=\eta^{\mu\nu}-\bar{h}^{\mu\nu}$ together with the change of variables $g_{\mu\nu}=\sqrt{-\hat{g}}\hat{g}_{\mu\nu}$ and recognising the field $\mathcal{G}^\rho_{\mu\nu} $ as being an independent connection \`a la Palatini, one obtains the Einstein Hilbert action already after one iteration
\begin{equation}
\mathcal{L}_h^{(1)}=\sqrt{-g}g^{\mu\nu}R_{\mu\nu}(\mathcal{G})\,.
\end{equation}
The power using the auxiliary field and the Palatini formalism is in the fact that the infinite series of self interactions are packaged together after only one iteration. The resulting self energy in the stress energy tensor satisfies the full diffeomorphism $g_{\mu\nu}\to g_{\mu\nu}+\nabla_{(\mu}\xi_{\nu)}$ and the consistency of the field equations is restored. This was the clever point that Deser realised and as we saw, it can indeed render the bootstrapping procedure much easier. See also \cite{Padmanabhan:2004xk,Butcher:2009ta,Deser:2009fq,Barcelo:2014mua} for related discussions on this.

\subsection{Lovelock invariants}\label{subsec_LovelockInvariant}

The Lagrangian that we constructed above for the massless spin-2 field has the property of being second order in derivatives at the level of the equations of motion and propagates only two physical degrees of freedom. From the initially ten degrees of freedom we obtained four constraints by making the $g_{00}$ and $g_{0i}$ components non-dynamical. Together with the diffeomorphism invariance we removed $2\times4$ degrees of freedom. In order to write down the consistent covariant and non-linear Lagrangian for $g_{\mu\nu}$ interactions with two propagating degrees of freedom, we made use of the bootstrapping procedure to resum the self interactions and the resulting theory satisfies the requirement that the equations of motion are second order in derivatives and that the massless spin-2 field couples to all type of matter and energy in the same manner as a consequence of the equivalence principle. It is well known that the Lovelock invariants are the only manifestly covariant terms that satisfy the condition of second order equations of motion and full diffeomorphisms. In four dimensions the only Lovelock invariants are the cosmological constant $\lambda$ and the Ricci scalar $R$. There is also the Gauss Bonnet Lovelock invariant but in four dimensions it corresponds to a total derivative. In terms of the antisymmetric Levi-Civita tensors, we can write the Lovelock invariants systematically as
\begin{eqnarray}
\lambda&\sim&\mathcal{E}^{\mu\nu\alpha\beta} \mathcal{E}_{\mu\nu\alpha\beta}\nonumber\\
R&\sim& \mathcal{E}^{\mu\nu\alpha\beta} \mathcal{E}^{\rho\sigma}_{\;\;\;\;\;\;\;\;\alpha\beta}\partial_\mu\partial_{\rho} h_{\nu\sigma}\,.
\end{eqnarray}
In tight relation to the Lovelock invariants one can construct divergenceless tensors. For instance, the metric $g_{\mu\nu}$ itself is the divergeless tensor associated to the cosmological constant. And similarly, the Einstein tensor $G_{\mu\nu}$ is the divergenceless tensor associated to the Ricci scalar. They can also be written in terms of the Levi-Civita tensors systematically as
\begin{eqnarray}
g_{\mu\nu}&\sim&\mathcal{E}_{\mu}^{\;\;\alpha\beta\gamma} \mathcal{E}_{\nu \alpha\beta\gamma}\nonumber\\
G_{\mu\nu}&\sim& \mathcal{E}_{\mu}^{\;\;\alpha\beta\gamma} \mathcal{E}_{\nu\;\;\;\;\;\;\; \gamma}^{\;\;\rho\sigma}\partial_\alpha\partial_{\rho} h_{\beta \sigma}\,.
\end{eqnarray}
 The divergenceless tensor corresponding to the Gauss Bonnet Lovelock invariant vanishes identically since the Gauss Bonnet term is just a total derivative in four dimensions. The divergenceless tensors are nothing else but the quantities obtained from the equations of motion of the Lovelock invariants. Actually, besides them there is yet another divergeceless tensor in four dimensions and it corresponds to the double dual Riemann tensor $L_{\mu\alpha\nu\beta}$, which in terms of the Levi-Civita tensors can be written as 
\begin{eqnarray}
L_{\mu\alpha\nu\beta}&\sim& \mathcal{E}_{\mu\alpha}^{\;\;\;\;\delta\kappa} \mathcal{E}_{\nu\beta\;\;\;\;\;\;\; }^{\;\;\;\;\rho\sigma}\partial_\delta\partial_{\rho} h_{\kappa \sigma}\,.
\end{eqnarray}
All these non-trivial Lovelock invariants and the corresponding divergenceless tensors are linear in the curvature (linear in $h$). If we were in five dimensions, we could have constructed yet another additional divergenceless tensor with six indices $\bar{L}_{\mu\alpha\nu\beta\rho\sigma}$. The equivalent Lovelock invariants and their divergenceless quantities in five dimensions would be constructed this time in terms of the five dimensional antisymmetric Levi-Civita tensors
\begin{eqnarray}\label{Lovelock5dim}
\lambda&\sim&\mathcal{E}^{abcde} \mathcal{E}_{abcde}\nonumber\\
g_{\mu\nu}&\sim&\mathcal{E}_{\mu}^{\;\;abcd} \mathcal{E}_{\nu abcd}\nonumber\\
R&\sim& \mathcal{E}^{abcde} \mathcal{E}^{a'b'}_{\;\;\;\;\;\;\;\;cde}\partial_a\partial_{a'} h_{bb'}\nonumber\\
G_{\mu\nu}&\sim& \mathcal{E}_{\mu}^{\;\;abcd} \mathcal{E}_{\nu\;\;\;\;\;\;\; cd}^{\;\;a'b'}\partial_a\partial_{a'} h_{bb'}\nonumber\\
L_{\mu\alpha\nu\beta}&\sim& \mathcal{E}_{\mu\nu}^{\;\;\;\;abc} \mathcal{E}_{\alpha\beta \;\;\;\;\;\;\;\; c}^{\;\;\;\;\;\;\;a'b'}\partial_a\partial_{a'} h_{bb'}\nonumber\\
\bar{L}_{\mu\alpha\nu\beta\rho\sigma}&\sim& \mathcal{E}_{\mu\nu\rho}^{\;\;\;\;\;\;\;ab} \mathcal{E}_{\alpha\beta\sigma}^{\;\;\;\;\;\;\;a'b'}\partial_a\partial_{a'} h_{bb'}\,,
\end{eqnarray}
which are all again linear in the curvature.
As we mentioned above, the Gauss Bonnet Lovelock invariant is a total derivative in four dimensions, but this is not the case anymore in five dimensions. It is a quantity that is second order in the curvature and in terms of the field $h_{\mu\nu}$ can be written as
\begin{eqnarray}
\mathcal L_{GB}= \mathcal{E}^{abcde} \mathcal{E}^{\;a'b'c'd'}_{\;\;\;\;\;\;\;\;\;\;\;\;\;\;\;\;e}\partial_a\partial_{a'} h_{bb'}\partial_c\partial_{c'} h_{dd'}\,.
\end{eqnarray}
Since it is not a total derivative in five dimensions anymore, there is also the divergenceless tensor associated to the Gauss Bonnet Lovelock invariant
\begin{eqnarray}
\mathcal{L}^{GB}_{\mu\nu}= \mathcal{E}_\mu^{\;\;abcde} \mathcal{E}^{\;a'b'c'd'}_{\nu}\partial_a\partial_{a'} h_{bb'}\partial_c\partial_{c'} h_{dd'}\,.
\end{eqnarray}
The antisymmetric structure of all these interactions in terms of the Levi-Civita tensors guarantees that the equations of motion are at most second order in derivatives. Thus, 
based on the argument of Lovelock invariants, if we want to have a consistent theory for a spin-2 field with second order equations of motion, then this uniquely leads to General Relativity as well 
\begin{eqnarray}\label{action_GR}
\mathcal{S}_{\rm EH}=\int \mathrm{d} ^4 x \left(\mathcal{L}_{\rm GR} +\mathcal{L}_{\rm matter}\right)
=\frac{M_{\rm Pl}^2}{2} \int \mathrm{d} ^4 x\sqrt{-g} R +\int \mathrm{d} ^4 x \mathcal{L}_{\rm m}\,,
\end{eqnarray}
with $\mathcal{L}_{\rm matter}$ being the matter field Lagrangian. As we saw above, General Relativity is invariant under full general coordinate transformations $g_{\mu\nu}\to g_{\mu\nu}+\nabla_{(\mu}\xi_{\nu)}$. In the linearized limit it translates to the invariance under 
\begin{equation}\label{lin_gauge_symm}
h_{\mu\nu}\to h_{\mu\nu}+\partial_\mu \xi_\nu+\partial_\nu \xi_\mu.
\end{equation}
We can expand the metric and its inverse in terms of perturbations around flat space-time
\begin{equation}
g_{\mu\nu}=\eta_{\mu\nu}+\frac{2}{M_{\rm Pl}}h_{\mu\nu}\, \;\;\;\; {\rm and} \;\;\;\; g^{\mu\nu}=\eta^{\mu\nu}-\frac2{M_p}h^{\mu\nu}+\frac4{M^{2}_p}h^{\mu\alpha}h_\alpha^\nu+\cdots. 
\end{equation}
Then the Lagrangian in (\ref{action_GR}) to second order in $h$ becomes
\begin{equation}
\mathcal{L}_h=-h^{\mu\nu}\hat{\mathcal{E}}_{\mu\nu}^{\alpha\beta}h_{\alpha\beta}+\frac{1}{M_{\rm Pl}}h_{\mu\nu} T^{\mu\nu}+\frac{1}{2M_{\rm Pl}^2}h_{\mu\nu} h_{\alpha\beta}T^{\mu\nu\alpha\beta}+\cdots\,,
\end{equation}
with the Lichnerowicz operator $\hat{\mathcal{E}}$ as defined in (\ref{eq:epsilondef})
and the stress-energy tensor $T_{\mu\nu}$ and its derivative with respect to the metric $T^{\mu\nu\alpha\beta}$ given as
\begin{eqnarray}
T_{\mu\nu}=\frac{-2}{\sqrt{-g}}\frac{\delta \sqrt{-g}\mathcal L_m}{\delta g^{\mu\nu}} \;\;\;\;\;\;\; {\rm and}\;\;\;\;\;\;
T_{\mu\nu\alpha\beta}=-\frac{2}{\sqrt{-g}}\frac{\delta \sqrt{-g}T_{\mu\nu}}{\delta g^{\alpha\beta}}\,.
\end{eqnarray}
Thus, we obtain the same linear theory as we constructed above but this time starting from the covariant Lovelock invariants and the requirement of second order equations of motion and full diffeomorphism.

The propagator of a massless spin-2 particle is given by
\begin{equation}\label{GRpropagator}
D_{\mu\nu\alpha\beta}=\frac{1}{\Box}\left( g_{\alpha(\mu}g_{\beta\nu)}-\frac12\frac{2}{d-2}g_{\mu\nu}g_{\alpha\beta} \right)  \,.
\end{equation}
Between two conserved sources the exchange amplitude of a massless spin-2 particle in four dimensions results in
\begin{equation}
W=-\frac{1}{2}\int \frac{d^4k}{(2\pi^4)}T^{*\mu\nu}\frac{1}{k^2}\left( 2g_{\alpha(\mu}g_{\beta\nu)}-g_{\mu\nu}g_{\alpha\beta} \right)T^{\alpha\beta}  \,.
\end{equation}
It means that for two mass densities it simply reads
\begin{equation}\label{exhangeAmp_masslessSpin2}
W=-\frac{1}{2}\int \frac{d^4k}{(2\pi^4)}T^{*00}\frac{1}{k^2}T^{00}  \,.
\end{equation}
When we compare this expression with what we had for the exchange amplitude of a scalar field in equation (\ref{exchange_ampl_scalar}), we see that the amplitudes carry the same sign. Hence, a massless spin-2 field as an exchange particle would also give rise to an attractive force between two sources as well. Based on the attractive nature of the gravitational forces, gravity could be attributed to both scalar and tensor field, but additional observations, for instance the bending of light, exclude the pure scalar candidate.

\subsection{Gibbons-Hawking-York boundary term}\label{subsection_GHY}
We have seen in different ways, that the consistent non-linear, covariant and local theory of a massless spin-2 field is uniquely given by the Einstein-Hilbert action (\ref{action_GR}).
The Einstein equations obtained by varying the Einstein-Hilbert action with respect to the metric gives Einstein's famous field equations $G_{\mu\nu}=T_{\mu\nu}/M_{\rm Pl}^2$. One important property that is worth mentioning here is the fact, that the stress energy tensor as a source term has dimension four, which therefore requires the presence of a coupling constant of dimension -2 in form of $M_{\rm Pl}^{-2}$. This scale dictates up to which scale we can use General Relativity as an effective field theory. Even though the form of the field equations is quite well known, some readers might not be fully aware of the necessity of a boundary term for the well-posedness of the variational principle and the associated ambiguity of the boundary term. Therefore, we shall quickly review the variation of the Einstein-Hilbert action here. For the variation we will need the relations
\begin{eqnarray}
\delta \sqrt{-g} &=& -\frac12 \sqrt{-g} g_{\alpha\beta}\delta g^{\alpha\beta} \\
\delta R_{\alpha\beta} &=&\nabla_\mu (\delta \Gamma^\mu_{\alpha\beta})-\nabla_\beta(\delta\Gamma^\mu_{\alpha\mu})\,.
\end{eqnarray}
With this at hand, we can easily perform the variation of the Einstein-Hilbert action as
\begin{eqnarray}\label{variationEH_GHY}
\delta\mathcal{S}_{\rm EH}&=&\int_{\mathcal{M}}d^4x \delta\left( \sqrt{-g}g^{\alpha\beta}R_{\alpha\beta} \right)\nonumber\\
&=&\int_{\mathcal{M}}d^4x \left(  \sqrt{-g}R_{\alpha\beta}\delta g^{\alpha\beta}+g^{\alpha\beta}R_{\alpha\beta}\delta \sqrt{-g}+\sqrt{-g} g^{\alpha\beta}\delta R_{\alpha\beta} \right)\nonumber\\
&=&\int_{\mathcal{M}}d^4x \sqrt{-g}\left( R_{\alpha\beta}-\frac12 g_{\alpha\beta}R\right)\delta g^{\alpha\beta}+\int_{\mathcal{M}}d^4x \sqrt{-g}g^{\alpha\beta}\delta R_{\alpha\beta}\nonumber\\
&=&\int_{\mathcal{M}}d^4x \sqrt{-g}G_{\alpha\beta}\delta g^{\alpha\beta}+\int_{\mathcal{M}}d^4x \sqrt{-g}g^{\alpha\beta}\delta R_{\alpha\beta}\,.
\end{eqnarray}
In the last expression we recognise how the Einstein tensor arises $G_{\alpha\beta}=R_{\alpha\beta}-\frac12 g_{\alpha\beta}R$ in the first part of the variation. However, as we can see there is a second term in the last line of (\ref{variationEH_GHY}), that we can not neglect and will give a significant contribution. At a closer inspection, it will contribute in the following form
\begin{equation}\label{GHY_integral}
\int_{\mathcal{M}}d^4x \sqrt{-g}g^{\alpha\beta}\delta R_{\alpha\beta}=\int_{\mathcal{M}}d^4x \sqrt{-g}g^{\alpha\beta} \nabla_\mu \left( g^{\alpha\beta}\delta\Gamma^\mu_{\alpha\beta}-g^{\alpha\mu}\delta\Gamma^\beta_{\alpha\beta}\right)\,.
\end{equation}
This integral represents nothing else but an integration of an exterior derivative of a differential form and hence we can apply directly Stockes theorem, which states that the integral of the exterior derivative of a differential form equals the integral of the differential form over the boundary $\int_{\mathcal{M}}dA=\int_{\partial\mathcal{M}}A$. Applied to our above integral, it means that we can rewrite (\ref{GHY_integral}) as
\begin{eqnarray}\label{GHY_integral2}
\int_{\mathcal{M}}d^4x \sqrt{-g}g^{\alpha\beta}\delta R_{\alpha\beta}&=&\int_{\mathcal{M}}d^4x \sqrt{-g}g^{\alpha\beta} \nabla_\mu \left( g^{\alpha\beta}\delta\Gamma^\mu_{\alpha\beta}-g^{\alpha\mu}\delta\Gamma^\beta_{\alpha\beta}\right)\nonumber\\
&=&\int_{\partial\mathcal{M}}d\Sigma_\mu\left(  g^{\alpha\beta}\delta\Gamma^\mu_{\alpha\beta}-g^{\alpha\mu}\delta\Gamma^\beta_{\alpha\beta}\right)
\end{eqnarray}
where we are now performing the integral on the three dimensional boundary surface 
\begin{equation}
d\Sigma_\mu=\epsilon n_\mu d\Sigma=\epsilon n_\mu \sqrt{|\gamma|}d^3y\,,
\end{equation}
with $n_\mu$ denoting the unit normal to the boundary $\partial\mathcal{M}$, $y_a$ the coordinates on the boundary, $\gamma_{\mu\nu}$ the induced metric on the boundary $\gamma_{\mu\nu}=g_{\mu\nu}-\epsilon n_\mu n_\nu$ and $\epsilon=n^\mu n_\mu=\pm1$ ($+1$ for timelike $\partial\mathcal{M}$ and $-1$ for spacelike $\partial\mathcal{M}$). Thus, in order to finalise the variation of the Einstein-Hilbert action, we have to perform the surface integral
\begin{equation}
\int_{\mathcal{M}}d^4x \sqrt{-g}g^{\alpha\beta}\delta R_{\alpha\beta}=\int_{\partial\mathcal{M}} d^3y\epsilon n_\mu \sqrt{|\gamma|}\left(  g^{\alpha\beta}\delta\Gamma^\mu_{\alpha\beta}-g^{\alpha\mu}\delta\Gamma^\beta_{\alpha\beta}\right)\,.
\end{equation}
As next, we can express the variation of the Christoffel symbols in terms of the metric and contract them with the unit normal. By doing this, the integral becomes
\begin{equation}
\int_{\partial\mathcal{M}} d^3y\epsilon n_\mu \sqrt{|\gamma|}\left(  g^{\alpha\beta}\delta\Gamma^\mu_{\alpha\beta}-g^{\alpha\mu}\delta\Gamma^\beta_{\alpha\beta}\right)=\int_{\partial\mathcal{M}} d^3y\epsilon \sqrt{|\gamma|} n^\mu \gamma^{\alpha\beta}\left( \partial_\alpha \delta g_{\mu\beta}-\partial_\mu \delta g_{\alpha\beta} \right) \,.
\end{equation}
The variation principle is such that the variation vanishes, $\delta g_{\mu\nu}=0$, everywhere on the boundary. However, in the above expression we have the derivatives of the variation. Do they also vanish? No, only the tangential derivatives will vanish, i.e. only the derivatives that are projected on the boundary via the contraction with the induced metric will vanish. Thus, the first term in the above expression will vanish $ \gamma^{\alpha\beta}\partial_\alpha \delta g_{\mu\beta}=0$, but not the second term. In the first term $\gamma^{\alpha\beta}$ projects the derivative $\partial_\alpha$ on the boundary, but in the second term the derivative $\partial_\mu$ is not projected. This results in
\begin{equation}
\int_{\partial\mathcal{M}} d^3y\epsilon n_\mu \sqrt{|\gamma|}\left(  g^{\alpha\beta}\delta\Gamma^\mu_{\alpha\beta}-g^{\alpha\mu}\delta\Gamma^\beta_{\alpha\beta}\right)=-\int_{\partial\mathcal{M}} d^3y\epsilon \sqrt{|\gamma|} n^\mu \gamma^{\alpha\beta}\partial_\mu \delta g_{\alpha\beta} \,.
\end{equation}
Summarising, the variation of the Einstein-Hilbert term gives the two contributions
\begin{equation}\label{totalVar_EHaction}
\delta\mathcal{S}_{\rm EH}=\int_{\mathcal{M}}d^4x \sqrt{-g}G_{\alpha\beta}\delta g^{\alpha\beta}-\int_{\partial\mathcal{M}} d^3y\epsilon \sqrt{|\gamma|} n^\mu \gamma^{\alpha\beta}\partial_\mu \delta g_{\alpha\beta}\,.
\end{equation}
As we can see in this expression, the surviving of the second term jeopardises the well-posedness of the variation principle of the Einstein-Hilbert action if the considered space-time has a boundary. One has to supplement the initial action by a boundary term put by hand in order to make the variational principle well-defined. This is done by the addition of the Gibbons-Hawking-York (GHY) boundary term
\begin{equation}
\mathcal{S}_{\rm GHY}= 2\int_{\partial\mathcal{M}} d^3y\epsilon \sqrt{|\gamma|} K\,,
\end{equation}
where $K$ represents the trace of the extrinsic curvature $K=\nabla_\mu n^\mu$. When we perform the variation of the GHY boundary term, we only need to vary the extrinsic curvature since the induced metric is fixed on the boundary. We can express the extrinsic curvature in terms of the Christoffel symbols and the induced metric as
\begin{equation}
K=\gamma^{\mu\nu}\left( \partial_\nu n_\mu-\Gamma_{\mu\nu}^\rho n_\rho \right)=-\gamma^{\mu\nu}\Gamma_{\mu\nu}^\rho n_\rho\,.
\end{equation}
We can further express the Christoffel symbols in terms of the metric and perform the variation of $K$
\begin{equation}
\delta K=-\gamma^{\mu\nu}\delta\Gamma_{\mu\nu}^\rho n_\rho=\frac12 \gamma^{\mu\nu} \partial_\rho \delta g_{\mu\nu} n^\rho\,,
\end{equation}
where we again used the fact that only tangential derivatives vanish. Thus, the variation of the GHY boundary term yields
\begin{equation}
\delta \mathcal{S}_{\rm GHY}= 2\int_{\partial\mathcal{M}} d^3y\epsilon \sqrt{|\gamma|} \frac12 \gamma^{\mu\nu} \partial_\rho \delta g_{\mu\nu} n^\rho\,.
\end{equation}
Comparing this expression with the second term in equation (\ref{totalVar_EHaction}) shows that they will exactly cancel each other. Thus, if the considered spacetime is not closed, then we have to consider the augmented action
\begin{equation}
 \mathcal{S}_{\rm EH}+ \mathcal{S}_{\rm GHY}=\int_{\mathcal{M}}d^4x \sqrt{-g} R+2\int_{\partial\mathcal{M}} d^3y\epsilon \sqrt{|\gamma|} K\,,
\end{equation}
and the variation of this total action
\begin{equation}
 \delta\mathcal{S}_{\rm EH}+ \delta\mathcal{S}_{\rm GHY}=\int_{\mathcal{M}}d^4x \sqrt{-g}G_{\alpha\beta}\delta g^{\alpha\beta}-\int_{\partial\mathcal{M}} d^3y\epsilon \sqrt{|\gamma|} n^\mu \gamma^{\alpha\beta}\partial_\mu \delta g_{\alpha\beta}+\int_{\partial\mathcal{M}} d^3y\epsilon \sqrt{|\gamma|}  \gamma^{\mu\nu} \partial_\rho \delta g_{\mu\nu} n^\rho\,,
\end{equation}
will be well-defined and result in $G_{\alpha\beta}$ since the last two terms in the above expression will compensate each other. The inclusion of the boundary term by hand introduces an ambiguity in the variational principle, which might be received as something not fundamental and unique. 

In some applications, for instance in the computation of the black hole entropy based on the Euclidean action, it is necessary to make a further inclusion of a counter term in order to make the action finite. The thermodynamical properties of a black hole are encoded in the path integral $\mathcal{Z}=\int \mathcal{D}[g]\mathcal{D}[\psi]e^{-\mathcal{S}_E(g,\psi)}$, where the Euclidean action $\mathcal{S}_E$ is obtained from the action $\mathcal{S}_E(g,\psi)=-i\mathcal{S}(g,\psi)$ after Wick rotation $t=-i\tau$. In vacuum $\psi=0$ with $g=\bar{g}+\delta g$, the dominant contribution to the partition function becomes 
\begin{equation}
\ln \mathcal{Z}= -\mathcal{S}_E(\bar{g})+\ln \left( \mathcal{D}[g]e^{-S^{(1)}(\delta g)} \right)\,.
\end{equation}
This can be directly used in order to compute the associated entropy of the black hole $S=\beta M+\ln \mathcal{Z} $. Using the Euclidean Schwarzschild solution $ds^2=(1-2M/r)d\tau^2+(1-2M/r)^(-1)dr^2+r^2d\Omega^2$, we can compute the Euclidean action. Since the Ricci scalar vanishes for the vacuum $\mathcal{S}_{\rm EH}=\int \sqrt{-g} R=0$, we need to include the GHY boundary term and an additional counter term
\begin{eqnarray}
\mathcal{S}_{\rm E}=\mathcal{S}_{\rm EH}+\mathcal{S}_{\rm GHY}+\mathcal{S}_{\rm CT}&=& \frac{1}{8\pi}\int_{\partial\mathcal{M}} d^3y\epsilon \sqrt{|\gamma|}(K-K_0)\nonumber\\
&=& \frac{1}{8\pi}\int_0^\beta d\tau \int_0^{2\pi}\int_0^\pi d\theta \sqrt{|\gamma|}(K-K_0)\nonumber\\
&=&\frac{\beta}{2}M=\frac{\beta^2}{16\pi}=4\pi M^2\,.
\end{eqnarray}
The computation of the entropy of a black hole requires the inclusion of the GHY boundary term $K$ and an additional counter term $K_0$ in order to make the integral converging.

In section \ref{section_GeomGrav} we have seen three equivalent formulations of General Relativity in the geometrical framework. Even if we do not consider the geometrical formulations in this section anymore, we can view torsion and non-metricity as some classical fields without the need of relating them to some geometrical interpretation. The formulation of General Relativity in form of TEGR improves the computation of the entropy in the sense that no boundary term is needed since $\mathcal{S}=\int\sqrt{-g}\mathring{\mathbb{T}}\ne0$ for the Schwarzschild action. However, one still needs to introduce a counter term so that the integral becomes converging. In the case of TEGR one obtains for the Euclidean action
\begin{eqnarray}
\mathcal{S}_{\rm E}^{\rm TEGR}&=& \frac{1}{8\pi}\int_0^\beta d\tau \int_0^{2\pi}\int_0^\pi d\theta \sqrt{|\gamma|}n_r(\mathcal{T}^r-\mathcal{T}^r_0)\nonumber\\
&=&\frac{\beta}{2}M=\frac{\beta^2}{16\pi}=4\pi M^2\,.
\end{eqnarray}
However, the ambiguity and lack of uniqueness of the GHY and the counter term might be avoided in the formulation of General Relativity based on non-metricity. Not only is the action of CGR non-vanishing $\mathcal{S}=\int\sqrt{-g}\mathring{\mathcal{Q}}\ne0$ since $\mathring{\mathcal{Q}}=-2/r^2$, but one also does not need to introduce any arbitrary counter term. The resulting action is directly finite. 
One can use the gauge freedom to choose freely an appropriate coordinates
\begin{equation}
d s^2 =  \frac{\left( 1-\frac{M}{2R}\right)^2}{\left( 1+\frac{M}{2R}\right)^2}d \tau^2 
 + \left( 1+\frac{M}{2R}\right)^4 d \ell^2\,, \label{isotropic}
\end{equation}
where $R=\sqrt{x^2+y^2+z^2}$ and $d \ell^2 = d x^2 +  d y^2 +  d z^2$. The Euclidean action in CGR becomes
\ba \label{stgraction}
\mathcal{S}_{\rm E}^{\rm CGR}&=&\frac{1}{16\pi}\int d^4x\sqrt{g}\mathring{\mathcal{Q}}\\
&=&\frac{M^2}{8\pi}\int_0^\beta d \tau\int d x d y d z\frac{1}{R^4}\nonumber\\
&=&\frac{M^2}{8\pi}\int_0^\beta d\tau\int d\Omega\int_{M/2}^{R_0} R^2 d R\frac{1}{R^4}\nonumber\\
&=&\frac{M^2}{2}\beta \left(\frac{2}{M}-\frac{1}{R_0}\right)\underset{R_0\to\infty}{\longrightarrow}\beta M=8\pi M^2.\nonumber
\ea
Thus, the formulation of General Relativity in terms of non-metricity allows to avoid the introduction of the GHY boundary term with well-posed variational principle and finite Euclidean entropy without any counter terms
(see for instance \cite{BeltranJimenez:2017tkd,BeltranJimenez:2018vdo} for detailed discussion on this). 

\subsection{ADM decomposition and the system of constraints}
\begin{center}
\begin{figure}[h]
\begin{center}
 \includegraphics[width=9.0cm]{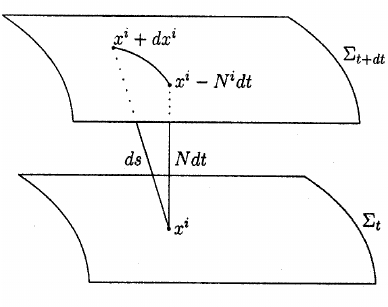}
\end{center}
  \caption{A pictorial representation of the foliation of the spacetime into a family of space-like surfaces. It is taken from \url{https://www.researchgate.net/figure/The-line-element-expressed-in-the-ADM-decomposition_fig4_238674029}.}
 \label{ADMdecomposition}
\end{figure}
\end{center}

We have seen that the Einstein-Hilbert action in equation (\ref{action_GR}) constructed out of the Lovelock invariant term $R$ is the unique consistent theory for a massless spin-2 field at the non-linear level, that gives rise to second order equations of motion. One crucial requirement for the correct number of physical degrees of freedom was the realisation of the full diffeomorphism. In order to convince ourselves that it really propagates the right number of degrees of freedom beyond the linear order we can study the Hamiltonian formulation in the ADM variables together with the realisation of the associated class of constraints. For this, consider the decomposition of the metric in the following form
\begin{eqnarray}\label{metric_ADMansatz}
g_{\mu\nu}=
\begin{pmatrix}
-N^2+N_iN_j \gamma^{ij}&N_j \\
N_i&\gamma_{ij}
 \end{pmatrix}\,,
\end{eqnarray}
with the spatial metric $\gamma_{ij}$, the lapse function $N$ and the shift $N_i$. This decomposition corresponds to a foliation of the spacetime into a family of space-like surfaces. See figure \ref{ADMdecomposition} for a pictorial representation of the foliation of the spacetime in terms of the ADM variables.

The Einstein-Hilbert action in equation (\ref{action_GR}) in terms of these variables reads
\begin{eqnarray}\label{GR_ADMvariables}
\mathcal{S}=M_p^2\int \mathrm{d}^4 \Pi^{ij} \dot{\gamma}_{ij}-NR^0 -N_i R^i)\,,
\end{eqnarray}
where $\Pi^{ij}$ is the conjugate momenta of the spatial metric $\Pi^{ij}=\partial \mathcal{L}/\partial \dot{\gamma}_{ij}$ and $R^0$ and $R^i$ stand for
\begin{eqnarray}
R^i&=& -2 \partial_j \Pi^{ij} \nonumber\\
R^0&=&\frac{-\sqrt{\gamma}}{\gamma} \left( \frac12 (\Pi^i{}_i)^2-\Pi^{ij}\Pi_{ij}\right)-\sqrt{\gamma}R_\gamma\,.
\end{eqnarray}
One immediate observation of (\ref{GR_ADMvariables}) is that the Lagrangian does not contain any dynamics for the lapse and shift and more importantly that it is linear in both. They act as Lagrange multipliers since $R^0$ and $R^i$ do not depend on the lapse and shift. Due to the absence of any dynamics for the lapse and shift, the associated conjugate momenta for $N$ and $N_i$ vanish identically 
\begin{equation}
\Pi^0=\partial \mathcal{L}/\partial \dot{N}=0 \qquad \text{and} \qquad \Pi^i=\partial \mathcal{L}/\partial \dot{N}_i=0. 
\end{equation}
Since the Lagrangian is linear in the lapse and shift, their equations of motion generate four constraint equations 
\begin{equation}
\mathcal{C}^0=R^0=0 \qquad \text{and} \qquad \mathcal{C}^i=R^i=0
\end{equation}
on the spatial metric $\gamma_{ij}$ and its momenta. These are first class constraints associated to the gauge symmetry of the theory since the Poisson brackets of the constraints $\{ \mathcal{C}^\mu,\mathcal{C}^\nu\}$ vanish on the constraint surface 
\begin{equation}
\{ \mathcal{C}^\mu,\mathcal{C}^\nu\}=0
\end{equation}
In this way we remove $2\times8$ dynamical variables leaving only two dynamical variables together with their conjugate momenta ($g_{\mu\nu}$ (10 components)-$2\times4$ (Diffs) leaving 2 degrees of freedom). \\

At the linear order we had seen that the construction of the Fierz-Pauli action strongly relied on the tuning of the coefficients which rendered $h_{00}$ and $h_{0i}$ non-dynamical and the resulting theory satisfied the linearised diffeomorphism. Instead of focusing on the dynamics of the equations of motion in (\ref{EOMh0i_cond}) and  (\ref{EOMh00_cond}), we could have on an equal footing computed the Hamiltonian of the general linear theory and obtained the conditions for the lapse and shift to appear linearly. This would have resulted in the same conclusion. Here, we have seen that this property is maintained at the fully non-linear level, when the Fierz-Pauli action is promoted to the Einstein-Hilbert action. The crucial requisite is that the Hamiltonian is linear in the lapse and the shifts, which on the other hand guarantees the presence of first class constraints that remove the non-physical degrees of freedom. The linearity in the lapse and the shifts is a virtue of the full diffeomorphism.

\section{Massive Spin-2 field}\label{sec_MassiveSpin2}
General Relativity describes the theory of gravitation with a massless spin-2 particle as an exchange particle and the interactions are built out of the tensor $g_{\mu\nu}$ and its derivatives. We have seen that the diffeomorphism invariance was crucial in order to ensure the presence of first class constraints that removed the unphysical degrees of freedom. From a theoretical point of view, it would be crucial to investigate whether the graviton itself could be a massive particle, and if so what kind of new interactions could be allowed. 
\subsection{Consistent construction of a massive spin-2 field}
For an exhaustive review of the consistent construction of a massive spin-2 particle we refer to \cite{deRham:2014zqa}. Here, we shall only give a quick introduction and summarise some of the fundamental properties of massive gravity and review some of the uncovered recent studies.
A first naive attempt for a mass term might consist in considering a simple Ansatz as
\begin{equation}
\sqrt{-g}\left[\frac{m^2}{2}\left(c_1g_{\mu\nu}g^{\mu\nu}+c_2g_{\mu}^{\mu}g_{\nu}^{\nu}\right)\right].
\end{equation}
However, in this way one would be writing nothing else but a cosmological constant, which does not correspond to a mass term. One could also consider an Ansatz like $m^2 R^2$. However, these interactions contain derivatives of the metric and can not be a mass term either. Very soon one realises that it is indispensable to introduce a second fiducial metric in order to construct a mass term. 
In terms of metric fluctuations on top of flat Minkowski metric (being the fiducial metric to contract the indices), one could try with the promising Ansatz at linear order
\begin{equation}\label{detunedFP}
\frac{m^2}{2}\left(c_1h_{\mu\nu}h^{\mu\nu}+c_2h_{\mu}^{\mu}h_{\nu}^{\nu}\right)\,.
\end{equation}
The inclusion of this mass term breaks explicitly the diffeomorphism invariance and hence the constraints related to the lapse and shift of the metric are not first class anymore and would remove only 4 variables, leaving potentially six propagating degrees of freedom. However, we know that the massive spin-2 representation of the Lorentz group carries only five physical degrees of freedom ($2s+1$), rendering the sixth mode unavoidably a ghost degree of freedom.
In fact, Fierz and Pauli found the first successful construction of a linear mass term without giving rise to any ghostly degree of freedom by applying the restriction $c_2=-c_1$ to the linear Ansatz (\ref{detunedFP}) \cite{Fierz:1939ix}. Away from this detuning the theory of massive gravity would contain six propagating degrees of freedom, one being a ghost and that would render the Hamiltonian unbounded from below. The mass of this ghost degree of freedom would scale as $m_g^2=-\frac{c_1(c_1+4c_2)m^2}{4(c_1+c_2)}$. As one can see immediately, the mass of the ghost goes to infinity for the Fierz-Pauli tuning. The consistent theory for gravitation with a massive spin-2 particle as an exchange particle has to be of the following form in the linearised regime
\begin{equation}
\mathcal{L}=-h^{\mu\nu}\hat{\mathcal{E}}_{\mu\nu}^{\alpha\beta}h_{\alpha\beta}-\frac{m^2}{2}\left(h_{\mu\nu}h^{\mu\nu}-h_{\mu}^{\mu}h_{\nu}^{\nu}\right)+\frac{1}{M_{\rm Pl}}h_{\mu\nu} T^{\mu\nu}.\label{linearMG}
\end{equation}
Variation with respect to $h$ gives the equations of motion
\begin{equation}
-2\hat{\mathcal{E}}_{\mu\nu}^{\;\;\;\alpha\beta}h_{\alpha\beta}-m^2\left(h_{\mu\nu}-\eta_{\mu\nu} h\right)=-\frac{1}{M_p}T_{\mu\nu}\,.
\end{equation}
The divergence of the equations of motion results in 
\begin{equation}
m^2(\partial^\mu h_{\mu\nu}-\partial_\nu h)=\frac{1}{M_p}\partial^\mu T_{\mu\nu},
\end{equation}
which is equivalent to the statement $\partial^\mu h_{\mu\nu}=\partial_\nu h$ for a conserved source. Similarly, the trace of the equation of motion imposes $-3m^2h=\frac{T}{M_p}$. These constraints remove five unphysical degrees of freedom. Hence, a consistent theory for massive gravity consists only of five propagating physical degrees of freedom, namely the two helicity-2, two helicity-1 and one helicity-0 degrees of freedom. The field equations together with these constraints become
\begin{equation}
(\Box-m^2)h_{\mu\nu}=-\frac{1}{M_p}\left(T_{\mu\nu}-\frac13\eta_{\mu\nu}T-\frac{1}{3m^2}\partial_\mu\partial_\nu T\right)\,.
\end{equation}
Thus, adding a mass term to the action of General Relativity breaks the invariance under general coordinate transformations and hence five physical degrees of freedom propagate instead of two, which will have important consequences on large cosmological scales, comparable to the Compton wavelength of the graviton. 
The broken diffeomorphism invariance can be restored by applying the Stueckelberg trick. For this purpose, one can introduce the following field redefinitions 
\begin{eqnarray}
h_{\mu\nu}\to h_{\mu\nu}+\partial_\mu A_\nu +\partial_\nu A_\mu  \;\;\;\; {\rm and} \;\;\;\; A_\mu \to A_\mu + \partial_\mu \phi.
\end{eqnarray}
The linearized Lagrangian (\ref{linearMG}) then becomes
\begin{eqnarray}
\mathcal L=-h^{\mu\nu}\hat{\mathcal{E}}_{\mu\nu}^{\alpha\beta}h_{\alpha\beta}-\frac{m^2}{2}\left(h_{\mu\nu}h^{\mu\nu}-h_{\mu}^{\mu}h_{\nu}^{\nu}\right)+\frac{1}{M_{\rm Pl}}h_{\mu\nu} T^{\mu\nu}\nonumber\\
-\frac{m^2}{2}F_{\mu\nu}^2-2m^2(h_{\mu\nu}\partial^\mu A^\nu-h\partial A)\nonumber-\frac{2}{M_{\rm Pl}}A_{\mu}\partial_{\nu} T^{\mu\nu}\\
-2m^2(h_{\mu\nu}\partial^\mu\partial^\nu \phi-h\partial^2 \phi)+\frac{2}{M_{\rm Pl}}\phi\partial_\mu\partial_{\nu} T^{\mu\nu}.
\end{eqnarray}
Since we introduced the Stuckelberg fields $A_\mu$ and $\phi$ the Lagrangian is now invariant under the simultaneous transformations 
\begin{eqnarray}
h_{\mu\nu}\to h_{\mu\nu}+\partial_\mu \xi_\nu+\partial_\nu \xi_\mu \;\;\;\; &&{\rm and} \;\;\;\; A_\mu \to A_\mu - \xi_\mu \nonumber\\
A_\mu \to A_\mu + \partial_\mu \theta \;\;\;\; &&{\rm and} \;\;\;\;  \phi \to \phi-\theta \,.
\end{eqnarray}
We can further canonically normalize the fields by $A_\mu\to\frac1m A_\mu$ and $\phi \to \frac{1}{m^2} \phi$. After taking the $m\to0$ limit, the linearized Lagrangian simplifies to
\begin{eqnarray}
\mathcal L=-h^{\mu\nu}\hat{\mathcal{E}}_{\mu\nu}^{\alpha\beta}h_{\alpha\beta}-\frac{1}{2}F_{\mu\nu}^2+\frac{1}{M_{\rm Pl}}h_{\mu\nu} T^{\mu\nu}
-2(h_{\mu\nu}\partial^\mu\partial^\nu \pi-h\partial^2 \pi)\label{undiagdecoupl}.
\end{eqnarray}
As it can be seen clearly, in this limit only very specific interactions survive. The scalar field is kinetically mixed with the tensor field and the vector field decouples completely. Note, that a crucial property of massive gravity is that the helicity-0 mode does not have an own kinetic term and arises only through derivative mixing.

In our Ansatz for the linear mass term in equation (\ref{detunedFP}) we made use of the flat metric $\eta_{\mu\nu}$ in order to raise and lower the space-time indices since we very soon realised that we can not use the field $g_{\mu\nu}$ for this purpose. This means that the presence of a second rank-2 tensor is unavoidable. Let us call this tensor $f_{\mu\nu}$. The next natural question that arises is how these two tensors interact at the non-linear level in order to construct consistent non-linear theory for massive gravity. Can we consider any potential interaction between these two $U(g^{\mu\alpha}f_{\alpha\nu})$? It turns out that we have to impose strong restrictions on these interactions in order to have the right non-linear theory with five physical degrees of freedom. 
For an arbitrary potential at the non-linear order, one will again face the fact that there will be $10-4=6$ degrees of freedom. This means that one has to construct the non-linear interactions such that one additional constraint arises removing one more degree of freedom. To appreciate this statement, we can again apply the Hamiltonian formulation as we did in equation (\ref{GR_ADMvariables}) for the massless spin-2 field. We make a similar ADM Ansatz for the second metric $f$
\begin{eqnarray}\label{fmetric_ADMansatz}
f_{\mu\nu}=
\begin{pmatrix}
-M^2+M_iM_j \tilde{\gamma}^{ij}&M_j \\
M_i&\tilde{\gamma}_{ij}
 \end{pmatrix}\,,
\end{eqnarray}
with its own lapse $M$, shift $M_i$ and spatial metric $\tilde{\gamma}_{ij}$. With this in mind the general potential interactions will have the general dependence 
\begin{eqnarray}\label{constraint_pot_MG_Hess}
m^2M_p^2\sqrt{-g}U(g^{\mu\alpha}f_{\alpha\nu})=m^2M_p^2 N\sqrt{\gamma} U(N, M, N_i, M_i, \gamma_{ij}, \tilde{\gamma}_{ij}).
\end{eqnarray}
In order to guarantee the presence of an additional constraint equation, we have to impose that the determinant of the Hessian matrix with respect to $N$ and $N_i$ vanishes identically \footnote{The degeneracy condition of the Hessian matrix was investigated in \cite{deRham:2011rn}. Even though the helicity-0 mode might look like having higher order equations of motion, the vanishing of the determinant of the Hessian matrix guarantees the presence of a primary constraint that removes the Boulware-Deser ghost.}. It means that combining the equations of motion for $N$ and $N_i$ should result in an additional constraint equation. The linearity in the lapse $N$ (with the right definition of the shift) together with the absence of the lapse in the shift equations of motion, will be the right requirements in order to satisfy the vanishing of the Hessian matrix. It turns out that these requirements uniquely fix the allowed potential interactions to be given by the fundamental tensor
\begin{equation}
\mathcal{K}^\mu_\nu=\left(\sqrt{g^{-1}f}\right)^\mu_\nu.
 \end{equation}
Similarly, as we were using the Levi-Civita construction scheme for the Galileon in equation (\ref{eqsGal_LeviC}) in terms of $\Pi_{\mu\nu}$, we can construct all the allowed non-linear interactions for the massive graviton as
\begin{align}\label{potentialsdRGT}
\mathcal{U}_1[\mathcal{K}] &= \epsilon^{\mu\nu\rho\sigma}  \epsilon^{\alpha}_{\;\;\;\; \nu\rho\sigma} \K_{\mu\alpha} = 6[K]\nonumber\\
\mathcal{U}_2[\mathcal{K}] &= \epsilon^{\mu\nu\rho\sigma}  \epsilon^{\alpha\beta}_{\;\;\;\;\; \rho\sigma} \K_{\mu\alpha} \K_{\nu\beta} = 2\left( [\K]^2-[\K^2]\right), \nonumber\\
\mathcal{U}_3[\mathcal{K}] &=\epsilon^{\mu\nu\rho\sigma}  \epsilon^{\alpha\beta\kappa}_{\;\;\;\;\;\;\; \sigma} \K_{\mu\alpha} \K_{\nu\beta}  \K_{\rho\kappa}=[\K]^3-3[\K][\K^2]+2[\K^3],  \nonumber\\
\mathcal{U}_4[\mathcal{K}] &=\epsilon^{\mu\nu\rho\sigma} \epsilon^{\alpha\beta\kappa\gamma} \K_{\mu\alpha} \K_{\nu\beta}  \K_{\rho\kappa}  \K_{\sigma\gamma} =[\K]^4-6[\K]^2[\K^2]+3[\K^2]^2+8[\K][\K^3]-6[\K^4]\,.
\end{align}
The action for a massive spin-2 field with only five propagating degrees of freedom then becomes
\begin{equation}\label{action_MG_effcoupl}
\mathcal{S} = \int \mathrm{d}^4x \Big[ \frac{\mpl^2}{2} \sqrt{-g}\left(R[g]-\frac{m^2}{2}\sum_n \alpha_n{\cal U}_n[\cal K]   \right)+\mathcal{L}_{\rm matter} \Big]
\end{equation}
with arbitrary parameters $\alpha_n$. The ghost free interactions of massive gravity can be also expressed in a more compact form in terms of a deformed determinant \cite{Hassan:2011vm}
\begin{eqnarray}
{\rm det}( \delta^\mu_\nu+K^\mu_\nu)=\sum_{i=0}^4\frac{-\alpha_i}{i!(4-i)!}\mathcal{E}_{\mu_1\cdots \mu_i \alpha_{i+1}\cdots \alpha_4}\mathcal{E}^{\nu_1\cdots \nu_i \alpha_{i+1}\cdots \alpha_4}
K^{\mu_1}_{\nu_1}\cdots K^{\mu_i}_{\nu_i}
\end{eqnarray}
On the other hand, the determinant of a matrix  $\sqrt{g^{\mu\alpha}f_{\alpha\nu}}$ can be expressed in terms of the elementary symmetric polynomials  
\begin{eqnarray}\label{deformedDet_MG}
{\rm det}( \delta^\mu_\nu+K^\mu_\nu)=\sum_{n=0}^4\alpha_n e_n(K)\,,
\end{eqnarray}
where they satisfy the following combination of traces
\begin{eqnarray}
e_0(K) &=& 1 \nonumber\\
e_1(K) &=& [K] \nonumber\\
e_2(K) &=&\frac12( [K]^2-[K^2]) \nonumber\\
e_3(K) &=&\frac16( [K]^3-3[K][K^2]+2[K^3]) \nonumber\\
e_4(K) &=&\frac1{24}( [K]^4-6[K]^2[K^2]+3[K^2]^2+8[K][K^3]-6[K^4]) ]\,.
\end{eqnarray}
With these potential interactions the Hamiltonian of massive gravity is of the form $H=\int  \mathrm{d}^3x (-N\mathcal{C}_1+\bar{\mathcal{H}})$ with the lapse independent part $\bar{\mathcal{H}}$ and the constraint equation $\mathcal{C}_1$ generated by the lapse equation of motion. This corresponds to a primary constraint, that will yield a secondary constraint $\mathcal{C}_2=\dot{\mathcal{C}}_1=\{\mathcal{C}_1, H\}=0$. This constitutes a second class constraint $\{ \mathcal{C}_1,\mathcal{C}_2 \}\ne0$ since there is not any gauge symmetry anymore and it removes only one physical degree of freedom (see \cite{deRham:2010ik,deRham:2010kj,Hassan:2011vm,Hassan:2011hr,Hassan:2011ea} for more details). One can also promote the $f$ metric to be dynamical by adding a kinetic term for it. This results in bigravity \cite{Hassan:2011zd}. It is the most natural extension of massive gravity.

Some of the fundamental physical properties of this theory manifest themselves already in the leading order interactions of the decoupling limit. This incorporates particularly the leading order contributions of the helicity-2 and helicity-0 polarisations of the graviton. The limit corresponds to taking $m\to 0, M_{\rm Pl}\to \infty$ while keeping $\Lambda_3\equiv (M_{\rm Pl} m^2)^{1/3}$ fixed. The Lagrangian in the decoupling limit simplifies to
\begin{equation}
\mathcal{L}=-\frac{1}{2}
h^{\mu\nu}\mathcal{E}^{\alpha\beta}_{\mu\nu} h_{\alpha\beta}+
 h^{\mu\nu}\sum_{n=1}^3 \frac{a_{n}}{\Lambda^{3(n-1)}_3} X^{(n)}_{\mu\nu}\!\left(\Pi\right),
\label{lagr1}
\end{equation}
where the three matrices $X$'s are expressed in terms of the fundamental tensor $\Pi_{\mu\nu}=\partial_\mu\partial_\nu \pi$ and the Levi-Civita symbol $\mathcal{E}^{\mu\nu\alpha\beta}$
\begin{eqnarray}
X^{(1)}_{\mu\nu}\left(\Pi\right)&=&{\mathcal{E}_{\mu}}^{\alpha\rho\sigma}
{{\mathcal{E}_\nu}^{\beta}}_{\rho\sigma}\Pi_{\alpha\beta}, \quad  \nonumber \\
X^{(2)}_{\mu\nu}\left(\Pi\right)&=&{\mathcal{E}_{\mu}}^{\alpha\rho\gamma}
{{\mathcal{E}_\nu}^{\beta\sigma}}_{\gamma}\Pi_{\alpha\beta}
\Pi_{\rho\sigma}, \nonumber \\
X^{(3)}_{\mu\nu}\left(\Pi\right)&=&{\mathcal{E}_{\mu}}^{\alpha\rho\gamma}
{{\mathcal{E}_\nu}^{\beta\sigma\delta}}\Pi_{\alpha\beta}
\Pi_{\rho\sigma}\Pi_{\gamma\delta}\, .
\label{Xs}
\end{eqnarray}
One immediately recognises that the decoupling limit interactions are very similar to the Galileon interactions with the difference that the $\pi$ field in front of the interactions in equations (\ref{eqsGal_LeviC}) is replaced by the $h_{\mu\nu}$ field. These interactions become highly strong at the energy scale $E\sim \Lambda_3$. Following symmetries determine the properties of the underlying interactions
\begin{itemize}
\item  global field-space Galilean transformations $\pi\to\pi+b_\mu x^\mu+b$
\item  linearized diffeomorphisms $h_{\mu\nu}\to h_{\mu\nu}+\partial_{(\mu}\xi_{\nu)}$ up to total derivatives.
\end{itemize}
For the consistent construction of the ghost free massive gravity theory, the successful development of the decoupling limit and understanding its properties was a very crucial step forward\cite{deRham:2010ik}.

\subsection{Vainshtein mechanism}\label{vainshtein_Galileon}
If one computes the graviton exchange amplitude between two sources in the linear ghost free massive spin-2 field (\ref{undiagdecoupl}), one does not recover the General Relativity result In the limit of vanishing graviton mass. This is due to the fact that the helicity-0 mode does not decouple in that limit and there remains derivative mixing with the helicity-2 modes. The mixing can be got rid of by a proper field redefinition $h_{\mu\nu}=\tilde h_{\mu\nu}+\pi\eta_{\mu\nu}$ at the prize of introducing couplings of the scalar field to the stress energy tensor $\pi T$. 
Exactly this coupling is at the origin of the vDVZ discontinuity \cite{vanDam:1970vg,Zakharov:1970cc}. However, the vDVZ discontinuity is just an artifact of the linear approximation and one can recover properly General Relativity in the vanishing mass limit by taking into account the non-linear interactions of the helicity-0 mode, known as the Vainshtein mechanism.

\begin{center}
\begin{figure}[h!]
\begin{center}
 \includegraphics[width=4.7cm, angle =-90 ]{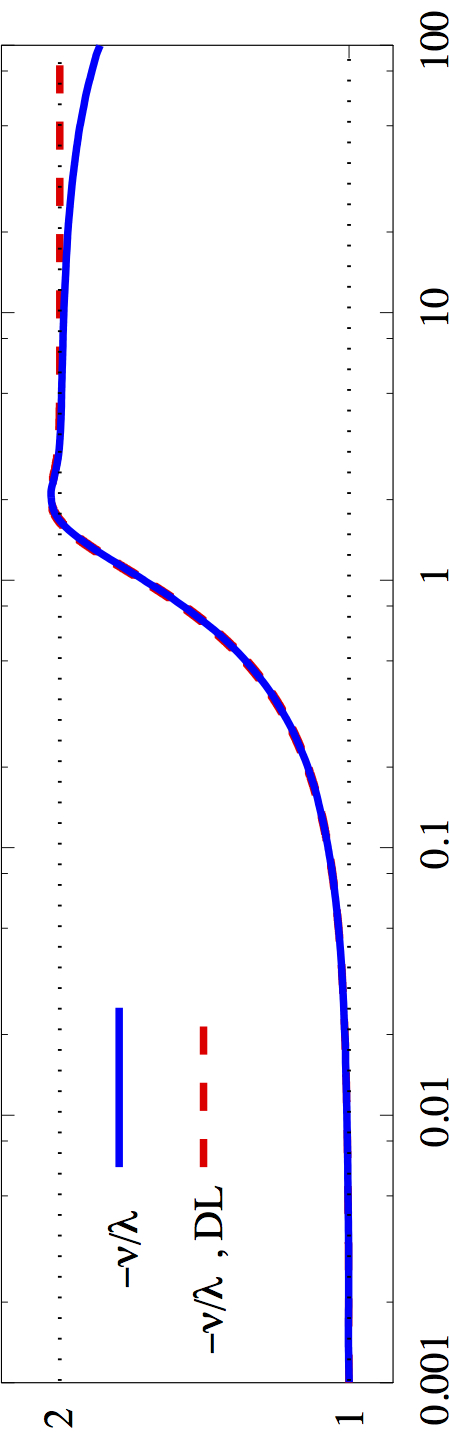}
\end{center}
  \caption{This figure is taken from \cite{Babichev:2009jt}, where for a static spherically symmetric solution $ds^2=-e^{\nu(R)}DT^2+e^{\lambda(R)}dR^2+R^2d\Omega^2$ the ratio between the functions $\nu(R)$ and $\lambda(R)$ versus radius (normalised by the Vainshtein radius) is plotted. The x-axis is $R/R_V$ and the $y$-axis is $\nu/\lambda$. One sees clearly the transition at the Vainshtein radius $R/R_V=1$, where the GR regime is recovered $\nu\sim-\lambda$. The corresponding decoupling limit result is also shown.}
   \label{Vainshtein_MG}
\end{figure}
\end{center}

The essence of the Vainshtein mechanism comes from the derivative self-interactions of the helicity-0 mode (of the scalar Galileon). To illustrate that schematically, let us assume a background field configuration with a localized source $T=M\delta^{(3)}(r)+\delta T$. The helicity-0 mode acquires the decomposition $\pi=\bar\pi(r)+\delta\pi(x^\mu)$. Even if the non-linearities are large on the background $\Lambda_3^{-3}\box\pi(\partial\pi)^2\gg1$, the perturbations are still weakly coupled. On top of this background the kinetic matrix symbolically takes the following form
\begin{equation}\label{modified_kin_Vainshtein}
	\mathcal{L}_\pi=-\frac12\left(1+\frac{\partial^2\bar\pi}{\Lambda^3}+\frac{(\partial^2\bar\pi)^2}{\Lambda^6}+\cdots\right)(\partial\delta\pi)^2+\frac{1}{\mpl}\delta\pi\delta T \,.
\end{equation}
The crucial step to remind now is that the effective coupling to the matter field will be redressed and couple to the self-interactions of the helicity-0 mode once we canonically normalise the field
\cite{Vainshtein:1972sx,Deffayet:2001uk}
\begin{equation}
	\mathcal{L}_\pi=-\frac12(\partial\delta\pi)^2	+\frac{1}{\mpl}\frac{\delta\pi\delta T}{\sqrt{(1+\frac{\partial^2\bar\pi}{\Lambda^3}+\frac{(\partial^2\bar\pi)^2}{\Lambda^6}+\cdots)}} \,.
\end{equation}
As we can see from this expression, the effective coupling to the matter field becomes small for a strongly self-interacting field $(1+\frac{\partial^2\bar\pi}{\Lambda^3}+\frac{(\partial^2\bar\pi)^2}{\Lambda^6}+\cdots)\gg1$. This additional coupling of the helicity-0 mode to the matter field has to be orders of magnitude weaker than the standard coupling of gravity in order to be in agreement with the constraints imposed by the absence of fifth forces. Thanks to the Vainshtein mechanism the helicity-0 mode is screened on small scales, while still being able to have cosmological effects \cite{Babichev:2009jt,Koyama:2011xz,Babichev:2013usa}. This is shown in figure \ref{Vainshtein_MG} taken from the pioneering work \cite{Babichev:2009jt} (see also the exhaustive review \cite{Joyce:2014kja} for a detail exposure to screening mechanisms).

\subsection{Quantum stability of massive spin-2 field}
In the previous subsection we have seen that for a consistent non-linear theory of massive gravity the square-root structure of the potential interactions is crucial. One fundamental question in this context is whether this specific structure gets detuned by quantum corrections and if so at what scale. Another essential question is also whether or not the mass of the graviton receives large quantum corrections. In other words we have to investigate the radiative stability of the overall and relative coefficients of the theory. For a technically natural theory it is required that quantum corrections remain small. One encounters technically natural tunings also in the Standard Model of Particle Physics. For instance, the electron mass is tuned to a small value relative to the electroweak scale. However, this tuning is technically natural due to the enhanced symmetry in the mass going to zero limit. The presence of a chiral symmetry in this limit protects the electron mass from large quantum corrections. In the case of massive gravity, we recover a gauge symmetry in the mass going to zero limit. Even if this is not a global symmetry, we expect a similar protection from large quantum corrections. 

This promising expectation indeed reveals itself when studying the quantum corrections in the decoupling limit of massive gravity. As we have seen in the previous subsection, the decoupling limit of massive gravity corresponds to the massless limit of the theory, where the interactions of the helicity-2 and the helicity-0 mode of the graviton become transparent and decouple from the rest. If we consider any Feynman diagram with the decoupling limit interactions acting at the vertices, then any external particle attached to this type of vertices receives at least two derivatives acting on it. Thus, we will have a similar non-renormalization theorem as we saw in section \ref{subsec_GalQuantum} for the scalar Galileon interactions. In order to illustrate this property, consider the decoupling limit interaction $h^{\mu\nu} \epsilon_{\mu}^{\;\;\alpha\rho\gamma} \epsilon_{\nu\;\;\;\;\gamma}^{\;\;\beta\sigma}\Pi_{\alpha\beta}\Pi_{\rho\sigma}$ in equation (\ref{lagr1}). In the non-trivial case we let the two $\pi$-particles with derivatives run in the loop with momenta $k_\mu$ and $(p+k)_\mu$ whereas contract the helicity-2 field without derivatives with an external helicity-2 particle. The contribution of this vertex in an arbitrary Feynman diagram is again proportional to  
\begin{eqnarray}
\mathcal A
 \propto  \int \frac{\mathrm{d}^4 k}{(2\pi)^4} D_k \,
 D_{p+k}\  \mathcal{E}^{\mu\nu}\, \epsilon_{\mu}^{\;\;\alpha\rho\gamma} \,
\epsilon_{\nu\;\;\;\;\gamma}^{\;\;\beta\sigma}
\, \, k_\alpha \, k_\beta \, (p+k)_\rho \, (p+k)_\sigma \cdots \,,
\end{eqnarray}
with $\mathcal{E}^{\mu\nu}$ representing the spin-2 polarization tensor and $D_k= k^{-2}$ the Feynman massless propagator of the $\pi$-particle. 
Due to the antisymmetric structure of the interactions, the only non-vanishing contribution to the scattering amplitude
 comes in with at least two powers of the external helicity-2 momentum $p_\rho p_\sigma$ corresponding to two derivatives in coordinate space. Hence, the quantum corrections do not take the same structure as the classical interactions of the massive gravity theory in the decoupling limit. This means that the overall and relative coefficients of the theory receives
 no quantum corrections in the decoupling limit as we anticipated above based on the symmetry argument. 
 
\begin{center}
\begin{figure}[h]
\begin{center}
 \includegraphics[width=11cm]{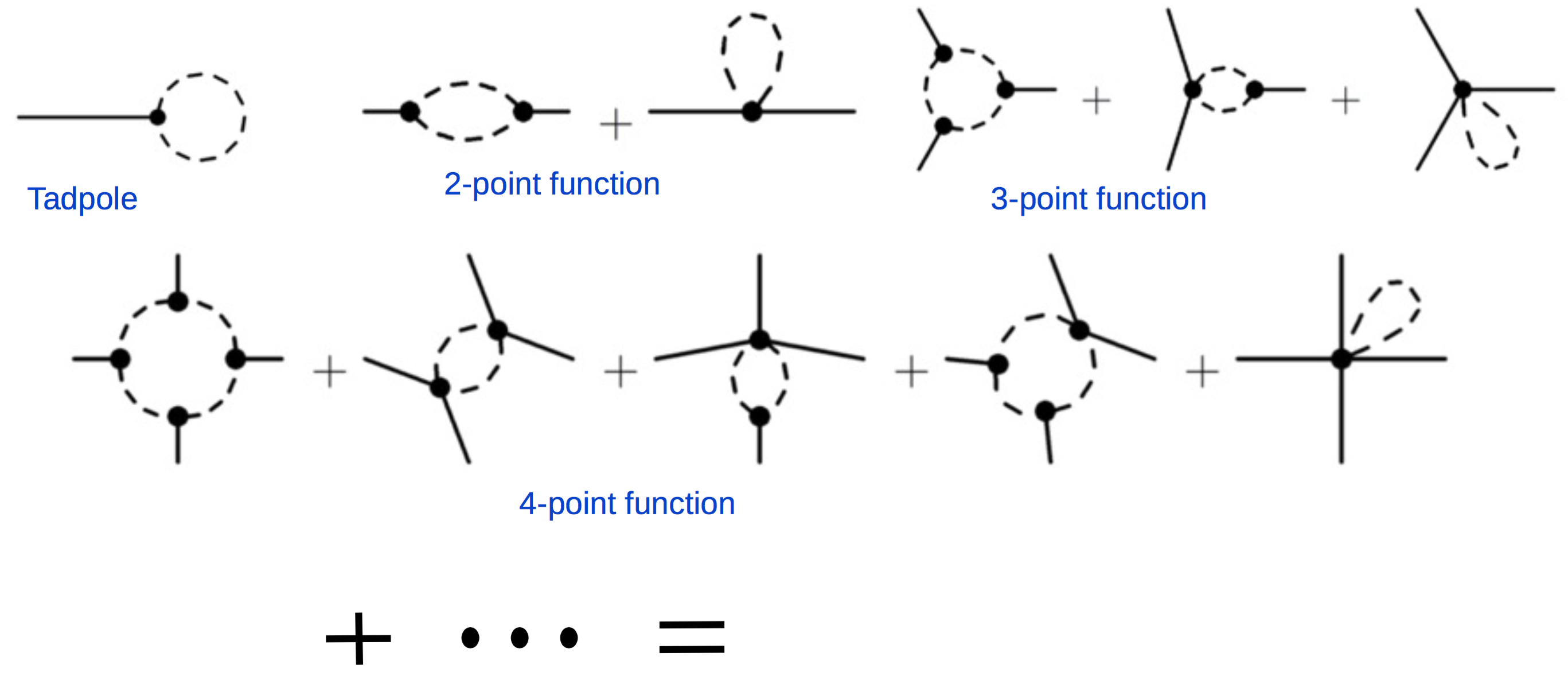}
\end{center}
  \caption{Contributions from matter loops.
Dashes denote the matter field propagator,
whereas solid lines denote the graviton. Shown are all the Feynman diagrams up to the 4-point function.
}
 \label{Feynman_diagrams_1}
\end{figure}
\end{center}

 These results in the decoupling limit indicate that the quantum corrections beyond this limit in the full theory should scale proportional to the mass of the graviton, since it vanishes in the massless limit. To be precise, the contributions to the graviton mass beyond the decoupling limit should scale as $\delta m^2 \lesssim m^2 (m/\mpl)^{2/3}$. This means that we expect a small graviton mass to be technically natural. For the purpose of studying the quantum corrections beyond the decoupling limit, let us assume that a scalar field plays the role of a matter field that couples minimally to massive gravity. In the next section we will investigate in more detail the possible consistent matter couplings, but for now we assume that the matter fields couple only to the $g$ metric. With these assumptions we will have two type of contributions at one loop: either matter fields or gravitons will run in the loops but not both simultaneously. The propagator of the massive graviton is given by
\begin{eqnarray}
\label{mGR_propagator}
D^{m}_{\mu\nu\alpha\beta}&=&\langle h_{\mu\nu}(x_1) h_{\alpha \beta}(x_2)\rangle=\tilde f_{\mu\nu\alpha\beta} \int \frac{\mathrm{d}^4 k}{(2\pi)^4}\frac{e^{i k_\mu \left(x_1^ \mu - x_2^\mu\right)}}{k^2+m^2-i\varepsilon} ,
\end{eqnarray}
with $\tilde f$ standing for
 \begin{equation}
\tilde f_{\mu\nu\alpha\beta}=\left(\tilde \eta_{\mu(\alpha}\tilde \eta_{\nu \beta)}-\frac 13 \tilde \eta_{\mu\nu}\tilde  \eta_{\alpha\beta}\right)
\hspace{15pt}{\rm and}\hspace{15pt}
\tilde \eta_{\mu\nu}=\eta_{\mu\nu}+\frac{k_\mu k_\nu}{m^2}\,,
\end{equation}
and similarly the Feynman propagator of the scalar matter field with mass $M$ reads
\begin{eqnarray}
D_{\chi}=\langle \chi(x_1) \chi(x_2)\rangle=\int \frac{\mathrm{d}^4 k}{(2\pi)^4}\frac{e^{i k_\mu \left(x_1^ \mu - x_2^\mu\right)}}{k^2+M^2-i\varepsilon}\,.
\end{eqnarray}
If we compare the propagator of the massive graviton (\ref{mGR_propagator}) with the expression we had for the massless graviton in equation (\ref{GRpropagator}), we see that there is a difference between the factors $1/3$ and $1/2$, which is at the origin of the vDVZ  discontinuity \cite{vanDam:1970vg,Zakharov:1970cc}. 
Let us consider first the one loop contributions where the matter fields run in the loop. At the level of the one-point function there is one tadpole contribution and at the two-point function level there are two contributions with three and four fields acting on the vertices respectively etc., as depicted in Figure \ref{Feynman_diagrams_1}.
The summation of all these one loop matter contributions or similarly the explicit computation of the one-loop effective action results in a contribution in form of a cosmological constant \cite{deRham:2013qqa}
\begin{eqnarray}
\mathcal{L}^{({\rm matter-loops})}_{1,\textrm{eff}}
\supset \frac {M^4}{64 \pi^2} \sqrt{- g} \log(\mu^2 )\,,
\end{eqnarray}
with $\mu$ standing for the regularization scale. This result is not surprising after taking into account the fact, that the graviton propagator does not have any effect on the vertices since we consider the same covariant matter coupling as in General Relativity and has only effect on the graviton propagator. These quantum corrections involve only matter fields inside the loop and hence are unaware of the graviton mass. 

\begin{center}
\begin{figure}[h]
\begin{center}
 \includegraphics[width=9cm]{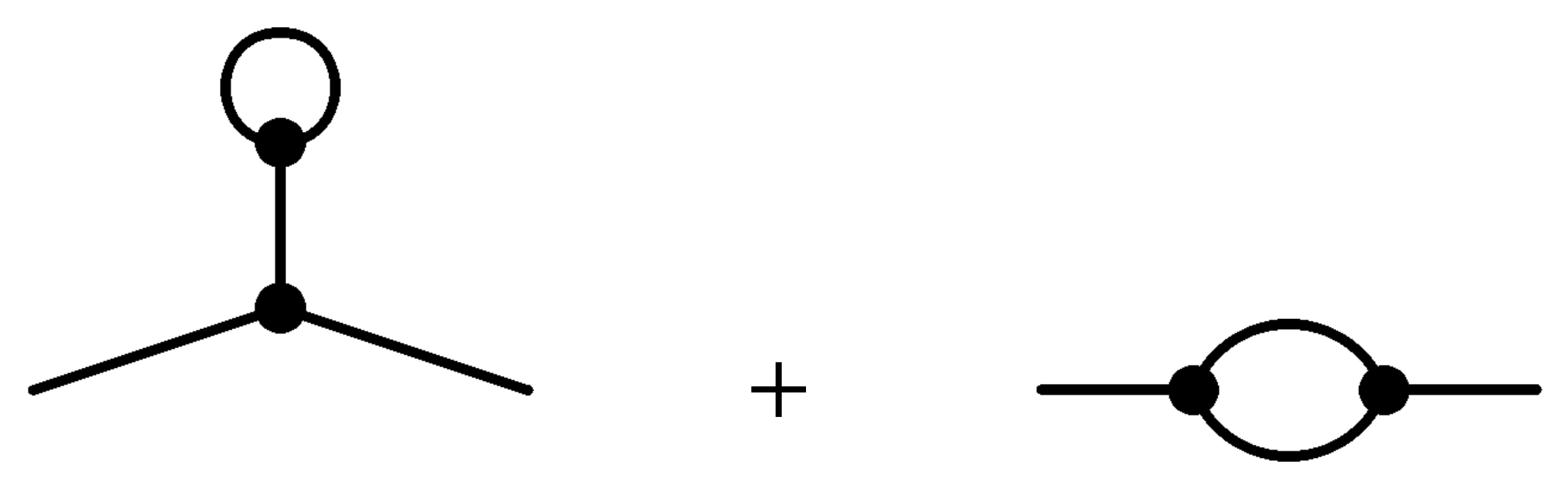}
\end{center}
  \caption{Contribution from graviton loops with the cubic interactions ${\cal U}_3$ acting on the vertices.
}
 \label{Feynman_diagrams_2}
\end{figure}
\end{center}

As next, let us consider the one loop contributions with the graviton running in the loop. Considering the quantum corrections with the vertices where the quartic potential ${\cal U}_4$ in (\ref{potentialsdRGT}) acts, one observes that the special structure of the potential interactions are preserved. However, the contributions with the cubic potential interactions ${\cal U}_3$ acting on the vertices as depicted in Figure \ref{Feynman_diagrams_2} reveals a detuning of the potential interactions.
 The second Feynman diagram in figure \ref{Feynman_diagrams_2} results in a detuning of the Fierz-Pauli interaction. For the higher n-point functions similar detuning of the potential interactions arise with a scaling 
 \begin{eqnarray}
\mathcal{L}^{({\rm graviton-loops})}_{1,\textrm{eff}}
\supset \frac {m^4}{\mpl^n} h^n \qquad \textrm{with} \qquad n\ge 2.
\end{eqnarray}
with the detuned structure of the $h^n$. As one can see, for small background values, these detunings are irrelevant below the Planck scale. The worry might come for large background values, since then the mass of the introduced ghost through the detuning could be lowered down within the regime of validity of the theory. Large background field configurations are typically encountered in the vicinity of compact objects like in Solar System, and since for the recovery of General Relativity the Vainshtein mechanism is crucial on these small scales, we need to make sure that no ghost is introduced at or below the strong coupling scale. Even in this case, once the Vainshtein mechanism is also properly taking into account at the level of the redressed interactions of the one loop effective action, one finds that the large background configuration suppresses further the quantum corrections. The one loop effective action in the vicinity of dense environments leads symbolically to contributions of the form
 \begin{eqnarray}
\mathcal{L}^{({\rm graviton-loops})}_{1,\textrm{eff}}
\sim \frac {1}{1+\frac{\bar{h}}{\mpl}}  \frac{m^4}{\mpl^2}(\delta h)^2
\end{eqnarray}
with the large background encoded in $\bar{h}$. Thus, the detuning of the potential interactions and consequently the introduced ghost would be irrelevant below Planck scales. Summarizing, at one loop we have:
 \begin{itemize}
 \item All the one loop matter contributions only give rise to a cosmological constant and leave the structure of the graviton potential unaffected.
  \item The graviton loop contributions detune the special structure of the potential interactions but this detuning is irrelevant below the Planck scale.
  \item The graviton mass receives small quantum corrections $\delta m^2 \lesssim m^2 (m/M_p)^{2/3}$ and is therefore technically natural. 
 \end{itemize}

\begin{center}
\begin{figure}[h!]
\begin{center}
  \includegraphics[width=7.3cm]{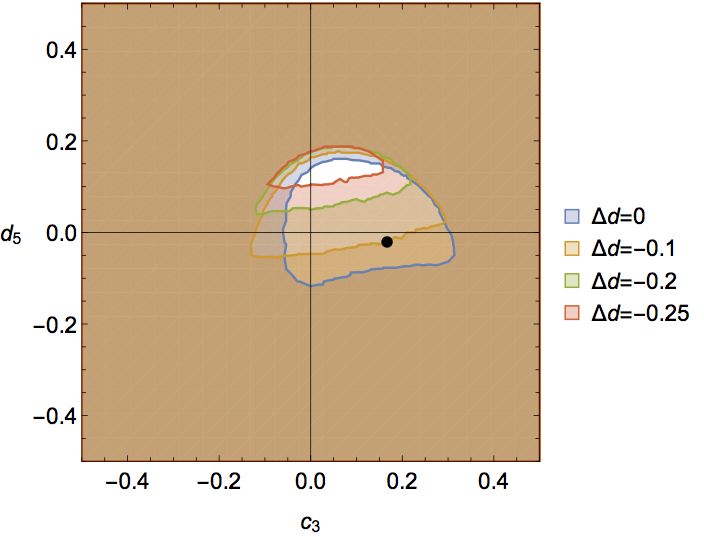}
   \includegraphics[width=7.5cm]{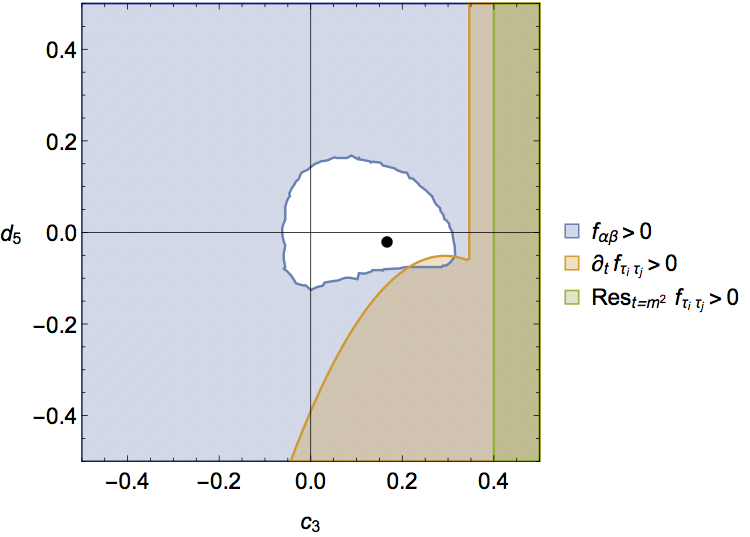}
\end{center}
  \caption{The surviving island: The analyticity/positivity requirements tune the coefficients of a $\Lambda_5$ theory to be exactly of the form of the $\Lambda_3$ massive gravity theory, which is encoded in the parameter $\Delta d$. The case with $\Delta d\ne0$ is ruled out from the requirement of a UV completion. $c_3$ and $d_5$ denote the theory parameters. The transversity amplitudes $f_{\alpha\beta}$ and their derivatives (for both the forward limit any beyond) can be obtained in section 7 of \cite{deRham:2018qqo}.}
 \label{MG_positivityBounds}
\end{figure}
\end{center}

\textbf{Implications of the UV completion}: \\
In section \ref{subsec_GalQuantum} we have quickly discussed the implications of the requirement to have Lorentz invariant UV completion on the low energy effective field theory for spin-0 fields with Galileon type of interactions. This requirement can be translated into some positivity bounds on the tree level scattering amplitudes. Using the positivity of the derivatives with respect to the momentum transfer $t$ of the imaginary part of the scattering amplitude away from the forward limit, the consequences of these bounds can be derived for the effective field theory of a scalar field, specially for the massive Galileon \cite{deRham:2017imi}. The same theoretical testing ground can be applied to non-zero spin fields, for instance to massive gravity, even though more caution is needed due to crossing relations and the standard helicity formalism has to be replaced by the transversity formalism. This was performed in \cite{Cheung:2016yqr,deRham:2017xox,deRham:2018qqo} and we refer the reader to them for a more detailed discussion. One of their remarkable result is, that even if one starts with a generic massive gravity theory with the scale of perturbative unitarity breaking at  $\Lambda_5=(\mpl m^4)^{1/5}$, imposing the positivity bounds filters out the specific structure of ghost-free interactions of massive gravity with the raised scale $\Lambda_3=(\mpl m^2)^{1/3}$. Thus, the $\Lambda_3$ massive gravity theory may have a local, Lorentz invariant Wilsonian UV completion. The surviving island after imposing the positivity bounds is shown in figure \ref{MG_positivityBounds}, taken from \cite{deRham:2018qqo}.

\subsection{Consistent couplings to matter fields}\label{CouplingsToMatter_MG}
In the previous section we have seen the behaviour of one-loop quantum corrections and shown the stability of the theory below the Planck scale. Concerning the matter loops, we had assumed a standard matter coupling as in General Relativity. In this section we will investigate whether or not there exist alternative consistent couplings to the matter fields. We will go through all the possible couplings of the matter fields to the $g$ and $f$ metrics and study case by case their consistency. We will see that apart from the standard minimal coupling to $g$, there is a unique consistent non-minimal coupling to $g$ and $f$, which is constructed out of an effective metric.\\

$\bullet$ The first possibility of coupling the matter fields consists of coupling one specific matter field to either $g$ or to $f$, but not to both simultaneously
 \begin{eqnarray}
\mathcal{L}_{\rm matter}= \mathcal{L}_g[g,\chi_g] + \mathcal{L}_f[f,\chi_f]\,,
\end{eqnarray}
where $\chi_g$ and $\chi_f$ represent two separate matter fields. At the classical level, since the matter fields are minimally coupled to either the $g$ or the $f$ metric, this way of coupling is stable and does not introduce any ghost degree of freedom. Studying the quantum corrections shows also that this coupling does not destroy the special structure of the graviton potential. In fact, matter loops only give rise to contributions in form of cosmological constants for the $g$ and $f$ metric
\begin{eqnarray}
\mathcal{L}^{({\rm matter-loops})}_{1,\textrm{eff}}
\sim M^4_g \sqrt{-g} \log(M_g/\mu )+M^4_f \sqrt{- f} \log(M_f/\mu )\,,
\end{eqnarray}
where $M_g$ and $M_f$ refer here to the mass of the mater field that couples to $g$ and $f$ respectively. Thus, this way of coupling the matter sector is consistent at the classical as well as at the quantum level. \\

 $\bullet$ The second possibility consists of coupling the same matter sector to both metrics simultaneously
  \begin{eqnarray}
\mathcal{L}_{\rm matter}= \mathcal{L}_g[g,\chi] + \mathcal{L}_f[f,\chi].
\end{eqnarray}
 As it can be seen in the Lagrangian, the same matter field $\chi$ couples to both metrics. First of all, one immediate result is that this way of coupling actually results in a Hamiltonian which is not linear in the lapse even after an allowed redefinition of the shift \cite{Yamashita:2014fga,deRham:2014naa}. It means that with this coupling the ghost degree of freedom as the sixth degree of freedom of the graviton reappears in the spectrum already at the classical level. Therefore, this coupling reintroduces the ghost issue. The attempt of trying to argue that one could use it as an effective theory with a cutoff given by the ghost mass is doomed to fail since the quantum corrections of this coupling introduces also ghostly operator with a scaling that could be made arbitrarily small given by the mass $M$ of the matter field. A closer look at the one loop matter contributions reveals
 \begin{eqnarray}\label{second_coupling}
\mathcal{L}^{({\rm matter-loops})}_{1,\textrm{eff}}
\sim M^4 \sqrt{-g_{\rm eff}} \log(M/\mu )  \qquad \textrm{with} \qquad g_{\rm eff}^{\mu\nu}=\frac{\sqrt{-g}g^{\mu\nu}+\sqrt{-f}f^{\mu\nu}}{\sqrt{-g}+\sqrt{-f}}\,.
\end{eqnarray}
The one loop matter contributions results in a cosmological constant for an effective metric that has the specific form as in equation (\ref{second_coupling}). However, the square root of the determinant of this effective metric does not correspond to the right structure of the healthy potential interactions of massive gravity. This means that the quantum corrections generate ghostly operator with the mass $m^2_{\rm ghost}=\Lambda^6/M^4$. Hence, the mass of the ghost could be made arbitrarily small by considering very massive matter fields. This detuning of the potential structure at an arbitrarily low scale makes this coupling unacceptable as an effective field theory. \\

 $\bullet$ The third possibility of coupling matter field is a rather unusual setup. Imagine that the same matter sector couples to both metrics, but the coupling is such that the kinetic term of the matter field is connected to the $g$ metric whereas the potential term to the $f$ metric
   \begin{eqnarray}
\mathcal{L}_{\rm matter}= \mathcal{L}^{\rm kin}_g[g,\partial\chi] + \mathcal{L}^{\rm pot}_f[f,\chi].
\end{eqnarray}
Since the $f$ metric is absent in the kinetic term of the matter field, this way of coupling does not alter the linearity in the lapse of the Hamiltonian and hence there is no ghostly degree of freedom introduced by this coupling and it is classically stable \cite{Yamashita:2014fga,deRham:2014naa}. Nevertheless, it shares the same problem of quantum instability as the previous matter coupling. The one loop matter contributions take the form
 \begin{eqnarray}\label{third_coupling}
\mathcal{L}^{({\rm matter-loops})}_{1,\textrm{eff}}
\sim M^4 \sqrt{-g_{\rm eff}} \log(M/\mu )  \qquad \textrm{with} \qquad \sqrt{-g_{\rm eff}}=\frac{\det(f)}{\sqrt{-g}}\,.
\end{eqnarray}
As it can be seen in the above equation, the resulting effective metric of this way of coupling does not correspond to the right potential structure and the introduced ghost scales again as $m^2_{\rm ghost}=\Lambda^6/M^4$, which can be made arbitrarily small by the corresponding choice of the matter field. Even if this way of coupling is consistent at the classical level, it is not stable under quantum corrections. Thus, this option can be disregarded as well. \\

 $\bullet$ If one insists on coupling the same matter sector to both metrics without introducing any ghost at the classical level below the strong coupling scale of the theory, it turns out that there is a unique effective metric that can be constructed\footnote{In the vielbein language this correspond to the linear combination $e_{\rm eff}=\alpha e+\beta f$ \cite{Noller:2014sta}.} \cite{deRham:2014naa,deRham:2014fha}
 \begin{eqnarray}\label{effective_metric}
g_{\mu\nu}^{\rm eff}=\alpha^2g_{\mu\nu}+2\alpha\beta g_{\mu\rho}\left( \sqrt{g^{-1}f} \right)^\rho_\nu+\beta^2 f_{\mu\nu}.
\end{eqnarray}
If we couple the matter sector to both metric at the same time via this effective metric
   \begin{eqnarray}
\mathcal{L}_{\rm matter}= \mathcal{L}_{g_{\rm eff}}[g_{\rm eff},\chi]\,,
\end{eqnarray}
then the theory is free of any ghostly degree of freedom up to the strong coupling scale. This effective metric is unique in the sense that the quantum corrections generate contributions only in form of the potential interactions
 \begin{eqnarray}
\mathcal{L}^{({\rm matter-loops})}_{1,\textrm{eff}}
\sim M^4 \sqrt{-g_{\rm eff}} \log(M/\mu )  \quad \textrm{with} \quad \sqrt{-g_{\rm eff}}=\sqrt{-g}\det\left( \alpha+\beta\sqrt{g^{-1}f}\right).
\end{eqnarray}
Therefore, this way of matter coupling is consistent at the quantum level as well. Any effective metric built out of the two metrics that does not have this exact form as in equation (\ref{effective_metric}), will unavoidably detune the classical potential structure. \\

\begin{center}
\begin{tabular}{|c|c|c|c| }
  \hline
  \multicolumn{4}{|c|}{\color{black}{different couplings to matter}} \\
  \hline
  \color{green}{minimal} & \color{blue}{both} & \color{orange}{kinetic} &\color{yellow}{effective}   \\
   \hline
    $ \mathcal{L}_g[g,\chi_g] + \mathcal{L}_f[f,\chi_f]$ & $\mathcal{L}_g[g,\chi] + \mathcal{L}_f[f,\chi]$ & $\mathcal{L}^{\rm kin}_g[g,\partial\chi] + \mathcal{L}^{\rm pot}_f[f,\chi]$ & $\mathcal{L}_{g_{\rm eff}}[g_{\rm eff},\chi]$   \\
   \hline
    classically stable & unstable & stable &stable  \\
   \hline
     5 dof & 6 dof & 5 dof & 5 dof +1 (above strong \\
       &  &  & coupling scale)  \\
   \hline
       $M^4_g \sqrt{-g}+M^4_f \sqrt{- f}$ & $M^4 \sqrt{-g_{\rm eff}}$ &$M^4 \sqrt{-g_{\rm eff}}$& $M^4 \sqrt{-g_{\rm eff}}$  \\
   \hline
    \hline
        & $\frac{\sqrt{-g}g^{\mu\nu}+\sqrt{-f}f^{\mu\nu}}{\sqrt{-g}+\sqrt{-f}}$ &$\frac{\det(f)}{\sqrt{-g}}$& $\sqrt{-g}\det\left( \alpha+\beta\sqrt{g^{-1}f}\right)$  \\
   \hline
     quantum stable & unstable & unstable &stable  \\
   \hline
     & $m^2_{\rm ghost}=\Lambda^6/M^4$ & $m^2_{\rm ghost}=\Lambda^6/M^4$ &  \\
   \hline
     \hline
     \color{green}{ EFT \checkmark} &   \xcancel{\color{blue}{ EFT}} &\xcancel{\color{orange}{ EFT}}   & \color{yellow}{ EFT \checkmark} \\
   \hline
  \multicolumn{4}{|c|}{only the first and the last option are stable and consistent EFT couplings} \\
  \hline
\end{tabular}
\end{center}

Summarising, the only consistent and viable way of coupling the matter fields within the regime of validity of the theory is either to couple separate sectors of matter fields to $g$ or to $f$ but not simultaneously to both, or if one insists on coupling to both metrics then only via the effective metric in equation (\ref{effective_metric}).
This does not depend on whether the $f$ metric is promoted to be dynamical or not. The same result applies also to bigravity \cite{Heisenberg:2014rka}. This effective metric is unique in the sense that it is the only composite metric that does not yield any Boulware-Deser ghost within the decoupling limit and does not detune the potential structure through quantum corrections \cite{Heisenberg:2015iqa,Heisenberg:2015wja,Huang:2015yga,Matas:2015qxa}.

\section{Scalar-Tensor theories}\label{sec_scalar-tensor}
In the previous sections we saw how one can construct consistent theories for scalar and tensor fields separately. In this section we would like to investigate how the constructed theories for the scalar field in flat space-time can be generalised to curved space-time. In other words, we would like to study the consistent couplings between scalar and tensor fields without introducing ghostly unphysical degrees of freedom. The first attempt that we could do would consist of promoting the flat Minkowski metric to a general non-flat metric $\eta_{\mu\nu}\to g_{\mu\nu}$ and similarly the partial derivative to a covariant derivative $\partial_\mu\to \nabla_\mu$. For instance, the k-essence type of scalar field in curved space-time would become \cite{ArmendarizPicon:2000dh}
\begin{eqnarray}\label{kessenceLagrangian}
\mathcal{S}=\int d^4x\sqrt{-g}P(X,\pi) \quad \text{with} \quad X=\frac12 g^{\mu\nu}\nabla_\mu\pi\nabla_\nu\pi\,.
\end{eqnarray}
The stress energy density obtained by varying the action with respect to the metric $T_{\mu\nu}=P_{X}\nabla_\mu\pi\nabla_\nu\pi-g_{\mu\nu}P$ would not violate the Null Energy Condition if one imposes $P_{X}\ge0$, where $P_{X}$ stands for the partial derivative with respect to $X$. The equation of motion for the scalar field takes the form
\begin{eqnarray}
(P_{X}g^{\mu\nu}+P_{XX}\nabla^\mu\pi\nabla^\nu\pi)\nabla_\mu\nabla_\nu\pi+2XP_{X\pi}-P_{\pi}=0
\end{eqnarray}
As we can see the equations of motions are second order in derivatives in curved space-time as well. If we consider small perturbations on top of a background configuration, for the stability of the theory we have to impose $1+2XP_{XX}/P_{X}>0$ and the sound speed is given by $c_s^2=(1+2XP_{XX}/P_{X})^{-1}$. For the absence of Laplacian instabilities, we have to require $c_s^2>0$. For detailed studies of the k-essence field, see \cite{ArmendarizPicon:2000dh,ArmendarizPicon:2000ah,Mukhanov:2005bu,Copeland:2006wr,Babichev:2007dw,Tsujikawa:2013fta}.

\subsection{Horndeski interactions}
The derivative interaction of the scalar field arising from the decoupling limit of the DGP model, that corresponds to the cubic Galileon can be simply generalised to curved spacetime in a similar way 
\begin{eqnarray}
\mathcal{S}=\int d^4x\sqrt{-g}\left(-\frac12(\nabla\pi)^2+\frac{\alpha_1}{\Lambda^3}(\nabla\pi)^2\Box\pi\right)\,.
\end{eqnarray}
Since the interaction term is only linear in the connection, one does not need to add non-minimal couplings to gravity in order to maintain the equations of motion at most second order in derivatives. Thus, the equations of motion of the cubic Galileon on curved spacetimes still remain second order. How about the remaining Galileon interactions? Can we simply promote them to curved spacetime? These questions were first investigated in \cite{Deffayet:2009wt}. The naive covariantisation of the quartic and quintic Galileon interactions results in higher order equations of motion. If we take for instance the quartic Lagrangian $\mathcal{L}_4$ in (\ref{Gal_interactions}) and promote it to curved spacetime by $\eta_{\mu\nu}\to g_{\mu\nu}$ and $\partial_\mu\to \nabla_\mu$ $(\Pi_{\mu\nu}=\nabla_\mu\partial_\nu\pi)$, then the resulting equation of motion is no longer given by $\mathcal{E}_4$ in equation (\ref{EOMgalileons}), but rather contains higher order terms
\begin{equation}
\mathcal{E}_4\supset 2(\partial\pi)^2(\nabla_\nu\nabla^\nu\Pi_\rho^\rho-\nabla_\nu\nabla_\rho\Pi^{\nu\rho})+10\Box\pi\partial_\mu\pi(\nabla^\nu\Pi_{\mu\nu}-\nabla_\mu\Pi^\nu_\nu)+\cdots
\end{equation}
These higher order contributions at the level of the equations of motion can be compensated by introducing non-minimal couplings at that order
\begin{equation}
\mathcal{L}_4\supset (\partial\pi)^2G^{\mu\nu}\partial_\mu\pi \partial_\nu\pi\,.
\end{equation}
After integration by parts the covariant version of the quartic Galileon interactions takes the form
\begin{equation}
\mathcal{L}_4=(\partial\pi)^2\left( 2(\Box\pi)^2-2\Pi_{\mu\nu}\Pi^{\mu\nu}-\frac12(\partial\pi)^2R \right)\,.
\end{equation}
Hence, since the quartic Galileon interactions are non-linear in the connection, one has to introduce non-minimal coupling to the Ricci scalar in order to maintain the equations of motion second order.
Exactly the same happens also for the quintic Galileon interactions \cite{Deffayet:2009wt}. This gave rise to the rediscovery of the Horndeski interactions \cite{Horndeski:1974wa}. 

\begin{center}
\begin{figure}[h!]
\begin{center}
  \includegraphics[width=6.0cm]{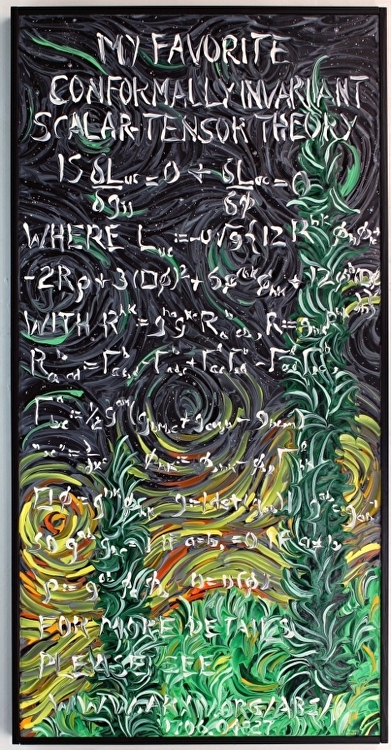}
\end{center}
  \caption{This picture is taken from Horndeski contemporary art gallery with the dedication "my favorite conformally invariant scalar-tensor field theory, completed 7/11/17". It is taken from \url{http://horndeskicontemporary.com/workszoom/2534281}.}
 \label{Horndeski_art}
\end{figure}
\end{center}

The question Horndeski was after was:  what is the most general Lorentz and diffeomorphism invariant scalar-tensor theory with second order equations of motion. The answer to this question led uniquely to the Horndeski interactions
\begin{eqnarray}
\mathcal{S}=\int d^4x\sqrt{-g}\left(\sum_{i=2}^5\mathcal{L}_i+\mathcal{L}_{\rm matter}\right)
\end{eqnarray}
with the individual Lagrangians expressed as
\begin{eqnarray}\label{HorndeskiGal}
\mathcal{L}_2&=&P(\pi,X)\nonumber\\
\mathcal{L}_3&=&-G_3(\pi,X)[\Pi]\nonumber\\
\mathcal{L}_4&=&G_4(\pi,X)R+G_{4,X}\left([\Pi]^2-[\Pi^2]\right)\nonumber\\
\mathcal{L}_5&=&G_5(\pi,X)G_{\mu\nu}\Pi^{\mu\nu}-\frac16G_{5,X}\left([\Pi]^3-3[\Pi][\Pi^2]+2[\Pi^3]\right)\,,
\end{eqnarray}
where the arbitrary functions $P$, $G_3$, $G_4$ and $G_5$ depend on the scalar field $\pi$ and its derivatives $X=-\frac12(\partial\pi)^2$ and furthermore $G_{i,X}=\partial G_i/\partial X$ and $G_{i,\pi}=\partial G_i/\partial \pi$. As it can be seen in the above expressions, the quartic and quintic interactions require the presence of non-minimal couplings to gravity via the Ricci scalar and the Einstein tensor. The relative tuning of the interactions in the quartic and quintic Lagrangians guarantee the second order nature of the equations of motion. Unfortunately, Horndeski quitted mathematics and became an artist. A quick look at the citations of his masterpiece at the time might let you wonder in silence and is livened up in its own right by now. The greater is the joy of seeing him becoming active again in physics \cite{Horndeski:2016bku}, and so {\it no one shall go beyond Horndeski without him}! Marveling at his gallery leaves no doubt about his greatness in arts as well (see figure \ref{Horndeski_art}).

Clearly, promoting the Galileon interactions to the curved spacetime breaks explicitly the Galileon symmetry, even though generalised Galileon symmetry can be constructed on maximally symmetric backgrounds \cite{Goon:2011qf,Burrage:2011bt}. Furthermore, the non-renormalisation theorem applies only on flat spacetime. However, in the case where the Galileon invariance is weakly broken, it has been shown that one can construct quasi de Sitter backgrounds, which are insensitive to loop corrections \cite{Pirtskhalava:2015nla}.

\subsection{A proxy theory to massive gravity}
As we saw in section \ref{sec_MassiveSpin2}, the decoupling limit of massive gravity contains Galileon interactions.
A subclass of these Horndeski interactions can be put in relation with massive gravity. This is achieved by covariantising the decoupling limit interactions of massive gravity. Consider the lowest order interaction between the helicity-0 and helicity-2 modes in the decoupling limit of massive gravity in equation (\ref{lagr1}) and perform an integration by part
\begin{eqnarray}\label{intDLX1}
h^{\mu\nu}X^{(1)}_{\mu\nu}=h^{\mu\nu}(\partial_\alpha\partial^\alpha\pi\eta_{\mu\nu}-\partial_\mu\partial_\nu\pi)
=(\Box h-\partial_\mu\partial_\nu h^{\mu\nu})\pi\,.
\end{eqnarray}
The last expression reminds us of the Ricci scalar in the weak field limit $R=\partial_\mu\partial_\nu h^{\mu\nu}-\Box h$. The direct covariantisation of (\ref{intDLX1}) would result in 
\begin{eqnarray}
(\Box h-\partial_\mu\partial_\nu h^{\mu\nu})\pi \;\;\;\; &\underbrace{\to}& -R\pi\,.\\
&{\rm cov.}&\nonumber
\end{eqnarray}
In a similar way we can apply the same procedure of integration by parts and covariantisation to the other two interactions of the decoupling limit Lagrangian of massive gravity, which gives the following correspondence
\begin{eqnarray}
h^{\mu\nu}X^{(1)}_{\mu\nu} &\longleftrightarrow& -\pi R\\
h^{\mu\nu}X^{(2)}_{\mu\nu} &\longleftrightarrow& -\partial_\mu\pi\partial_\nu\pi G^{\mu\nu}\\
h^{\mu\nu}X^{(3)}_{\mu\nu}  &\longleftrightarrow& -\partial_\mu\pi\partial_\nu\pi \Pi_{\alpha\beta} L^{\mu\alpha\nu\beta}\,.
\end{eqnarray}\label{eq:corres}
Hence, from the decoupling limit Lagrangian we construct a specific scalar-tensor theory
which relate the decoupling limit of massive gravity to a subclass of Horndeski interactions \cite{deRham:2011by}
\begin{equation} \label{eq:TotalAction}
\mathcal{L}=\sqrt{-g}\left(M_{\rm Pl}^2 R+M_{\rm Pl} \left(-\pi R-\frac{a_2}{\Lambda^3}\partial_\mu\pi\partial_\nu\pi G^{\mu\nu}-\frac{a_3}{\Lambda^6}\partial_\mu\pi\partial_\nu\pi \Pi_{\alpha\beta} L^{\mu\alpha\nu\beta}\right)+\mathcal L^{\rm matter}\right)\,,
\end{equation}
with $\Pi_{\mu\nu}=\nabla_\mu\partial_\nu\pi$.
 In terms of the Horndeski functions, this theory corresponds to the specific choice of the functions
 \begin{eqnarray}
&&P(\pi,X)=0, \quad G_3(\pi,X)=0,\nonumber\\
&&G_4(\pi,X)=M_{\rm Pl}^2-M_{\rm Pl}\pi-\frac{M_{\rm Pl}}{\Lambda^3}a_2X \quad \text{and} \quad G_5(\pi,X)=3\frac{M_{\rm Pl}}{\Lambda^6}a_3X
\end{eqnarray}
 in equation (\ref{HorndeskiGal}). Hence, in the same way that one can relate the decoupling limit of massive gravity with the Galileon interactions, one can relate its covariant version with a subclass of Horndeski interactions after covariantization (see \cite{deRham:2011by,Heisenberg:2014kea} for more detail). \\

\subsection{Beyond Horndeski}\label{sec_beyondHorndeski}
So far these interactions had the restriction of yielding second order equations of motion in order to avoid any Ostrogradski instability related to higher time derivatives. However, as initiated in \cite{Zumalacarregui:2013pma,Gleyzes:2014dya} this restriction is not necessary. Even in the presence of higher order equations of motion one can avoid the Ostrogradski instabilities if the presence of a constraint equation is still maintained. This resulted in new type of interactions. The interactions $\mathcal{L}_2$ and $\mathcal{L}_3$ remain the same but for the last two Lagrangians in $\mathcal{L}_4$ and $\mathcal{L}_5$ one obtains new contributions of the form \cite{Gleyzes:2014dya} 
\begin{eqnarray}\label{eq:beyondHorndeski}
\mathcal{L}_4^{N}&=&F_4\epsilon^{\alpha_1 \alpha_2\alpha_3\gamma_4}
\epsilon^{\beta_1 \beta_2 \beta_3}{}_{\gamma_4}\nabla_{\alpha_1}\pi\nabla_{\beta_1}\pi \Pi_{\alpha_2 \beta_2}\Pi_{\alpha_3 \beta_3} \nonumber\\
\mathcal{L}_5^{N}&=&F_5\epsilon^{\alpha_1 \alpha_2\alpha_3\alpha_4}
\epsilon^{\beta_1 \beta_2 \beta_3 \beta_4}\nabla_{\alpha_1}\pi\nabla_{\beta_1}\pi \Pi_{\alpha_2 \beta_2}\Pi_{\alpha_3 \beta_3} \Pi_{\alpha_4 \beta_4} \,.
\end{eqnarray}
These new interactions written in terms of the Levi-Civita tensors correspond simply to the covariant version of the quartic and quintic Galileons on flat space-time. Therefore, the resulting equations of motions are third order in derivatives but still avoiding unwanted ghostly degrees of freedom. That this is the case can be easily understood in the ADM formulation of the interactions. The Hamiltonian in terms of the ADM variables is given by
\begin{equation}
H=\int d^3x \left( \mathcal{P}^{ij}\dot{\gamma}_{ij}-\mathcal{L} \right)=\int d^3x \left( N\mathcal{H}_0+N^i\mathcal{H}_i \right)\,,
\end{equation}
where the introduced quantities $\mathcal{H}_0$ and $\mathcal{H}_i$ are given by
\begin{eqnarray}
\mathcal{H}_0&=& -\sqrt{\gamma} \left(  A_2-\frac38 \frac{A_3^2}{A_4}+\frac{A_3[\mathcal{P}]}{2\sqrt{\gamma}A_4}+\frac{2[\mathcal{P}^2]-[\mathcal{P}]^2}{2\gamma A_4}+B_4R \right)\nonumber\\
\mathcal{H}_i&=&-2D_j \mathcal{P}i^j{}_i\,,
\end{eqnarray}
with $D_j$ denoting the spatial covariant derivative, $\mathcal{P}^{ij}$ the conjugate momenta and the functions $A_i$ and $B_i$ representing 
\begin{eqnarray}
A_2&=& G_2-(-X)^{1/2}\int\frac{G_{3,\pi}}{2\sqrt{-X}}dX \nonumber\\
A_3&=&-\int G_{3,X}\sqrt{-X}dX-2\sqrt{-X}G_{4,\pi} \nonumber\\
A_4&=& -G_4+2XG_{4,X}+\frac{X}{2}G_{5,\pi}-X^2F_4 \nonumber\\
B_4&=&G_4+\sqrt{-X}\int \frac{G_{5,\pi}}{4\sqrt{-X}}dX \nonumber\\
A_5&=& -\frac{(-X)^{3/2}}{3}G_{5,X}+(-X)^{5/2}F_5\nonumber\\
B_5&=& -\int G_{5,X}\sqrt{-X}dX\,.
\end{eqnarray}
In the unitary gauge with the constant scalar field hypersurfaces, the time diffeomorphism is broken whereas the spatial ones are still being intact. The first class constraint related to the spatial diffeomorphism $\mathcal{H}_i$ removes six degrees of freedom, whereas the broken time diffeomorphism yields only a second class constraint $\mathcal{H}_0+N\partial \mathcal{H}_0/\partial N=0$ removing one additional degree of freedom. Hence, there are only three propagating degrees of freedom, where one of them is the scalar field. Note, that in the presence of only $\mathcal{L}_4^{N}$ one can map the new interaction into Horndeski by means of a disformal transformation. The same is true if only $\mathcal{L}_5^{N}$ is present, but not if both new terms are present. 

As we have imposed in the case of massive gravity (\ref{constraint_pot_MG_Hess}), the crucial point to remember is that the associated Hessian matrix has a vanishing determinant, which guarantees the presence of a primary constraint. In this case, these degenerate theories can avoid to have an Ostrogradski instability but still have higher order equations of motion. The construction of this type of beyond Horndeski interactions has created a lot of activities in the literature (see for instance \cite{Gleyzes:2014qga,Langlois:2015cwa,Langlois:2015skt,BenAchour:2016fzp}). The Hamiltonian analysis of the constraints system is often performed in the unitary gauge, where the constant scalar field hypersurfaces coincide with the constant time hypersurfaces. The results concluded in this particular gauge might not be applicable beyond this gauge \cite{Deffayet:2015qwa}.

\subsection{DHOST theories}\label{secDHOSTtheories}
Motivated by the Beyond Horndeski interactions, one can aim at constructing the most general higher order scalar-tensor theories with an additional primary constraint, that ensures the propagation of only three physical degrees of freedom. These are the Degenerate Higher-Order Scalar-Tensor (DHOST) theories \cite{Langlois:2015cwa,Achour:2016rkg} up to cubic order in second-order derivatives of the scalar field, satisfying the degeneracy condition, or in other words the vanishing of the determinant of the Hessian matrix\footnote{In fact the requirement of the degeneracy conditions is the very definition of first- and second-class constrained Hamiltonian systems with and without gauge symmetries introduced by Dirac himself and was used in 2011 to construct the degenerate conditions for massive gravity \cite{deRham:2011rn} and in 2014 to construct the degenerate conditions for generalised Proca interactions \cite{Heisenberg:2014rta}.}.

We can start with the most general Lagrangian up to cubic order in $\Pi_{\mu\nu}$ \cite{Langlois:2015cwa,Achour:2016rkg,BenAchour:2016fzp}
 \begin{eqnarray}
\label{action}
\mathcal{S} &=& \int d^4 x \, \sqrt{- g }
\left[ f_0(X,\pi) + f_1(X,\pi) \Box \pi
+
f_2(X,\pi) \, R+ C_{(2)}^{\mu\nu\rho\sigma} \,  \Pi_{\mu\nu} \, \Pi_{\rho\sigma}+
\right.
\cr
&&
\left. \qquad \qquad\qquad + f_3(X, \pi) \, G_{\mu\nu} \Pi^{\mu\nu}  +  
C_{(3)}^{\mu\nu\rho\sigma\alpha\beta} \, \Pi_{\mu\nu} \, \Pi_{\rho\sigma} \, \Pi_{\alpha \beta} \right]  \,.
\end{eqnarray}
The terms quadratic in $\Pi_{\mu\nu}$ are given by $C_{(2)}^{\mu\nu\rho\sigma} \,  \Pi_{\mu\nu} \, \Pi_{\rho\sigma} =\sum_{A=1}^{5}a_A(X,\pi)\,   \mathcal{L}^{(2)}_ A$, where the individual Lagrangians $\mathcal{L}^{(2)}_ A$ satisfy 
\be
\label{QuadraticDHOST}
\begin{split}
& \mathcal{L}^{(2)}_1 = \Pi_{\mu \nu} \Pi^{\mu \nu} \,, \qquad
\mathcal{L}^{(2)}_2 =(\Box \pi)^2 \,, \qquad
\mathcal{L}^{(2)}_3 = (\Box \pi) \partial^{\mu}\pi \Pi_{\mu \nu} \partial^{\nu}\pi \,,  \\
& \mathcal{L}^{(2)}_4 =\partial^{\mu}\pi  \Pi_{\mu \rho} \Pi^{\rho \nu} \partial_{\nu}\pi \,, \qquad
\mathcal{L}^{(2)}_5= (\partial^{\mu}\pi \Pi_{\mu \nu} \partial^{\nu}\pi)^2\,.
\end{split}
\ee
In a similar way, the cubic order interactions can be constructed as $C_{(3)}^{\mu\nu\rho\sigma\alpha\beta} \, \Pi_{\mu\nu} \, \Pi_{\rho\sigma} \, \Pi_{\alpha \beta} = \sum_{A=1}^{10} b_A(X,\pi)\,  L^{(3)}_A$, with the Lagrangians $\mathcal{L}^{(3)}_ A$ this time given by 
\be
\label{CubicDHOST}
\begin{split}
& L^{(3)}_1=  (\Box \pi)^3  \,, \quad
L^{(3)}_2 =  (\Box \pi)\, \Pi_{\mu \nu} \Pi^{\mu \nu} \,, \quad
L^{(3)}_3= \Pi_{\mu \nu}\Pi^{\nu \rho} \Pi^{\mu}_{\rho} \,,   \quad
 L^{(3)}_4= \left(\Box \pi\right)^2 \partial_{\mu}\pi \Pi^{\mu \nu} \partial_{\nu}\pi \,, \\
&L^{(3)}_5 =  \Box \pi\,  \partial_{\mu}\pi  \Pi^{\mu \nu} \Pi_{\nu \rho} \partial^{\rho}\pi \,, \quad
L^{(3)}_6 = \Pi_{\mu \nu} \Pi^{\mu \nu} \partial_{\rho}\pi \Pi^{\rho \sigma} \partial_{\sigma}\pi \,,   \quad
 L^{(3)}_7 =  \partial_{\mu}\pi \Pi^{\mu \nu} \Pi_{\nu \rho} \Pi^{\rho \sigma}  \partial_{\sigma}\pi \,, \\
&L^{(3)}_8 = \partial_{\mu}\pi   \Pi^{\mu \nu} \Pi_{\nu \rho} \partial^{\rho}\pi\, \Pi_{\sigma} \Pi^{\sigma\lambda} \partial_{\lambda}\pi \,,   \quad
 L^{(3)}_9 = \Box \pi \left(\partial_{\mu}\pi  \Pi^{\mu \nu} \partial_{\nu}\pi\right)^2  \,, \quad
L^{(3)}_{10} = \left(\partial_{\mu}\pi  \Pi^{\mu \nu} \partial_{\nu}\pi\right)^3 \,.
\end{split}
\ee
These general Lagrangians with the arbitrary functions propagate four modes, among which one unavoidably gives rise to an Ostrogradsky instability. One has to impose constraints on the free function in order to reduce the system to three propagating degrees of freedom. 

\begin{center}
\begin{figure}[h!]
\begin{center}
  \includegraphics[width=15.0cm]{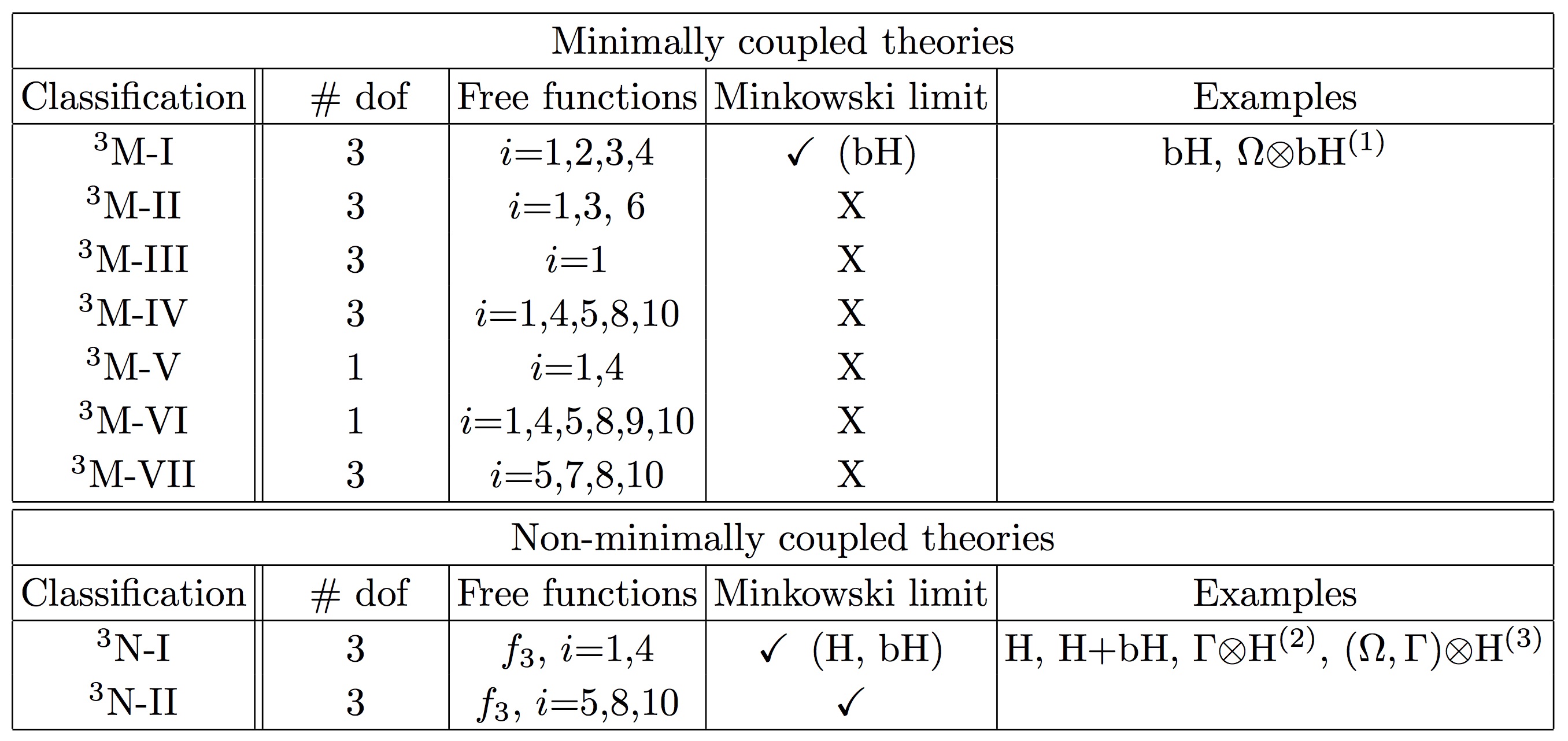}
\end{center}
  \caption{This table shows all the possible cubic degenerate classes. The subscript $i$ stands for the free $b_i$ functions, H for Horndeski and bH for beyond Horndeski, $\Omega$ for conformal transformations and $\Gamma$ for disformal transformations. One sees the 9 degenerate subclasses, where 7 correspond to the case with $f_3$ and 2 with $f_3\ne0$. See \cite{Achour:2016rkg} for a more detailed information on the classification.}
 \label{DHOSTclasses}
\end{figure}
\end{center}

Concentrating only on quadratic DHOST theories with the six arbitrary functions $f_2$ and $a_A$ (the functions $f_0(X,\pi)$ and $f_1(X,\pi)$ can be kept arbitrary since they are at most linear in the connection), the degeneracy condition translates into restrictions on $f_2$, $f_{2,X}$ and $a_A$, which classify the interactions into seven subclasses (three with the special case $f_2=0$ and four with $f_2\ne0$) \cite{Langlois:2015cwa}. For the particular choices of the functions $f_2=G_4$, $a_1=-a_2=2G_{4,X}+XF_4$ and $a_3=-a_4=2F_4$ one obtains the Horndeski interactions $\mathcal{L}_4$ in (\ref{HorndeskiGal}) and the beyond Horndeski interactions $\mathcal{L}_4^N$ in (\ref{eq:beyondHorndeski}).
Including the cubic DHOST theories enlarge the possible degenerate combinations significantly. There are nine degenerate subclasses, out of which seven correspond to the case with $f_3$ and two with $f_3\ne0$. Allowing the quadratic and cubic theories to coexist results in 25 independent degenerate combinations \cite{Achour:2016rkg}. The classification is shown in figure \ref{DHOSTclasses} taken from \cite{Achour:2016rkg}.

It is worth to emphasise that the DHOST theories are invariant under the most general disformal transformations $\tilde{g}_{\mu\nu}=C(\pi,X)g_{\mu\nu}+D(\pi,X)\partial_\mu\pi\partial_\nu\pi$. This is not the case for Horndeski and beyond Horndeski theories. In fact, the Horndeski theories are only invariant for the restricted functions $C_{,X}=0$ and $D_{,X}=0$ and the beyond Horndeski for the functions $C(\pi)$ and $D(\pi,X)$. Needless to say, in the presence of matter fields, the disformally related theories are physically not equivalent. In terms of the disformal transformations, DHOST theories can be divided into two main classes, those that are related to Horndeski and beyond Horndeski by means of disformal transformations and those which can not be obtained from Horndeski and beyond Horndeski by this type of transformations. The first type of theories (related by disformal transformations) are denoted as DHOST I and the latter ones by DHOST II in the literature.

\section{Vector Fields}

The Standard Model of Particle Physics contains as important force carriers abelian and non-abelian vector fields. This might motivate the consideration of vector fields in the cosmological evolution of the universe apart from scalar fields.
For this purpose, we shall consider a field that carries the spin-1 representation of the Lorentz group. 
 In four dimensions a vector field carries four numbers and in principal all of the four degrees of freedom could propagate, one of them always being a ghost. 
 How many real physical degrees of freedom propagate will depend on the presence or breaking of gauge symmetry in the same way as we had for the tensor fields.
 We have seen that the fundamental theory for a massless spin-2 field carries a gauge symmetry with two physical degrees of freedom. Introducing a mass term
 broke explicitly the gauge symmetry and a massive spin-2 field propagates three degrees of freedom. We will see that in an analogous way the presence of a gauge symmetry enforces the spin-1 field to carry only two physical modes, whereas the massive spin-1 field contains three after the gauge symmetry breaking.
 
 \subsection{Massless spin-1 field: Maxwell}
 Naively, we could start with the most general Lagrangian quadratic in the field on flat space-time
\begin{equation}\label{Lag_Maxwell}
\mathcal L_{A_\mu}=\frac{\alpha_1}2\partial_\mu A_\nu \partial^\mu A^\nu+\frac{\alpha_2}2\partial_\mu A_\nu \partial^\nu A^\mu,
\end{equation}
with the two arbitrary dimensionless parameters $\alpha_1$ and $\alpha_2$ (the possible contraction $(\partial_\mu A^\mu)^2$ is equivalent to the second term after integration by parts). In order to reduce the number of four degrees of
freedom down to two, we need to impose the presence of constraint equations. This on the other hand is guaranteed only if we 
ensure that the determinant for the Hessian vanishes
\begin{eqnarray}
\det{H_{\mathcal L_{A_\mu}}^{\mu\nu}}=\det{\frac{\partial^2\mathcal{L}_{A_\mu}}{\partial \dot A_\mu \partial \dot A_\nu}}=0.
\end{eqnarray}
The vanishing of the determinant should be achieved by making the temporal and longitudinal modes non-dynamical. Their
dynamics become transparent by looking at their conjugate momenta 
\begin{equation}
\Pi^0=\frac{\partial \mathcal{L}_{A_\mu}}{\partial \dot{A}_0}=(\alpha_1+\alpha_2)\dot{A}^0 \qquad 
\text{and} \qquad \Pi^i=\frac{\partial \mathcal{L}_{A_\mu}}{\partial \dot{A}_i}=\alpha_1\dot{A}^i+\alpha_2\partial^i A^0\,.
\end{equation} 
Hence, in order to make the temporal
component non-dynamical we need to impose $\alpha_2=-\alpha_1$. Without imposing this condition, the Hamiltonian density would
be in the general case
\begin{eqnarray}
\mathcal{H}_{\mathcal L_{A_\mu}}&=&\frac{\Pi_0^2}{2(\alpha_1+\alpha_2)}+\frac{\Pi_i \Pi^i}{\alpha_1}-\partial_i A_0 \Pi^i \left(\frac{2\alpha_2}{\alpha_1}+1\right)+\partial_i A_0\partial^iA^0\left(\alpha_1+\alpha_2\right)\frac{\alpha_2}{\alpha_1} \nonumber\\
&&-\frac12(\alpha_1\partial_i A_j\partial^iA^j+\alpha_2\partial_i A_j\partial^jA^i).
\end{eqnarray}
We see that the After fixing the condition $\alpha_2=-\alpha_1$, we immediately observe that the Hamiltonian simplifies to
\begin{eqnarray}\label{Hamiltonian_vector_gauge}
H_{\mathcal L_{A_\mu}}&=&\int d^3x \left(-\frac{\vec{\Pi}^2}{\alpha_1}-\nabla \vec{\Pi} A_0  -\frac{1}{4}F_{ij}F^{ij} \right).
\end{eqnarray}
In order for the transverse modes not to be a ghost degree of freedom, we have to further impose that $\alpha_1<0$. Otherwise, the Hamiltonian would become unbounded from below. After canonically normalizing the
vector field $\alpha_1=-1$, this uniquely leads to Maxwell theory 
\begin{equation}
\mathcal L_{A_\mu}=-\frac14F_{\mu\nu}^2 \qquad \text{with} \qquad F_{\mu\nu}=\partial_\mu A_\nu -\partial_\nu A_\mu \,.
\end{equation} 
The vanishing of the temporal momentum $\Pi^0=0$ is an important ingredient in the construction of a consistent kinetic term. It constitutes a primary constraint $\mathcal C_1=\Pi^0=0$. Related to it, one can generate a secondary constraint by computing the Poisson bracket between the primary constraint and the Hamiltonian density
\begin{equation}\label{eq_gaugesymm}
\mathcal C_2=\dot{\Pi}^0=\left\{ \mathcal C_1,\mathcal{H}_{\mathcal L_{A_\mu}}\right\}=\frac{\partial \mathcal{H}_{\mathcal L_{A_\mu}}}{\partial A_\mu}\frac{\partial \mathcal C_1}{\partial \Pi^\mu}-\frac{\partial \mathcal{H}_{\mathcal L_{A_\mu}}}{\partial \Pi^\mu}\frac{\partial \mathcal C_1}{\partial A_\mu}=\partial_i \Pi^i\sim 0.
\end{equation}
This is basically the time evolution of the primary constraint.
Computing the Poisson bracket of the primary constraint with the secondary constraint reveals
\begin{equation}\label{Poisson_gaugesymm}
\left\{ \mathcal C_1, \mathcal C_2 \right\}=0.
\end{equation}
This reflects the fact that we have a first class constraint, generating a gauge symmetry and hence removing one more degree of freedom, namely the longitudinal mode. One particularity of the Hamiltonian in equation (\ref{Hamiltonian_vector_gauge}) is the linear dependence on the temporal component of the vector field. This means that $A_0$ becomes a Lagrange multiplier after imposing the condition $\alpha_2=-\alpha_1$. The constraint associated to $A_0$ is first class and hence, one does not only get rid of the $A_0$ dependence but also the $\nabla \vec{\Pi}$ dependence.
Thus, a manifestly Lorentz invariant and local theory for a massless spin-1 field has to be invariant under gauge transformations
\begin{equation}\label{eq_gaugesymm}
A_\mu \to A_\mu + \partial_\mu \theta.
\end{equation}
This on the other hand guarantees the propagation of only two vector degrees of freedom, which is the case for the Maxwell term.

The vector field might also couple to other fields in form of a current. This uniquely gives Maxwell theory
 \begin{equation}\label{Lag_Maxwell}
\mathcal L_{A_\mu}=-\frac14F_{\mu\nu}^2-J_\mu A^\mu,
\end{equation}
with the external source $J_\mu$. The variation with respect to the vector field yields the equations of motion
\begin{equation}
\partial_\nu F^{\mu\nu}=J^\mu.
\end{equation}
Taking the divergence of the equation of motion yields $\partial_\mu J^\mu=0$, i.e. the external source is conserved. The equations of motion are gauge-invariant. We can fix the gauge by imposing the Lorenz condition $\partial_\mu A^\mu=0$. By doing so the equations of motion reduce to $\Box A^\mu=J^\mu$. The gauge choice together with the residual gauge $\Box \theta =0$ gets rid of the two unphysical modes. 

Even if one does not impose directly the $U(1)$ symmetry on the matter coupling, the consistency of the equations of motion inevitably introduces the gauge symmetry into the matter coupling. 
For a quick illustration of this statement, imagine that we have a complex scalar field as a matter field $\mathcal{L}_{(0)}=-\partial_\mu \phi \partial^\mu\phi^*$ with the global symmetries $\phi\to e^{i\alpha}\phi$ and $\phi^*\to e^{-i\alpha}\phi^*$. This global symmetry of the matter Lagrangian has an associated conserved Noether current that is obtained by varying the matter action
\begin{equation}
\delta S_{(0)}=\int d^4x \left( \Box\phi\delta\phi^*+\Box\phi^*\delta\phi-\partial_\mu(\partial^\mu\phi^*\delta\phi+\partial^\mu\phi\delta\phi^*) \right)\,,
\end{equation}
where the first two terms are the corresponding equations of motion and the last term in the parenthesis is the Noether current $j^\mu_{(0)}=ie(\phi\partial^\mu\phi^*-\phi^*\partial^\mu\phi)$ which is conserved on-shell with respect to the initial matter Lagrangian $\mathcal{L}_{(0)}$. The most natural coupling of the complex scalar field to the massless vector field would be via $A_\mu j^\mu_{(0)}$
\begin{equation}
\mathcal{L}_{(1)}=-\partial_\mu \phi \partial^\mu\phi^*+A_\mu j^\mu_{(0)}\,.
\end{equation}
However, this coupling will also contribute to the equations of motion of the vector field $\delta \mathcal{L}_{A_\mu}=\partial_\nu F^{\mu\nu}=j^\mu_{(0)}$ and yields an inconsistency. Since the field strength tensor $F_{\mu\nu}$ is antisymmetric, the divergence of the equations of motion of the vector field
\begin{equation}
\partial_\mu \delta \mathcal{L}_{A_\mu}=0=\partial_\mu\partial_\nu F^{\mu\nu}=\partial_\mu j^\mu_{(0)}\ne0\,
\end{equation}
gives a vanishing identity on the left hand side but a non-zero term on the right hand side, since $j^\mu_{(0)}$ is only conserved on-shell, but with respect to $\mathcal{L}_{(0)}$ and not with respect to the scalar equations of motion from $\mathcal{L}_{(1)}$, that contains new contributions from $A_\mu j^\mu_{(0)}$. The crucial point to bear in mind is that the Noether current is only conserved on-shell. Thus, we have to compute the associated Noether current with respect to $\mathcal{L}_{(1)}$. This now gives
\begin{eqnarray}
\delta S_{(1)}&=&\int d^4x \Big( (\Box\phi-ieA_\mu\partial^\mu\phi)\delta\phi^*+(\Box\phi^*-ieA_\mu\partial^\mu\phi^*)\delta\phi \nonumber\\
&&-\partial_\mu(j^\mu_{(0)}-ieA^\mu\phi^*\delta\phi-ieA^\mu\phi\delta\phi^*) \Big)\,,
\end{eqnarray}
with the next leading contribution of the Noether current in the form
\begin{equation}
j^\mu_{(1)}=j^\mu_{(0)}-2e^2A^\mu\phi\phi^*\,.
\end{equation}
The promoted Lagrangian 
\begin{equation}
\mathcal{L}_{(2)}=-\partial_\mu \phi \partial^\mu\phi^*+A_\mu j^\mu_{(1)}
\end{equation}
has now a local gauge symmetry $A_\mu\to A_\mu+\partial_\mu \theta$, $\phi\to \phi+e^{i\theta}\phi$ and $\phi^*\to \phi+e^{-i\theta}\phi^*$. Thus, even if we do not impose a local gauge symmetry on the matter coupling at the beginning, the consistency of the equations of motion enforces this upon us. By defining the derivative as $D_\mu\phi=(\partial_\mu+iA_\mu)\phi$ and $D_\mu\phi^*=(\partial_\mu-iA_\mu)\phi^*$, we can write the interacting Lagrangian simply as $\mathcal{L}=-\frac14F\mn F^{\mu\nu}-D_\mu\phi D^\mu\phi^*$, that has manifestly the local gauge symmetry.

 \subsection{Massive spin-1 field: Proca}
As in the case of the spin-2 field, adding a mass term to the Maxwell action breaks explicitly the gauge symmetry (\ref{eq_gaugesymm}). We can construct a mass term simply by contracting its space-time indices $A_\mu A^\mu$. The Lagrangian after the inclusion of the mass term becomes 
\begin{equation}\label{LagSimpleProca}
\mathcal L_{A_\mu}=-\frac14F_{\mu\nu}^2-\frac12m_A^2A_\mu^2-J_\mu A^\mu\,,
\end{equation}
which represents a massive spin-1 field with three propagating degrees of freedom. In order to understand this statement, let us have a look at the Hamiltonian. 
The above Hamiltonian of the massless vector field in equation (\ref{Hamiltonian_vector_gauge}) with $\alpha_1=-1$ now becomes through the inclusion of the mass term
\begin{eqnarray}\label{Hamiltonian_simpleProca}
H^m_{\mathcal L_{A_\mu}}&=&\int d^3x \left(\vec{\Pi}^2-\nabla \vec{\Pi} A_0  -\frac{1}{4}F_{ij}F^{ij}+\frac{m^2}{2}(A_0^2-\vec{A}^2) \right).
\end{eqnarray}
The secondary constraint obtains an additional term coming from the mass term $\mathcal C_2=\partial_i \Pi^i-m^2A_0\sim 0$ and consequently the Poisson bracket of the primary constraint with the secondary constraint does not vanish any longer $\left\{ \mathcal C_1, \mathcal C_2 \right\}=m^2$. In the presence of the mass term we have a second class constraint that removes only one degree of freedom in difference to the massless case. Thus, the massive spin-1 field propagates three degrees of freedom. If we compare the expression of the Hamiltonian for the Proca field (\ref{Hamiltonian_simpleProca}) with the Hamiltonian of the massless case (\ref{Hamiltonian_vector_gauge}), we immediately observe that the temporal component is no longer a Lagrange multiplier in the Proca theory, since it appears quadratic in the Hamiltonian. However, it is still not dynamical and plays the role of an auxiliary field, and its associated equation of motion imposes a constraint equation.

The equation of motion of the vector field modifies into
\begin{equation}
\partial_\nu F^{\mu\nu}-m_A^2 A^\mu=J^\mu\,.
\end{equation}
The divergence of the equation of motion results in the constraint
\begin{equation}
-m_A^2\partial_\mu A^\mu=\partial_\mu J^\mu
\end{equation}
The equation of motion simply becomes a Klein-Gordon equation $(\Box-m_A^2) A^\mu=J^\mu$, together with the condition $\partial_\mu A^\mu=0$ for conserved currents. Similarly as for the spin-2 field the gauge invariance can be restored
using the Stueckelberg trick. For this purpose we can perform the field redefinition
\begin{equation}
A_\mu \to A_\mu + \partial_\mu \pi.
\end{equation}
The Lagrangian for the massive spin-1 field (\ref{LagSimpleProca}) takes the following form after this change of variables
\begin{equation}
\mathcal L_{A_\mu}=-\frac14F_{\mu\nu}^2-\frac12m_A^2(A_\mu+\partial_\mu\pi)^2-J_\mu (A^\mu+\partial_\mu \pi)\,,
\end{equation}
which is now invariant under the simultaneous transformations $A_\mu \to A_\mu + \partial_\mu \theta$ and $\pi \to \pi-\theta$. After canonically normalizing the additional field $\pi\to \frac{1}{m_A}\pi$ the interactions read 
\begin{equation}
\mathcal L_{A_\mu}=-\frac14F_{\mu\nu}^2-\frac12m_A^2A_\mu^2-\frac12(\partial\pi)^2-m_AA_\mu\partial^\mu\pi-J_\mu \left(A_\mu+\frac{\partial_\mu\pi}{m_A} \right).
\end{equation}
We can also take the $m_A\to0$ limit in a continuous way. The resulting theory in this limit describes a decoupled massless scalar field and 
a massless vector field
\begin{equation}
\mathcal L_{A_\mu}=-\frac14F_{\mu\nu}^2-\frac12\partial\pi^2-J_\mu A^\mu
\end{equation}
This reflects the fact, that the $m_A\to 0$ limit of the massive spin-1 field does not give rise to any vDVZ discontinuity as it is the case for a massive spin-2 field.\\

In the massless case we have a first class constraint system, where the primary and secondary constraints form a closed algebra with $\{\mathcal{C}_1,\mathcal{C}_2 \}=0$. The temporal component $A_0$ plays the role of a Lagrangemultiplier. Only two propagating degrees of freedom remain after imposing the condition associated to $A_0$. 
\begin{center}
\begin{tabular}{ |l|l| }
  \hline
  \multicolumn{2}{|c|}{\color{blue}{first class constraints}} \\
  \hline
  primary constraint & $\mathcal{C}_1=0$ \\
   \hline
  secondary constraint & $\mathcal{C}_2=\dot{\mathcal{C}}_1=0$ \\
    \hline
  \multicolumn{2}{|c|}{\color{blue}{ gauge} $\;\;\; \{\mathcal{C}_1,\mathcal{C}_2 \}=0$} \\
  \hline
\end{tabular}
\end{center}

In the massive case we have a second class constraint system, where the primary and secondary constraints do not form a closed algebra anymore $\{\mathcal{C}_1,\mathcal{C}_2 \}\ne0$. The temporal component $A_0$ plays the role of an auxiliary field. Therefore, three propagating degrees of freedom survive after imposing the condition associated to $A_0$. 
\begin{center}
\begin{tabular}{ |l|l| }
  \hline
  \multicolumn{2}{|c|}{\color{blue}{second class constraints}} \\
  \hline
  primary constraint & $\mathcal{C}_1=0$ \\
   \hline
  secondary constraint & $\mathcal{C}_2=\dot{\mathcal{C}}_1=0$ \\
    \hline
  \multicolumn{2}{|c|}{ \xcancel{\color{blue}{ gauge}}$\;\;\; \{\mathcal{C}_1,\mathcal{C}_2 \}=m^2$} \\
  \hline
\end{tabular}
\end{center}

Let us finish this subsection by commenting on the repulsive nature of a vector field. If we have two conserved currents, we can compute the exchange amplitude of a spin-1 particle between them. The propagator of the vector field obeys
\begin{equation}
D_{\mu\nu}=\frac{\eta_{\mu\nu}}{\Box-m_{A_\mu}^2}\,.
\end{equation}
Hence, the exchange amplitude reads
\begin{equation}
Q(j)=-\frac12\int \frac{d^4k}{(2\pi)^4}j^{*\mu}\frac{\eta_{\mu\nu}}{k^2-m_{A_\mu}^2}j^\nu\,.
\end{equation}
For the charges, the $00$- component of the above expression gives
\begin{equation}
Q(j)=+\frac12\int \frac{d^4k}{(2\pi)^4}j^{*0}\frac{1}{k^2-m_{A_\mu}^2}j^0\,.
\end{equation}
If we compare this with what we had for the spin-0 and spin-2 fields in equations (\ref{exchange_ampl_scalar}) and (\ref{exhangeAmp_masslessSpin2}) respectively, we see that the exchange of a spin-1 particle gives rise to a repulsive force. The exchange amplitudes comes with a sign difference compared to what we had for the spin-0 and spin-2 fields.

\subsection{Massless Vector Galileons}\label{sec_MasslessVectorGal}
We have seen how the requirement to construct a healthy ghost-free theory for a massless vector field uniquely leads to the Maxwell theory.
One immediate question that comes in mind is whether one can construct derivative self-interactions for a massless vector field in the same spirit as the Galileon scalar interactions without altering the number of propagating degrees of freedom. There has been already some attempts in the literature to construct vector Galileon interactions beyond the Maxwell kinetic term with second order interactions at the level of the Lagrangian but still with second order equations of motion on flat space-times \cite{Deffayet:2013tca}. The idea was to construct the derivative self-interactions such that the gauge symmetry would remain preserved and only two physical degrees of freedom would propagate. The conclusion was that the Maxwell kinetic term is the unique term fulfilling these conditions for an abelian massless vector field. This No-go result can be understood easily. The interactions that one is after are of the following schematic form
\begin{equation}
\mathcal L=\epsilon^{\mu_1\mu_2\cdots} \epsilon^{\nu_1\nu_2\cdots} F_{\mu_1\mu_2}\cdots F_{\nu_1\nu_2}\cdots \left(\partial_{\mu_k}F_{\nu_l \nu_{l+1}}\cdots \right) \left(\partial_{\mu_j}F_{\nu_m \nu_{m+1}}\cdots \right)
\end{equation}
In four dimensions since the two Levi-Civita tensors do not have sufficient space-time indices to be contracted, one can not construct such terms. On the other hand, even if in five dimensions such a contraction becomes possible $\epsilon^{\mu\nu\rho\sigma\kappa}\epsilon^{\alpha\beta\gamma\delta\xi}F_{\mu\nu}F_{\mu\nu}\partial_\rho F_{\gamma\delta}\partial_\xi F_{\sigma\kappa}$, the resulting interactions correspond to just total derivatives. The construction in terms of the Levi-Civita tensors makes this No-go result apparent immediately. One can obtain the same conclusion by adapting to the same approach as Horndeski applied to construct the consistent scalar-tensor theories with second order equations of motion, which was also pursued in \cite{Deffayet:2013tca}. Starting from the equations of motion we can write down all the possible terms as polynomial functions of $\partial F$ and $F$ with arbitrary coefficients and demand the second order nature of the equations. Once this is accomplished one can then construct the Lagrangian from the equations of motions. The equations of motion of the vector field should be of the form
\begin{equation}\label{gelAnsatz_EOM_vecGal}
 E^\alpha=E^\alpha(A_\mu, \partial_\nu A_{\mu}, \partial_\nu\partial_\rho A_{\mu}, \eta_{\mu\nu},\mathcal{E}_{\mu\nu\rho\sigma} )\,.
\end{equation}
The question is whether these equations of motion can be constructed from a variational principle of an action $\delta \mathcal{S}=\int d^4x E^\mu \delta A_\mu$ avoiding any contributions with $\partial^3A$ or higher order. The integrability conditions of the equations of motion are \cite{Deffayet:2013tca}
\begin{eqnarray}
&&\frac{\partial E^{\mu}}{\partial A_\nu}-\frac{\partial E^{\nu}}{\partial A_\mu}+\partial\rho\left(\frac{\partial E^{\nu}}{\partial A_{\mu,\rho}} -\partial_\sigma \frac{\partial E^{\nu}}{\partial A_{\mu,\rho\sigma}} \right) =0 \nonumber\\
&&\frac{\partial E^{\nu}}{\partial A_{\mu,\rho}} +\frac{\partial E^{\mu}}{\partial A_{\nu,\rho}} -2\partial_\sigma \frac{\partial E^{\nu}}{\partial A_{\mu,\rho\sigma}}=0 \nonumber\\
&&\frac{\partial E^{\mu}}{\partial A_{\nu,\rho\sigma}} -\frac{\partial E^{\nu}}{\partial A_{\mu,\rho\sigma}}=0
\end{eqnarray}
Furthermore, imposing the invariance under gauge symmetries brings the form of (\ref{gelAnsatz_EOM_vecGal}) into (removes the explicit dependence on $A_\mu$)
\begin{equation}\label{gaugeAnsatz_EOM_vecGal}
 E^\alpha=E^\alpha( \partial_\nu A_{\mu}, \partial_\nu\partial_\rho A_{\mu}, \eta_{\mu\nu}, \mathcal{E}_{\mu\nu\rho\sigma} )
\end{equation}
and enforces given conditions on the dependence of $E^\mu$ on $A_\mu$ and its derivatives
\begin{eqnarray}
&&\frac{\partial E^{\mu}}{\partial A_{\nu,\rho\sigma}}+\frac{\partial E^{\mu}}{\partial A_{\rho,\nu\sigma}}+\frac{\partial E^{\mu}}{\partial A_{\sigma,\nu\rho}}=0\\
&&\frac{\partial E^{\mu}}{\partial A_{\nu,\rho}}+\frac{\partial E^{\mu}}{\partial A_{\rho,\nu}}=0 \\
&&\frac{\partial E^{\mu}}{\partial A_{\nu}}=0
\end{eqnarray}
Combining the properties of these conditions results in the final condition \cite{Deffayet:2013tca}
\begin{equation}\label{gauge_vecGal_nogo}
\frac{\partial^2 E^{\mu}}{\partial A_{\nu_1,\rho_1\sigma_1} \partial A_{\nu_2,\rho_2\sigma_2}}=0\,.
 \end{equation}
 This means that the equations of motion of the vector field are at most linear in $A_{\nu,\rho\sigma}$ and have to have the following form
\begin{equation}\label{gaugeAnsatz_EOM_vecGal}
 E^\mu=K_1^{\mu\alpha\beta\gamma}A_{\alpha,\beta\gamma}+K_2^\mu\,,
 \end{equation}
 where $K_1$ and $K_2$ depend only on $(\partial_\nu A_{\mu}, \eta_{\mu\nu}, \mathcal{E}_{\mu\nu\rho\sigma})$. Hence, one can not have vector Galileon interactions since the equations of motion are at most linear in the second derivatives of the vector field and is a clear No-go result. A similar analysis can be also performed for an arbitrary p-form Galileon \cite{Deffayet:2010zh,Deffayet:2016von,Deffayet:2017eqq}.

\subsection{Generalized Proca theories}\label{sec_GeneralizedProcaTheories}
In the previous subsection we have seen that one can not construct consistent and non-trivial vector Galileon interactions with second order equations of motion for a gauge invariant vector field.
A natural question that immediately arises is whether we can generalize the interactions of a massive vector field, the Proca field, without changing the propagating number of degrees of freedom yielding second order equations of motion. The idea is to avoid the previous No-go result by abandoning the gauge invariance of the vector field in the hope to construct new derivative self-interactions for a vector field \cite{Heisenberg:2014rta}. We would like to construct more general interactions and maintain three propagating degrees of freedom, namely two transverse and one longitudinal mode of the vector field. For this purpose, we will impose the two following conditions:
\begin{itemize}
\item The equations of motion are second order.
\item The temporal component of the vector field $A_0$ should not carry any dynamics.
\end{itemize} 
Imposing only the first condition is not enough since second order equation of motion for the temporal component would mean that $A_0$ is a dynamical ghost degree of freedom. We know that the massive spin-1 representation of the Lorentz group should only carry three dynamical fields and hence the inclusion of derivative self-interactions should not alter this property.

First of all, we can promote the mass term to a general potential term $V(A^2)$. Since there is not any derivative of the vector field involved, this trivially satisfies our requirement of not modifying the spectrum of propagating degrees of freedom. 
Similarly, any gauge invariant interactions constructed out of $F\mn$ and its dual $\tilde{F}\mn$ will not cause the propagation of a forth mode. Moreover, interactions which do not contain any dynamics for the zeroth component of the vector field $A_0$ will not change this property, like for instance $A^\mu A^\nu F_{\mu}{}^{\alpha}F_{\nu\alpha}$. We can collect all these type of interactions into the function
\begin{equation}
\mathcal L_2  = f_2(A_\mu, F_{\mu\nu}, \tilde{F}_{\mu\nu}).
\end{equation}
This function contains all the contractions between $A_\mu$, $F_{\mu\nu}$ and $\tilde{F}_{\mu\nu}$. For instance, it contains the gauge invariant interactions like $f_2 \supset (F^2)^n$, or even parity violating terms $f_2 \supset (F\tilde{F})^n$, and terms that have dependence on $A_0$ but without time derivatives applied on it like in $f_2 \supset A_\mu A_\nu F^{\rho\mu}F_{\rho}^\nu$... etc. Thus, the independent contractions will be in form of\footnote{Note, that $X$ in (\ref{kessenceLagrangian}) would correspond to the longitudinal scalar mode and hence we kept the same notation for vector. In this section we will refer $X$ to the vector norm. However, whenever there might be a confusion, we will use distinguished notations.} $X=-A_\mu A^\mu/2$, $F=-F_{\mu\nu}F^{\mu\nu}/4$ and $Y=A^\mu A^\nu F_{\mu}{}^{\alpha}F_{\nu\alpha}$ \cite{Heisenberg:2014rta,Fleury:2014qfa} and we can also rewrite the function in terms of these three scalar quantities (ignoring the partiy violating scalar quantity $\tilde{F}=F_{\mn} \tilde{F}^{\mn}$)
\begin{equation}
\mathcal L_2  =f_2(X,F,Y)\,.
\end{equation}
At this order, these are the only unique interactions that do not carry any dynamics for the temporal component of the vector field and give rise to second order equations of motion.

As next, in first order in derivatives of $A_\mu$, we can only have the following interaction
\begin{equation}
\mathcal L_3  = f_3(X)\; \partial_\mu A^\mu \label{vecGalL3}\,,
\end{equation}
where $f_3$ is an arbitrary function of the vector field norm $X=-A_\mu A^\mu/2$. The presence of this function makes that the interaction does not correspond to a total divergence. The temporal component of the vector field does not obtain any dynamics from this interaction, even in the presence of the Maxwell kinetic term. The corresponding Hessian matrix vanishes identically 
\begin{eqnarray}
H_{\mathcal L_3}^{\mu\nu}=0.
\end{eqnarray}
The form of the interaction and why it is the unique interaction at this order becomes transparent when it is written in terms of the Levi-Civita tensor
 \begin{equation}\label{GenProcaL3}
\mathcal{L}_3=-\frac{f_3(X)}{6}\mathcal{E}^{\mu\nu\rho\sigma}\mathcal{E}^{\alpha}_{\;\;\nu\rho\sigma}\partial_\mu A_\alpha=f_3(X)\partial_\mu A^\mu \,.
 \end{equation}
 At this order, there is only one way of contracting the indices of the Levi-Civita tensors with $\partial_\mu A_\alpha$ and this term gives rise to the cubic Galileon interaction for the longitudinal mode in the decoupling limit ${\mathcal{E}}^{\mu\alpha\rho\sigma}{{\mathcal{E}}^{\nu\beta}}_{\rho\sigma}\Pi_{\mu\nu}\partial_{\alpha}\pi\partial_\beta\pi$ given in (\ref{eqsGal_LeviC}). Instead of contracting the indices of the Levi-Civita tensors with the hidden metrics, we could also contract them with the vector field in the schematic form $\mathcal{E}\mathcal{E} A^2 \partial A $. To be more precise, we can construct interactions as $\tilde{f}_3(X)\mathcal{E}^{\mu\nu\rho\sigma}\mathcal{E}^{\alpha\beta}{}_{\rho\sigma}\partial_\mu A_\alpha A_{\nu}A_{\beta}$,
which would result in a contraction of the form 
 \begin{equation}\label{GenProcaL3al}
\tilde{\mathcal{L}}_3=\tilde{f}_3(X)A^\mu A^\nu (\partial_\mu A_\nu),
 \end{equation}
which is related to the above one in equation (\ref{GenProcaL3}) by means of a disformal transformation of the metric $\eta_{\mu\nu}\rightarrow \eta_{\mu\nu}+\tilde{f}_3(X)A_\mu A_\nu$. We will not distinguish between these type of interactions, since they are equivalent after integration by parts and can be related to each other by disformal transformations. If we try to construct terms with two more additional vector fields in the form $\mathcal{E}\mathcal{E} A^4 \partial A $, where all of the vector fields would be contracted with the space-time indices of the Levi-Civita tensors, we see immediately that these interactions vanish identically, since two of the $A's$ are symmetric whereas the Levi-Civita tensors are antisymmetric (of course if some of the vector fields are contracted among themselves then they could again be absorbed into the disformal transformation in $\tilde{f}_3(X)A_\mu A_\nu$).

In the next order in derivatives of the vector field we have the following three possible ways of contracting the indices
\begin{equation}
\mathcal L_4  =  f_4(X) \; \left[c_1(\partial\cdot A)^2+c_2\partial_\rho A_\sigma \partial^\rho A^\sigma+c_3\partial_\rho A_\sigma \partial^\sigma A^\rho\right]  \label{vecGalL4gen}
\end{equation}
where $f_4$ is again an arbitrary function of the vector norm and $c_1$, $c_2$ and $c_3$ are free parameters (even though $c_1$ and $c_3$ give rise to the same contributions after integration by parts). We have to fix them in a way, that guarantees that only three physical degrees of freedom are dynamical. For this, we have to enforce the presence of a second class constraint.
This on the other hand is ensured, if the determinant of the Hessian matrix is zero \cite{Heisenberg:2014rta}. It is given by
\begin{eqnarray}
H_{\mathcal L_4}^{\mu\nu}=\frac{\partial^2\mathcal{L}_4}{\partial \dot A_\mu \partial \dot A_\nu} =f_4
\begin{pmatrix}
2(c_1+c_2+c_3)&0&0&0 \\
0&-2c_2 &0&0 \\
0&0&-2c_2 &0 \\
0&0&0&-2c_2
 \end{pmatrix}\,.
\end{eqnarray}
As can be clearly seen from the eigenvalues of the Hessian matrix, the determinant vanishes for two choices. We either choose $c_2=0$ or $c_1+c_2+c_3=0$. In the first case, three of the eigenvalues automatically vanish with only the zero component of the vector field propagating. In the latter case, only one eigenvalue vanishes and three of the vector components propagate. The for us interesting case is the second one. Choosing $c_1=1$, this condition is equal to $c_3=-(1+c_2)$, hence the Lagrangian with three propagating degrees of freedom at this order becomes
\begin{equation}
\mathcal L_4  =  f _4\; \left[(\partial\cdot A)^2+c_2\partial_\rho A_\sigma \partial^\rho A^\sigma-(1+c_2)\partial_\rho A_\sigma \partial^\sigma A^\rho\right]  \label{vecGalL4}
\end{equation}

The form of these interactions at the quadratic order in derivatives of the vector field becomes transparent again when we write them in terms of the Levi-Civita tensors. In the symmetric scalar Galileon $\Pi_{\mu\nu}$ in the interactions (\ref{eqsGal_LeviC}) we had only one way of contacting the indices of the antisymmetric Levi-Civita tensors with $\Pi_{\mu\nu}$. On the contrary, for the Proca vector field we have the symmetric and antisymmetric parts of $\partial_\mu A_\nu$ and hence we have two different independent ways of contracting the indices with the Levi-Civita tensors
\begin{eqnarray}
\mathcal{L}_4&=&-\frac{1}{2}\mathcal{E}^{\mu\nu\rho\sigma}\mathcal{E}^{\alpha\beta}_{\;\;\;\;\rho\sigma}(f_4(X)\partial_\mu A_\alpha\partial_\nu A_\beta+c_2\tilde{f}_4(X)\partial_\mu A_\nu\partial_\alpha A_\beta)\nonumber\\
&=&f _4\; \left[(\partial\cdot A)^2-\partial_\rho A_\sigma \partial^\sigma A^\rho\right]+c_2\tilde{f} _4(\partial_\rho A_\sigma \partial^\rho A^\sigma-\partial_\rho A_\sigma \partial^\sigma A^\rho) \,.
\end{eqnarray}
The terms proportional to $c_2$ are just the field strength tensor $F$. Therefore, we can rewrite the Lagrangian as
\begin{equation}\label{quartic_vecGal}
\mathcal L_4  = f _4 \; \left[(\partial\cdot A)^2-\partial_\rho A_\sigma \partial^\sigma A^\rho\right] + c_2 \tilde{f} _4F_{\rho\sigma}^2\,.
\end{equation}
We can simply reabsorb the contribution in form of the field strength tensor in $\mathcal L_2$ in the function $f_2$ instead of introducing redundancies at the present order. Hence, we shall neglect the contribution $c_2 \tilde{f} _4F_{\rho\sigma}^2$ in $\mathcal L_4$.

Instead of the metric, we could have contracted the indices of the Levi-Civita tensors with the vector field $A_\mu$ in form of $\mathcal{E}^{\mu\nu\rho\delta}\mathcal{E}^{\alpha\beta\sigma}{}_{\delta}\partial_\mu A_\alpha\partial_\nu A_\beta A_\rho A_\sigma$. In this way, we would have constructed terms of the form 
\begin{equation}
\mathcal L_4  =\hat{f} _4(X)A^\mu A^\nu(\partial_\nu A_\mu(\partial\cdot A)-\partial_\nu A_\rho \partial^\rho A_\mu).
\end{equation} 
These interactions are again at the same footing as the interactions in equation (\ref{quartic_vecGal}), since they correspond to just a disformal transformation of the previous ones and their equivalence becomes apparent after integrations by part. Similarly, the other possible contraction $\mathcal{E}^{\mu\nu\rho\delta}\mathcal{E}^{\alpha\beta\sigma}{}_{\delta}\partial_\mu A_\nu\partial_\alpha A_\beta A_\rho A_\sigma$ will give rise to contributions of the form $[FFAA]$ that are already included in $\mathcal L_2$. The attempt to contract two more additional vector fields with the Levi-Civita tensors in the schematic form $\mathcal{E}\mathcal{E} A^4 \partial A \partial A$ fails again due to the antisymmetric structure of the interactions.

As we mentioned before, the vanishing of the Hessian matrix indicates the existence of a constraint that can get rid of the unwanted temporal component of the vector field. Its presence reveals itself after computing the corresponding conjugate momentum $\Pi^\mu_{\mathcal L_4}=\frac{\partial \mathcal L_4}{\partial\dot A_\mu}$. The zero component of it $\Pi^0_{\mathcal L_4}=-2f_4\; \vec{\nabla}\vec{A}$ does not have any time derivatives and hence yields the constraint equation 
\begin{equation}
\mathcal C_1=\Pi^0_{\mathcal L_4}+2f_4\; \vec{\nabla}\vec{A}. 
\end{equation}
Associated to this constraint equation, there is a secondary constraint imposed by
 \begin{equation}
\{ H, \mathcal C_1\}=\frac{\partial H}{\partial A_\mu}\frac{\partial \mathcal C_1}{\partial \Pi^\mu}-\frac{\partial H}{\partial \Pi^\mu}\frac{\partial \mathcal C_1}{\partial A_\mu}\,.
\end{equation}
These constraint equations ensure, that $\mathcal L_4$ possesses only three dynamical degrees of freedom and the temporal component enters only as an auxiliary field.\\

Continuing the same strategy as in the previous orders, we can construct the corresponding interactions at cubic order in derivatives of the vector field. We start again with
all the possible contractions at that order
\begin{eqnarray}
\mathcal L_5  &=&  f_5(X) \left[ d_1(\partial\cdot A)^3-3d_2(\partial\cdot A)\partial_\rho A_\sigma \partial^\rho A^\sigma-3d_3(\partial\cdot A)\partial_\rho A_\sigma \partial^\sigma A^\rho \right. \nonumber\\
&&\left. +2d_4\partial_\rho A_\sigma \partial^\gamma A^\rho\partial^\sigma A_\gamma+2d_5\partial_\rho A_\sigma \partial^\gamma A^\rho\partial_\gamma A^\sigma\right] \label{vecGalL5gen},
\end{eqnarray}
where $f_5$ is a general function of the norm of the vector field and $d_1$, $d_2$, $d_3$, $d_4$ and $d_5$ are free parameters, which have to be restricted in a way that guarantees that the Hessian matrix has one vanishing eigenvalue. This is achieved by demanding the conditions 
\begin{eqnarray}
d_3=1-d_2, \;\;\;\;\;\; \;\; d_4=1-\frac{3d_2}{2},  \;\;\;\;\;\;\;d_5=\frac{3d_2}{2}
\end{eqnarray}
with $d_1=1$. The quintic Lagrangian then becomes
\begin{eqnarray}
\mathcal L_5  &=&  f_5(X) \Big[(\partial\cdot A)^3-3d_2(\partial\cdot A)\partial_\rho A_\sigma \partial^\rho A^\sigma-3(1-d_2)(\partial\cdot A)\partial_\rho A_\sigma \partial^\sigma A^\rho  \nonumber\\
&& +2\left(1-\frac{3d_2}{2}\right)\partial_\rho A_\sigma \partial^\gamma A^\rho\partial^\sigma A_\gamma+2\left(\frac{3d_2}{2}\right)\partial_\rho A_\sigma \partial^\gamma A^\rho\partial_\gamma A^\sigma\Big] \label{vecGalL5}
\end{eqnarray}
The terms proportional to $d_2$ actually can carry an independent function and does not need to be restricted to $ f_5$, in the same was as it was the case for the $c_2$ term in $\mathcal L_4$. In terms of the field strength tensor we can rewrite these interactions also as \cite{Heisenberg:2014rta,Allys:2015sht}
\begin{eqnarray}
\mathcal L_5  &=&  f_5(X) \left[(\partial\cdot A)^3-3(\partial\cdot A)\partial_\rho A_\sigma \partial^\sigma A^\rho+2\partial_\rho A_\sigma \partial^\gamma A^\rho\partial^\sigma A_\gamma \right] \nonumber\\
&&+d_2 \tilde{f}_5(X) \left[ \left(\frac{1}2(\partial\cdot A)F_{\rho\sigma}^2-\partial_\sigma A_\gamma F_\rho^{\;\;\sigma}F^{\rho\gamma}\right)\right].
\end{eqnarray}

This exact form imposed by the vanishing of the determinant of the Hessian matrix naturally arises when we construct the interactions using the Levi-Civita tensors. In terms of the Levi-Civita tensors, we can again contract the indices in two independent ways
\begin{eqnarray}
\mathcal{L}_5&=&-\mathcal{E}^{\mu\nu\rho\sigma}\mathcal{E}^{\alpha\beta\delta}_{\;\;\;\;\;\;\sigma}(f_5(X)\partial_\mu A_\alpha\partial_\nu A_\beta \partial_\rho A_\delta +d_2\tilde{f}_5(X)\partial_\mu A_\nu\partial_\rho A_\alpha \partial_\beta A_\delta) \nonumber\\
& =&f_5(X)\left[(\partial\cdot A)^3-3(\partial\cdot A)\partial_\rho A_\sigma \partial^\sigma A^\rho 
+2\partial_\rho A_\sigma \partial^\gamma A^\rho\partial^\sigma A_\gamma \right] \nonumber\\
&+&d_2\tilde{f}_5(X)\left[( \partial^\beta A^\alpha(\partial^\alpha A_\gamma \partial^\gamma A_\beta - \partial_\gamma A_\beta \partial^\gamma A_\alpha)+(\partial\cdot A)(\partial_\gamma A_\beta-\partial_\beta A_\gamma)\partial^\gamma A^\beta\right] \,.
\end{eqnarray}
In fact the term proportional to $d_2$ can be expressed in terms of the dual of the strength tensor $\tilde{F}_{\mn}$. In this way the quantic Lagrangian can be rewritten in a more compact form
\begin{eqnarray}
\mathcal L_5  &=&f_5(A^2)\;\left[(\partial\cdot A)^3-3(\partial\cdot A)\partial_\rho A_\sigma \partial^\sigma A^\rho +2\partial_\rho A_\sigma \partial^\gamma A^\rho\partial^\sigma A_\gamma \right] \nonumber\\
&&  +\tilde{f}_5(A^2)\tilde{F}^{\alpha\mu}\tilde{F}^\beta_{\;\;\mu}\partial_\alpha A_\beta\,.
\end{eqnarray}
The longitudinal part of the vector field belongs to the Galileon interactions. If we would impose the condition that the longitudinal mode should not have any trivial total derivative interactions, then the series for the longitudinal mode would stop here \cite{Heisenberg:2014rta}. Relaxing this condition, we can construct the interactions forth order in derivates of the vector field \cite{Allys:2015sht,Jimenez:2016isa}. We could again write down all the possible contractions with arbitrary coefficients and then fix them by imposing the vanishing of the determinant of the Hessian matrix. Since we familiarised ourselves with this procedure already in the previous orders, we can simply obtain the desired consistent structure of the interactions by the contractions with the Levi-Civita tensors. At this order in $\mathcal L_6$, we again have the two independent possible contractions \cite{Allys:2015sht,Jimenez:2016isa}
\begin{align}
\mathcal{L}_6=&-\mathcal{E}^{\mu\nu\rho\sigma}\mathcal{E}^{\alpha\beta\delta\kappa}(f_6(X)\partial_\mu A_\alpha\partial_\nu A_\beta \partial_\rho A_\delta \partial_\sigma A_\kappa +e_2\tilde{f}_6(X)\partial_\mu A_\nu\partial_\alpha A_\beta \partial_\rho A_\delta  \partial_\sigma A_\kappa)\nonumber\\
&=f_6(X)\left[3\partial^\beta A^\alpha(\partial_\alpha A_\beta \partial_\mu A_\nu \partial^\nu A^\mu-2\partial_\alpha A_\nu \partial^\mu A_\beta \partial^\nu A_\mu)+8(\partial\cdot A)\partial_\beta A_\nu \partial^\mu A^\beta \partial^\nu A_\mu \right. \nonumber\\
&\left.-6 (\partial\cdot A)^2\partial_\mu A_\nu \partial^\nu A_\mu+(\partial\cdot A)^4 \right] \nonumber\\
&+e_2\tilde{f}_6(X)\Big[ (\partial\cdot A)(2\partial^\gamma A^\beta(\partial_\beta A_\delta \partial^\delta A_\gamma-\partial_\delta A_\gamma \partial^\delta A_\beta)+(\partial\cdot A)(\partial_\delta A_\gamma - \partial_\gamma A_\delta)\partial^\delta A^\gamma)) \nonumber\\
&+\partial^\beta A^\alpha(-2\partial_\alpha A_\delta \partial^\gamma A_\beta \partial^\delta A_\beta+\partial_\gamma A_\delta(2\partial^\gamma A_\alpha \partial^\delta A_\beta+(\partial_\alpha A_\beta -\partial_\beta A_\alpha)\partial^\delta A^\gamma))\Big]
\end{align}
The terms independent of $e_2$ correspond to just total derivatives and vanish identically at the level of the equations of motion, which reflects the fact that the interactions for the longitudinal mode stop at the previous order. Neglecting the terms that depend purely on $F_{\mn}$ and $\tilde{F}_{\mn}$ (since they are already incorporated in $f_2$), the non-trivial sixth order interaction becomes \cite{Jimenez:2016isa}
\begin{eqnarray}
\mathcal L_6  =e_2f_6(X) \tilde{F}^{\alpha\beta}\tilde{F}^{\mu\nu}\partial_\alpha A_\mu \partial_\beta A_\nu \,.
\end{eqnarray}
Thus, demanding that the vector field possesses only three propagating degrees of freedom with second order equations of motion and that its longitudinal part is purely Galileon type of interactions results in the total Lagrangian for the vector field 
\begin{equation}\label{generalizedProfaField}
\mathcal L_{\rm gen. Proca} = -\frac14 F_{\mu\nu}^2 +\sum^5_{n=2}\alpha_n \mathcal L_n \,,
\end{equation}
where the self-interactions of the vector field are \cite{Heisenberg:2014rta,Jimenez:2016isa}
\begin{eqnarray}\label{vecGalProcaField}
\mathcal L_2 & = &f_2(X,F,Y)\nonumber\\
\mathcal L_3 & = &f_3(X) \;\; \partial_\mu A^\mu \nonumber\\
\mathcal L_4  &=&  f _4(X)\;\left[(\partial\cdot A)^2-\partial_\rho A_\sigma \partial^\sigma A^\rho\right]   \nonumber\\
\mathcal L_5  &=&f_5(X)\;\left[(\partial\cdot A)^3-3(\partial\cdot A)\partial_\rho A_\sigma \partial^\sigma A^\rho \right. \nonumber\\
&&\left.+2\partial_\rho A_\sigma \partial^\gamma A^\rho\partial^\sigma A_\gamma \right]  +\tilde{f}_5(X)\tilde{F}^{\alpha\mu}\tilde{F}^\beta_{\;\;\mu}\partial_\alpha A_\beta \nonumber\\
\mathcal L_6  &=&f_6(X) \tilde{F}^{\alpha\beta}\tilde{F}^{\mu\nu}\partial_\alpha A_\mu \partial_\beta A_\nu \,.
\end{eqnarray}
These are the most general derivative self-interactions for a massive vector field with second order equations of motion and three propagating degrees of freedom on flat space-time (up to disformal transformations). Note, that the series stops at this order and there are not any higher order interactions.
Following the systematic construction of the generalized Proca interactions in terms of the two antisymmteric Levi-Civita tensors, the series has to stop after the sixth order of interactions, since there are no indices left in the two Levi-Civita tensors to be contracted in order to construct a possible $\mathcal L_7$ term in four dimensions. Any naive attempt to construct such terms for instance as  $\mathcal L_7^{\rm Perm,2}=M^\mu{} _{\nu} (\partial_\mu A^\nu+\partial^\nu A_\mu)S_\mu{}^\nu$ \cite{Allys:2015sht} with
\be
M=F^4-\frac12 [F^2]F^2+\frac18\Big([F^2]^2-2[F^4]\Big)\Id
\label{CHtheoremL7}
\ee
would vanish in four dimensions by the virtue of the Cayley-Hamilton theorem applied on $F^\mu{}_\nu$ and, hence, such interactions trivialise in four dimensions \cite{Jimenez:2016isa}.
In five dimensions with the Levi-Civita tensors carrying five indices, we could indeed construct such seventh order Lagrangians, for instance
\begin{align}
\mathcal{L}_7^{\rm Gal}=\mathcal{E}^{\mu\nu\rho\sigma\tau}\mathcal{E}^{\alpha\beta\delta\kappa\omega}\partial_\mu A_\alpha\partial_\nu A_\beta \partial_\rho A_\delta \partial_\sigma A_\kappa \partial_\tau A_\omega \,.
\end{align}
and the above Lagrangian that was denoted by $\mathcal L_7^{\rm Perm,2}$
\begin{align}
\mathcal L_7^{\rm Perm,2}=-\frac{1}{2}\mathcal{E}^{\mu\nu\rho\sigma\tau}\mathcal{E}^{\alpha\beta\delta\kappa\omega}\partial_\mu A_\nu\partial_\alpha A_\beta \partial_\rho A_\sigma  \partial_\delta A_\kappa  \partial_\tau A_\omega\; .
\end{align}
This is in the same spirit as the Lovelock invariants of the massless spin-2 field in equations (\ref{Lovelock5dim}) or the higher dimensional Galileon interactions.
For more detailed explanations on the consistent constructions of the generalised Proca theories see \cite{Heisenberg:2014rta,Allys:2015sht,Jimenez:2016isa}.

\subsection{Construction of the generalised Proca from the decoupling limit}
We could have obtained the systematic form of these interactions from the previous subsection by starting from the consistent construction of the interactions in the decoupling limit. As we mentioned previously, in the case of the vector field, the derivative of the vector field $\partial_\alpha A_\beta$ has its symmetric and antisymmetric part. It will be convenient to
introduce its symmetric part by $S_{\mu\nu}=\partial_\mu A_\nu+\partial_\nu A_\mu$ and its antisymmetric part as usual by $F_{\mn}$. In terms of $A$, $F$ and $S$, we can write the interactions systematically as an expansion
\be
\mathcal L\sim\sum_{m,n,p}c_{m,n,p}\left(\frac{A}{\Lambda_M}\right)^m\left(\frac{F}{\Lambda_F^2}\right)^n\left(\frac{S}{\Lambda_S^2}\right)^p \,,
\ee
with their corresponding scales $\Lambda_M$, $\Lambda_F$ and $\Lambda_S$ of the interactions and $c_{m,n,p}$ being some coefficients. 
Using the Stueckelberg trick we can restore the broken gauge invariance of the vector field by performing the change of variables  $A_\mu \to A_\mu+\partial_\mu\pi/M$ with $\pi$ the Stueckelberg field and $M$ denoting the mass of the vector field. Written in this way, the additional scalar field is nothing else but the longitudinal mode of the original massive vector field. When we take the decoupling limit by sending the mass of the vector field to zero, then the leading part of the vector field becomes $A_\mu \to \partial_\mu\pi/M$ and similarly $S_{\mu\nu}\to\partial_\mu\partial_\nu\pi/M$. The above expansion of the interactions on the other hand simply becomes 
\be
\mathcal L_{\rm dec}\sim\sum_{m,n,p}c_{m,n,p}\left(\frac{\partial\pi}{M\Lambda_M}\right)^m\left(\frac{F}{\Lambda_F^2}\right)^n\left(\frac{\partial\partial\pi}{M\Lambda_S^2}\right)^p \,.
\label{Ldec}
\ee
Starting with the lowest order $p=0$, we immediately observe, that in this case the interactions contain only one derivative per scalar field
\be\label{intDLp0}
\mathcal L_{\rm dec}^{p=0}\sim\sum_{m,n}c_{m,n,0}\left(\frac{\partial\pi}{M\Lambda_M}\right)^m\left(\frac{F}{\Lambda_F^2}\right)^n \,.
\ee
After performing the resummations in $m$ and $n$, we would be obtaining exactly the same type of interactions that we were grouping into $\mathcal{L}_2$. We will have three types of interactions at this order. The interactions with $m=0$ would be just functions of $F^2$. Similarly, the interactions with $n=0$ would be just functions of the vector norm $A^2$. Last but not least, interactions with $m\ne0$ and $n\ne0$ would just correspond to functions of $Y=F_{\mu}{}^\alpha F_{\nu\alpha}A^\mu A^\nu$. Thus, the schematic interactions in (\ref{intDLp0}) are nothing else but the corresponding interactions in the decoupling limit originating from $f_2(A_\mu, F_{\mu\nu}, \tilde{F}_{\mu\nu})$. As we have seen, due to their tensor structure, the interactions in $f_2$ can be written as functions of $X=A^2$, $F=F^2$ and $Y=F_{\mu}{}^\alpha F_{\nu\alpha}A^\mu A^\nu$, so the appropiate resummations of $m$ and $n$ will have one to one correspondings with the functions $f_2(X,F,Y)$ beyond the decoupling limit. \\

Next, let us analyse the type of interactions we can have linear in $\partial\partial\pi$ with $p=1$
\be\label{intDLp1}
\mathcal L_{\rm dec}^{p=1}\sim\sum_{m,n}c_{m,n,1}\left(\frac{\partial\pi}{M\Lambda_M}\right)^m\left(\frac{F}{\Lambda_F^2}\right)^n\left(\frac{\partial\partial\pi}{M\Lambda_S^2}\right) \,.
\ee
First of all, the interactions with $n=0$ are the standard Galileon interactions at cubic order $f_3(\partial\pi^2)\Box\pi$ and here they correspond to the leading order contributions in the decoupling limit coming from the interactions of $f_3(A^2) \partial_\mu A^\mu $ in $\mathcal L_3$. The interactions with $n\ne0$ constitute the first non-trivial mixing between the gauge field and the scalar field, which has second derivatives acting on it. Therefore, these interactions require more caution. After resummation, the resulting symmetric rank-2 tensor $K^{\mu\nu}$ will be contracted with $\partial_\mu \partial_\nu\pi$, hence in order to guarantee second order derivatives field equations arising from $K^{\mu\nu}\partial_\mu \partial_\nu\pi$, the $(00)$ component of this tensor can not accommodate time derivatives other than $\dot{\pi}$. Thus, the symmetric rank-2 tensor $K^{\mu\nu}$ can only be built out of $f_3(\partial\pi^2)\tilde{F}^{\mu\alpha}\tilde{F}^\nu{}_\alpha$, since its magnetic part $\tilde{F}^{0\alpha}\tilde{F}^0{}_\alpha\propto B^2$ is purely potential without any time derivatives acting on $A$. Summarising, the decoupling limit interactions at order $p=1$ will be summed into
\be
\mathcal L_{\rm dec}^{p=1}\sim\left(c_{2,0,1}(\partial\pi)^2\eta^{\mu\nu}+c_{0,2,1}\tilde{F}^{\mu\alpha}\tilde{F}^\nu{}_\alpha\right)\frac{\partial_\mu\partial_\nu\pi}{M\Lambda_S^2} \,,
\label{Ldecp1}
\ee
where the coefficients can be arbitrary functions of $(\partial\pi^2)$. As mentioned above, the first type of interactions are the corresponding interactions of $f_3(A^2) \partial_\mu A^\mu $ in $\mathcal L_3$ in the decoupling limit, whereas the second type are the interactions that would be promoted to $\tilde{f}_5(A^2)\tilde{F}^{\alpha\mu}\tilde{F}^\beta_{\;\;\mu}\partial_\alpha A_\beta$ in $\mathcal L_5$ beyond the decoupling limit.
These interactions are the only interactions at that order in the decoupling limit, that guarantee second order equations of motion for $\pi$ and since $\tilde{F}^{\mu\nu}$ is divergence-free, also higher order equations of motion for the gauge field are directly avoided. \\

Now, let us study the next order interactions in $\partial\partial\pi$ with $p=2$
\be\label{intDLp2}
\mathcal L_{\rm dec}^{p=2}\sim\sum_{m,n}c_{m,n,2}\left(\frac{\partial\pi}{M\Lambda_M}\right)^m\left(\frac{F}{\Lambda_F^2}\right)^n\left(\frac{\partial\partial\pi}{M\Lambda_S^2}\right)^2 \,.
\ee
The purely scalar interactions with $n=0$ must have again the same structure as the Galileon interactions, which would ensure second order equations of motion for the purely longitudinal mode sector. For the mixing between the scalar field and the gauge field with $n\ne0$ we again have to construct them in a way, that only the magnetic part of the gauge field is directly coupled to the second time derivatives of $\pi$. Hence, we have to construct a tensor of rank-4 $K^{\mu\nu\alpha\beta}\partial_\mu \partial_\nu\pi\partial_\alpha \partial_\beta\pi$ that is purely built out of the magnetic part. This is again only satisfied by the dual field strength. These two requirements together with the tensor structure uniquely lead to 
\begin{align}
\mathcal L_{\rm dec}^{p=2}\sim &c_{2,0,2}(\partial\pi)^2\frac{(\Box\pi)^2-(\partial_\mu\partial_\nu\pi)^2}{M^2\Lambda_S^4}
+c_{0,2,2}\tilde{F}^{\mu\nu}\tilde{F}^{\alpha\beta}\frac{\partial_\mu\partial_\alpha\pi\partial_\nu\partial_\beta\pi}{M^2\Lambda_S^4} \,,
\label{Ldecp2}
\end{align}
where again the coefficients can be an arbitrary functions of  $(\partial\pi^2)$. The first part is nothing else but the quartic Galileon interactions, which would be related to $ f _4(A^2) \left[(\partial\cdot A)^2-\partial_\rho A_\sigma \partial^\sigma A^\rho\right] $ in $\mathcal L_4$ beyond the decoupling limit and the second part constitutes the unique mixing between the gauge field and the scalar field at that order, that would be promoted to $ f_6(A^2) \tilde{F}^{\alpha\beta}\tilde{F}^{\mu\nu}\partial_\alpha A_\mu \partial_\beta A_\nu$ in $\mathcal L_6$ beyond the decoupling limit. \\

Last but not least, we shall pay a special attention to the interaction cubic order in $\partial\partial\pi$ with $p=3$
\be\label{intDLp3}
\mathcal L_{\rm dec}^{p=3}\sim\sum_{m,n}c_{m,n,3}\left(\frac{\partial\pi}{M\Lambda_M}\right)^m\left(\frac{F}{\Lambda_F^2}\right)^n\left(\frac{\partial\partial\pi}{M\Lambda_S^2}\right)^3 \,.
\ee
Similarly, as before the purely scalar interactions with $n=0$ correspond to the Galileon interactions, the quintic Galileon interactions. For the mixed interactions with the gauge field, imposing the same conditions as above reveals that one can not construct any healthy mixed interaction between the gauge field and the Stueckelberg field at that order. Thus, we can not have any new interaction with $n\ne0$. The series stop here and the decoupling limit has a finite order of allowed interactions. Constructing the interactions beyond the decoupling limit is straightforward by means of promoting $\partial_\mu\pi\rightarrow A_\mu$ and $\partial_\mu\partial_\nu\pi\rightarrow S_{\mu\nu}$
\begin{equation}\label{generalizedProfaField_S}
\mathcal L_{\rm gen. Proca} = -\frac14 F_{\mu\nu}^2 +\sum^5_{n=2}\alpha_n \mathcal L^S_n \,,
\end{equation}
where the self-interactions of the vector field are \cite{Heisenberg:2014rta,Jimenez:2016isa}
\begin{eqnarray}\label{vecGalProcaFieldinS}
\mathcal L^S_2 & = &f_2(A_\mu, F_{\mu\nu}, \tilde{F}_{\mu\nu})\nonumber\\
\mathcal L^S_3 & = &f_3(A^2)[S] \nonumber\\
\mathcal L^S_4  &=&  f _4(A^2)\;\left( [S]^2-[S^2] \right)   \nonumber\\
\mathcal L^S_5  &=&f_5(A^2)\;\left( [S]^3-3[S][S^2]+2[S^3]\right)  
+\tilde{f}_5(A^2)\tilde{F}^{\alpha\mu}\tilde{F}^\beta_{\;\;\mu}S_{\mu\nu} \nonumber\\
\mathcal L^S_6  &=&f_6(A^2) \tilde{F}^{\alpha\beta}\tilde{F}^{\mu\nu}S_{\alpha\mu}S_{\beta\nu} \,.
\end{eqnarray}
This gives an additional proof from the decoupling limit that the constructed interactions of the generalised Proca (\ref{vecGalProcaField}) are the most general interactions with second order equations of motion with three propagating degrees of freedom. It is worth emphasising again that the generalized Proca interactions can be divided into two types of interactions, namely those that can be directly obtained from the scalar Galileon interactions by promoting $\partial_\mu \pi \to A_\mu$ and those that are genuinely new interactions with intrinsic vector modes, that do not have any scalar counter example. These are the interactions $\tilde{F}^{\alpha\mu}\tilde{F}^\beta_{\;\;\mu}S_{\mu\nu}$ and $\tilde{F}^{\alpha\beta}\tilde{F}^{\mu\nu}S_{\alpha\mu}S_{\beta\nu}$.

\subsection{Deformed determinant for the generalised Proca}

In the same way as we could express the scalar Galileon interactions and the massive gravity interactions in terms of a deformed determinant in equations (\ref{deformedDet_Gal}) and (\ref{deformedDet_MG}) respectively, we shall use the structure of the Proca interactions based on the antisymmetry of the Levi-Civita tensor in order to express them in terms of a determinantal formulation as well.
The generating determinant takes this time the form \cite{Jimenez:2016isa}
\begin{equation}\label{determinant}
f(A^2)\det(\delta^{\mu}{}_\nu+\mathcal{C}^\mu{}_\nu)\,,
\end{equation}
where the fundamental matrix reads
\be
\mathcal{C}^\mu{}_\nu=aF^\mu{}_\nu +bS^\mu{}_\nu +c A^\mu A_\nu
\ee
with $a$, $b$ and $c$ denoting some parameters of dimension $-2$. In massive gravity the fundamental matrix was $K^\mu{}_\nu$ and in Galileon interactions $\Pi^\mu{}_\nu$. In the case of generalised Proca interactions, this gets simply replaced by $\mathcal{C}^\mu{}_\nu$. In the same spirit as the interactions in massive gravity and scalar Galileon interactions, one can express the determinant in terms of the elementary symmetric polynomials 
\be
\det(\delta^{\mu}{}_\nu+\mathcal{C}^\mu{}_\nu)=\sum_{n=0}^4 e_n(\mathcal{C}^\mu{}_\nu)
\ee
Since the zeroth order symmetric elementary polynomial is trivial $e_0 = 1$ let us start with the first order
\begin{eqnarray}
e_1 &=&-\frac{1}{6}\epsilon_{\mu\nu\alpha\beta}\epsilon^{\rho\nu\alpha\beta}\mathcal{C}^\mu{}_\rho=[\mathcal{C}]
= b[S]+cA^2\,.
\end{eqnarray}
We recognise the $\mathcal L_3$ interaction in the first term, whereas a simple potential interaction as a part of $\mathcal L_2$ in the second term.
The second order symmetric elementary polynomial gives the following combination of traces
\begin{eqnarray}
e_2 &=&-\frac{1}{4}\epsilon_{\mu\nu\alpha\beta}\epsilon^{\rho\sigma\alpha\beta}\mathcal{C}^\mu{}_\rho\mathcal{C}^\nu{}_\sigma=\frac{1}{2}\Big([\mathcal{C}]^2-[\mathcal{C}^2]\Big) \nonumber\\
&=&\frac{1}{2}\Big( a^2[F^2]+b^2([S]^2-[S^2])
+  2bc(A^2\eta^{\mu\nu}-A^\mu A^\nu )S_{\mu\nu} \Big). 
\end{eqnarray}
The first two contributions correspond to the interactions in $\mathcal L_4$, whereas the last term proportional to $S\mn$ is nothing else but the interaction in $\mathcal L_3$, where the indices of the Levi-Civita tensors are contracted with two additional vector fields instead of the metrics, which would result as a disformal transformation of the original interactions in $\mathcal L_3$.
The cubic symmetric elementary polynomial reads
\begin{eqnarray}
e_3 &=&-\frac{1}{6}\epsilon_{\mu\nu\alpha\beta}\epsilon^{\rho\sigma\delta\beta}\mathcal{C}^\mu{}_\rho\mathcal{C}^\nu{}_\sigma \mathcal{C}^\alpha{}_\delta   \nonumber\\
&=& \frac{a^2c}{2} F_\alpha{}^\mu F_{\beta\mu}(A^2 \eta^{\alpha\beta}-2A^\alpha A^\beta )
+\frac{1}2 a^2b ([F^2][S]-2F^{\alpha\beta}F_{\alpha}{}^\mu S_{\beta\mu})  \nonumber\\
&+&\frac12cb^2 S_\alpha{}^\mu S_{\beta\mu}(2A^\alpha A^\beta -A^2\eta^{\alpha\beta})  
+\frac12cb^2[S]S_{\alpha\beta}(A^2 \eta^{\alpha\beta}-2A^\alpha A^\beta ) \nonumber\\
&+&\frac{b^3}{6}([S]^3-3[S][S^2]+2[S^3]) \,.
\end{eqnarray}
We can identify the interactions in $\mathcal L_5$ together with their altered version in terms of the disformal metric determined by the vector field and we also recognise the presence of the purely intrinsic vector mode interactions $-\frac12a^2b\tilde{F}^{\mu\alpha}\tilde{F}^{\nu}{}_\alpha S_{\mu\nu}$ in the second line.
Finally, the quartic elementary polynomial 
\begin{eqnarray}
e_4 =-\frac{1}{24}\epsilon_{\mu\nu\alpha\beta}\epsilon^{\rho\sigma\delta\gamma}\mathcal{C}^\mu{}_\rho\mathcal{C}^\nu{}_\sigma \mathcal{C}^\alpha{}_\delta \mathcal{C}^\beta{}_\gamma 
\end{eqnarray}
gives the remaining interactions in $\mathcal L_6$. Since we are in four dimensions, the series stops at that order $e_n=0$ for $n\ge 5$.

\subsection{Quantum stability of the vector interactions}
As we have seen in the previous subsections, the generalized Proca interactions can be divided into two groups, namely those that can be directly obtained from the extension of the scalar Galileon interactions by promoting $\partial_\mu \pi \to A_\mu$ and those that are genuinely new interactions with intrinsic vector modes. For the quantum stability these two groups will behave differently. For the first type of interactions, we expect a similar non-renormalization theorem as we saw in the scalar counter part in section \ref{subsec_GalQuantum} and hence they should be protected from quantum corrections. The second type interactions might receive quantum corrections, even though small. The different contractions of the antisymmetric Levi-Civita tensors with $\partial A$ will be essential for the presence of the non-renormalization theorem. In the following we would like to investigate this in more detail. For this purpose, we shall consider Feynman diagrams with purely vector modes. The propagator for a massive vector field in Fourier modes reads
\begin{equation}\label{propagator_massiveVector}
D_{\mu\nu}=\frac{1}{k^2-m^2}\left( -\eta_{\mu\nu}+\frac{k_\mu k_\nu}{m^2} \right)\,.
\end{equation}
Let us first have a closer look to the first type of interactions, which are the direct extension of the scalar Galileons. As an example consider the quartic interaction contracted as $A^\mu A^\nu \mathcal{E}_{\mu}{}^{\alpha\rho\gamma}\mathcal{E}_{\nu\;\;\;\; \gamma}^{\;\;\beta\sigma}\partial_\alpha A_\beta \partial_\rho A_\sigma$. The most worrisome contributions will come from the Feynman diagrams with the least power of external momentum. Therefore, we shall 
only pay attention to the Feynman diagrams with the vertex where the two vector fields without the derivatives run on the external legs with momentum $p$ and $q$ whereas the two with derivatives run in the loop with momentum $k$ and $p+q-k$. From this vertex, the transition amplitude receives a contribution in form of
\begin{equation}\label{Feynman1_massiveVector}
\mathcal{A}= \int \frac{d^4k}{(2\pi)^4} D^{\kappa\mu}(p) D^{\delta\nu}(q) \mathcal{E}_{\mu}{}^{\alpha\rho\gamma}\mathcal{E}_{\nu\;\;\;\; \gamma}^{\;\;\beta\sigma}k_\alpha(p+q-k)_\rho D_{\lambda\beta}(k) D_{\xi\sigma}(p+q-k) 
\end{equation}
where $D_{\mu\nu}(k)$ is the propagator of the massive vector field with momentum $k$ in (\ref{propagator_massiveVector}). We see immediately that from this contribution the only non-vanishing terms have at least two external momenta due to the antisymmetric structure of the vertex. This on the other hand means that the operators that will be generated by these quantum corrections will have two more derivatives than the original classical operators. Thus, the classical interaction $A^\mu A^\nu \mathcal{E}_{\mu}{}^{\alpha\rho\gamma}\mathcal{E}_{\nu\;\;\;\; \gamma}^{\;\;\beta\sigma}\partial_\alpha A_\beta \partial_\rho A_\sigma$ is not renormalized. This is in the same spirit as in the scalar Galileon interactions. The same non-renormalization theorem is also present for this first type of vector interactions as we anticipated and hence the Galileon vector interactions are radiatively stable. 

Let us now pay special attention to the second type of interactions, which do not have any scalar counter part and therefore correspond to genuinely new interactions with intrinsic vector modes. As an example consider again a quartic interaction but this time with the contraction $A^\mu A^\nu \epsilon_{\mu}{}^{\alpha\rho\gamma}\epsilon_{\nu\;\;\;\; \gamma}^{\;\;\beta\sigma}\partial_\alpha A_\rho \partial_\beta A_\sigma$. We again let the two vector fields without the derivatives run on the external legs and the other two with derivatives run in the internal legs. The contribution to the transition amplitudes becomes this time
\begin{equation}\label{Feynman2_massiveVector}
\mathcal{A}= \int \frac{d^4k}{(2\pi)^4} D^{\kappa\mu}(p) D^{\delta\nu}(q) \epsilon_{\mu}{}^{\alpha\rho\gamma}\epsilon_{\nu\;\;\;\; \gamma}^{\;\;\beta\sigma}k_\alpha(p+q-k)_\beta D_{\lambda\rho}(k) D_{\xi\sigma}(p+q-k) \,.
\end{equation}
In difference to the previous contribution we can have now non-vanishing terms proportional to $k_\alpha k_\beta$ without increasing the number of external momenta. It means that we will have quantum corrections of the same structure as the classical interactions. Hence, this second type of vector interactions are unfortunately not stable under radiative corrections. It would be interesting to study their exact scaling and whether they always remain small.\\

\textbf{Implications of the UV completion}: \\
In section \ref{subsec_GalQuantum} we had a quick look at the implications of a Lorentz invariant UV completion for a spin-0 field on the low energy effective field theory. This was put into effect by the positivity bounds on the tree level scattering amplitudes. The positivity bounds on the derivatives with respect to the momentum transfer $t$ of the imaginary part of the scattering amplitude away from the forward limit can yield significant restriction on the parameters of the considered theory \cite{deRham:2017imi}. This theoretical testing ground can be also directly applied to the spin-1 field \cite{Bonifacio:2016wcb,deRham:2018qqo}. 

Assuming the existence of a standard UV completion of the Proca field, the parameter space of the generalised Proca theories can be restrained from the Froissart-Martin bound on top of the phenomenological bounds. For the implications of the 2-2 scattering amplitudes it is sufficient to consider interactions up to quartic order in the field. For illustration purposes, let us consider for the interactions up to quartic Lagrangian in equation (\ref{vecGalProcaField}) the functions $\mathcal{L}_2=F+m^2X$, $f_3=\frac{-2g_3m^3}{\Lambda^3}X$ in $\mathcal{L}_3$ and $f_4=\frac{-2g_4m^4}{\Lambda^6}X$ in $\mathcal{L}_4$. The requirement of a Wilsonian UV completion translates into the bound $3g_3^2<-g_4<g_3^2$, which would enforce a free theory with $g_3=g_4=0$. Thus, for this particular selection of the functions there is no UV completion for non-vanishing parameters $g_3$ and $g_4$.

One remarkable result is that the UV completion can be accommodated if an additional operator is included at this order and this is exactly the other interaction $Y$ in $\mathcal{L}_2$. In the above example of quartic interactions we had included all of the quartic interactions up to $\mathcal{L}_4$ without the $Y$ contribution in $\mathcal{L}_2$. Now, including the purely intrinsic vector interaction $Y$ in $\mathcal{L}_2$ as $\mathcal{L}_2=F+m^2X+\frac{C_1m^2}{\Lambda^6}Y$, the positivity bounds change to $3g_3^2-C_1<-g_4<g_3^2$ or alternatively $C_1>2g_3^2>-2g_4$, which can be realised with the coefficients of the same order \cite{deRham:2018qqo}. 

It is very interesting that the vector Galileon interactions proportional to $g_3$ and $g_4$ alone can not satisfy the positivity bounds and the genuine vector interactions in $Y$ (which does not have the corresponding scalar Galileon counterpart) are forced upon us for a Wilsonian UV completion. It would be very interesting to consider the bounds arising from higher order scattering amplitudes and the role of the genuine purely intrinsic vector interactions proportional to $\tilde{f}_5(X)$ and $f_6(X)$ in (\ref{vecGalProcaField}). The example of the 2-2 scattering amplitudes suggest that these intrinsic vector interactions will be crucial ingredients for the high energy field theory. Note that these are exactly the interactions which do not satisfy the non-renormalisation theorem in the decoupling limit that we discussed above.
\begin{center}
\begin{figure}[h]
\begin{center}
 \includegraphics[width=11.5cm]{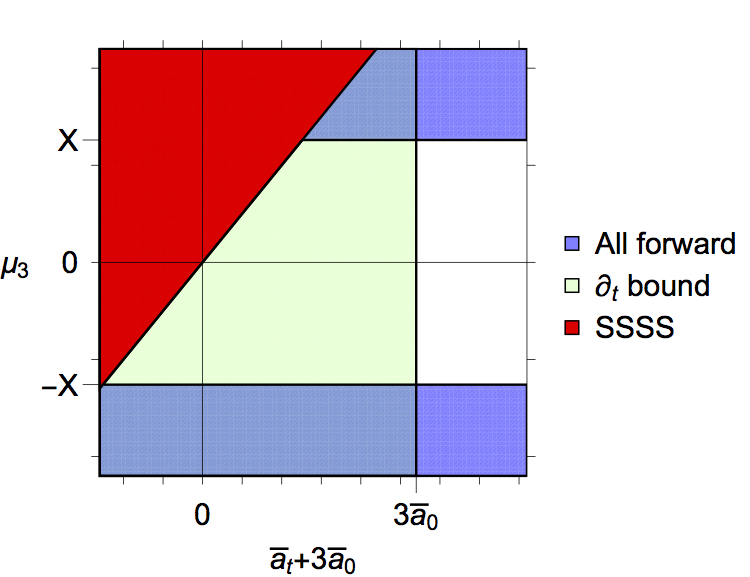}
\end{center}
  \caption{This figure is taken from \cite{deRham:2018qqo}, which illustrates the allowed region by analyticity. The parameters are defined as $3\bar{a}_0=3a_0-a_5-C_1-4C_2$, $2\bar{a}_t=a_3+a_4-2a_5$ and $\mu_3=a_4-a_5-a_3/2-C_1/2-2C_2$. These coefficients originate from their equation (3.5) and can be easily compared to our notation with restricted functions $f_n(X)$ in our equation (\ref{vecGalProcaField}). }
   \label{PositivityboundsProca}
\end{figure}
\end{center}

\subsection{Vector dualities}
In subsection \ref{subsec_GalDual} we saw that we can construct dualities between Galileon theories. For a specific choice of the parameters we could map a quintic Galileon theory to a free theory. Similar dualities can be also constructed for the vector fields. The corresponding Legendre transformation satisfies this time 
$|\eta_{\mu\nu}+s\partial_{\mu}A_\nu|=|\eta_{\mu\nu}-s\tilde\partial_{\mu}\tilde{A}_\nu|$. The $U(1)$ invariant Maxwell action $\mathcal S_{A_\mu}^{m=0}=-\frac12\int d^4x(\partial_\mu A_\nu \partial^\mu A^\nu-\partial_\mu A_\nu \partial^\nu A^\mu )$ can be rewritten in the matrix notation as $\mathcal S_{A_\mu}^{m=0}=-\frac12\int d^4x(\Tr[B^TB]-\Tr[B^2])$ with $B_\mu{}^\nu(x)=\partial_\mu A^\nu(x)$ and its dual $\tilde B_\mu{}^\nu(\tilde x)=\tilde \partial_\mu \tilde{A}^\nu(\tilde x)$. Hence, performing the change of coordinates $|\partial x^a/\partial \tilde x^b|=\det(1-s\tilde{B}(\tilde x))$ gives the following dual theory to Maxwell \cite{deRham:2014lqa}
\begin{equation}
\mathcal S_{\rm Maxwell}^{\rm dual}=-\frac12\int d^4x\det(1-s\tilde{B})\left(\Tr\left[\left(\frac{\tilde B}{1-s\tilde B}\right)^T\frac{\tilde B}{1-s\tilde B}\right]-\Tr\left[\frac{\tilde B^2}{(1-s\tilde B)^2}\right] \right)\,.
\end{equation}
Similarly, the duality can be applied on the free Lagrangian of the Proca field. So, let us consider now $S_{A_\mu}^{m\ne0}=-\frac12\int d^4x(\Tr[B^TB]-\Tr[B^2]+m^2A^2)$. The change of coordinates results in the following dual theory 
\begin{equation}
\mathcal S_{\rm Proca}^{\rm dual}=\mathcal S_{\rm Maxwell}^{\rm dual}-\frac12m^2\int d^4x\det(1-s\tilde{B})\tilde{A}_\mu^2\,.
\end{equation}
If we compare this with the determinantal structure we introduced in (\ref{determinant}), we see that the fundamental matrix of the generalised Proca $\mathcal{C}^\mu{}_\nu=aF^\mu{}_\nu +bS^\mu{}_\nu +c A^\mu A_\nu$ is much general than the Legendre transformation, but we still expect to recover some of the subclasses of the interactions of the generalised Proca.

\section{Vector-Tensor theories}\label{sec_VectorTensorTheories}
So far the vector interactions that we discussed were on flat space-time. How can one generalize them to the curved space-time case?
If we insist on the invariance of gauge symmetry, we have seen that one can not construct the analogues of Galileon interactions for the massless vector field. The only allowed terms are the Maxwell kinetic term or higher order operators of $F_{\mu\nu}$ but no vector Galileons with the systematic form $(\partial F)^n$. 
For the massive Proca field, we could successfully establish derivative self-interactions with second order equation of motion. In difference to the Galileon interactions we were able to construct sixth order interactions and we also saw that there are purely intrinsic vector mode interactions without scalar corresponding. In the following we would like to investigate the curved version of these interactions for both gauge invariant and broken cases.
\subsection{Unique non-minimal coupling for a $U(1)$ gauge field}
In the scalar case, we have seen that promoting the flat space-time interactions to curved space-time requires the presence of non-minimal couplings with gravity. For the vector field, we are after a similar question: preserving the gauge symmetry what is the form of the 
most general action for a massless vector field with a non-minimal coupling to gravity leading to second order equations of motion for both, the vector field and the gravitational sector. In order to maintain the gauge symmetry, we can not consider potential terms or direct couplings of the vector field to curvature and the derivative couplings should be only via the field strength $F_{\mu\nu}$. 

For instance, one could construct terms with a coupling of $F_{\mu\nu}$ to the Riemann tensor. However, these couplings would yield gravitational equations, that are not second order. For second order equations of motion, one hast to couple $F_{\mu\nu}$ to a divergence-free tensor constructed out of the Riemann tensor. In section \ref{subsec_LovelockInvariant} we have seen that the only divergence-free tensors in four dimensions are the metric $g_{\mu\nu}$, the Einstein tensor $G_{\mu\nu}$ and the double dual Riemann tensor $L^{\alpha\beta\gamma\delta}= -\frac{1}{2}\mathcal{E}^{\alpha\beta\mu\nu}\mathcal{E}^{\gamma\delta\rho\sigma}R_{\mu\nu\rho\sigma}$. These divergenceless tensors are the equivalents of the non-trivial Lovelock invariants at the level of the equations of motion. 

The contractions of $F_{\mu\nu}$ with the first divergence free tensor, namely the inverse metric $g^{\mu\nu}$, gives rise to the Maxwell kinetic term or higher order operators of $F_{\mu\nu}$, which are not the interesting non-minimal couplings that we are after. On the other hand, since the Einstein tensor $G_{\mu\nu}$ is symmetric in the indices $\mu\nu$, we can not contract it with the antisymmetric tensor, hence $G_{\mu\nu}F^{\mu\nu}=0$. The attempt to contract the indices of $F$ with a combination of the Einstein tensor and the metric fails as well. The reason for that is that the product of two divergenceless tensors is not divergence-free. Thus, $G^{\mu\nu}g^{\alpha\beta}$ is not divergenceless even though $G^{\mu\nu}$ and $g^{\mu\nu}$ are. This means that terms of the form $G^{\mu\nu}g^{\alpha\beta}F_{\mu\alpha}F_{\nu\beta}$ will not satisfy our requirement of second order equations of motion. This is true for any combination of divergenceless tensors, therefore the product of two Einstein tensors $G^{\mu\nu}G^{\alpha\beta}F_{\mu\alpha}F_{\nu\beta}$ fails for the same reason. Thus, the only non-minimal coupling with second order equations of motion will be the coupling to the double dual Riemann tensor $L^{\alpha\beta\gamma\delta}$. 

The most general action for a massless vector field on curved space-time with second order equations of motion is therefore given by \cite{Horndeski:1976gi,Barrow:2012ay,Jimenez:2013qsa}
\begin{align}
S&=\int {\rm d}^4x\sqrt{-g}\left[\frac{1}{2}M_p^2R-\frac14F_{\mu\nu}F^{\mu\nu}+\frac{1}{4M^2}L^{\alpha\beta\gamma\delta}F_{\alpha\beta}F_{\gamma\delta} \right]\nonumber \\
&=\int {\rm d}^4x\sqrt{-g}\left[\frac{1}{2}M_p^2R-\frac14F_{\mu\nu}F^{\mu\nu}
+\frac{1}{2M^2}\Big(R F_{\mu\nu}F^{\mu\nu} - 4R_{\mu\nu}F^{\mu\sigma}F^{\nu}_{\;\;\sigma} + R_{\mu\nu\alpha\beta} F^{\mu\nu} F^{\alpha\beta} \Big) \right],
\label{actionLFFvector}
\end{align}
with $M$ representing the scale where this non-minimal coupling will be relevant. Note at this stage, that if we contract the indices of the two field strength tensors in the other way, namely $L^{\alpha\gamma\beta\delta}F_{\alpha\beta} F_{\gamma\delta}$, then we obtain the same type of interaction as above and it does not correspond to a new term. We also did not want to consider parity violating terms of the form $L^{\alpha\beta\gamma\delta} F_{\alpha\beta}\tilde F_{\gamma\delta}$. Another attempt might be to contract the double dual Riemann tensor with two duals of $F$ in the form $L^{\alpha\beta\gamma\delta}\tilde F_{\alpha\beta}\tilde F_{\gamma\delta}$. Since the dual of $F$ contains an additional Levi-Civita tensor, these contractions correspond to contracting $F$ with the Riemann tensor and therefore does not give rise to second order equations of motion. The crucial fact to note here is that $F$ is a closed form satisfying the Bianchi identities whereas its dual does not, which therefore allows for $LFF$ terms but not for $L\tilde{F}\tilde{F}$ terms. 

Summarizing, as Horndeski proved \cite{Horndeski:1976gi}, and \cite{Jimenez:2013qsa} rediscovered, the action (\ref{actionLFFvector}) is the most general action for the electromagnetic field with non-minimal coupling leading to second order equations of motion and recovering Maxwell theory in flat spacetime. This limit corresponds to taking spacetime curvature much smaller than $M^2$ and suppressing strongly the non-minimal coupling.

It is worth to mention that this non-minimal coupling can be also obtained from a Kaluza-Klein reduction of the Gauss-Bonnet terms in five dimensions. In this way one does not only generate $LFF$ type of terms but also a quartic interaction term $\frac{1}{M^2}[(F_{\alpha\beta}F^{\alpha\beta})^2-2F^\alpha_\beta F^\beta_\gamma F^\gamma_\delta F^\delta_\alpha]$. It is interesting to note that this non-minimal coupling involves the same type of terms that are obtained when introducing vacuum polarization corrections from a curved background in standard electromagnetism \cite{Drummond:1979pp}. In that case,  the non-minimal couplings are suppressed by the mass of the charged fermion running inside the loops (typically the electron mass in standard QED). 

Such radiative corrections will also arise if the vector field is coupled to some other charged field of mass $m$. This means that the lightest possible particle charged under the $U(1)$ field should satisfy $m\gg M$ for those quantum corrections not to spoil the Horndeski interaction. On the other hand, one might also wonder whether loop corrections involving gravitons could spoil the Horndeski structure as well, since they will produce radiative corrections of this type (i.e., terms linear in the curvature and quadratic in $F_{\mu\nu}$), but without the appropriate coefficients. However, these corrections will be suppressed by the Planck mass so that, for the action (\ref{actionLFFvector}) to make sense as an effective field theory, we would need to require $M\ll M_p$. This is a safe bound since we know that General Relativity breaks down at the Planck scale anyways. These type of terms were also considered in \cite{Turner:1987bw} as a possible mechanism for the generation of magnetic fields during inflation. This becomes a possibility because the non-minimal interactions break the conformal invariance of the Maxwell term in four dimensions.

The Bianchi identities for the Riemann tensor and the field strength tensor guarantee that the equations of motion are second order.
The possible dangerous terms could come from variation with respect to $A_\mu$ with derivatives applying on the Riemann tensor, namely $\frac{1}{2}\mathcal{E}^{\alpha\beta\mu\nu}\mathcal{E}^{\gamma\delta\rho\sigma}\nabla_\gamma R_{\mu\nu\rho\sigma}F_{\alpha\beta}\delta A_\delta$. However, these terms cancel by the virtue of the Bianchi identity for the Riemann tensor $R_{\mu\nu[\rho\sigma;\gamma]}=0$. Hence, the equation of motion for the vector field is given by
\begin{eqnarray}
\left[g^{\mu\rho}g^{\nu\sigma}-\frac{1}{M^2}L^{\mu\nu\rho\sigma}\right]\nabla_\nu F_{\rho\sigma}&=&0 \,,
\end{eqnarray}
which can be also rewritten as
\begin{eqnarray}
\nabla_\nu\left[F^{\mu\nu}-\frac{1}{M^2}L^{\mu\nu\rho\sigma}F_{\rho\sigma}\right]=0,
\end{eqnarray}
since the double dual Riemann tensor is divergence-free. The variation with respect to the metric on the other hand results in the energy momentum tensor of the form
\begin{eqnarray}
T_{\mu\nu}&=&
\frac{1}{2M^2}\left[-R_{\alpha\beta\gamma\delta}\tilde{F}^{\alpha\beta}\tilde{F}^{\gamma\delta}g_{\mu\nu}+2 R_{\mu\beta\gamma\delta}\tilde{F}_\nu^{\;\;\beta}\tilde{F}^{\gamma\delta}+4\nabla^\gamma\nabla^\beta\left(\tilde{F}_{\mu\beta}\tilde{F}_{\gamma\nu}\right)\right] \nonumber\\
&-&F_{\mu\alpha}F_\nu^{\;\;\alpha}
+\frac{1}{4}g_{\mu\nu}F_{\alpha\beta}F^{\alpha\beta}\,.
\label{TmunuLFF}
\end{eqnarray}
In the above expression, the last term in the first line looks like a dangerous term with higher order derivatives. However, on closer inspection one realizes its equivalent form
\begin{eqnarray}\label{e:ddff}
\nabla_\gamma\nabla_\beta\left(\tilde{F}^{\mu\beta}\tilde{F}^{\gamma\nu}\right)=R^\mu_{\lambda\beta\gamma}\tilde{F}^{\lambda\beta}\tilde{F}^{\gamma\nu}+R_{\lambda\gamma}\tilde{F}^{\mu\lambda}\tilde{F}^{\gamma\nu}+\nabla_\gamma\tilde{F}^{\mu\beta}\nabla_\beta\tilde{F}^{\gamma\nu} \,.
\end{eqnarray}
As it becomes clear from this expression, only second order derivatives are involved. As mentioned above, in four dimensions this Horndeski interaction is the unique interaction between a massless vector field and tensor field with second order equations of motion without introducing any ghostly degree of freedom.

\subsection{Generelised Proca theories in curved spacetime}\label{secProcacurved}

A natural fellow-up question is how we can couple in a similar way a massive vector field to gravity from the lessons that we learned for the massless case. Unless the latter case, for massive vector fields we have seen that we can construct Galileon type of derivative interactions on flat spacetime. We have to pay a special attention in the presence
of gravity in order to maintain the equations of motion second order. Promoting naively the partial derivatives of the interactions in section \ref{sec_GeneralizedProcaTheories} to covariant derivatives would yield higher order equations of motion. However, we can compensate for this by adding non-minimal couplings to gravity as counter-terms. For this, we shall use the allowed couplings to the divergenceless tensors multiplied with an overall function. Since the pure Stueckelberg field sector belongs to the scalar Horndeski interactions, we can borrow the same reasoning for the vector field. 

On the other hand, for the interactions with explicit couplings of the Stueckelberg to the transverse modes we have to find the corresponding non-minimal couplings. The generalisation of the derivative self-interactions in equation (\ref{vecGalProcaField}) to a curved spacetime results in
\begin{equation}\label{generalizedProfaField_curved}
\mathcal L^{\rm curved}_{\rm gen. Proca} = -\frac14 \sqrt{-g}F_{\mu\nu}^2 + \sqrt{-g}\sum^6_{n=2} \mathcal L_n
\end{equation}
where this time the interactions $\mathcal L_n$ become \cite{Heisenberg:2014rta,Jimenez:2016isa}
\begin{eqnarray}\label{vecGalcurv}
\mathcal L_2 & = & G_2(A_\mu,F_{\mu\nu}) \nonumber\\
\mathcal L_3 & = &G_3(X)\nabla_\mu A^\mu \nonumber\\
\mathcal L_4 & = & G_{4}(X)R+G_{4,X} \left[(\nabla_\mu A^\mu)^2-\nabla_\rho A_\sigma \nabla^\sigma A^\rho\right] \nonumber\\
\mathcal L_5 & = & G_5(X)G_{\mu\nu}\nabla^\mu A^\nu-\frac{1}{6}G_{5,X} \Big[
(\nabla\cdot A)^3 \nonumber\\
&+&2\nabla_\rho A_\sigma \nabla^\gamma A^\rho \nabla^\sigma A_\gamma -3(\nabla\cdot A)\nabla_\rho A_\sigma \nabla^\sigma A^\rho \Big] \nonumber \\
&-&g_5(X) \tilde{F}^{\alpha\mu}\tilde{F}^\beta_{\;\;\mu}\nabla_\alpha A_\beta  \nonumber \\
\mathcal L_6 & = & G_6(X)\mathcal{L}^{\mu\nu\alpha\beta}\nabla_\mu A_\nu \nabla_\alpha A_\beta 
+\frac{G_{6,X}}{2} \tilde{F}^{\alpha\beta}\tilde{F}^{\mu\nu}\nabla_\alpha A_\mu \nabla_\beta A_\nu
\end{eqnarray}
with $\nabla$ denoting the covariant derivative, $X=-\frac12 A^2$ and $\mathcal{L}^{\mu\nu\alpha\beta}$ the double dual Riemann tensor with the notation $\mathcal{L}^{\mu\nu\alpha\beta}=-\frac12L^{\mu\nu\alpha\beta}=\frac14\mathcal{E}^{\mu\nu\rho\sigma}\mathcal{E}^{\alpha\beta\gamma\delta}R_{\rho\sigma\gamma\delta}$. We can rewrite the second Lagrangian $G_2(A_\mu,F_{\mu\nu})$ as a function of three independent variables $X$, $Y=A^\mu A^\nu F_\mu{}^\alpha F_{\nu\alpha}$ and $F=-\frac14F_{\mu\nu}F^{\mu\nu}$ and we shall ignore parity violating terms of the form $F_{\mu\nu}\tilde{F}^{\mu\nu}$. Hence, the function $G_2$ becomes $G_2(X,Y,F)$ \cite{Heisenberg:2014rta,Fleury:2014qfa}. In addition, one could also consider non-minimal couplings of the form $G^{\mu\nu}A_\mu A_\nu$ in $G_2$. This term is fine on its own due to the absence of any time derivative of the temporal component of the vector field. Nevertheless, this type of interactions can be included in  $\L_4$ after integration by parts. 

Note, that these interactions are the most general interactions giving rise to second order equations of motion up to disformal transformations. The non-minimal couplings $G_{4}(X)R$, $ G_5(X)G_{\mu\nu}\nabla^\mu A^\nu$ and $G_6(X)L^{\mu\nu\alpha\beta}\nabla_\mu A_\nu \nabla_\alpha A_\beta$ are needed in order to cancel the higher order contributions from the counterparts of the covariantization of the self-interactions. However, for the interaction 
$\tilde{G}_5(Y) \tilde{F}^{\alpha\mu}\tilde{F}^\beta_{\;\;\mu}\nabla_\alpha A_\beta$ in $\L_5$ there is no need to introduce a non-minimal coupling as counter-term, since this term is linear in the connection. In fact, the same is true also for the interaction $\L_3$. We can write both contributions as $(G_3g^{\mu\nu}-\tilde{G}_5\tilde{F} ^{\mu\alpha}\tilde{F}^\nu{}_\alpha)\nabla_\mu A_\nu$. 

The above interactions rewritten in terms of $F_{\mu\nu}$ and $S_{\mu\nu}$ reveals the different nature of the interactions, especially those that have the pure intrinsic vector contribution that does not have the scalar correspondence \cite{Jimenez:2016isa}
\begin{eqnarray}\label{vecGalcurvFS}
\mathcal L_2 & = & \hat{G}_2(X,F,Y) \\
\mathcal L_3 & = &\frac12 G_3(X)[S] \nonumber\\
\mathcal L_4 & = & G_{4}(X)R+G_{4,X} \frac{ [S]^2-[S^2]}{4} \nonumber\\
\mathcal L_5 & = & \frac{G_5(X)}{2}G^{\mu\nu} S_{\mu\nu}-\frac{G_{5,X}}{6}\frac{ [S]^3-3[S][S^2] +2[S^3]}{8} \nonumber\\
&+&g_5(X) \tilde{F}^{\mu\alpha}\tilde{F}^\nu_{\;\;\alpha} S_{\mu\nu} \nonumber \\
\mathcal L_6 & = & G_6(X)L^{\mu\nu\alpha\beta} F_{\mu\nu}F_{\alpha\beta}
+\frac{G_{6,X}}{2} \tilde{F}^{\alpha\beta}\tilde{F}^{\mu\nu}S_{\alpha\mu} S_{\beta\nu}\nonumber \,.
\end{eqnarray}
In these Lagrangians, the scalar Horndeski counter part corresponds to taking  $S_{\mu\nu}\rightarrow 2\nabla_\mu\nabla_\nu\pi$. The purely vector nature survives in the couplings $\pi$ with $\tilde{F}^{\mu\nu}$ in the interactions $g_5$ in $\mathcal L_5$ and $G_6$ in $\mathcal L_6$.
For the equations of motion to be of second order nature, the way how the derivatives apply on $S$ at the level of equations of motion plays a crucial role. As long as they satisfy the following relation
\begin{align}
2\nabla_{[\alpha} S_{\beta]\gamma}=&[\nabla_\alpha,\nabla_\beta]A_\gamma+[\nabla_\alpha,\nabla_\gamma]A_\beta-[\nabla_\beta,\nabla_\gamma]A_\alpha +\nabla_\gamma F_{\alpha\beta}
\end{align}
the equations of motion will be second order.\\

\subsection{Beyond Generalized Proca theories}
The previous construction relied on the restriction of second order equations of motion. In order to maintain this property on curved space-time, we have included non-minimal couplings to gravity. In fact, we can construct more general interactions by removing this restriction. Following the spirit of beyond Horndeski construction for scalar fields in section\ref{sec_beyondHorndeski}, we can construct similar beyond generalized Proca interactions. Even if higher order equations of motion arise, the interactions are such that the presence of a constraint equation guarantees the right number of propagating degrees of freedom. The crucial point is that the temporal component of the vector field does not receive dynamics. Including this type of beyond generalized Proca interactions, the action becomes \cite{Heisenberg:2016eld}
\begin{equation}
S=\int d^4x \sqrt{-g} \left( {\cal L}_F
+ \sum_{i=2}^{6}  {\cal L}_n
+{\cal L}^{\rm N}\right)\,, 
\label{beyond_gen_Proc_action}
\end{equation}
with the generalized Proca interactions $ {\cal L}_n$ from equation (\ref{vecGalcurv}) and the new beyond generalized Proca interactions
\be
{\cal L}^{\rm N}=
{\cal L}_4^{\rm N}+{\cal L}_5^{\rm N}+
\tilde{{\cal L}}_5^{\rm N}+
{\cal L}_6^{\rm N}\,,
\ee
where the introduced Lagrangians stand for
\ba
\hspace{-1.2cm}
& &{\cal L}_4^{\rm N}
=f_4 (X)\hat{\delta}_{\alpha_1 \alpha_2 \alpha_3 \gamma_4}^{\beta_1 \beta_2\beta_3\gamma_4}
A^{\alpha_1}A_{\beta_1}
\nabla^{\alpha_2}A_{\beta_2} 
\nabla^{\alpha_3}A_{\beta_3}\,, \label{L4N}\\
\hspace{-1.2cm}
& &{\cal L}_5^{\rm N}
=
f_5 (X)\hat{\delta}_{\alpha_1 \alpha_2 \alpha_3 \alpha_4}^{\beta_1 \beta_2\beta_3\beta_4}
A^{\alpha_1}A_{\beta_1} \nabla^{\alpha_2} 
A_{\beta_2} \nabla^{\alpha_3} A_{\beta_3}
\nabla^{\alpha_4} A_{\beta_4}\,,\label{L5N} \\
\hspace{-1.2cm}
& &\tilde{{\cal L}}_5^{\rm N}
=
\tilde{f}_{5} (X)
\hat{\delta}_{\alpha_1 \alpha_2 \alpha_3 \alpha_4}^{\beta_1 \beta_2\beta_3\beta_4}
A^{\alpha_1}A_{\beta_1} \nabla^{\alpha_2} 
A^{\alpha_3} \nabla_{\beta_2} A_{\beta_3}
\nabla^{\alpha_4} A_{\beta_4}\,,
\label{L5Nd} \\
\hspace{-1.2cm}
& &
{\cal L}_6^{\rm N}
=f_{6}(X)
 \hat{\delta}_{\alpha_1 \alpha_2 \alpha_3 \alpha_4}^{\beta_1 \beta_2\beta_3\beta_4}
\nabla_{\beta_1} A_{\beta_2} \nabla^{\alpha_1}A^{\alpha_2}
\nabla_{\beta_3} A^{\alpha_3} \nabla_{\beta_4} A^{\alpha_4}\,,
\label{L6N}
\ea
with $\hat{\delta}_{\alpha_1 \alpha_2\gamma_3\gamma_4}^{\beta_1 \beta_2\gamma_3\gamma_4}=\mathcal{E}_{\alpha_1 \alpha_2\gamma_3\gamma_4}
\mathcal{E}^{\beta_1 \beta_2\gamma_3\gamma_4}$, and 
the functions $f_{4,5,6}, \tilde{f}_{5}$ depend on $X=-1/2A_\mu A^\mu$. The detuning
between the relative coefficients of the non-minimal couplings to gravity and the
derivative self interactions through the presence of ${\cal L}^{\rm N}$ results in higher order
equations of motion. However, the number of propagating degrees of freedom still remain five in total.\\

\begin{center}
\begin{figure}[h!]
\begin{center}
 \includegraphics[width=16.0cm]{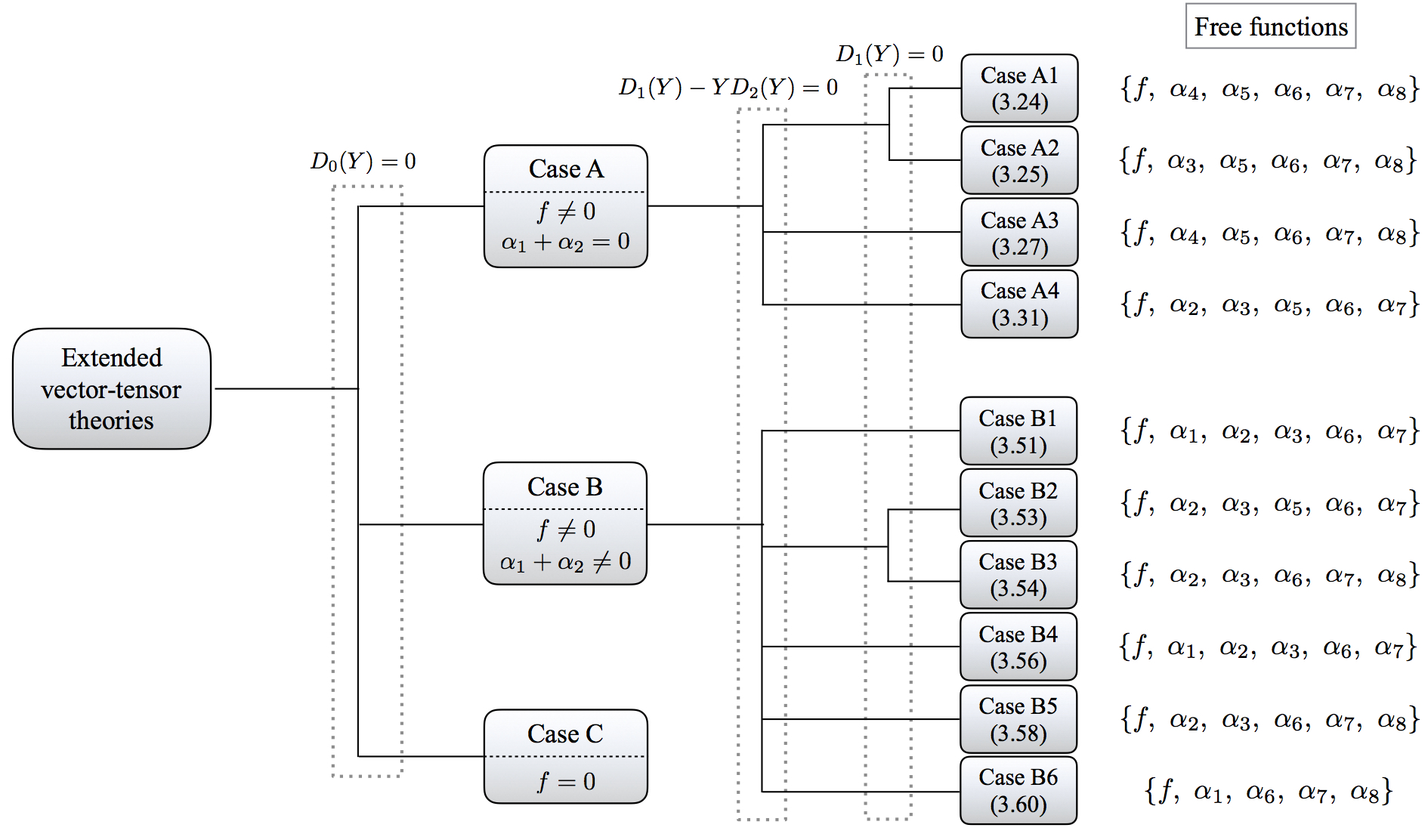}
\end{center}
\caption{\label{figExtendedVectorTheories}
This figure is taken from \cite{Kimura:2016rzw} and shows the classification of different interactions fulfilling the degeneracy condition.}
\end{figure}
\end{center}

Besides the systematical construction in terms of the Levi-Civita tensors, one can on a similar footing consider all the possible contractions between the fields and their derivatives with arbitrary coefficients but constrain those afterwards such that the determinant of the Hessian matrix vanishes and similarly the Hamiltonian possesses a constraint equation. In fact exactly this reasoning was applied in section \ref{sec_GeneralizedProcaTheories} to construct the generalized Proca interactions $\mathcal{L}_{4,5,6}$ as an alternative to the construction with the Levi-Civita tensors. In \cite{Kimura:2016rzw} this approach was applied to construct the corresponding interactions up to $\mathcal{L}_4$ on curved space-time with new contributions
\begin{equation}
S=\int d^4x \sqrt{-g} \left( f R+G_2+G_3\nabla_\mu A^\mu+ \mathcal{C}_4 \right)\,,
\end{equation}
where $\mathcal{C}_4=4C^{\mu\nu\rho\sigma}\nabla_\mu A_\nu \nabla_\rho A_\sigma$ and the tensor $C^{\mu\nu\rho\sigma}$ is the most general tensor constructed out of the metric $g^{\mu\nu}$, the vector field $A^\mu$ and the Levi-Civita tensor and the contraction with $\nabla A$ and can be written in terms of $S_{\mu\nu}$ and $F_{\mu\nu}$ as
\begin{eqnarray}
\mathcal{C}_4&=&a_1 S_{\mu\nu} S^{\mu\nu}+a_2 [S]^2+a_3 A^{\mu}A^{\nu} S_{\mu\nu}[S]+a_4 A^{\mu}A^{\nu} S_{\mu\rho}S_\nu{}\rho+a_5(A^{\mu}A^{\nu} S_{\mu\nu})^2 \nonumber\\
&+& a_6 F_{\mu\nu}F^{\mu\nu}+a_7A^{\mu}A^{\nu} F_{\mu\rho}F_\nu{}\rho+a_8A^{\mu}A^{\nu} F_{\mu\rho}S_\nu{}\rho+a_9 F_{\mu\nu}\tilde{F}^{\mu\nu}\,.
\end{eqnarray}
The last term is parity violating and we can ignore it in the following. The parameters $a_i$ and $f$ are functions of $X$ and are arbitrary. We have to impose the necessary conditions on these parameters in order to guarantee the presence of a constraint equation, that would remove the temporal component of the vector field. For this purpose, let us assume the ADM form for the metric $g_{\mu\nu}=\gamma_{\mu\nu}-n_{\mu}n_\nu$ with the induced metric $\gamma_{\mu\nu}$ and the normal vector $n_{\mu}$. Similarly, we can decompose the vector field as $A_\mu=-n_\mu n^\nu A_\nu+\gamma_{\mu\nu}A^\nu$. The associated extrinsic curvature is given by $K_{\mu\nu}=\gamma_\mu{}^\rho\gamma_\nu{}^\sigma\nabla_\rho n_\sigma $. In terms of these variables we can work out the conditions for the vanishing of the Hessian matrix or similarly the degeneracy condition of the kinetic matrix $\mathcal{K}$. It turns out that the determinant of the kinetic matrix can be brought into the following form \cite{Kimura:2016rzw} 
\begin{equation}
\det \mathcal{K}= D_0(X)+D_1(X)(n^\mu A_\mu)^2+D_2(X)(n^\mu A_\mu)^4\,,
\end{equation}
with the $D_0$ function given by
\begin{equation}
D_0(X)=\frac{(a_1+a_2)}{16} \mathcal{D}(f,a_1,a_2,a_4,a_8,b_1) \qquad \text{with} \qquad b_1=-2a_6-a_7X.
\end{equation}
where the function $ \mathcal{D}$ depends on all the variables in the parenthesis. Similarly, the function $D_1$ can be decomposed into three parts 
\begin{equation}
D_1(X)=\mathcal{E}_1(f,a_1,a_2,X) a_8^2+\mathcal{E}_2(f,f_X,a_1,a_2,a_3,a_4,a_5,X)a_8+\mathcal{E}_3(f,f_X,a_1,a_2,a_3,a_4,X,\beta)\,.
\end{equation}
The expression of $D_2$ on the other hand is such that it satisfies the following relation
\begin{eqnarray}
D_1-XD_2&=&-\frac{(f-a_1X)}{16X}(2a_1+X(a_4+a_8)-\beta)\Big[4(a_1+a_2+(a_3+a_4)X+a_5X^2) \nonumber\\
&&\times(2f+(a_1+3a_2)X)-3X(2a_2+4f_X+a_3X)^2\Big]+\frac{D_0}{X}\,.
\end{eqnarray}
In order to have vanishing determinant of the kinetic matrix, we have to impose the degeneracy condition
\begin{equation}
D_0=D_1=D_2=0\,.
\end{equation}
From the vanishing of $D_0$, we immediately observe that there are two options. The first option is setting $a_2=-a_1$, which we were obtaining from the systematical construction in terms of the Levi-Civita tensors and corresponds to $[S^2]-[S]^2$. The second option is to set the function $\mathcal{D}=0$, which represents new type of interactions. A subclass of the generalized Proca interactions in (\ref{vecGalcurv}) is equivalent to choosing $f=G_4$, $a_1=-a_2=2G_{4,X}$ and $a_3=a_4=a_5=a_8=0$ and satisfies the degeneracy condition. In fact, the kinetic matrix in this case takes the simple form
\begin{eqnarray}
K_{\mu\nu}=
\begin{pmatrix}
0&0&0&0 \\
0&K_{22} &0&K_{24} \\
0&0&0 &K_{34} \\
0&K_{24}&K_{34}&K_{44}
 \end{pmatrix}\,.
\end{eqnarray}
The exact expression for the non-zero components is irrelevant here (see \cite{Kimura:2016rzw} for more detail). The crucial point is that the determinant of the kinetic matrix vanishes representing the presence of a constraint that removes the fourth degree of freedom of the vector field. The same property is also observed if we include the fourth order beyond generalized Proca interaction ${\cal L}_4^{\rm N}$. In this case, the parameters correspond to $f=G_4$, $a_1=-a_2=2G_{4,X}+f_4X$, $a_3=-a_4=2f_4$ and $a_5=a_8=0$ and the kinetic matrix modifies into
\begin{eqnarray}
K_{\mu\nu}=
\begin{pmatrix}
0&0&0&K_{14} \\
0&K_{22} &0&K_{24} \\
0&0&0 &K_{34} \\
K_{14}&K_{24}&K_{34}&K_{44}
 \end{pmatrix}\,.
\end{eqnarray}
The determinant of this matrix vanishes as well and this proves that the beyond generalized Proca interactions up to quartic order are degenerate. These two examples show nicely, how the condition of vanishing Hessian matrix or equivalently the vanishing of the kinetic matrix permits to construct the allowed vector interactions also on curved space-time. The set of allowed interactions up to quartic order fulfilling the degeneracy condition has been classified in \cite{Kimura:2016rzw} and summarized into a nice figure shown in figure \ref{figExtendedVectorTheories}\,. Similar ADM and Hamiltonian analysis can be done for the quintic and sixth order generalised Proca and beyond generalised Proca interactions.

\subsection{Vainshtein mechanism in generalised Proca}
Theories with derivative self-interactions as the Galileon or helicity-0 mode of massive gravity strongly rely on the successful implementation of the Vainshtein mechanism, as we saw in section \ref{vainshtein_Galileon}. The longitudinal mode of the massive vector field also carries derivative self-interactions, therefore it becomes crucial to study how
the longitudinal mode gravitates on spherically symmetric backgrounds. This was exactly performed in \cite{DeFelice:2016cri} and it was shown that the cubic-order self-interactions lead to the suppression of the
longitudinal mode through the Vainshtein mechanism even in the presence of the auxiliary temporal component.

Let us consider the line element of the spherically symmetric and static background $ds^{2}=-e^{2\Psi(r)}dt^{2}+e^{2\Phi(r)}dr^{2}+r^{2} \left( d\theta^{2}+\sin^{2}\theta\,d\varphi^{2}\right)$ and matter field configuration with density $\rho_m(r)$ and pressure $P_m(r)$, and similarly the vector field configuration $A^{\mu}=\left(\phi(r), e^{-2\Phi}\chi'(r),0,0 \right)$. As a proof of concept it will be sufficient to consider only the interactions up to cubic order $\mathcal{L}_2$ and $\mathcal{L}_3$ in equation (\ref{vecGalcurv}) with $G_4=\mpl^2/2$. The vector field equations are given by 
\ba
& &
\frac{1}{r^2} \frac{d}{dr} (r^2 \phi')-e^{2\Phi}G_{2,X} \phi
-G_{3,X}\phi\, \frac{1}{r^2} \frac{d}{dr} (r^2 \chi') +2\phi \left( \Psi''+\Psi'^2-\Psi'\Phi' \right)\nonumber \\
& &
-\left( \phi \chi' G_{3,X}-3\phi'-\frac{4\phi}{r} \right)
\Psi'
+\left( \phi \chi' G_{3,X}-\phi' \right)\Phi'=0\,,
\label{Vainshtein_eqphi} \\
& &
\chi' G_{2,X}+\left( e^{2\Psi} \phi \phi'+\frac{2}{r}
e^{-2\Phi} \chi'^2 \right) G_{3,X}
+\left( e^{2\Psi} \phi^2+e^{-2\Phi}\chi'^2 \right)G_{3,X}\Psi'=0\,.
\label{Vainshtein_eqpsi}
\ea
In order to facilitate the computation, we can focus on the functions $G_2=m^2X$ and $G_3(X)=\beta_3 X$ with an arbitrary parameter $\beta_3$. The matter field equation is simply the continuity equation $P_m'+\Psi'(\rho_m+P_m)=0$. In order to be in agreement with local gravity experiments, we have to impose that the gravitational potentials $\Phi$ and $\Psi$ are close to those in General Relativity. 

Let us approximate the matter density as $\rho_m(r)\sim\rho_0$ for $r<r_*$ and $\rho_m(r)\sim0$ for $r>r_*$, thus, with an abrupt change at $r_*$. In the regime of weak gravity, the interior solutions $r<r_*$ for $\Phi$ and $\Psi$ are $\Phi_{\rm GR} \simeq \frac{\rho_0 r^2}{6\mpl^2}$ and $\Psi_{\rm GR} \simeq \frac{\rho_0}{12\mpl^2} \left( r^2-3r_*^2 \right)$, respectively, and the exterior solutions are $\Phi_{\rm GR} \simeq \frac{\rho_0r_*^3}{6\mpl^2r}$ and $\Psi_{\rm GR} \simeq -\frac{\rho_0r_*^3}{6\mpl^2r}$ for $r>r_*$. Plugging these interior and exterior solutions in the vector field equations \eqref{Vainshtein_eqphi}-\eqref{Vainshtein_eqpsi}, and restricting the solutions to the form $\phi(r)=\phi_0+f(r)$, we can find analytical approximative solutions for $\phi$ and $\chi$
\ba
\phi(r)
=\phi_0 \left[ 1
-{\cal B}\frac{\rho_0}{6\mpl^2}
r^2 \right]\, \qquad \text{and} \qquad
\chi'(r)
=\sqrt{\frac{\rho_0\phi_0^2}{6\mpl^2}
\left[ {\cal B}-\frac12\right]}\,r\,.
\label{chiphiinL3}
\ea
in the interior regime $r<r_*$, where we introduced the short-cut notations ${\cal B}\equiv (1+b_3)\left( 1-\sqrt{\frac{b_3}{1+b_3}}\right)$ and $b_3=\frac{3(\beta_3 \phi_0\mpl)^2}{4\rho_0}$. From these solutions, it becomes clear that the screening effect is efficient to suppress the propagation of the longitudinal mode for large coupling $|\beta_3|$. 

\begin{center}
\begin{figure}[h!]
\begin{center}
\includegraphics[width=2.9in]{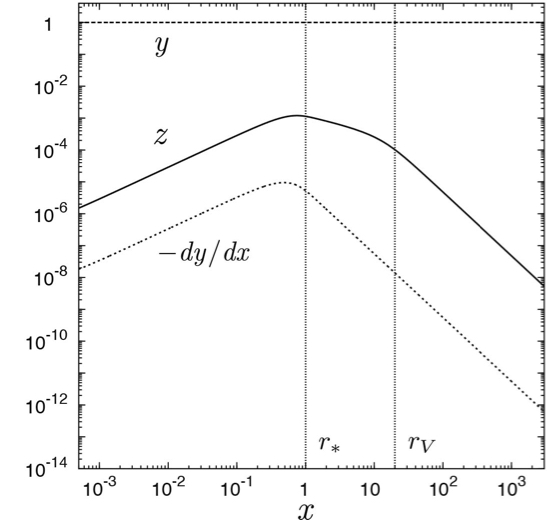}
\includegraphics[width=2.9in]{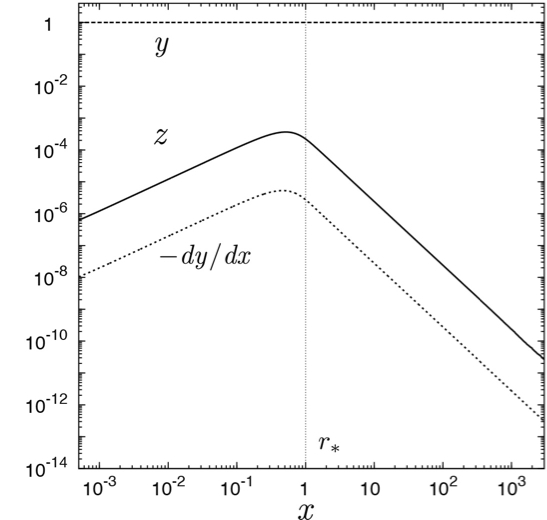}
\caption{This figure is taken from \cite{DeFelice:2016cri} and shows the numerical solutions to $y=\phi/\phi_0$,
$-dy/dx$, and $z=\chi'/\phi_0$ as a function of $x=r/r_*$
for the matter profile $\rho_m=\rho_0 e^{-4r^2/r_*^2}$
with $\Phi_*=10^{-4}$.
The left panel corresponds to $b_3=10^{-4}$ and the right panel to
$b_3=1$, and the two vertical
lines denote the scales $r=r_*$ and $r_V=20r_*$, respectively. The numerical solutions
confirm the analytical estimations made in the main text.
\label{ScreeningProca}}
\end{center}
\end{figure}
\end{center}

Similarly, using the weak gravity exterior solutions for $\Phi_{\rm GR}$ and $\Psi_{\rm GR}$, we can also estimate the solutions for $\phi$ and $\chi$ in the exterior regime for $r>r_*$
\ba
\phi'(r)=-\frac{\rho_0\phi_0r_*^3}{3\mpl^2r^2}
{\cal B}_{\tilde{b}_3}\, \qquad \text{and} \qquad
\chi'(r)=\sqrt{\frac{\rho_0r_*^3\phi_0^2}{6\mpl^2r}
\left[ {\cal B}_{\tilde{b}_3}-\frac12 \right]}\,.
\label{chiphiinL3bigr}
\ea
where ${\cal B}_{\tilde{b}_3}$ is ${\cal B}$ with respect to the rescaled $\tilde{b}_3 \equiv b_3 \frac{r^3}{r_*^3}$. Now, we have to distinguish between the strength of the coupling constant. For $b_3\gg1$ (or equivalently $\tilde{b}_3\gg1$), the two vector modes acquire $\phi'(r) \simeq -\frac{\rho_0\phi_0r_*^3}{6\mpl^2r^2}$ and $\chi'(r) \simeq \frac{\rho_0r_*^3}{6\beta_3 \mpl^2 r^2}$ for $r>r_*$. For $b_3 \lesssim 1$ there is a transition radius $r_V=r_*/b_3^{1/3}$, the Vainshtein radius, where the longitudinal mode changes its $r$ dependence. For distances $r_*<r \ll r_V$, the profiles of the two vector fields change to $\phi'(r) \simeq -\frac{\rho_0\phi_0r_*^3}{3\mpl^2r^2}$ and $\chi'(r)\simeq\sqrt{\frac{\rho_0r_*^3\phi_0^2}{12M_{\rm pl}^2r}}$, respectively. For distances $r \gg r_V$ we have $\tilde{b}_3 \gg 1$ and the longitudinal mode decreases faster than in the regime $r_*<r \ll r_V$ with a suppressed amplitude. Above the Vainshtein radius $\chi'(r)$ decays quickly. Summarising, for large coupling constants $b_3\gg1$ the longitudinal mode is strongly suppressed both in the interior and exterior regimes due to a efficient Vainshtein mechanism. On the other hand, for $s_{\beta_3} \lesssim 1$, the Vainshtein screening of the longitudinal mode becomes efficient for large distances $r>r_V$. This is an intriguing property of the vector Galileon. It is very interesting that the suppression of the longitudinal mode
happens outside the radius $r_V$ for small $|\beta_3|$ and gives a unique opportunity to test this feature of the
vector Galileons. For more detail we refer the reader to \cite{DeFelice:2016cri}. These analytical estimations have been also numerically confirmed there, which is shown in figure \ref{ScreeningProca}.

\section{Scalar-Vector-Tensor theories}
In the previous sections we have discussed the consistent constructions of scalar-tensor and vector-tensor theories constituting the scalar Horndeski and the generalised Proca theories (and beyond). These theories initiated a lot of activities in the literature. A crucial step towards unifying these two 
important classes of gravity theories, the Horndeski and generalised Proca theories, was attempted in \cite{Heisenberg:2018acv}. The resulting Lagrangian corresponds to the most general scalar-vector-tensor theories with second order equations of motion, for both the gauge invariant and the gauge broken cases. This opens a new area of research and has also important implications for a broad spectrum of different research areas, including the application to inflation and generation of primordial magnetic fields, new black hole and neutron star solutions, dark matter and dark energy. The presence of both the scalar and vector field allows for richer phenomenology. We shall start with the construction of the scalar-vector-tensor theories with gauge invariance.

\subsection{Gauge invariant scalar-vector-tensor theories}
We can start the analysis with the restricted case, where the vector field is gauge invariant. We ask ourselves the question: what is the most general Lagrangian for a scalar field, a diffeomorphism invariant tensor field and a $U(1)$ gauge field with derivative interactions, that gives rise to second order equations of motion with five physical degrees of freedom. The restriction of gauge invariance shrinks the number of allowed interactions considerably. Our knowledge about the consistent construction of generalized Proca theories will be very useful for figuring out the allowed interactions for the gauge invariant scalar-vector-tensor theories. 

The construction of the generalized Proca interactions is based on the breaking of gauge invariance of the vector field, which nonetheless can be reintroduced by a scalar Stueckelberg field. In this way, the same interactions can be alternatively written as scalar-vector theories. By construction, the scalar Stueckelberg field appears only via derivatives and therefore has the shift symmetry. Promoting these interactions to the more general case with broken shift symmetry will allow us to construct the desired scalar-vector-tensor theories with the genuine new couplings between the scalar field and the vector field. The relative tuning of the interactions with derivative dependence of the Stueckelberg field does not require any additional adjustment for those, which break the shift symmetry.
The arising action of the scalar-vector-tensor theories is given by \cite{Heisenberg:2018acv}
\begin{equation}
\mathcal{S}=\mathcal{S}_{\rm ST}+\mathcal{S}_{\rm SVT},
\end{equation}
with the well known scalar-tensor interactions in $\mathcal{S}_{\rm ST}$ being the Horndeski interactions that we saw in section \ref{sec_scalar-tensor}. For the sake of easier presentation we recall them here again. Note that we will treat the quadratic Horndeski $\mathcal{L}^2_{\rm ST}$ separately from the rest and hence the pure Horndeski part starts from $i=3$, $\mathcal{S}_{\rm ST}=\int d^4x \sqrt{-g}\sum_{i=3}^5\mathcal{L}^i_{\rm ST}$ with
\begin{eqnarray}\label{HorndeskiL3toL5}
\mathcal{L}^3_{\rm ST}&=&G_3[\Pi]\\
\mathcal{L}^4_{\rm ST}&=&G_4R+G_{4,X}\left([\Pi]^2-[\Pi^2]\right)\nonumber\\
\mathcal{L}^5_{\rm ST}&=&G_5G_{\mu\nu}\Pi^{\mu\nu}-\frac{G_{5,X}}{6}\left([\Pi]^3-3[\Pi][\Pi^2]+2[\Pi^3]\right)\,. \nonumber
\end{eqnarray}
As usual, the functions $G_{3,4,5}(\pi,X)$ depend on the scalar field $\pi$ and its derivatives $X=-\frac12(\partial\pi)^2$, with $G_{i,X}=\partial G_i/\partial X$ and $G_{i,\pi}=\partial G_i/\partial \pi$. Since the vector field $A_\mu$ is gauge invariant, it will only appear via the gauge invariant field strength $F_{\mu\nu}=\nabla_\mu A_\nu-\nabla_\nu A_\mu$ and its dual $\tilde{F}^{\mu\nu}=\frac12\epsilon^{\mu\nu\alpha\beta}F_{\alpha\beta}$. 

We can construct the corresponding scalar-vector-tensor Lagrangians order by order as $\mathcal{S}_{\rm SVT}=\int d^4x \sqrt{-g}\sum_{i=2}^4\mathcal{L}^i_{\rm SVT}$. We can start with the simplest quadratic Lagrangian, which does not carry any second derivative for the scalar Stueckelberg field. The second order Lagrangian of the scalar-vector-tensor theories is given by an arbitrary function of the form $\mathcal{L}^2=f_2(\pi, \partial_\mu\pi, F_{\mu\nu}, \tilde{F}_{\mu\nu})$ \cite{Heisenberg:2014rta,Fleury:2014qfa}. We can use these four arguments in order to build independent scalar quantities. These are given by $X$, $F=F_{\mu\nu}F^{\mu\nu}$, $\tilde{F}=F_{\mu\nu}\tilde{F}^{\mu\nu}$ and $Y=\partial_\mu\pi \partial_\nu \pi F^{\mu\alpha}F^\nu{}_\alpha$. Therefore, we can write the second order Lagrangian of the scalar-vector-tensor theories as
 \begin{equation}\label{L2SVTgauge}
\mathcal{L}^{2}_{\rm SVT}=f_2(\pi,X,F,\tilde{F},Y) \,.
\end{equation}
The dependence on $\pi$ and $X$ allows us to include the second order Lagrangian of the Horndeski $\mathcal{L}^2_{\rm ST}=G_2(\pi,X)$ directly here. For this reason we had omitted it in (\ref{HorndeskiL3toL5}). The scalar quantities $\partial_\mu\pi \partial_\nu \pi \tilde{F}^{\mu\alpha}\tilde{F}^\nu{}_\alpha$  and $\partial_\mu\pi \partial_\nu \pi F^{\mu\alpha}\tilde{F}^\nu{}_\alpha$ do not represent independent contractions and are already included in the scalar quantities in $f_2$. 

So far, the interactions carry only one derivative per field. More interesting interactions arise when we allow two derivatives per field, but they require more caution. New genuine couplings between the scalar Stueckelberg field and the gauge field can be constructed imposing consistent conditions on their symmetries and coupling tensors. First of all, let us start with the linear interaction in $\nabla\partial\pi$. This object can be coupled to a symmetric rank-2 tensor via $\mathcal{M}_3^{\mu\nu}\nabla_\mu\partial_\nu\pi$, which can only depend on $\partial_\mu\pi$, $\tilde{F}_{\mu\nu}$ and $g_{\mu\nu}$. For the absence of Ostrogradski instabilities it is crucial to avoid higher time derivatives, therefore $\mathcal{M}^{00}_3$ should not have any other time derivatives than $\dot\pi$ and contain the gauge field only via $\tilde{F}^{0\alpha}\tilde{F}^0{}_\alpha\sim B^2$. These requirements uniquely determines the third order Lagrangian of the scalar-vector-tensor interactions to be of the form
 \begin{equation}
\mathcal{L}^3_{\rm SVT}=\mathcal{M}_3^{\mu\nu}\nabla_\mu\partial_\nu\pi
\end{equation}
with the symmetric rank-2 tensor $\mathcal{M}_3^{\mu\nu}$ given by
\begin{equation}
\mathcal{M}^{\mu\nu}_3=\left( f_3(\pi,X)g_{\rho\sigma}+\tilde{f}_3(\pi,X)\partial_\rho\pi\partial_\sigma\pi\right) \tilde{F}^{\mu\rho}\tilde{F}^{\nu\sigma}\,.
\end{equation}
with the arbitrary free functions $f_3$ and $\tilde{f}_3$, that depend on $X$ and $\pi$.

The next order interactions quadratic in $\partial^2\pi$ can be constructed in a similar way. The forth order Lagrangian of the scalar-vector-tensor interactions results in
\begin{equation}
\mathcal{L}^{4}_{\rm SVT}=\left( \mathcal{M}_4^{\mu\nu\alpha\beta}\nabla_\mu\partial_\alpha\pi\nabla_\nu\partial_\beta\pi+f_4(\pi,X)L^{\mu\nu\alpha\beta}F_{\mu\nu}F_{\alpha\beta}\right)
\end{equation}
with the double dual Riemann tensor $L^{\mu\nu\alpha\beta}$ and the four-rank tensor $\mathcal{M}_4^{\mu\nu\alpha\beta}$ of the form
\begin{equation}
\mathcal{M}^{\mu\nu\alpha\beta}_4=\left( \frac12f_{4,X}+\tilde{f}_4(\pi)\right)\tilde{F}^{\mu\nu}\tilde{F}^{\alpha\beta}\,.
\end{equation}
The part of the $X$ dependence requires the introduction of the non-minimal coupling to have second order equations of motions, but not the $\pi$ dependence, therefore the additional $\tilde{f}_4$ dependence can be included. 

Summarising, the general scalar-vector-tensor theories with gauge invariance and second order equations of motion can be constructed as $\mathcal{S}=\int d^4x \sqrt{-g}\left(\sum_{i=3}^5\mathcal{L}^i_{\rm ST}+\sum_{i=2}^4\mathcal{L}^i_{\rm SVT}\right)$ with the pure scalar sector $\mathcal{L}^i_{\rm ST}$ and the genuine scalar-vector-tensor interactions $\mathcal{L}^i_{\rm SVT}$
\begin{eqnarray}\label{genLagrangianSVT}
\mathcal{L}^2_{\rm SVT}&=&f_2(\pi,X,F,\tilde{F},Y) \\
\mathcal{L}^3_{\rm SVT}&=&\mathcal{M}_3^{\mu\nu}\nabla_\mu\partial_\nu\pi \nonumber\\
\mathcal{L}^{4}_{\rm SVT}&=& \mathcal{M}_4^{\mu\nu\alpha\beta}\nabla_\mu\partial_\alpha\pi\nabla_\nu\partial_\beta\pi+f_4(\pi,X)L^{\mu\nu\alpha\beta}F_{\mu\nu}F_{\alpha\beta}\nonumber\,.
\end{eqnarray}
These interactions give rise to second order equations of motion, for both the scalar and gauge field and there are five propagating degrees of freedom.
It is worthwhile to mention that the pure gauge invariant vector-tensor theories are recovered in the limit of a constant scalar field. The standard Horndeski vector interactions $\sqrt{-g}L^{\mu\nu\alpha\beta}F_{\mu\nu}F_{\alpha\beta}$ arises in the case $f_4=\text{constant}$. In \cite{Heisenberg:2018vti} it has been shown, that these gauge invariant scalar-vector-tensor theories admit interesting black hole solutions with scalar and vector hair and stable perturbations \cite{Heisenberg:2018mgr}.

\subsection{Scalar-vector-tensor theories with broken gauge invariance}
\label{Nongauge}
So far the constructed interactions were scalar-vector-tensor theories with $U(1)$ gauge invariance. Removing this restriction allows for the construction of more general interactions. In the broken gauge case, the vector field enters also via the tensor $S_{\mu\nu}=\nabla_\mu A_\nu+\nabla_\nu A_\mu$ apart from the field strength tensor $F_{\mu\nu}$ and its dual $\tilde{F}_{\mu\nu}$. We also introduce an effective tensor given by
 \begin{eqnarray}\label{SVTeff}
\mathcal{G}^{\mu\nu}_{f_{nj}} = f_{n1}(\pi,X_i)g_{\mu\nu}+f_{n2}(\pi,X_i)\partial_\mu\pi\partial_\nu\pi 
+f_{n3}(\pi,X_i)A_\mu A_\nu+f_{n4}(\pi,X_i)A_\mu \partial_\nu\pi
\end{eqnarray}
where $i=1,2,3$ and $j=1,2,3,4$ and the subscript $f_{nj}$ denotes the four different arbitrary functions of $\pi$ and $X_i$, where the latter represent the scalar quantities $X_1=-\frac12(\partial\pi)^2$, $X_2=-\frac12\partial_\mu\pi A^\mu$, $X_3=-\frac12A_\mu A^\mu$, respectively. The quadratic Lagrangian can be an arbitrary function of the following form 
\begin{equation}
\mathcal{L}^{2,ng}_{\rm SVT}=f_2(\pi,X_1,X_2,X_3,,F,\tilde{F},Y_1,Y_2,Y_3) \,,
\end{equation}
where $Y_i$ represent the scalar quantities $Y_1=\partial_\mu\pi \partial_\nu \pi F^{\mu\alpha}F^\nu{}_\alpha$, $Y_2=\partial_\mu\pi A_\nu F^{\mu\alpha}F^\nu{}_\alpha$, $Y_3=A_\mu A_\nu F^{\mu\alpha}F^\nu{}_\alpha$. This is a natural extension of the gauge invariant quadratic scalar-vector-tensor theories in (\ref{L2SVTgauge}) to the gauge broken case. The crucial difference comes in the $i$-dependence of the functions $X$ and $Y$. With the help of the effective tensor introduced in (\ref{SVTeff}), we can construct the cubic Lagrangian using some of its functional dependence. The interactions correspond to scalar quantities constructed out of $S_{\mu\nu}$ and the disformal part of the effective metric. For the cubic interactions, the general form of the effective metric gets restricted to the disformal transformation $g^{\mu\nu}+A^{\mu}A^{\nu}$ and the $X_3$ dependence only
\begin{equation}
\mathcal{L}^{3,ng}_{\rm SVT}=\Big(f_{31}(\pi,X_3)g^{\mu\nu}+f_{32}(\pi,X_3)A^{\mu}A^{\nu}\Big) S_{\mu\nu}
\end{equation}
with the two different functions $f_{31}$ and $f_{32}$ depending on $\pi$ and $X_3$. As usual the cubic interactions do not require any non-minimal coupling since the connection appears linear. More caution will be needed for the quartic Lagrangian, since it again requires the presence of non-minimal coupling to the Ricci scalar
\begin{equation}
\mathcal L^{4,ng}_{\rm SVT}  =  f_{4}(\pi,X_3)R+f_{4,X_3}S_2
\end{equation}
where $S_2=([S]^2-[S^2])/4$ is the standard quadratic elementary polynomial of $S_{\mu\nu}$. Note, that the $X_3$ dependence again surfaces as the unique scalar quantity out of the $X_i$ and requires the relative tuning to the non-minimal coupling to the gravity sector as we saw for the generalised Proca interactions. In a similar way we can construct the next order interactions. Again, the $X_3$ dependent part of the interactions needs the introduction of a counterpart in form of a non-minimal coupling to the Einstein tensor
\begin{eqnarray}
\mathcal L^{5,ng}_{\rm SVT}& = & \frac{f_5(\pi,X_3)}{2}G^{\mu\nu} S_{\mu\nu}-\frac{f_{5,X_3}}{6}S_3\nonumber\\
&+&\mathcal{M}_5^{\mu\nu}\nabla_\mu\partial_\nu\pi+\mathcal{N}_5^{\mu\nu}S_{\mu\nu}
\end{eqnarray}
where $S_3=( [S]^3-3[S][S^2] +2[S^3])/8$ and the rank-2 tensors are given by
\begin{eqnarray}\label{svt_h5andtildeh5}
\mathcal{M}^{\mu\nu}_5= \mathcal{G}_{\rho\sigma}^{h_{5j}} \tilde{F}^{\mu\rho}\tilde{F}^{\nu\sigma} \qquad \text{and} \qquad
\mathcal{N}^{\mu\nu}_5= \mathcal{G}_{\rho\sigma}^{\tilde{h}_{5j}} \tilde{F}^{\mu\rho}\tilde{F}^{\nu\sigma}\,.
\end{eqnarray}
 Thus, the $\nabla_\mu\partial_\nu\pi$ and $S_{\mu\nu} $ dependent parts of the Lagrangian come in with different independent functions $h_{5j}$ and $\tilde{h}_{5j}$, respectively. As it was pointed out 
in \cite{Heisenberg:2018acv}, the naive explicit dependence on all the $h_{nj}$ functions would introduce dynamics for the temporal component of the vector field on a general background and therefore needs an additional caution. In order for this not to happen, the dependence of $\mathcal{M}_5^{\mu\nu}$ would need to be restricted to $X_1$ only and similarly the dependence of $\mathcal{N}_5^{\mu\nu}$ to $X_3$. Thus, the forms of the rank-2 tensors in equation \eqref{svt_h5andtildeh5} have to be further restricted down to
\begin{eqnarray}
&&\mathcal{M}^{\mu\nu}_5= \Big(h_{51}(\pi,X_1)g_{\rho\sigma}+h_{52}(\pi,X_1)\partial_{\rho}\pi\partial_{\sigma}\pi\Big) \tilde{F}^{\mu\rho}\tilde{F}^{\nu\sigma} \\
&&\mathcal{N}^{\mu\nu}_5= \Big(\tilde{h}_{51}(\pi,X_3)g_{\rho\sigma}+\tilde{h}_{52}(\pi,X_3)A_{\rho}A_{\sigma}\Big)  \tilde{F}^{\mu\rho}\tilde{F}^{\nu\sigma}\,.
\end{eqnarray}

Last but not least, we can construct the sixth order Lagrangian with genuine scalar-vector-tensor interactions and non-minimal coupling to gravity as
\begin{eqnarray}
\mathcal{L}^{6,ng}_{\rm SVT}&=&f_6(\pi,X_1)L^{\mu\nu\alpha\beta}F_{\mu\nu}F_{\alpha\beta}+ \mathcal{M}_6^{\mu\nu\alpha\beta}\nabla_\mu\partial_\alpha\pi\nabla_\nu\partial_\beta\pi \nonumber\\
&+&\tilde{f}_6(\pi,X_3)L^{\mu\nu\alpha\beta}F_{\mu\nu}F_{\alpha\beta}+ \mathcal{N}_6^{\mu\nu\alpha\beta}S_{\mu\alpha}S_{\nu\beta}
\end{eqnarray}
with the tensors $\mathcal{M}_6$ and $\mathcal{N}_6$ given by
\begin{equation}
\mathcal{M}^{\mu\nu\alpha\beta}_6=2f_{6,X_1}\tilde{F}^{\mu\nu}\tilde{F}^{\alpha\beta} \qquad \text{and} \qquad \mathcal{N}^{\mu\nu\alpha\beta}_6=\frac12\tilde{f}_{6,X_3}\tilde{F}^{\mu\nu}\tilde{F}^{\alpha\beta}\qquad  \,.
\end{equation}
with the two different functions $f_6$ and $\tilde{f}_6$ respectively. 

Summarising, the general scalar-vector-tensor theories with broken gauge invariance is given by $\mathcal{S}=\int d^4x \sqrt{-g}\left(\sum_{i=3}^5\mathcal{L}^i_{\rm ST}+\sum_{i=2}^4\mathcal{L}^i_{\rm SVT}\right)$ with the pure scalar sector $\mathcal{L}^i_{\rm ST}$ as before and the new scalar-vector-tensor interactions $\mathcal{L}^i_{\rm SVT}$
\begin{eqnarray}\label{genLagrangianSVTnoGauge}
\mathcal{L}^{2,ng}_{\rm SVT}&=&f_2(\pi,X_1,X_2,X_3,F,\tilde{F},Y_1,Y_2,Y_3) \\
\mathcal{L}^{3,ng}_{\rm SVT}&=&\Big(f_{31}(\pi,X_3)g^{\mu\nu}+f_{32}(\pi,X_3)A^{\mu}A^{\nu}\Big) S_{\mu\nu}\nonumber\\
\mathcal L^{4,ng}_{\rm SVT} & = & f_{4}(\pi,X_3)R+f_{4,X_3}S_2 \nonumber\\
\mathcal L^{5,ng}_{\rm SVT}& = & \frac{f_5(\pi,X_3)}{2}G^{\mu\nu} S_{\mu\nu}-\frac{f_{5,X_3}}{6}S_3\nonumber\\
&+&\mathcal{M}_5^{\mu\nu}\nabla_\mu\partial_\nu\pi+\mathcal{N}_5^{\mu\nu}S_{\mu\nu} \nonumber\\
\mathcal{L}^{6,ng}_{\rm SVT}&=&f_6(\pi,X_1)L^{\mu\nu\alpha\beta}F_{\mu\nu}F_{\alpha\beta}+ \mathcal{M}_6^{\mu\nu\alpha\beta}\nabla_\mu\partial_\alpha\pi\nabla_\nu\partial_\beta\pi \nonumber\\
&+&\tilde{f}_6(\pi,X_3)L^{\mu\nu\alpha\beta}F_{\mu\nu}F_{\alpha\beta}+ \mathcal{N}_6^{\mu\nu\alpha\beta}S_{\mu\alpha}S_{\nu\beta}\nonumber\,.
\end{eqnarray}
Since the vector sector of the interactions does not carry any gauge invariance, there are six propagating degrees of freedom in this case (two tensors, two vectors and two scalars). Note, that we can obtain the same scalar-vector-tensor interactions from the decoupling limit of the multi-Proca theories, that we will discuss in the next section. We believe that these scalar-vector-tensor theories will have very rich applications to cosmology, specially to the early universe cosmology and dark matter phenomenology. The first application to cosmology (inflation and dark energy) has been studied in \cite{Heisenberg:2018mxx}.

\section{Set of vector fields}
In the previous constructions of the vector field interactions, we were considering a single vector field. For the gauge invariant vector field, we saw that one can not construct vector Galileon derivative self-interactions for a single gauge field and the unique consistent non-minimal coupling on curved space-time is via the double dual Riemann tensor. Breaking explicitly the gauge invariance allowed us to construct the generalized Proca interactions in the same spirit as the Galileon/Horndeski interactions. In this section we would like to consider a set of vector fields and their interactions, for both non-abelian gauge invariant and broken cases.

\subsection{Non-abelian vector fields}
One immediate interesting question worth investigating is how massless vector fields can interact in a consistent way. In fact, some of the fundamental particles of the Standard Model of Particle Physics are non-abelian gauge fields. These are the very well known Yang-Mills theories. Consider a set of massless vector fields $A_\mu^a$, where $a$ denotes the index of the field space. The consistent Lagrangian arises after the successful extension of the abelian vector field to the case of non-abelian vector field
\begin{equation}\label{non-abelianGaugeFields}
\mathcal{L}=-\frac{1}{4}\mathcal{G}_{ab}\mathcal{F}^{a}_{\mu\nu}\mathcal{F}^{b}_{\mu\nu}g^{\mu\alpha}g^{\nu\beta}\,,
\end{equation}
where  $\mathcal{G}_{ab}$ is the metric of the field space and $\mathcal{F}^a_{\mu\nu} =F^a_{\mu\nu} + g_c f^{abc}A^b_\mu A^c_\nu$ with $F^a_{\mu\nu} = \partial_\mu A^a_\nu -\partial_\nu A^a_\mu$ and $g_c$ is the coupling constant of the non-abelian field and $f^{abc}$ are the structure constants of the Lie algebra of the isometry group of $\mathcal{G}_{ab}$.

The generators $T_a$ of the Lie algebra satisfy $[T_a, T_b]=i f_{ab}{}^cT_c$. The non-abelian vector fields correspond to nothing else but given values in the Lie algebra of the group and under an isometry transformation with parameters $\theta^a$ they transform in the adjoint representation $A_\mu^a \to A_\mu^a + f_{bc}{}^a\theta^b A^c_\mu-\partial_\mu \theta^a/g_c$, i.e., representing a connection. With this one can then define the covariant derivative as $D_\mu\equiv \partial_\mu I-ig_c A^a_\mu T_a$, with the commutation relation $[D_\mu,D_\nu]=-ig_c\mathcal{F}^a_{\mu\nu}T_a$.  

Apart from the standard Lagrangian of the Yang-Mills theories (\ref{non-abelianGaugeFields}), a natural question is whether these theories could be extended further in order to have derivative self-interactions of the non-abelian gauge fields. As it was shown in \cite{Deffayet:2010zh}, unfortunately the first non-trivial interactions arise in five dimensions and hence in four dimensions one can not construct derivative self-interactions while still retaining the non-abelian gauge symmetry. Hence, there is a similar No-go result in four dimensions as for the abelian gauge fields (as we saw in section \ref{sec_MasslessVectorGal}). This is easily understood in terms of the Levi-Civita tensors. If we want to construct Galileon type of interactions for the non-abelian gauge field, because of the space-time index of the derivative in $\partial_\mu F_{\alpha\beta}^a$, we need to contract the terms with the Levi-Civita tensors with five indices 
\begin{equation}
\mathcal{S}=\int d^5x\,\epsilon^{\mu\nu\alpha\beta\rho}\epsilon^{\kappa\delta\eta\chi\sigma}F_{\mu\nu}^aF_{a\kappa\delta}\partial_\alpha F_{\eta\chi}^b\partial_\sigma F_{b\beta\rho}\,.
\end{equation} 
Such non-abelian Galileon interactions could be useful for constructing higher dimensional gravity theories.

\subsection{Multi-Proca fields}\label{sec_multiProca}
How about if we break the non-abelian gauge symmetry? Can we then construct derivative self-interactions for a set of massive vector fields in four dimensions in the same spirit as generalised Proca theories? When we add a mass term to the above Lagrangian in equation (\ref{non-abelianGaugeFields}), we break the non-abelian gauge symmetry explicitly 
 \begin{equation}
\mathcal{L}=-\frac{1}{4}\mathcal{G}_{ab}\mathcal{F}^{a\mu\nu}\mathcal{F}^b_{\mu\nu}-\frac12\mathcal{G}_{ab} A_\mu^a A^{b\mu} .
\end{equation}
For the set of these massive vector fields one can indeed construct Galileon type of interactions. The mass term breaks the non-abelian gauge symmetry but it keeps the original global symmetry. In the following we shall consider only terms that have a global rotational symmetry, the $SO(3)$ invariance, as a remnicient of the original $SU(2)$ gauge symmetry with $\mathcal{G}_{ab}=\delta_{ab}$ and $f_{abc}=\epsilon_{abc}$. As in the abelian Proca case, we will make use of $F_{\mu\nu}^a$ and $S_{\mu\nu}^a\equiv \partial_\mu A_\nu^a+\partial_\nu A_\mu^a$ to distinguish between different interactions. 

For the construction of the consistent derivate self-interactions for a set of massive vector fields, we can again make use of the antisymmetric Levi-Civita tensors and consider all the possible contractions order by order in $\partial_\mu A_\nu^a$ and in vector fields without derivatives in the schematic form $\mathcal{E}^{\mu\nu\alpha\beta}\mathcal{E}^{\rho\sigma\gamma\delta}\partial_\mu A^a_\rho\cdots  A^b_\nu A^c_\sigma\cdots$ \cite{Jimenez:2016upj}. Since we want to maintain the internal global $SO(3)$ symmetry, we will also consider all the possible contractions of the internal indices with $\delta_{ab}$ and  $\epsilon_{abc}$. For now, we shall not consider parity violating terms. As it was the case for the single Proca field, we can collect all the non-abelian gauge invariant terms or terms that do not have any dynamics for the temporal component of the vector fields into the function
\begin{equation}\label{multiProca_L2}
\mathcal{L}_2=f_2(A_\mu^a, F^a_{\mu\nu})\,.
\end{equation}

To first order in $\partial_\mu A_\nu^a$ without any additional vector fields, we can not construct a Lorentz scalar with global $SO(3)$ symmetry since the internal index remains without contruction. It means that at this order we have to add an even number of vector fields without derivatives in the schematic form $[{\mathcal{E}}{\mathcal{E}}\partial A A^{2n}]$ with $n\ne0$ in order to construct such Galileon interactions. This is different from the single Proca field case where we could construct $\partial_\mu A^\mu$ (or equivalently $[S]$) at this order due to the absence of the internal index. 
For $n=1$ on the other hand we can construct the terms
\begin{eqnarray}\label{Testeq_L32A}
\mathcal{L}_3^{(2A)}&=&f_{3,1} {\mathcal{E}}^{\mu\nu\alpha\beta}{\mathcal{E}}^{\rho\sigma}{}_{\alpha\beta}\partial_\mu A^a_\rho A^b_\nu A^c_\sigma  \epsilon_{abc} 
+f_{3,2} {\mathcal{E}}^{\mu\nu\alpha\beta}{\mathcal{E}}^{\rho\sigma}{}_{\alpha\beta}\partial_\mu A^a_\nu A^b_\rho A^c_\sigma \epsilon_{abc}
\end{eqnarray}
with $f_{3,i}$ being scalar functions of the vector fields. Since we have three group indices, we can only contract the possible interactions with the antisymmetric $ \epsilon_{abc} $ tensor. As a consequence, these terms simply correspond to $F_{\mu\nu}^a A^\mu_b A^\nu_c \epsilon_a{}^{bc}$ and hence are already included in equation (\ref{multiProca_L2}). This on the other hand means that we can not generate terms of the form $[SAA]$, which is very different with respect to the single Proca field case, where this was possible. 
In order to be able to construct the first non-trivial interactions at this order, we need to consider four vector fields without derivatives with $n=2$. However, if these four vector fields indices are contracted with the Levi-Civita tensors, then we will only generate $[FA^4]$ terms. To be more precise, if we consider the following interactions
\begin{equation}
\mathcal{L}_3^{(4A)}\supset g_{3,1} {\mathcal{E}}^{\mu\nu\alpha\beta}{\mathcal{E}}^{\rho\sigma\delta}{}_{\beta}\partial_\mu A^a_\rho A^b_\nu A^c_\sigma  A^d_\alpha A^e_\delta  \epsilon_{abc} \delta_{de}
+g_{3,2} {\mathcal{E}}^{\mu\nu\alpha\beta}{\mathcal{E}}^{\rho\sigma\delta}{}_{\beta}\partial_\mu A^a_\nu A^b_\alpha A^c_\rho  A^d_\sigma A^e_\delta  \epsilon_{abc} \delta_{de}\,,
\end{equation}
where the four vector fields without derivatives are contracted with the two Levi-Civita tensors as well, then all the generated terms contain only $F$ contracted with the vector fields, hence we again would only have terms of the form $[FA^4]$, which already belong to \eqref{multiProca_L2}. We have now five group indices and we can contract them only with $ \epsilon_{abc} $ and $\delta_{de}$. For the construction of interactions of the type $[SA^4]$, we need to contract the  Lorentz indices of two vector fields among themselves, like so
\be
\mathcal{L}_3^{(4A)}\supset g_{3,3} {\mathcal{E}}^{\mu\nu\alpha\beta}{\mathcal{E}}^{\rho\sigma}{}_{\alpha\beta}\partial_\mu A^a_\rho A^b_\nu A^c_\sigma  A^d_\delta A^{e \delta}  \epsilon_{abe} \delta_{cd}.
\ee
Besides the contributions $[FA^4]$, the above contraction generates a new interaction term with the structure
\begin{eqnarray}
\mathcal{L}_3^{(4A)}= g_{3} S^{a\mu\nu} A_\mu^b A_\nu^d A_\alpha^c A^{e \alpha} \delta_{de} \epsilon_{abc},
\end{eqnarray}
 with $g_3$ again standing for a scalar function of the vector fields. 
We can continue with the next interactions with six vector fields without derivatives corresponding to  $n=3$, where the interactions have the schematic form 
$[{\mathcal{E}} {\mathcal{E}} \partial A A^6]$. In the same way as previously if we contract all of the Lorentz indices of the vector fields with the two Levi-Civita tensors, like so $ {\mathcal{E}}^{\mu\nu\alpha\beta}{\mathcal{E}}^{\rho\sigma\delta\gamma}\partial_\mu A^a_\rho A_\nu^b A_\sigma^c A_\alpha^d A^{e}_{\delta} A^{f\beta} A_\gamma^g$, then we will be only constructing terms of the form $[FA^6]$, which we had grouped into $\mathcal{L}_2$.  For generating terms of the form $[SA^6]$ we again have to consider contractions of the Lorentz indices of the vectors with no derivatives among themselves
 \begin{align}
\mathcal{L}_3^{(6A)} =  {\mathcal{E}}^{\mu\nu\alpha\beta}{\mathcal{E}}^{\rho\sigma}{}_{\alpha\beta} \partial_\mu A^a_\rho A_\nu^b A_\sigma^c A_\lambda^d A^{e\lambda} A^{f\delta} A_\delta^g
\Big(h_{3,1}\epsilon_{abd}\delta_{cf}\delta_{eg}+h_{3,2}\epsilon_{bdf}\delta_{ae}\delta_{cg}+h_{3,3}\epsilon_{abe}\delta_{cd}\delta_{fg}\Big)\,,
\end{align}
with $h_{3i}$ being a scalar function of the vector fields. Up to terms $[FA^6]$, the new interactions arising from these contractions are
\begin{align}
\mathcal{L}_3^{(6A)} = S^{a\mu\nu} A_\mu^b A_\nu^c A_\alpha^d A^{e\alpha} A^{f\beta} A_\beta^g 
\Big(h_{3,1}\epsilon_{abd}\delta_{cf}\delta_{eg}+h_{3,2}\epsilon_{bdf}\delta_{ae}\delta_{cg}+h_{3,3}\epsilon_{abe}\delta_{cd}\delta_{fg}\Big).
\end{align}
At this point it is worth to emphasize that we can always add additional pairs of vector fields with the Lorentz indices contracted among themselves but not the internal indices, since this would correspond to simply promoting $\delta_{ab}\to\delta_{ab}+ A_{a\alpha} A_b^\alpha$ or $\eta_{\mu\nu}\to\eta_{\mu\nu}+A^a_\mu A_{a\nu}$. In this way one can construct terms of the form $[SA^8]$ and $[SA^{10}]$... etc.

These techniques used to construct the interactions in first order of $(\partial A)$ in a systematic way, can be easily extended to the higher order terms in $(\partial A)$. As next, let us construct the corresponding interactions to second order in the derivative $[{\mathcal{E}}{\mathcal{E}}\partial A\partial A A^{2n}]$. Since we have now two internal indices, that can be simply contracted with $\delta_{ab}$, we can construct terms with $n=0$
\begin{eqnarray}
\mathcal{L}_4^{(0A)}= f_{4,1}{\mathcal{E}}^{\mu\nu\alpha\beta}{\mathcal{E}}^{\rho\sigma}{}_{\alpha\beta} \partial_\mu A^a_\rho \partial_\nu A^b_\sigma \delta_{ab}  
+f_{4,2}{\mathcal{E}}^{\mu\nu\alpha\beta}{\mathcal{E}}^{\rho\sigma}{}_{\alpha\beta} \partial_\mu A^a_\nu \partial_\rho A^b_\sigma \delta_{ab} .  
\end{eqnarray}
The last interaction is just $F^2$, which is again included in $\mathcal{L}_2$. On the contrary, from the first interaction we obtain the direct extension of the single Proca case to the multi-Proca case preserving the global symmetry
\begin{eqnarray}\label{LagL4nonAproca}
\mathcal{L}_4^{(0A)}= f_4 \Big(S_{\mu\nu}^aS^{b\mu\nu}-S_\mu^{a\mu}S_{\nu }^{b\nu}\Big)\delta_{ab}.
\end{eqnarray}
Similarly, one can construct the interactions with $n=1$ using the corresponding independent contractions with the Levi-Civita tensors. By doing so, one obtains
\begin{align}
\mathcal{L}_4^{(2A)} =& 
+g_{4,1}\Big(S_\lambda^{a\lambda}S^{a\mu\nu}-S^{a\mu}{}_\lambda S^{a\lambda\nu}\Big)A^b_\mu A^b_\nu 
+g_{4,2}\Big(S_\lambda^{a\lambda}S^{b\mu\nu}-\frac12S^{a(\mu}{}_\lambda S^{b\nu )\lambda}\Big)A^a_\mu A^b_\nu \nonumber\\
&+g_{4,3}{\mathcal{E}}^{\alpha\beta\gamma\delta}\tilde{F}^a_{\alpha\lambda} S^{b\lambda}{}_\beta A^a_\gamma A^b_\delta 
+g_{4,4}\Big(S_{\mu\nu}^aS^{b\mu\nu}-S_\mu^{a\mu}S_{\nu }^{b\nu}\Big)A^{a\lambda}A^b_\lambda.
\end{align}
The third interaction is genuine of the multi-Proca theory and does not exist for the single vector field case. Continuing with the interactions with $n=2$, we can consider the following contractions
\begin{align}
 \mathcal{L}_4^{(4A)} &\supset h_{4,1} {\mathcal{E}}^{\mu\nu\gamma\alpha}{\mathcal{E}}^{\rho\sigma\delta\beta} \partial_\mu A^a_\rho \partial_\nu A^c_\sigma A_\gamma^bA_\delta^d A_\alpha^e A_\beta^f \epsilon_{abe}\epsilon_{cdf} \nonumber\\
    &+h_{4,2}{\mathcal{E}}^{\mu\nu\gamma\alpha}{\mathcal{E}}^{\rho\sigma\delta\beta} \partial_\mu A^a_\nu \partial_\gamma A^c_\rho A_\alpha^bA_\sigma^d A_\delta^e A_\beta^f \epsilon_{abe}\epsilon_{cdf} .
\end{align}
The interactions proportional to $h_{4,1}$ are of the form $[SSA^4]$ and $[FFA^4]$, whereas the interactions proportional to $h_{4,2}$ yield contributions of the mixed form $[FSA^4]$. There will be again also terms originating from interactions where the Lorentz indices of the vector fields are contracted among themselves 
\begin{align}
 \mathcal{L}_4^{(4A)} &\supset h_{4,3}{\mathcal{E}}^{\mu\nu\gamma\alpha}{\mathcal{E}}^{\rho\sigma\delta}{}_\alpha \partial_\mu A^a_\rho \partial_\nu A^c_\sigma A_\gamma^bA_\delta^d A^{e\beta} A_\beta^f \epsilon_{abe}\epsilon_{cdf} \nonumber\\
&+h_{4,4} {\mathcal{E}}^{\mu\nu\gamma\alpha}{\mathcal{E}}^{\rho\sigma\delta}{}_\alpha \partial_\mu A^a_\nu \partial_\gamma A^c_\rho A_\sigma^bA_\delta^d A^{e\beta} A_\beta^f \epsilon_{abe}\epsilon_{cdf} \,.
\end{align}
These contractions also give rise to terms of the form $[SSA^4]$, $[FFA^4]$ and $[FSA^4]$. In the same way as in the previous order, we can always add additional pairs of vector fields with the contracted Lorentz indices and construct terms $[SSA^6]$ and  $[FSA^6]$ ... etc. We can summarize the interactions at this order as contributions in the form
\begin{equation}
\L_4=\mathcal{K}_{ab}^{\mu_1\mu_2\nu_1\nu_2} S^a_{\mu_1\nu_1}S^b_{\mu_2\nu_2}
+\mathcal{M}_{ab}^{\mu_1\mu_2\nu_1\nu_2} F^a_{\mu_1\nu_1}S^b_{\mu_2\nu_2}\;,\nonumber\\
\end{equation}
with the objects $\mathcal{K}_{a\dots}^{\mu_1\dots\nu_1\dots}$, $\mathcal{M}_{a\dots}^{\mu_1\dots\nu_1\dots}$ built out of $A_\mu^a$, space-time metric, group metric and the Levi-Civita tensors and being completely antisymmetric in the indices $\mu_i$ and $\nu_i$ separately. 

As next, we can use the same construction scheme to generate the next order interactions in derivatives with the general form $[\epsilon\epsilon\partial A\partial A\partial A A^{2n}]$. The same difficulty encountered in the first order derivative interactions for $n=0$ also appears here and makes it impossible to construct terms cubic order in $S$. Consider the following contractions
\begin{eqnarray}
\mathcal{L}_5^{(0A)}= f_{5,1}{\mathcal{E}}^{\mu\nu\alpha\beta}{\mathcal{E}}^{\rho\sigma\delta}{}_{\beta} \partial_\mu A^a_\rho \partial_\nu A^b_\sigma \partial_\alpha A^c_\delta \epsilon_{abc} 
+ f_{5,2}{\mathcal{E}}^{\mu\nu\alpha\beta}{\mathcal{E}}^{\rho\sigma\delta}{}_{\beta} \partial_\mu A^a_\nu \partial_\rho A^b_\sigma \partial_\alpha A^c_\delta \epsilon_{abc}.
\end{eqnarray}
On closer look, one realises that the first term just vanishes and hence one can not construct the known cubic interactions from the single vector field case and from the second term one only obtains $F^3$ terms, where again it is not possible to construct a $\tilde{F} \tilde{F} S$ term, that we are familiar with from the single Proca case.

New interactions can be constructed for $n=1$. For this, we write down the possible contractions as
\begin{eqnarray}
\mathcal{L}_5&=& 
 g_{5,1}{\mathcal{E}}^{\mu\nu\alpha\beta}{\mathcal{E}}^{\rho\sigma\delta\gamma} \partial_\mu A^a_\rho \partial_\nu A^b_\sigma \partial_\alpha A^c_\delta A^d_\beta A^e_\gamma \epsilon_{dea} \delta_{bc}\nonumber\\
&+&  {\mathcal{E}}^{\mu\nu\alpha\beta}{\mathcal{E}}^{\rho\sigma\delta\gamma}\partial_\mu A^a_\nu \partial_\rho A^b_\sigma \partial_\alpha A^c_\delta A^d_\beta A^e_\gamma
 (g_{5,2} \epsilon_{dea} \delta_{bc}+g_{5,3}\epsilon_{acd} \delta_{be}+g_{5,4}\epsilon_{abd} \delta_{ce}) .
\end{eqnarray}
Interactions of the form $[FSSAA]$ are generated from the first line, whereas terms with the structure $[FFSAA]$ are obtained from the second line. Unfortunately, at this order we can not construct terms of the form $[SSSAA]$. We can pack all of these new contributions into the following form
\begin{align}
\L_5=&\mathcal{M}_{abc}^{\mu_1\mu_2\mu_3\nu_1\nu_2\nu_3}F^a_{\mu_1\nu_1}S^b_{\mu_2\nu_2} S^c_{\mu_3\nu_3} 
+\mathcal{N}_{abc}^{\mu_1\mu_2\mu_3\nu_1\nu_2\nu_3}F^a_{\mu_1\nu_1}F^b_{\mu_2\nu_2} S^c_{\mu_3\nu_3} \nonumber\\
&+\hat{\mathcal{N}}_{abc}^{\mu_1\mu_2\mu_3\nu_1\nu_2\nu_3}F^a_{\mu_1\mu_2}F^b_{\nu_1\nu_2} S^c_{\mu_3\nu_3}\;, \nonumber\\
\end{align}
where $\mathcal{M}_{a\dots}^{\mu_1\dots\nu_1\dots}$, $\mathcal{N}_{a\dots}^{\mu_1\dots\nu_1\dots}$ and $\hat{\mathcal{N}}_{a\dots}^{\mu_1\dots\nu_1\dots}$ represent again quantities built out of the spacetime and group metrics, their Levi-Civita tensors and $A_\mu^a$ and are antisymmetric in the indices $\mu_i$ and $\nu_i$.

Last but not least, we can construct the terms with four derivatives in a similar way. Consider now the following contractions
\begin{eqnarray}\label{eq_L60A}
\mathcal{L}_6^{(0A)}&=&{\mathcal{E}}^{\mu\nu\alpha\beta}{\mathcal{E}}^{\rho\sigma\delta\gamma} \partial_\mu A^a_\rho \partial_\nu A^b_\sigma \partial_\alpha A^c_\delta \partial_\beta A^d_\gamma 
(f_{6,1}\delta_{ab}\delta_{cd}+f_{6,2}\delta_{ac}\delta_{bd}) \nonumber\\
&+&{\mathcal{E}}^{\mu\nu\alpha\beta}{\mathcal{E}}^{\rho\sigma\delta\gamma} \partial_\mu A^a_\nu \partial_\rho A^b_\sigma \partial_\alpha A^c_\delta \partial_\beta A^d_\gamma 
 (f_{6,3}\delta_{ab}\delta_{cd}+f_{6,4}\delta_{ac}\delta_{bd}) \,.
\end{eqnarray}
Up to total derivatives in the form $[S^4]$ and the interactions $[F^4]$ belonging to $\mathcal{L}_2$, we generate new terms of the form $[F^2S^2]$ from the first and second line. They correspond to the interactions that we are familiar with from the single Proca case. However, due to the internal index we construct two types of interactions at this order, namely $\tilde{F}^{a\mu\nu}\tilde{F}^{\alpha\beta}_a S^b_{\mu\alpha} S_{b\nu\beta}$ and $\tilde{F}^{a\mu\nu}\tilde{F}^{b\alpha\beta} S_{a\mu\alpha} S_{b\nu\beta}$. The interactions at this order can be summarised into
\begin{align}
\L_6=&\mathcal{N}_{abcd}^{\mu_1\mu_2\mu_3\mu_4\nu_1\nu_2\nu_3\nu_4}F^a_{\mu_1\mu_2}F^b_{\nu_1\nu_2} S^c_{\mu_3\nu_3} S^d_{\mu_4\nu_4}\nonumber\\
&+\hat{\mathcal{N}}_{abcd}^{\mu_1\mu_2\mu_3\mu_4\nu_1\nu_2\nu_3\nu_4}F^a_{\mu_1\nu_1}F^b_{\mu_2\nu_2} S^c_{\mu_3\nu_3} S^d_{\mu_4\nu_5}\,.
\end{align}

So far these interactions were constructed on flat space-time. In order to build the analog on curved space-time, we can follow the same strategy of adding non-minimal couplings as in the single Proca case. In fact, the first non-trivial interactions on curved space-time can be obtained by a direct extension of the single Proca interactions to the multi-Proca interactions with global $SO(3)$ symmetry \cite{Jimenez:2016upj}
\begin{eqnarray}\label{vecMGalcurv}
\mathcal L_2 & = & G_2(A_\mu^a,F^a_{\mu\nu}) \\
\mathcal L_4 & = & G_{4}R+G'_{4}\delta_{ab} \frac{ S^a_{\mu\nu}S^{b\mu\nu}-S^{a\mu}_\mu S^{b\nu}_\nu}{4}
\nonumber\\
\mathcal L_6 & = & G_6L^{\mu\nu\alpha\beta} F^a_{\mu\nu}F_{a\alpha\beta}
+\frac{G'_{6}}{2} \tilde{F}^{a\alpha\beta}\tilde{F}^{\mu\nu}_aS^b_{\alpha\mu} S_{b\beta\nu}. \nonumber
\end{eqnarray} 
The restriction of the global $SO(3)$ symmetry shrinks the allowed interactions and therefore it is not possible to construct the analog of $\mathcal L_3$ and $\mathcal L_5$ interactions of the single Proca, as we have also seen in the flat space-time case in the above construction procedure. There are also genuine new interactions, that can not be constructed by the direct extension of single Proca interactions. For instance, the following terms exist only due to the additional internal index of the vector field \cite{Jimenez:2016upj}
\begin{eqnarray}
\mathcal L_3 & = & S^{a\mu\nu} A_\mu^b A_\nu^d A_\alpha^c A^{e \alpha} \delta_{de} \epsilon_{abc} \nonumber\\
\mathcal L_4&=&{\mathcal{E}}^{\alpha\beta\gamma\delta}\tilde{F}^a_{\alpha\lambda} S^{b\lambda}{}_\beta A^a_\gamma A^b_\delta \nonumber\\
\mathcal L_5 & = &\epsilon_{abc} A^a_\mu A^{\mu d} \tilde{F}_d^{\alpha\nu}\tilde{F}^{b\beta}_\nu S^c_{\alpha\beta}
\end{eqnarray} 
These terms do not require the presence of non-minimal couplings since they are only linear in $S$ or in other words linear in the connection. See \cite{Jimenez:2016upj,Allys:2016kbq} for more detail on the construction of this type of multi-Proca interactions, and \cite{Emami:2016ldl,Rodriguez:2017wkg} for their applications.

\section{Cosmology}

Cosmology is the elaborate study of the entire universe. One of the tenacious efforts is the throughout deep understanding of the largest scale structures and the dynamics of the universe. These studies enable us to find the answers for the persistent fundamental questions about the formation and evolution of the universe. The whole picture of the universe underwent drastic changes during history. The origin of cosmology as an empirical scientific discipline was starting with the Copernican principle and Newtonian mechanics. Modern scientific cosmology began with the advancement of Albert Einstein's theory of General Relativity in the twentieth century. 

Studies in the different areas of scientific disciplines like Big Bang Nucleosynthesis, the very early universe, Cosmic Microwave Background (CMB), formation and evolution of large-scale structure, dark matter and dark energy conducted the standard model of cosmology according to which the universe began approximately 13 billion years ago with a Big Bang of hot, dense state and arrived now at a state of energy density consisting of 70 percent of dark energy, 25 percent of cold dark matter (CDM) and 5 percent of baryonic matter. This Standard Model of Big Bang cosmology is the prevailing cosmological model that satisfactorily describes the physics on cosmological scales. Neither dark matter, which does not appear in the Standard Model of Particle Physics, nor the even more "exotic" dark energy are fully understood. 

The main goals of modern cosmology are to answer the current open questions about the composition and the history of the universe. According to the standard cosmological model the history of the universe is divided into different periods depending on the dominant forces and processes in each period. The Big Bang theory proclaims that the universe began about 13 billion years ago in a violent explosion. For an incomprehensibly small fraction of a second, the universe was an infinitely dense and hot fireball. A peculiar unknown form of energy, suddenly pushed out the fabric of spacetime in a process called "inflation". In this period the universe's original lumpiness was smoothed out and the universe was left with homogeneity and isotropy as we see today. Quantum mechanical fluctuations during this process were imprinted on the universe as density fluctuations, which later seeded the formation of structure. Afterwards the universe continued to expand but in a slower manner than before. The process of phase transition formed out the most basic forces in the nature: first gravity, then the strong nuclear forces followed by the weak nuclear and electromagnetic forces. 

Among the four forces gravity governs the large scale dynamics of the universe. Gravity is described by General Relativity which describes space-time as a four dimensional manifold whose metric tensor $g_{\mu\nu}$ is a dynamical field which is described by Einstein's field equations. These equations couple the metric to the matter-energy content of space time. As the structure of spacetime dictates the motion of matter and energy, which determine the structure of spacetime, General Relativity is inevitably non-linear. Application of the theory of General Relativity to the large scale structure of the universe leads to various cosmological theories. The starting point for these theories is what is termed cosmological principle.

The cosmological principle is a fundamental assumption about the large scale structure of the cosmos. Application of the theory of General Relativity to the large scale structure of the universe leads to various cosmological theories and the cosmological principle strictly restrains these possible cosmological theories. Making a handful of assumptions one can elaborate very powerful predictions that can be compared with the observations. As a matter of fact cosmology relies on two fundamental assumptions: Averaging over sufficiently large scales, the observable properties of the universe are isotropic, i.e. independent of direction and there are no preferred places in the universe, i.e. homogeneous. Nevertheless, it is a matter to be clarified what sufficiently large scales are. The distribution of galaxies on the sky is not uniform or random, rather they form clusters and groups of galaxies. Also clusters of galaxies are not distributed uniformly, but their positions are correlated, grouped together in superclusters.

However, no evidence of structures with linear dimension $>100{\rm h}^{-1}$ Mpc have been found. Hence, the universe seems to be basically homogeneous if averaged over scales of $200$ Mpc. Additionally, the CMB is almost perfectly isotropic (the observed anisotropies are of the order of $10^{-5}$). The second assumption of isotropy is a reflection of the Copernican revolution, stating that the earth does not occupy a privileged location in the universe. If the universe is isotropic around all of its points, it is also homogeneous. By the second assumption, the first must hold for every observer in the universe. Thus, these two assumptions can be summarised on large spatial scales as the universe is homogeneous and isotropic. 

Compatible with the symmetries of homogeneity and isotropy the metric tensor takes the Friedman-Lemaitre-Robertson-Walker (FLRW) form, i.e. the line element is given by
\begin{equation}\label{FLRWmetric}
\mathrm{d} s^2=-N(t)^2\mathrm{d} t^2+a^2(t)\mathrm{d} \vec{x}^2\,,
\end{equation}
with $N$ denoting the lapse, $a$ the scale factor and $t$ the cosmic time. In this section, we will discuss in detail the cosmological implications of scalar-tensor, vector-tensor and tensor-tensor theories and point out their distinctive features. Before doing so, we shall first introduce the standard cosmological framework within General Relativity.

\subsection{Cosmology in standard General Relativity} \label{BackgroundCosmologyGR}
As we mentioned above, the background metric is given by the FLRW metric. Similarly, the homogeneity and isotropy enforce the stress energy tensor of the matter fields to take the form of the ideal fluid $T_{\mu\nu}=(\rho+p)u_\mu u_\nu+pg_{\mu\nu}$, with the four-velocity $u_\mu$, the mean energy density $\rho$ and the pressure $p$. Einstein's covariant field equations $G_{\mu\nu}+\Lambda g_{\mu\nu}=8\pi GT_{\mu\nu}$ translate into the Friedmann and acceleration equations
\begin{eqnarray}
3\left( \frac{\dot{a}}{a} \right)^2&=&8\pi G\rho+\Lambda \nonumber\\
3\left( \frac{\ddot{a}}{a}\right)&=&-4\pi G(\rho+3p)+\Lambda \,.
\end{eqnarray}
There is also the continuity equation of the matter fields $\dot\rho=-3\left( \frac{\dot{a}}{a} \right)(\rho+p)$. A brief remark is in order: the continuity equation is equivalent to the first law of thermodynamics, $dQ = dU-pdV = d(\rho a^3)-pd(a^3)$, where the left hand side can be interpreted as the change in internal energy and the right hand side as the pressure work. Hereby the absence of heat flow was assumed, since this would define a preferred direction and therefore violate the assumption of isotropy. However, these three equations are not independent. In fact, differentiating the Friedmann equation and then replacing the expression for $\ddot a$ using the acceleration equation gives exactly the continuity equation. In order to close the system of equations for $a$, $\rho$ and $p$, an equation of state of the form $p=p(\rho)$ is assumed. One customary assumption is $p=w\rho$ with the equation of state parameter $w$, where the continuity equation immediately yields $\rho\sim a^{-3(1+w)}$. In general the non-relativistic matter (or dust), ultra-relativistic matter (or radiation) and the cosmological constant correspond to $w=0,1/3,$ and $-1$, respectively. 

For making the equations more managable, it is common to introduce dimensionless density parameters $\Omega_i$, which express the energy densities in units of the critical density $\rho_{\rm crit}$, where the latter is defined as $\rho_{\rm crit}=\frac{3H^2}{8\pi G}$ with $H=\dot{a}/a$. This reflects the fact that in a sphere filled with matter of critical density, the gravitational potential is exactly balanced by the specific kinetic energy, which is obvious when rewriting the above equation as $\frac{4\pi G}{3}\left( \frac{\rho_{\rm crit}a^3}{a} \right)=\frac{\dot{a}^2}{2}$. Today, the critical density has a value of $\rho_{{\rm crit},0}=1.88\times 10^{-29}h^2\;g\;cm^{-3}$, corresponding to a proton mass in approximately $10^5cm^3$ of the cosmic volume, or about one galaxy mass per $Mpc^3$. 

In modern cosmology, it is now a tradition to parameterize the source terms in the Friedmann equation by means of four $\Omega$-parameters, where the various parameters follow from the definition of the critical density. The $\Omega$-parameters for matter and radiation at the present epoch, are in general given by $\Omega_{i0}\equiv\frac{\rho_{i0}}{\rho_{\rm crit}}=\frac{8\pi G}{3H_0^2}\rho_{i0}$. In the same way one can also introduce a density parameter $\Omega_\Lambda$ for the cosmological constant, as well as an $\Omega_K$ for the curvature term. The matter energy density $\Omega_{m0}$ corresponds to the sum of the contributions of the baryonic matter, the cold dark matter and the massive neutrinos. The main contribution of the radiation density $\Omega_{r0}$ comes from CMB, and can therefore be neglected nowadays. In contrast, when in the earlier period the universe was much smaller and hotter, it was dominated by the radiation density. The sum over the different density parameters adds up to unity $\Omega_{r0}+\Omega_{m0}+\Omega_{\Lambda}+\Omega_{K}=1$. 

The Friedmann equation can be also alternatively written as
\begin{eqnarray}
H^2&=&H_0^2\Big(\Omega_{r0}a^{-4}+\Omega_{m0}a^{-3}+\Omega_\Lambda+\Omega_{K}a^{-2}\Big) \nonumber\\
H^2&=&H_0^2\Big(\Omega_{r0}(1+z)^{-4}+\Omega_{m0}(1+z)^{-3}+\Omega_\Lambda+\Omega_{K}(1+z)^{-2}\Big)\,,
\end{eqnarray}
with the redshift satisfying $a=1/(1+z)$. Owing to the different scaling behavior of the density contributions in the above equation, one can differentiate several cosmological epochs, dominated by a certain cosmic fluid. Today, the radiation density is negligible compared to the matter density and the contribution from the cosmological constant. Nevertheless, going back in time, matter starts dominating after the time of matter-$\Lambda$ equality and the universe can be described by the Einstein-de Sitter limit, when going further back in time. Finally radiation takes over, since it grows faster than the matter density. Thus, there exists a time of matter-radiation equality $a_{\rm eq}=\Omega_{r0}/\Omega_{m0}$, which is very important for structure formation, since the two regimes differ significantly, as far as the evolution of density perturbations is concerned. Driven by several different types of measurements, we are able to pin down the cosmological parameters with unpreceded precision. Using CMB and supernova redshift measurements, galaxy surveys, Big Bang Nucleosynthesis, weak lensing we can put joint constraints on the Hubble constant and the density parameters. All these different probes are independent from each other as far as the systematics and the underlying physical model is concerned. Therefore, one is able to constrain the cosmological parameters by combining the different results.

The most accepted model of structure formation acts on the assumption that right after the Big Bang initial quantum fluctuations were magnified to macroscopic scales, where they provide the seeds for the formation of structure in the later universe. Characteristicaly, these fluctuations grow gravitationally with time and may finally collapse to a luminous object like a galaxy. As the measurements of the CMB anisotropy of $10^{-5}$ at a redshift of $z\sim 1100$ show, the fluctuations in the early universe are indeed very small, but must have forced to the rich structure we see nowadays in the universe. Due to the fact that the universe is homogeneous and isotropic, and therefore obeying the Friedmann equation when averaged over sufficiently large scales ($>100{\rm h}^{-1}$ Mpc), it is well founded to consider the universe as made of a uniform smooth background with inhomogeneities like galaxies etc. superimposed on it. This simplification makes the study of structure formation reduce to the study of the growth of inhomogeneities in an otherwise smooth universe, which can in turn be divided into two parts:
\begin{itemize}
\item \textbf{\textit{Linear regime:}} While the inhomogeneities are small, it is justified to describe their growth by linear perturbations to the otherwise smooth background. Due to its mathematical simplicity this approach is very convenient in the early stages of clustering.

\item \textbf{\textit{Non-Linear regime:}} As soon as the deviations from the background become large, linear perturbation theory fails, and the system enters the stage of non-linear evolution. Unfortunately, this regime can not be treated analytically\footnote{Even though in section \ref{sectionKFT} we will discuss a new analytical microscopic non-equilibrium field theory, kinetic field theory, for non-linear cosmic structure formation, that might change completely our perception of cosmological structure formation.}, except for some idealized special cases, so that numerical simulations are the standard way to study non-linear evolution.
\end{itemize}

Averaging the matter density and solving the Einstein equations with the smooth density distribution lead to results comparable to those obtained by averaging the exact solutions based on a inhomogeneous matter distribution. Lamentably, the exact solutions are not known with any degree of confidence for a realistic universe, so that there is no way to justify the assumption theoretically. If wrong, there could be a correction term on the right hand side of Einstein's equation arising from the averaging of the energy density distribution. At the moment there is no observational evidence for such a correction and it is not assumed to exist. \\

\textbf{Standard perturbations in General Relativity}: \\
The first task is to develop the mathematical machinery capable of describing the growth of structures. The basic procedure is to understand the growth of perturbations in the framework of General Relativity. For that reason one bears in mind perturbations of the metric $g_{\mu\nu}$ and the stress-energy tensor $T_{\mu\nu}$, which may be defined as $g_{\mu\nu}=\bar{g}_{\mu\nu}+\delta g_{\mu\nu}$ and $T_{\mu\nu}=\bar{T}_{\mu\nu}+\delta T_{\mu\nu}$, where the bar denotes the background quantities. When assuming that the perturbations $(\delta g_{\mu\nu}, \delta T_{\mu\nu})$ are small compared to the smooth background $( \bar{g}_{\mu\nu}, \bar{T}_{\mu\nu})$ in a suitable manner, then one can linearise Einstein's equations to obtain a second order differential equation of the form
\begin{equation}\label{linearisedFLRW_EinsteinPertEquations}
\hat{\mathcal{L}}(\bar{g}_{\mu\nu})\delta g_{\mu\nu}=\delta T_{\mu\nu}\,,
\end{equation}
where $\hat{\mathcal{L}}$ represents a linear differential operator depending on the background space-time. In general the metric perturbations $\delta g_{\mu\nu}$ and matter perturbations $\delta T_{\mu\nu}$ can be further decomposed into scalar, vector and tensor modes, according to standard Lifshitz classification of cosmological perturbations arising from the irreducible $SO(3)$ representation \cite{Mukhanov:1990me}.
\begin{itemize}
\item \textbf{\textit{scalar modes:}} This kind is very important for structure formation, since ordinary density fluctuations belong to this type of perturbations. We will denote them by $\phi$, $B$, $\psi$, $E$, $\delta\rho$ (or equivalently $\delta=\delta \rho/\bar{\rho}$) and $v$ (or equivalently $\theta=-vk^2$), where the last two are the density perturbations and fluid velocity of the matter fields.
\item \textbf{\textit{vector modes:}} This kind describes vortical modes that are quickly damped away by cosmic expansion, and are therefore negligible for structure formation applications. They will be represented by $B_i$, $E_i$ and $v_i$, with the latter representing the matter perturbations.
\item \textbf{\textit{tensor modes:}} This kind represents the gravitational waves, including primordial ones, that may be detected by CMB polarization probes in the form of the famous B-modes. We have witnessed the detection of gravitational waves produced by merging of black holes and neutron stars \cite{Abbott:2016blz,TheLIGOScientific:2017qsa}. They will be indicated by $h_{ij}$.
\end{itemize}
In terms of these irreducible representations, the metric perturbations take the form
\begin{eqnarray}\label{perturbed_metric}
\delta g_{00} &=& -2a^2\,\phi\,,\nonumber\\
\delta g_{0i} &=& a^2\,\left(\partial_i B+B_i\right)\,,\nonumber\\
\delta g_{ij} &=& a^2 \left[2\,\delta_{ij}\psi +\left(\partial_i\partial_j-\frac{\delta_{ij}}{3}\partial^k\partial_k\right)E+\partial_{(i}E_{j)}+h_{ij}\right]\,,
\end{eqnarray}
where the corresponding modes are functions of time and space and accord to the transformations under spatial rotations. Furthermore, following divergenceless and traceless conditions are applied $\delta^{ij}h_{ij} = \partial^ih_{ij} = \partial^i E_i = \partial^i B_i=0$. 
Due to the fact that equation (\ref{linearisedFLRW_EinsteinPertEquations}) is linear, it is accommodating to expand the solutions in terms of some appropriate mode functions. For the flat FLRW cosmology, this mode functions are given by plane waves. For the simplification it is adequate to work in Fourier space, gaining a set of equations $\hat{\mathcal{L}}_{\vec{k}}\delta g_{\vec{k}}=\delta T_{\vec{k}}$ for each mode, labeled by a wave vector $\vec{k}$. Here $\hat{\mathcal{L}}_{\vec{k}}$ is a second order linear differential operator with time. The results of this set of ordinary differential equations, with respect to some initial conditions, reflects the evolution of each mode separately, solving the problem of linear gravitational clustering completely.

It is worth mentioning that the biggest problem of this treatment is the gauge ambiguity of General Relativity, which is a consequence of invariance under general coordinate transformations $x_\mu\to x'_\mu$ of the theory, that change the form and numerical value of the metric $g_{\mu\nu}$ completely. Trivial changes of coordinates can make perturbations grow at a different rate or even decay, so it is justified to arise the question which one of the infinite choice is the more appropriate one. An adequate example for this problem is given by the cosmological dipole, i.e. the dipole term in the angular expansion of the CMB temperature, which is caused by a Doppler shift due to the peculiar velocity $v$ of our local group with respect to the CMB frame. The perturbations to the CMB temperature, and therefore our peculiar velocity, are measured to be rather small in units of the light velocity $(\delta T/T)_{\rm CMB dipole}\sim 10^{-3}$, so that linear cosmological perturbation theory can be applied. Indeed, in a coordinate frame moving with $-v$ with respect to our local group, no dipole would be seen. This example shows that cosmological perturbations, as represented here by the CMB temperature, do depend on the chosen gauge and may be zero in some gauges, and non-zero in others. 

To be able to talk meaningfully about the growth of inhomogeneities, it is inevitable to fix the gauge problem in the framework of General Relativity. Principally, there are two methods to solve this problem. The first one solves the problem by force, by choosing a particular coordinate system and computing everything within this frame. The remaining problem then is to physically interpret the computed quantities (two prominent gauges are the synchronous and the conformal Newtonian gauge). Another way is to introduce gauge invariant quantities (Bardeen potentials), which are linear combinations of various perturbed physical variables. This approach guarantees gauge invariance, but it is more complicated than the first one, and the gauge invariant objects possess in general no simple physical interpretation. 

In most of the perturbation analysis, the tenacious part is the scalar perturbations analysis. So we shall concentrate on the scalar perturbations here. For simplicity, let us assume that the stress energy tensor consists only of a dark matter component with the four velocity $u^\mu=\left(\Phi/a, \partial^i v/a \right)$ and pressure $p=0$. Therefore, the stress energy density of the dark matter component simply corresponds to $T^{\mu\nu}=\rho u^\mu u^\nu$, with the energy density $\rho=\bar{\rho}(1+\delta)$ where $\delta=\delta \rho/\bar{\rho}$ and in terms of the Fourier modes we have $v=-\theta/k^2$ where $\theta$ represents the divergence of the fluid velocity. \\

\textbf{Conformal Newtonian gauge:}\\
In the conformal Newtonian gauge choice, the scalar perturbations $B$ and $E$ are set to zero from the beginning. Hence, out of the six variables $\phi$, $\psi$, $B$, $E$, $\delta$ and $\theta$, we are left with four. In terms of the remaining perturbations, the Einstein's field equations in the conformal Newtonian gauge become \cite{Ma:1995ey}
\begin{eqnarray}\label{EinsteinEqsNewtGaug}
&&k^2\psi+3\mathcal{H}(\dot\psi+\mathcal{H}\phi)=-4\pi Ga^2 \rho\delta  \nonumber\\
&&k^2(\dot\psi+\mathcal{H}\phi)=4\pi Ga^2\bar\rho \theta \nonumber\\
&&\ddot{\psi}+\mathcal{H}(2\dot{\psi}+\dot{\phi})+\frac{k^2}{3}(\psi-\phi)+\phi(\mathcal{H}^2+2\dot{\mathcal{H}})=0\nonumber\\
&&k^2(\psi-\phi)=0\,,
\end{eqnarray}
where we used the fact that $T^0_0=-\rho\delta$ and $T_i^i=T_i^j=0$ with $\bar{p}=0$ and $\mathcal{H}$ is the conformal Hubble function. On the other hand, we have the Bianchi identity, which results from the divergenceless nature of the Einstein tensor. They result for cold dark matter in the following equations:
\begin{eqnarray}\label{BianchiEqsNewtGaug}
&&\dot{\delta}=-\theta+3\dot\psi \nonumber\\
&&\dot{\theta}=-\mathcal{H}\theta+k^2\phi \,.
\end{eqnarray}
Thus, we have in total six equations for four variables $\phi$, $\psi$, $\delta$ and $\theta$. However, the equations are redundant, i.e. not all of them are independent. We can use the forth equation in (\ref{EinsteinEqsNewtGaug}) in order to express $\phi$ in terms of $\psi$ and in this way we get rid of the explicit dependence on $\phi$. Furthermore, we can combine the first two equations in (\ref{EinsteinEqsNewtGaug}) to integrate out also the $\psi$ field in terms of $\delta$ and $\theta$. This gives the algebraic relation
\begin{equation}
\psi=-\frac{3\mathcal{H}^2(k^2\delta+3\mathcal{H}\theta)}{2k^4}\,,
\end{equation}
where we made use of the background equation $\bar{\rho}=3\mathcal{H}^2/(8\pi Ga^2)$. The resulting final equations for the remaining fields $\delta$ and $\theta$ are then two coupled first order differential equations 
\begin{eqnarray}\label{BianchiEqsNewtGaugFinal}
&&\dot{\delta}=-\theta+\frac{9\mathcal{H}^2(k^2\mathcal{H}\delta+(k^2+3\mathcal{H}^2)\theta)}{2k^4}  \nonumber\\
&&\dot{\theta}=-\frac{\mathcal{H}}{2}\left( 3\mathcal{H}\delta+2\theta+\frac{9\mathcal{H}^2\theta}{k^2} \right)\,.
\end{eqnarray}
Of course we can equivalently write these two first order differential equations as one second order differential equation. This result is consistent with the expectation that the scalar modes of the metric perturbations are not dynamical and the only dynamical degree of freedom comes from the matter perturbations. \\

\textbf{Synchronous gauge:}\\
Another widely used gauge choice is the synchronous gauge, where one puts the perturbations in the temporal components of the metric to zero. In the synchronous gauge one has $\phi=0$ and $B=0$. Thus, out of the six perturbations one has this time $\psi$, $E$, $\delta$ and $\theta$. Since with this gauge choice the diffeomorphism invariance is not completely fixed, one still has the freedom of the remaining gauge in order to fix the velocity perturbation of the dark matter component $\theta=0$ \cite{Ma:1995ey}. After integrating out the non-dynamical degrees of freedom, one remains with the equivalent first order differential equations in this gauge choice
\begin{eqnarray}\label{EqsSynGaugFinal}
&&\dot{E}=-H(E+3H\delta)  \nonumber\\
&&\dot{\delta}=-\frac{1}{2}E \,.
\end{eqnarray}
We again remain with the same amount of propagating degrees of freedom. We have one second order differential equation for the matter perturbation. We can indeed directly map the equations in the synchronous gauge (\ref{EqsSynGaugFinal}) to the equations in the conformal Newtonian gauge (\ref{BianchiEqsNewtGaugFinal}) by means of gauge transformations. \\

\textbf{Flat gauge:}\\
Another gauge choice, which is rather orthogonal to the synchronous gauge, corresponds to fixing the perturbations in the spatial components of the metric to zero. In this gauge choice with $\psi=0$ and $E=0$, the Einstein's field equations become 
\begin{eqnarray}\label{EinsteinEqsFlatGaug}
&&2k^2B-3\mathcal{H}(\delta+2\phi)=0  \nonumber\\
&&\frac32\mathcal{H}^2(k^2B-\theta)+k^2\mathcal{H}\phi=0 \nonumber\\
&&-\frac{k^2}3(2B\mathcal{H}+\phi+\dot{B})+\mathcal{H}\dot{\phi}=0\nonumber\\
&&2B\mathcal{H}+\phi+\dot{B}=0 \,.
\end{eqnarray}
Similarly, the Bianchi identity yields the following equations in this gauge choice:
\begin{eqnarray}\label{BianchiEqsFlatGaug}
&&\dot{\delta}=-\theta \nonumber\\
&&\dot{\theta}=k^2B\mathcal{H}-\mathcal{H}\theta+k^2(\phi+\dot{B})\,.
\end{eqnarray}
We again have six equations of motion for four variables and some of them are redundant. We can directly use the first equation in (\ref{EinsteinEqsFlatGaug}) in order to integrate out the $\phi$ field in terms of the other fields $\phi=\frac{k^2B}{3\mathcal{H}}-\frac{\delta}2$. From the fourth equation one obtains $\dot{B}=-2B\mathcal{H}-\phi$, which plugged back into the third equation yields $\dot{\phi}=0$. Thus, also in this gauge choice one ends up with two first order differential equations of the matter field perturbations
\begin{eqnarray}\label{BianchiEqsFlatGaugFinal}
&&\dot{\delta}=-\theta \nonumber\\
&&\dot{\theta}=-\mathcal{H}(k^2B+\theta)\,,
\end{eqnarray}
where $B$ satisfies the algebraic equation $B=\frac{3\mathcal{H}(k^2\delta+3\mathcal{H}\theta)}{2k^4+9k^2\mathcal{H}^2}$. From this we see that the physical properties of the theory does not depend on the gauge choice as it should be. We always obtain one dynamical scalar mode in the perturbation spectrum after integrating out the non-dynamical fields. One just has to be careful with the physical interpretation of the computed quantities. This became very transparent for the synchronous gauge for instance, where the velocity of the dark matter fluid was set to zero from the beginning. \\

\textbf{Without fixing any gauge:}\\
One can also obtain the right physical properties without fixing any gauge. One can leave the purely gauge fields in the equations of motion, but once the non-dynamical modes are replaced by means of algebraic equations, one obtains the right combination of gauge invariant quantities in the final differential equations. To illustrate this, let us consider the metric perturbations as they stand in (\ref{perturbed_metric}) without fixing any gauge. In this case, the Einstein's field equations take the more general form
\begin{eqnarray}\label{EinsteinEqsWFGaug}
&&3\mathcal{H}(\dot\psi+\mathcal{H}\phi)+k^2(\psi-\mathcal{H}(B-\dot{E}))=-4\pi Ga^2 \rho\delta  \nonumber\\
&&k^2(\dot\psi+\mathcal{H}\phi)=4\pi Ga^2\bar\rho (-Bk^2\theta) \nonumber\\
&&\ddot{\psi}+\mathcal{H}(2\dot{\psi}+\dot{\phi})-\frac{k^2}{3}(\phi-\psi+2\mathcal{H}(B-\dot{E})+\dot{B}-\ddot{E})+\phi(\mathcal{H}^2+2\dot{\mathcal{H}})=0\nonumber\\
&&\psi-\phi+2\mathcal{H}(B-\dot{E})+\dot{B}-\ddot{E}=0\,.
\end{eqnarray}
Correspondingly, the matter field equations arising from the Bianchi identity without gauge fixing read
\begin{eqnarray}\label{BianchiEqsWFGaug}
&&\dot{\delta}=-\theta+3\dot\psi+k^2\dot{E} \nonumber\\
&&\dot{\theta}=-\mathcal{H}\theta+k^2(\phi +\dot{B})+k^2\mathcal{H}B \,.
\end{eqnarray}
Without fixing any gauge we have six equations for six variables but again not all of the equations of motions are independent. Two of the fields are purely gauge artifacts and we can not evolve them in time. However, we can use the equations in order to integrate out some of the non-dynamical fields. From the second equation of (\ref{EinsteinEqsWFGaug}) we can obtain the algebraic equation for $B$, which corresponds to 
\begin{equation}
B=\frac{\theta}{k^2}-\frac{2(H\phi+\dot\psi)}{3H^2}\,.
\end{equation} 
Similarly, we can integrate out the $\phi$ field using the first equation 
\begin{equation}
\phi=\frac{-9H^2\delta+6H(\theta-k^2\dot{E})-6k^2\psi}{4k^2+18H^2}-\frac{\dot{\psi}}{H}\,.
\end{equation}  
The third equation in (\ref{EinsteinEqsWFGaug}) and the first equation in (\ref{BianchiEqsWFGaug}) give the same contribution after plugging back the expressions for $B$ and $\phi$. The equation reads $\theta-k^2\dot{E}+\dot{\delta}-3\dot{\psi}=0$. Similarly, the remaining two equations depend on these quantities. The nature of the equations is such that only the combinations $(\theta-k^2\dot{E})$ and $(\delta-3\psi)$ appear together. This is not surprising since these quantities are gauge invariant. Combining these quantities into the variables $\Theta=\theta-k^2\dot{E}$ and $\Delta=\delta-3\psi$, we end up with the two first order differential equations
\begin{eqnarray}\label{BianchiEqsWFGaugFinal}
&&\dot{\Delta}+\Theta=0 \nonumber\\
&&4k^2\Theta+6\mathcal{H}^2(k^2\Delta+3\dot{\Theta})-2k^2\mathcal{H}(\Theta+3\dot{\Theta})+9H^3(\Theta-3\dot{\Theta})=0\,.
\end{eqnarray}
These are the resulting equations of motion if one performs the computation without fixing any gauge. The gauge invariant combinations of the fields arise naturally. One could also alternatively start with Einstein's field equations as they stand in (\ref{EinsteinEqsWFGaug}) and perform the following gauge transformations in terms of gauge invariant quantities
\begin{eqnarray}\label{WFGaugTrafo}
&&\phi \to \Phi-\dot{B}+\ddot{E}-H(B-\dot{E}) \nonumber\\
&&\psi \to \Psi+H(B-\dot{E})\nonumber\\
&&\delta \to \Delta+3H(B-\dot{E})\nonumber\\ 
&&\theta \to \Theta +k^2 \dot{E}
\end{eqnarray}
After plugging this change of variables back into Einstein's field equations (\ref{EinsteinEqsWFGaug}), one observes two things: 
\begin{itemize}
\item The explicit dependence on the fields $B$ and $E$ disappears in the equations of motions.
\item The resulting equations have exactly the same form as the equations in the conformal Newtonian gauge.
\end{itemize}
We have seen that independently of the gauge choice one always encounters two first order differential equations (or equivalently one second order differential equation) representing one propagating scalar degree of freedom originating from the matter perturbations. Depending on the purposes of the undertaking, one gauge choice might be more appropriate than others.\\

\textbf{Inhomogeneities:}\\
As we mentioned above, even if the previous assumptions made in this section allows to derive many features of the universe, for an accurate understanding of the universe on all scales, it is not satisfactory to assume isotropy and homogeneity, since observations proof that this is not the case on small scales. In order to describe the universe on small scales, like galaxies and galaxy clusters, it is necessary to abandon the homogeneity and isotropy and instead of that develop a new machinery to take deviations from the assumed homogeneity into account. As mentioned before, the measurements of the CMB anisotropy of $10^{-5}$ at a redshift of $z\sim 1100$ already indicate, that the early universe yet inhibits small deviations from total isotropy growing with time, due to gravitational clustering. 

It therefore became an indispensable task in modern cosmology to develop the necessary mathematical tools to deal with perturbations theory. If the analysis fails to be treated analytically, the required numerical tools have to be developed. The main goal of most cosmological model prediction is a successful and fast solution of the Einstein Boltzmann equations governing the linear evolution of perturbations in the universe. There has been several attempts for such codes, that produce fast and accurate solutions to this set of first order linear homogeneous differential equations. Once these codes are developed, their immediate predictions have to be then compared to measurements from different cosmological surveys to put constrains on the parameters of the underlying cosmological model using Monte-Carlo Markov Chain techniques.

The previous assumptions were mainly based on the validity of linear perturbation theory, which is equal to the condition $\delta\ll1$. But when the density contrast reaches unity, the linear approximation breaks down, and has to be replaced by other approaches. To accentuate the need for this, one just has to look to the interesting structures in the universe, like galaxies or clusters of galaxies because they are highly non-linear. A galaxy cluster, for example, corresponds to a value of up to $\delta\sim 10^5$, so that there will be no way of applying linear theory on such objects. Since every theory of structure formation must be capable of describing the formation and evolution of non-linear objects, the major developments in these directions were modelling of gravitational collapse and also N-body simulations. 

One of the prominent model is the spherical collapse model. The dark matter distribution in the universe can be considered to consist of spherical overdense clouds of dark matter, which are also widely known as halos. These halos can reach highly non-linear densities in their centres. One can achieve a rough understanding of the properties of such halos by investigating the evolution of a homogeneous overdense sphere, the so called spherical collapse or top-hat model, in the framework of the expanding Friedmann model. This model tells us that as soon as the linear extrapolated density contrast of a matter concentration exceeds $\delta_{\rm lin}\sim 1.69$, it collapses and gets virialised. A virialised region can be defined as a region where the inner density is given by $\rho\ge 200\bar{\rho}$, where $\bar{\rho}$ denotes the mean density of the background universe. 

Even though the spherical collapse model helps to describe and comprehend some of the basic physics of halo formation, it only provides a snapshot of the whole process. It is evident from the fact that small structures go non-linear first, that at early times most of the dark matter is in low mass halos. The further evolution of these halos is determined by accretion and merging processes, so that present high-mass halos developed out of the merging process of numerous low-mass halos. Whereas numerical simulations automatically reproduce this behaviour, this fact has to be somehow taken into account for any kind of analytical model of halo formation.

Apart from the formation of halos, their internal structure is a very active field of research, because the density and the mass profiles of dark matter halos are still theoretically and observationally not well understood. Generally speaking the virial theorem demands that any system of self-gravitating particles does not have an equilibrium state. Furthermore it states, that the total energy is minus half of its potential energy, implying that any energy loss produces a deeper potential. So it urges a tighter binding of the halo, which in turn increases the energy loss. Thus, any density profile must reflect a potentially long-lived, but transient state. The target of the next generation of cosmological probes is to produce an extremely detailed map of the distribution of galaxies and matter in our universe. In order to properly analyse the measurements of those probes, the use of N-body codes will be unavoidable. N-body simulations will be also powerful to disentangle certain degeneracies between different cosmological models or modified gravities.


\subsection{Challenges of cosmological solutions in massive gravity}
In massive gravity the gravitational interactions are weaken on scales larger than the inverse mass of the graviton due to the Yukawa suppression. If the graviton mass is chosen to be small, the gravitational force would shut down on large scales. This on the other hand could naturally explain the observed accelerated expansion of the universe without invoking any exotic fluid for dark energy. One can understand this also from a different perspective. In the helicity decomposition of massive gravity, the presence of the helicity-0 mode could play the role of a condensate whose energy density could source self-acceleration. However, this scalar mode of the graviton does not correspond to a scalar field like in the quintessence and k-essence theories playing the role of a fluid on the right hand side of Einstein's field equations. In fact, this scalar mode descends from a full-fledged tensor field and there is not such a thing as an Einstein frame where one could decouple this scalar mode from the helicity-2 mode by means of local transformations. This is not surprising since one genuinely modifies the gravitational sector by giving a mass to the graviton. 

Actually, the original motivation for massive gravity was rather to tackle the cosmological constant problem. The mass term would then play the role of a high pass filter modifying the effect of long wavelength sources such as a cosmological constant. In other words, the vacuum energy would gravitate weakly. The first realisations of these ideas were pursued in the decoupling limit of massive gravity \cite{deRham:2010tw}. As we saw in section \ref{sec_MassiveSpin2}, in this limit the leading interactions have the schematic form
\begin{equation}
\mathcal{L}=-h\partial^2h+h(\partial^2\pi+\alpha_2(\partial^2\pi)^2+\alpha_3(\partial^2\pi)^3)+hT.
\end{equation}
For the self-accelerated solution, we can approximate the de Sitter metric locally as a small perturbations over flat Minkowski space-time $ds^2=\eta_{\mu\nu}(1-\frac12H^2x^\alpha x_{\alpha})dx^\mu dx^\nu$. The equations of motion in the decoupling limit becomes in this case $H^2f_1(\alpha_2,\alpha_3)=0$ and $H^2=m^2f_2(\alpha_2,\alpha_3)$, where $f_1$ and $f_2$ are functions of the two parameters of the theory.  As it can been seen, the values of the Hubble parameter is set by the graviton mass. The helicity-0 mode acts as a condensate with negative pressure. Furthermore, the fluctuations on top of this self-accelerating solution in the decoupling limit are stable and the scalar mode decouples from the matter fields in this approximation. For these background solutions, the helicity-1 mode of the massive graviton was consistently set to zero. However, there will still be non-zero fluctuations of the vector mode. Unfortunately, the kinetic term of the vector perturbations vanishes on this background and they become infinitely strongly coupled \cite{deRham:2010tw,Tasinato:2012ze}. 

For the degravitating solution on the other hand, one assumes the presence of a large cosmological constant $T_{\mu\nu}=-\lambda \eta_{\mu\nu}$. The decoupling limit equations of motion are of the form $H^2g_1(\alpha_2,\alpha_3)=0$ and $M_pH^2+g_2(\alpha_2,\alpha_3)=\lambda/(3M_p)$ with again $g_1$ and $g_2$ being functions of the two parameters of the theory. Even in the presence of the cosmological constant one can have flat solutions with $H=0$, since the helicity-0 mode plays the role of a condensate, whose energy density compensates the cosmological constant. This solution relates directly the function $g_2$ to the cosmological constant. The fluctuations on top of this degravitating solutions are stable. However, the strong coupling scale for the perturbations scales with the screened cosmological constant, which on the other hand lowers the allowed value of the vacuum energy significantly in order not to be in conflict within the solar system and accordingly for a successful Vainshtein mechanism \cite{deRham:2010tw}. 

The existing self-accelerating solutions motivated to consider the possible cosmological solutions beyond the decoupling limit in the full theory. Assuming a flat homogeneous and isotropic Ansatz for the dynamical metric $d_s^2=-N^2dt^2+a^2d\vec{x}^2$, we can study the possible class of cosmological solutions by computing the equations of motion with respect to the lapse $N$ and the scale factor $a$. One immediate observation is that the combination of these two equations (or equivalently the Bianchi identity) results in a condition of the form
\begin{equation}
m^2 \partial_0(a^3-a^2)=0,
\end{equation}
which means that this condition enforces the scale factor to be constant. Hence, there are not any flat FLRW solutions in massive gravity \cite{PhysRevD.84.124046}. This same No-go result applies also to the spatially closed FLRW solutions. However, one can construct open FLRW solutions, as it was shown in \cite{Gumrukcuoglu:2011ew}. In this case the Bianchi constraint modifies and the scale factor can evolve in time. The standard Friedman equation changes to
\begin{equation}
3H^2-3\frac{|K|}{a^2}=\rho + \rho_m\,,
\end{equation}
with the energy density of matter fields $\rho$ and the effective energy density from the graviton mass terms
\begin{equation}
\rho_m=-m^2(1-\tilde{k}) \Big(3(2-\tilde{k})+\alpha_3(1-\tilde{k})(4-\tilde{k})+\alpha_4(1-\tilde{k})^2 \Big)
\end{equation}
where $\tilde{k}=\sqrt{|K|}f/a$ and $f$ representing the Stueckelberg field associated with the restored time diffeomorphism. Even if these open FLRW solutions look promising at first sight, the study of perturbations on top of these solutions reveal either strong coupling or instabilities \cite{Gumrukcuoglu:2011zh}. In the linear perturbation analysis on top of these open FLRW backgrounds, the scalar and vector modes have vanishing kinetic terms, signalling strong couplings of these modes. The extension of this study to a small anisotropy limit of the Bianchi-I background revealed an unavoidable ghost mode in the even sector. Hence, massive gravity with a flat reference metric does not allow for stable FLRW solutions. To overcome this problem it was attempted to consider more general reference metrics, for instance de Sitter or FLRW metrics. However, in these more general cases the helicity-0 mode of the massive graviton becomes a Higuchi ghost for consistent backgrounds with Vainshtein mechanism implemented \cite{Fasiello:2012rw}.

A more promising attempt for the cosmological application was the bigravity extension of massive gravity \cite{Comelli:2011zm,vonStrauss:2011mq,Akrami:2012vf,Akrami:2013ffa}. Apart from the homogeneous and isotropic Ansatz for the $g$ metric $d_{s{_g}}^2=-N_g^2dt^2+a_g^2d\vec{x}^2$, one assumes a similar Ansatz for the $f$ metric $d_{s{_f}}^2=-N_f^2dt^2+a_f^2d\vec{x}^2$ with the corresponding lapse $N_f$ and scale factor $a_f$. In this case one has two Friedmann equations 
\begin{eqnarray}
3H_g^2&=&m^2\rho_m^g+\frac{\rho_g}{M_g^2} \nonumber\\
3H_f^2&=&m^2\frac{M_g^2}{M_f^2}\rho_m^f+\frac{\rho_f}{M_f^2} \,,
\end{eqnarray}
with the corresponding Hubble functions $H_g=\dot{a}_g/(N_ga_g)$ and $H_f=\dot{a}_f/(N_fa_f)$ and the corresponding effective energy densities of the mass terms, that depend on the parameters of the theory and on the ratio of the two scale factors $\chi=a_f/a_g$. There are also some matter fields that couple only to $g$, and other independent matter fields that couple to $f$ with their energy densities $\rho_g$ and $\rho_f$ (corresponding to the first option we discussed in section \ref{CouplingsToMatter_MG}). Besides the Friedmann equations, one has also two sets of acceleration equations and conservation equations of matter fields. The Bianchi identity, or similarly combining the background equations, imposes the constraint
\begin{equation}
J(H_g-\xi H_f)=0\,,
\end{equation}
where $J$ depends on the parameters of the theory and the scale factors. As it is clear from this constraint equation, there are two branches of solutions.
\begin{itemize}
\item  For instance, we can impose that J=0, this on the other hand means that the ratio of the scale factors is constant. Therefore, the potential interactions contribute only in form of cosmological constants for both metrics. Studying perturbations on top of these solutions reveal that one essentially has two copies of General Relativity in this branch, meaning that three degrees of freedom become strongly
coupled since they loose their kinetic terms (the kinetic terms of the vector and scalar perturbations is proportional to $J$ and vanishes on this branch with $J=0$). This is the analogue of the self-accelerating branch of massive gravity that we saw above, where the regime of validity of the effective field theory shrinks to zero due to strong coupling. Besides that, the perturbations beyond the linear order will show again the presence of ghostly degree of freedom as in massive gravity \cite{Comelli:2012db}. Thus, we can disregard this branch of solutions within the framework of standard matter field coupling.

\item The other branch of solutions consists of imposing the condition $H_g=\xi H_f$. In this case the ratio of the scale factors can evolve in time and there have been two cases studied in the literature extensively depending on the different evolution. There is the finite branch of solutions (with $\xi$ evolving from zero to a finite value) and the infinite branch of solutions (where the ratio evolves from infinity to a finite value) \cite{Konnig:2014xva}. Unfortunately, the perturbations in the infinite branch of solutions have shown ghost instability with a very low mass of the ghost, such that the theory can not be used as an effective field theory \cite{DeFelice:2014nja,Cusin:2014psa}. On the other hand, in the finite branch of solutions one found gradient instabilities (or crossing a singularity) \cite{Comelli:2011zm,Comelli:2012db,Comelli:2014bqa} and two ways out have been proposed in the literature 
\begin{itemize}
\item The condition on the graviton mass in order to push the instability outside the regime of validity of the low energy effective theory requires a very large mass $m \gg H $ \cite{DeFelice:2013nba,DeFelice:2014nja}. Hence, the gradient instability can be avoided at the prize of fine-tuning.
\item Another condition for the same purpose could be $M_g \gg M_f$ \cite{Akrami:2015qga}, even though one has to be careful with not lowering too much the strong-coupling scale.
\end{itemize}
\end{itemize}

So far, the above mentioned cosmological solutions were assuming minimal matter couplings to the $g$ and $f$ metrics. In fact, there are new type of cosmological solutions if one gives up this restriction. We have seen in section \ref{CouplingsToMatter_MG} that matter fields can couple to an effective metric $g_{\mu\nu}^{\rm eff}$ in a consistent way without introducing any ghost degrees of freedom within the decoupling limit. In the presence of a non-minimally coupled matter field via this effective metric, the Friendmann equation in massive gravity modifies into \cite{deRham:2014naa}
\begin{equation}
3H^2=m^2\rho_m+\frac{\alpha a_{\rm eff}^3}{M_g^2a^3} \rho\,,
\end{equation}
where $\rho$ represents the energy density of the non-minimally coupled matter field and $a_{\rm eff}=\alpha a+\beta$. As a consequence, the Stueckelberg field equation of motion does not correspond to $J=0$ anymore, but rather to the constraint
\begin{equation}
m^2J=\frac{\alpha\beta a_{\rm eff}^2}{M_P^2a^2} P\,,
\end{equation}
with $P$ being the pressure of the doubly coupled matter field. The presence of this new effective matter coupling not only evades the No-go result for the flat FLRW solutions in massive gravity, but also the essential properties of perturbations due to this changed constraint equation.  In the standard open FLRW solutions in massive gravity, that we reported on above, the vector and scalar perturbations were proportional to $J$ and hence the Stueckelberg constraint equation unavoidably resulted in vanishing kinetic terms for these modes. The effective coupling modifies this property and hence all the five physical degrees of freedom have non-vanishing kinetic terms. Furthermore, the study of perturbations has shown the absence of ghost and gradient instabilities upon the conditions $\alpha>0$, $\beta>0$ and $\dot{H}<0$ \cite{Gumrukcuoglu:2014xba}. Motivated by these positive results, one can push forward the study of this effective matter coupling for cosmological applications and compare with observations \cite{Heisenberg:2016spl,Heisenberg:2016dkj} (see also \cite{Lagos:2015sya,Brax:2016ssf}).

This same non-minimal coupling can be applied to bigravity. In this case, the two Friedmann equations with respect to $g$ and $f$ change into
\begin{eqnarray}
3H_g^2&=&m^2\rho_m^g+\frac{\alpha a_{\rm eff}^3}{M_g^2a^3}\rho \nonumber\\
3H_f^2&=&m^2\frac{M_g^2}{M_f^2}\rho_m^f+\frac{\beta a_{\rm eff}^3}{M_f^2a^3\xi^3}\rho \,.
\end{eqnarray}
Similarly, the two acceleration equations and the conservation equation of the matter field modify accordingly. The key change comes again in the constraint equation
\begin{equation}
\left(m^2J-\frac{\alpha\beta a_{\rm eff}^2}{M_g^2a^2} P \right) (H_g-\xi H_f)=0\,.
\end{equation}
In the branch of solutions with $H_g=\xi H_f$, there are unfortunately gradient instabilities in the vector sector in the early time regime, and even worse ghost instability in the scalar sector in the late time regime of the evolution. Therefore, this branch of solutions can be completely disregarded \cite{Gumrukcuoglu:2015nua,Comelli:2015pua}. A more promising branch of solutions is the branch with $J=\frac{\alpha\beta a_{\rm eff}^2}{m^2M_g^2a^2} P$. In this branch it was shown that all of the perturbations including the scalar modes are free of ghost instabilities \cite{Gumrukcuoglu:2015nua}. A more detail study of the perturbations performed in \cite{Gao:2016dwl} showed the necessary conditions also for the avoidance of gradient instabilities. Within the framework of bigravity with non-minimal matter couplings, this branch of solutions constitutes the unique consistent cosmological solutions without ghost and gradient instabilities. \\

Apart from the bimetric extension of massive gravity, there have been other extensions considered in the literature for a better cosmological application. Among them, the quasidilaton received much attention. In this extension, an additional scalar field is coupled in a specific way to massive graviton \cite{PhysRevD.87.064037} and the interactions are invariant under the quasidilaton global symmetry. Regretfully, in this formulation the self-accelerating solutions in the quasidilaton massive gravity are plagued by a ghost instability. In order to cure this instability a promising extension with a new coupling constant was proposed \cite{DeFelice:2013tsa}. The extension is such that this new coupling constant leaves the background dynamics untouched whereas it has large impact on the stability of the perturbations and their phenomenology in the Cosmic Microwave Background and in the effective mass of gravitational waves \cite{Kahniashvili:2014wua}. The scalar perturbations with the presence of the matter fields were investigated in the late-time asymptotic solution in \cite{Motohashi:2014una}. This analysis has been further generalized to the fully dynamical background equations of motion and the perturbations on top of this general background with the matter fields \cite{Heisenberg:2015voa,Gumrukcuoglu:2016hic}. Another promising extension of the quasidilaton theory is based on the effective metric. In this new formulation the quasidilaton lives on the effective metric and hence couples non-minimally to both metrics \cite{Mukohyama:2014rca}. This on the other hand has interesting consequences for the self-accelerating solutions and their stability. \\

The presence of the effective matter coupling is not only unique for the consistent cosmological solutions, but also for the application to dark matter phenomenology. One can build a relativistic model of dark matter within the framework of bigravity, where the phenomenology of dark matter at galactic scales resembles modified Newtonian dynamics (MOND) while reproducing $\Lambda$CDM behaviour at large cosmological scales. This is based on a formulation of dipolar dark matter with two different species of dark matter particles coupled to the two metrics and linked by an internal vector field, that lives on the effective metric. Through a mechanism of gravitational polarization one recovers MOND phenomenology. The model is based on the action
\begin{align}
S &= \int\mathrm{d}^{4}x\biggl\{ \sqrt{-g}\biggl(\frac{M_g^2}{2}R_g -
\rho_\text{bar}-\rho_g\biggr)
+\sqrt{-f}\biggl(\frac{M_f^2}{2}R_f-\rho_f\biggr)
\nonumber\\ &\qquad\qquad +\sqrt{-g_\text{eff}} \biggl[ \frac{m^2}{4\pi} +
  \mathcal{A}_\mu \left(j_g^\mu - \frac{\alpha}{\beta}j_f^\mu\right) +
  \frac{a_0^2}{8\pi}\,\mathcal{W}\bigl(X\bigr)
  \biggr]\biggr\}\,,\label{lagrangian}
\end{align}
with $j_g^{\mu}$ and $j_f^{\mu}$ standing for the mass currents of the two types of dark matter particles and $\mathcal{W}(\mathcal{X})$ for the non-canonical kinetic term of the vector field 
$\mathcal{X} = -g_\text{eff}^{\mu\rho} g_\text{eff}^{\nu\sigma}\mathcal{F}_{\mu\nu}
  \mathcal{F}_{\rho\sigma}/2a_{0}^{2}$. In the non-relativistic limit the Poisson equation for the Newtonian potentials $U_g$ and $U_f$
associated with the two metrics becomes
\begin{equation}\label{scalareq}
\Delta\left(\frac{2M_g^2}{\alpha}U_g - \frac{2M_f^2}{\beta}U_f\right)
= -\frac{1}{\alpha}\bigl(\rho^*_\text{bar} +
\rho^*_g\bigr)+\frac{1}{\beta} \rho^*_f\,,
\end{equation}
with $\rho^*_\text{bar}$, $\rho^*_g$ and $\rho^*_f$ representing the ordinary Newtonian densities of the matter fluids respectively. 
Furthermore, the two Newtonian potentials are linked together by the virtue of Bianchi identity $\bm{\nabla}\bigl(\alpha U_g + \beta U_f\bigr) = 0$. After using the vector field equations of motion together with the equations of motion of the baryons and the dark matter particles, 
the equation for the ordinary potential $U_g$ becomes
\begin{equation}\label{poissonpol}
\bm{\nabla}\cdot\Bigl[\bigl(1-\mathcal{W}_{X}\bigr)\bm{\nabla}
  U_g\Bigr] = - 4 \pi \rho^*_\text{bar} \,.
\end{equation}
This is exactly of the Bekenstein-Milgrom form~\cite{BekM84} with MOND interpolating function given by $\mu=1-\mathcal{W}_{X}$. For more informations on the realisation of this mechanism of gravitational polarization see \cite{Blanchet:2015sra,Blanchet:2015bia,Bernard:2015gwa,Blanchet:2017duj}.
\subsection{Cosmological implications of Scalar-Tensor Theories}\label{sec_cosmology_horndeski}
Among the modified gravity theories, models based on scalar fields are the most extensively explored models for cosmology. One immediate practical reason is that they can yield accelerated expansion without breaking the isotropy of the universe with a background field configuration $\pi(t)$. In order to act as a dark energy candidate, they need to be very light, of the order $m_\pi\sim 10^{-33}$ eV. There have been many works in the literature on specific models and instead of going through them one by one, we shall directly concentrate on the Horndeski theories given in equation \eqref{HorndeskiGal}. Since the Horndeski scalar-tensor theories already accommodate quintessence, k-essence, $f(R)$, Brans-Dicke and Galileon models, in this way these models will be automatically covered and their cosmological implications can then be studied by choosing some specific functions.

At the background level, we will again have an homogeneous and isotropic Ansatz for the metric $ds^{2}=-N(t)^2\, dt^{2}+a^{2}(t)\delta_{ij}dx^{i}dx^{j}$ and a field configuration $\pi=\pi(t)$ for the scalar field, respectively. The background equations of motion are obtained by varying with respect to $N$, $a$ and $\pi$. The corresponding Friedmann equation is given by \cite{DeFelice:2011hq}
\begin{eqnarray}
&&\mathcal{E}=-P+2XP_{,X}-2XG_{3,\pi}+6X\dot{\pi}HG_{3,X}-6H^{2}G_{4} \nonumber\\
&&-6H\dot{\pi}G_{4,\pi} -12HX\dot{\pi}G_{4,\pi X} +24H^{2}X(G_{4,X}+XG_{4,XX}) \nonumber\\
&&+ 2H^{3}X\dot{\pi}\left(5G_{5,X}+2XG_{5,XX}\right)-6H^{2}X\left(3G_{5,\pi}+2XG_{5,\pi X}\right)=-\rho\,.
\end{eqnarray}
The accelerating equation on the other hand corresponds to
\begin{eqnarray}
&&\mathcal{P}=P-2X\left(G_{3,\pi}+\ddot{\pi}\, G_{3,X}\right)+  2\left(3H^{2}+2\dot{H}\right)G_{4}-12H^{2}XG_{4,X} \nonumber\\
&&-4H\dot{X}G_{4,X}-8\dot{H}XG_{4,X}-8HX\dot{X}G_{4,XX}+2\left(\ddot{\pi}+2H\dot{\pi}\right)G_{4,\pi}\nonumber\\
&& -2X\left(2H^{3}\dot{\pi}+2H\dot{H}\dot{\pi}+3H^{2}\ddot{\pi}\right)G_{5,X}-4H^{2}X^{2}\ddot{\pi}\, G_{5,XX}+4HX\left(\dot{X}-HX\right)G_{5,\pi X} \nonumber\\
&&+2\left[2\left(\dot{H}X+H\dot{X}\right)+3H^{2}X\right]G_{5,\pi}+4HX\dot{\pi}\, G_{5,\pi\pi}=0
\end{eqnarray}
We also have the background equation of motion for the scalar field $\pi$, which can be compactly written as
\begin{eqnarray}\label{ScalarBGeq}
\frac{1}{a^3}\frac{d}{dt}(a^3J)&=&P_{,\pi}-2X\left(G_{3,\pi\pi}+\ddot{\pi}\, G_{3,\pi X}\right)+6\left(2H^{2}+\dot{H}\right)G_{4,\pi}\nonumber\\
&+&6H\left(\dot{X}+2HX\right)G_{4,\pi X}-6H^{2}XG_{5,\pi\pi}+2H^{3}X\dot{\pi}\, G_{5,\pi X}
\end{eqnarray}
where $J$ stands for
\begin{eqnarray}
J & \equiv&  \dot{\pi}P_{,X}+6HXG_{3,X}-2\dot{\pi}G_{3,\pi}+6H^{2}\dot{\pi}\left(G_{4,X}+2XG_{4,XX}\right)-12HXG_{4,\pi X}\nonumber \\
&  & +2H^{3}X\left(3G_{5,X}+2XG_{5,XX}\right)-6H^{2}\dot{\pi}\left(G_{5,\pi}+XG_{5,\pi X}\right)\,.
\end{eqnarray}
Of course there is also the equations of motion of the matter fields, but these are identical to what we have in General Relativity, since we will assume minimal couplings for them. For instance, the non-relativistic matter field has the equation of motion $\dot{\rho}+3H\rho=0$. As it becomes clear from the background equations of motion, the generous freedom of the four functions allows very easily a self-accelerating background, that could be relevant for dark energy and inflation.

For the purpose of the cosmological scalar perturbations analysis, let us consider the metric to be of the form
\begin{equation}
ds^{2}=-(1+2\Psi)\, dt^{2}-2\partial_{i} B dtdx^{i}+a^{2}(t)(1+2\Phi)\delta_{ij}dx^{i}dx^{j}\,,\label{pertmetr}
\end{equation}
and the scalar field as
\begin{equation}
\pi=\pi(t)+\delta\pi(t,{\vec{x}})\,.
\end{equation}
Let us also assume perturbation of the matter fields in form of density perturbation $\delta\rho$ (density contrast $\delta=\delta\rho/\rho$) and fluid velocity $v$.
The Einstein field equations in Fourier space get modified into
\begin{eqnarray}
& & A_{1}\dot{\Phi}+A_{2}\dot{\delta\pi}-\rho_{m}\dot{v}+A_{3}\frac{k^{2}}{a^{2}}\Phi+A_{4}\Psi+A_{5}\frac{k^{2}}{a^{2}}B+\left(A_{6}\frac{k^{2}}{a^{2}}-\mu\right)\delta\pi-\rho\delta=0\,,\label{eq:Psi}\\
 & & B_{1}\ddot{\Phi}+B_{2}\ddot{\delta\pi}+B_{3}\dot{\Phi}+B_{4}\dot{\delta\pi}+B_{5}\dot{\Psi}+B_{6}\frac{k^{2}}{a^{2}}\Phi+\left(B_{7}\frac{k^{2}}{a^{2}}+3\nu\right)\delta\pi\nonumber \\
 &  & {}+\left(B_{8}\frac{k^{2}}{a^{2}}+B_{9}\right)\Psi+B_{10}\frac{k^{2}}{a^{2}}\dot{B}+B_{11}\frac{k^{2}}{a^{2}}B+3\rho\dot{v}=0\,,\label{eq:Phi}\\
 & & C_{1}\dot{\Phi}+C_{2}\dot{\delta\pi}+C_{3}\Psi+C_{4}\delta\pi+\rho v=0\,,\\
 &  & D_{1}\ddot{\Phi}+D_{2}\ddot{\delta\pi}+D_{3}\dot{\Phi}+D_{4}\dot{\delta\pi}+D_{5}\dot{\Psi}+D_{6}\frac{k^{2}}{a^{2}}\dot{B} +\left(D_{7}\frac{k^{2}}{a^{2}}+D_{8}\right)\Phi \nonumber \\
 &  & +\left(D_{9}\frac{k^{2}}{a^{2}}-M^{2}\right)\delta\pi+\left(D_{10}\frac{k^{2}}{a^{2}}+D_{11}\right)\Psi+D_{12}\frac{k^{2}}{a^{2}}B=0\,,\label{eq:dPi}
\end{eqnarray}
Similarly, the two equations of motion of the matter fields in Fourier space get modified into
\begin{eqnarray}
&  & \dot{v}-\Psi=0\,,\label{eq:mat1}\\
 & & \dot{\delta}+3\dot{\Phi}+\frac{k^{2}}{a^{2}}v-\frac{k^{2}}{a^{2}}B=0\,.\label{eq:mat2}
\end{eqnarray}
The coefficients $A_i$, $B_i$, $C_i$ and $D_i$ are all defined in the \ref{AppendixHorndeski} and more detail can be found in \cite{DeFelice:2011hq}. On top of the scalar degree of freedom of the matter field, there is one additional propagating scalar field $\delta\pi$. There are no propagating vector modes in the gravity sector as in General Relativity. 

It is a common practice to work in the quasi-static approximation in order to study the evolution of the modes deep inside the Hubble radius $k_2/a^2\gg H^2$, assuming that the propagation speed of the scalar degree of freedom is not $c_s\ll1$, since in this case the approximation breaks down. In this approximation we can directly neglect the time derivatives of the scalar degree of freedom $\delta\pi$, which facilitates the computation significantly \cite{Tsujikawa:2007xu,Kobayashi:2009wr,DeFelice:2010as}. In this approximation the temporal component and the traceless part of the gravitational field equations together with the scalar field equation of motion simplify to (in the Newtonian gauge with $B=0$)
\begin{eqnarray}
&&B_6\Phi+B_7\delta\pi+B_8\Psi \approx0\nonumber\\
&& B_7\frac{k^2}{a^2}\Phi+\left(D_{9}\frac{k^{2}}{a^{2}}-M^{2}\right)\delta\pi+A_6\frac{k^2}{a^2}\psi \approx0\nonumber\\
&&B_8\frac{k^2}{a^2}\Phi+A_6\frac{k^2}{a^2}\delta\pi-\rho\delta \approx0\,.
\end{eqnarray}
A combination of these equations can be brought into the form of a modified Poisson equation
\begin{equation}
\frac{k^2}{a^2}\Psi\approx -4\pi G_{\rm eff}\rho\delta\,,
\end{equation}
with the effective gravitational coupling given by \cite{DeFelice:2011hq}
\begin{equation}
G_{\rm eff}=\frac{2\mpl^2((B_6D_9-B_7^2)(k/a)^2-B_6M^2)}{(A_6^2B_6+B_8^2D_9-2A_6B_7B_8)(k/a)^2-B_8^2M^2}G\,.
\end{equation}
The evolution of the matter perturbations in terms of the gauge invariant density contrast $\tilde\delta=\delta+3Hv$ in the quasi-static approximation is given by the simple differential equation
\begin{equation}
\ddot{\tilde{\delta}}+2H\dot{\tilde{\delta}}-4\pi G_{\rm eff}\rho\tilde\delta\approx0\,.
\end{equation} 
Another relevant quantity is the slip parameter, which represents the difference between the two gravitational potentials $\eta=-\Phi/\Psi$ and in the quasi-static approximation simplifies to
\begin{equation}
\eta\approx \frac{(B_8D_9-A_6B_7)(k/a)^2-B_8M^2}{(B_6D_9-B_7^2)(k/a)^2-B_6M^2}
\end{equation}
These expressions for the effective gravitational coupling and the slip parameter have been then studied for the specific sub-classes of theories, for instance $f(R)$, Branns-Dicke, kinetic gravity braidings and covariant Gelileon theories \cite{DeFelice:2011hq}.
The direct cosmological application of covariant Galileons witnessed a flurry of investigations concerning self-accelerating de Sitter solutions \cite{Silva:2009km,DeFelice:2010jn}, inflation \cite{Creminelli:2010ba,LevasseurPerreault:2011mw,Kobayashi:2010cm,Burrage:2010cu}, dark energy and its observational implications \cite{Gannouji:2010au,DeFelice:2010as,Brax:2011sv,DeFelice:2010pv,DeFelice:2010nf,deRham:2011by,Heisenberg:2014kea}.

Similarly, one can also study the linear perturbations of the more general classes of scalar-tensor theories, like for instance DHOST. Using the language of effective description of dark energy \cite{Gubitosi:2012hu,Gleyzes:2013ooa,Piazza:2013coa} in terms of lapse, shift and the extrinsic curvature, the quadratic action for DHOST can be written as \cite{Langlois:2017mxy}
\begin{eqnarray}\label{DHOSTeffectivecosmo}
\mathcal{S}_{\rm DHOST}=\int d^3x dta^3\frac{M^2}{2}\Big(\delta K_{ij}\delta K^{ij}-\left( 1+\frac23\alpha_L\right)\delta K^2
+(1+\alpha_T)\left( R\frac{\delta\sqrt{h}}{a^3}+\delta_2R \right)\nonumber\\
+H^2\alpha_K\delta_N^2+4H\alpha_B\delta K\delta N+(1+\alpha_H)R\delta N+4\beta_1\delta K\delta\dot{N}+\beta_2\delta\dot{N}^2+\frac{\beta_3}{a^2}(\partial_i\delta N)^2\Big)\,.
\end{eqnarray}
Horndeski scalar-tensor theories with second order equations of motion correspond to the restricted case with the only free functions $\alpha_K$, $\alpha_B$, $\alpha_T$ and $\alpha_M=d\ln M^2/(Hdt)$ (in terms of the notation used in \cite{Bellini:2014fua}) and the beyond Horndeski theories contain the additional function $\alpha_H$. The DHOST theories introduce the additional four functions $\alpha_L$, $\beta_1$, $\beta_2$ and $\beta_3$ \cite{Langlois:2017mxy}. As we mentioned in section \ref{secDHOSTtheories}, the DHOST theories can be divided into two groups, those that are related to Horndeski by means of a disformal transformation and those that are unrelated to Horndeski theories. The parameters of the first category $\mathcal{C}_I$ satisfy the relations $\alpha_L=0$, $\beta_2=-6\beta_1^2$ and $\beta_3=-2\beta_1(2(1+\alpha_H)+\beta_1(1+\alpha_T))$ and in the second category $\mathcal{C}_{II}$ the $\beta_i$ parameters are given in terms of $(\alpha_L, \alpha_T, \alpha_H)$ parameters. The stability analysis of cosmological perturbations in the DHOFT theories reveals that the second category of interactions $\mathcal{C}_{II}$ that are not related to Horndeski suffer from gradient instabilities either in the scalar or in the tensor sector and therefore can be disregarded. Thus, the only valuable remaining DHOST theories are those that are related directly to the Horndeski interactions by means of disformal transformations. 

\subsection{Cosmology with gauge invariant vector fields}
The Standard Model of Particle Physics contains both abelian and non-abelian vector fields as the fundamental fields of the gauge interactions. Therefore, this motivates an exploration of the role of bosonic vector fields in the cosmological evolution besides scalar fields. Similarly to the scalar field modifications, vector fields as part of the gravitational interactions could naturally yield an accelerated expanding universe on large scales while being screened on small scales. 

Vector fields can be also in a natural way applied to the early universe to realise an inflationary epoch. Starting with the standard Maxwell kinetic term, one immediately realises that the associated energy density quickly decays since it is conformally invariant. One way out is to consider a potential for the vector field $V(A_\mu A^\mu)$. This was for instance an idea that was followed in \cite{Ford:1989me}, where the slow-roll regime was realised by an almost constant potential despite the exponential variation of $A_\mu A^\mu$. One can also consider a non-minimal coupling to the gravity sector, which was pursued in \cite{Golovnev:2008cf} with a coupling of the form $R A_\mu A^\mu$. However, this would explicitly break the gauge invariance and we shall not comment on it any further in this subsection. 

In all these considerations in order to guarantee an isotropic expansion, one has to either assume a large number of randomly oriented vector fields, or a triplet of orthogonal vector fields \cite{ArmendarizPicon:2004pm} or a vector field with only the zero component non-vanishing \cite{Koivisto:2008xf} (massive vector field). In the case of a potential term for the vector field, the contribution to the $0i-$ component of the stress energy tensor goes as $T_{0i}=2A_0 A_i V'$. For this to be zero, one has to either have a vanishing time component of the vector field or vanishing spatial components. Furthermore, one has to impose that the off diagonal terms of the stress energy tensor vanish and the diagonal component to be equal in order to be compatible with the FLRW metric. This is possible with the triads. 

Another possibility was considered in \cite{Ackerman:2007nb} with a fixed-norm vector field and its impact on the CMB anisotropy was studied. In all of these considerations, it was later shown that it is very hard to maintain the stability requirements \cite{Himmetoglu:2008hx}. The encountered instabilities for some of these models make them cosmologically unviable \cite{Himmetoglu:2009qi}. Similar ideas of non-minimal couplings to the gravity sector were also considered in \cite{Jimenez:2008au, Jimenez:2008sq} for the purpose of dark energy. In this subsection we shall consider only the cosmological solutions of vector field models with the explicit gauge invariance preserved. The consequences of breaking this symmetry for cosmology will be investigated in the next subsection. Maintaining gauge invariance allows a homogeneous and isotropic background only in two ways: either considering $N$ randomly distributed vector fields or a triad configuration.

\begin{figure}
\begin{center} 
\includegraphics[width=11.5cm]{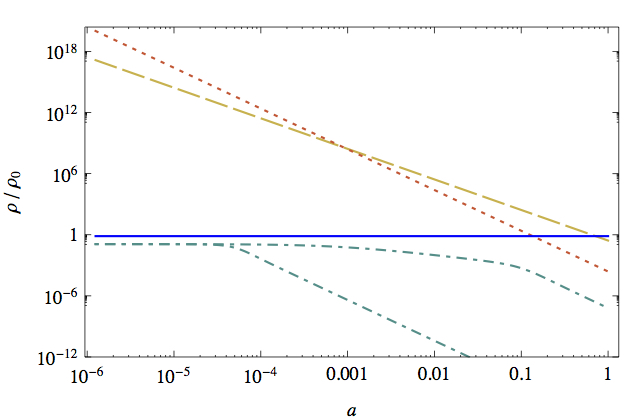}
\caption{The figure is taken from \cite{Jimenez:2013qsa}, where the evolution of the energy density $\rho_A$ (dotdashed) is shown for $M=10^7H_0$ and $M=25H_0$. Furthermore, one can compare it with the cosmological evolution for the standard components in form of radiation (dotted),matter (dashed), and a constant energy density of the cosmological constant (blue line).}
\label{cosmoevolLFF}
\end{center}
\end{figure}

As we have seen in section \ref{sec_VectorTensorTheories}, imposing the $U(1)$ symmetry allows only a unique coupling with the double dual Riemann tensor $LFF$. Let us consider the most general action for a massless vector field on curved space-time with second order equations of motion presented in equation (\ref{actionLFFvector})
\begin{align}
S=\int {\rm d}^4x\sqrt{-g}\left[\frac{1}{2}\mpl^2R-\frac14F_{\mu\nu}F^{\mu\nu}+\frac{1}{4M^2}L^{\alpha\beta\gamma\delta}F_{\alpha\beta}F_{\gamma\delta} \right]\,.\nonumber 
\end{align}
Assuming the metric to be of the homogeneous and isotropic FLRW type and considering a vector field with only non-vanishing spatial components $A_\mu=(0,A_i(t))$, the vector field equations of motion are of the form \cite{Jimenez:2013qsa}
\begin{eqnarray}
\left(1+\frac{4 \mathcal H^2}{M^2a^2}\right)\vec{A}''-\frac{8\mathcal H \left(\mathcal H^2-\mathcal H'\right)}{M^2a^2}\vec{A}'=0 \,,
\end{eqnarray}
where $\mathcal H$ is the comoving Hubble parameter and the derivatives are with respect to the conformal time $d\eta^2=dt^2/a^2$. Of course, as we mentioned above, a vector field with the spatial components is not compatible with the symmetries of the FLRW metric, therefore the vector field should be considered as subdominant. In the case of de Sitter expansion with $a=e^{ht}$ we have $\mathcal H'=\mathcal H^2$. This means that the second term of the vector field equations of motion disappears yielding $\vec{A}''=0$. This is not very interesting since it corresponds to the conventional electrodynamics. In the case of a power law expansion, the dependence on $M^2$ does not any longer factor out and we see the effect of the non-minimal coupling. The corresponding energy density in this case takes the simple form \cite{Jimenez:2013qsa}
\begin{eqnarray}
\rho_{A}=\frac{1}{2a^4}\left(1+\frac{12 \mathcal H^2}{M^2a^2}\right)  \vert\vec{A}'\vert^2\,.
\end{eqnarray}
One immediately observes, that if the non-minimal coupling is subdominant, the energy density decays as in standard electrodynamics $\rho_{A}\propto 1/a^4$. On the other hand, if the non-minimal coupling dominates, then one has $\rho_A \propto (a\mathcal H)^{-2}$. 
For instance in a radiation dominated universe, the energy density will be constant until the point where $\mathcal H^2\simeq a^2M^2$ and then it starts decaying. This behaviour can be seen in figure \ref{cosmoevolLFF}. In a similar way, the longitudinal and transverse pressures can be obtained from the stress energy tensor (\ref{TmunuLFF}) and one has to make sure that large scale anisotropies are not generated. It turns out that for this to be the case, one has to impose $M\gsim H_{eq}\simeq10^{-29}$ eV with $H_{eq}$ denoting the Hubble expansion rate at matter and radiation equality. In this case, a initially subdominant contribution of the vector field remains always subdominant. 

So far we have assumed that the vector field played the role of a subdominant component of the universe. What happens if the vector field contributes in a non-negligible way to the expansion. For this purpose let us consider a vector field pointing along the z-direction and an axisymmetric Bianchi I metric $ds^2=dt^2-a^2_\perp(t)\left(dx^2+dy^2\right)-a^2_\parallel(t) dz^2$. Corresponding to the two scale factors, we have the expansion rate along the transverse direction $H_{\perp}=\dot{a}_\perp/a_\perp$ and longitudinal direction $H_{\parallel}=\dot{a}_\parallel/a_\parallel$. We can plug in these Ansaetze into the metric field equations, which result in
\begin{eqnarray}
H_\perp^2+2H_\parallel H_\perp&=&8\pi G\rho,\\ \label{e:Eperp}
\dot{H}_\perp+\dot{H}_\parallel+H_\perp^2+H_\parallel^2+H_\perp H_\parallel&=&-8\pi Gp_\perp,\\
2\dot{H}_\perp+3H_\perp^2&=&-8\pi Gp_\parallel,    \label{e:Epap}
\end{eqnarray}
with the corresponding energy density and pressures
\begin{eqnarray}
\rho&=&\rho_\Lambda+\frac{1}{2a_\parallel^2}\left(1+\frac{12 H_\perp^2}{M^2}\right)\dot{A}_z^2,\\
p_\perp&=&-\rho_\Lambda+\frac{1}{2a_\parallel^2}\left[\left(
1-4\frac{H_\perp(H_\perp-H_\parallel)+\dot{H}_\perp}{M^2}
\right)\dot{A}_z^2-\frac{8 H_\perp}{M^2}\dot{A}_z\ddot{A}_z\right],\\
p_\parallel&=&-\rho_\Lambda-\frac{1}{2a_\parallel^2}\left(1+\frac{4 H_\perp^2}{M^2}\right)\dot{A}_z^2.
\end{eqnarray}
Similarly, we have the equation of motion of the vector field
\begin{eqnarray}
\left(1+\frac{4H_\perp^2}{M^2}\right)\ddot{A}_z+\left[\left(1+\frac{4 H_\perp^2}{M^2}\right)(2H_\perp-H_\parallel)+\frac{8 H_\perp\dot{H}_\perp}{M^2}\right]\dot{A}_z&=&0.
\end{eqnarray}
We have equations of motion that contain $\ddot{A}_z$, $\dot{H}_\perp$ and $\dot{H}_\parallel$. For convenience, one can define an average expansion rate $H$ and its departure as shear $R$ in terms of $H_\perp=H(1-R)$ and
$H_\parallel=H(1+2R)$. We can use the Friedmann equation in order to eliminate the $\dot{A}_z$ dependence and the vector field equations of motion to replace $\ddot{A}_z$. Plugging these expressions back into the remaining equations results in an autonomous system of first order differential equations
\begin{eqnarray}
\dot{H}&=& f_1(H,R,\rho_\Lambda) \nonumber\\
\dot{R}&=& f_2(H,R,\rho_\Lambda) \,,
\end{eqnarray}
with the two specific functions $f_1$ and $f_2$. Brought the equations in this autonomous form, we can easily find the critical points and study their stability. The critical points correspond to $f_1=0$ and $f_2=0$. It turns out that there are two critical points characterised by $I:\left(\;R=0,\;\; H=\sqrt{\frac {\rho_\Lambda}{3\mpl^2}}\right)$ and $II:\left(\;R=1,\;\; H=\frac{\sqrt{2\rho_\Lambda}}{3\mpl}\right)$.

\begin{center} 
\begin{figure}[h!]
\begin{center} 
\includegraphics[width=6.7cm]{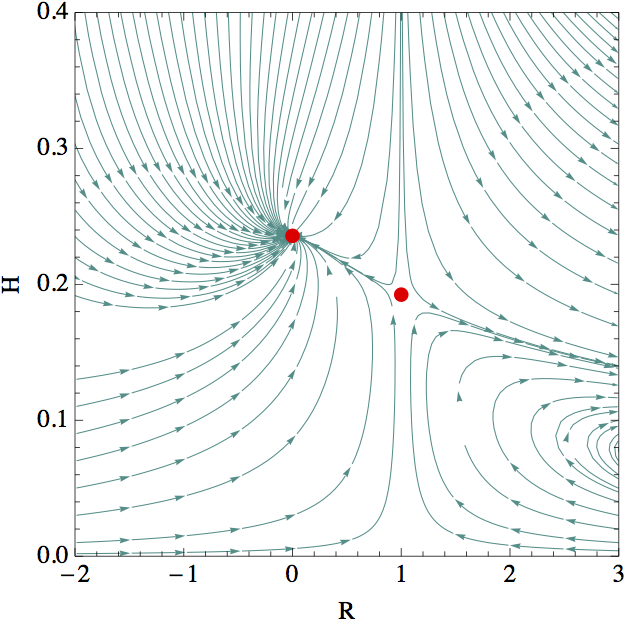}
\includegraphics[width=6.7cm]{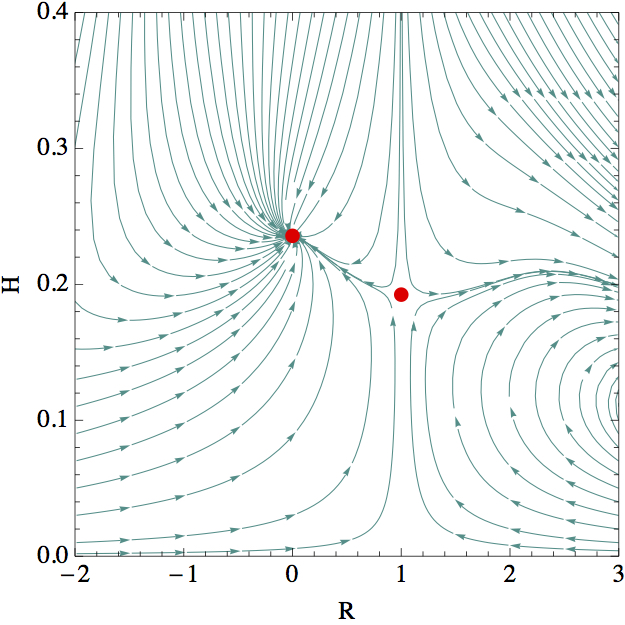}
\includegraphics[width=6.7cm]{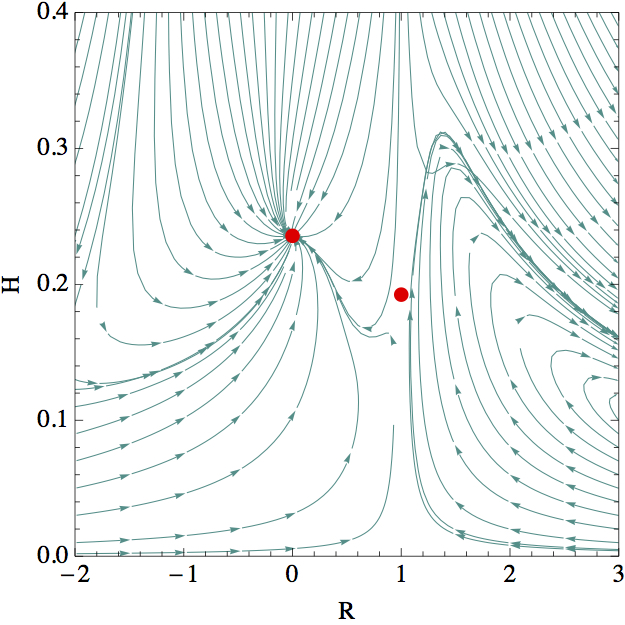}
\caption{In this figure three examples of the phase map for the autonomous system are shown with increasing values of $M^2$ from left to right \cite{Jimenez:2013qsa}. One can see that the de Sitter critical point ($R=0$) is an attractor and the electric Bianchi type I solution ($R=1$) is unstable.  }
\label{phasemapLFF}
\end{center}
\end{figure}
\end{center}

The critical point with ($R=0$) is nothing else but the isotropic de Sitter solution driven by the cosmological constant. On the other hand, the second critical point with ($R=1$) corresponds to the electric Bianchi type I solution, which is not physical since it yields a negative energy density of the vector field. 
It is interesting to note that the location of the critical points does not depend on $M^2$ so that the non-minimal coupling does not play any role. Moreover, we can consider small perturbations around the critical points in order to study their stability. The linearised perturbations have the form
\begin{eqnarray}
\frac{d}{dt}\left(\begin{array}{c}
\delta R \\ \delta  H\end{array}\right)  &=& M_{I,II} \left(
\begin{array}{c}
\delta R \\ \delta H\end{array}\right) \,.
\end{eqnarray}
From this, one obtains that the first critical point is stable whereas the second one is a saddle point. This can be explicitly seen in the figure \ref{phasemapLFF}. This means that the stability of the de Sitter universe is maintained despite the presence of the non-minimal interaction. 
Apart from the property of de Sitter attractor, one has to ensure the absence of ghost and Laplacian instabilities. In order to avoid ghost instabilities on cosmological backgrounds, one has to impose the condition
\begin{equation}
1+\frac{4 H^2}{M^2}>0.
\end{equation}
Similarly, for the avoidance of Laplacian instabilities, the propagation speed in the high frequency regime has to be positive
\begin{equation}
c_s^2= 1-\frac{\frac{4H^2}{M^2}(1-\frac{\mathcal H'}{\mathcal H^2})}{1+\frac{4H^2}{M^2}} >0.
\label{speed:flrw}
\end{equation}
This type of Horndeski vector-tensor theories, although very interesting on its own, quite generally suffer from ghosts or Laplacian instabilities in regions where the non-minimal coupling dominates over the Maxwell kinetic term \cite{Jimenez:2013qsa}. Unfortunately, this property does not alter by promoting the gauge field to a non-abelian gauge field. A non-abelian $SU(2)$ gauge field with a non-minimal Horndeski coupling to gravity can generate a de Sitter solution followed by a graceful exit to a radiation-dominated epoch \cite{Davydov:2015epx}. However, the second order action of perturbations for this Horndeski Yang-Mills theory reveals ghost and Laplacian instabilities in the tensor sector on top of the homogeneous and isotropic quasi de Sitter background \cite{BeltranJimenez:2017cbn}. In the case of gauge invariant vector theories, one can not apply them to dark energy maintaining the homogeneity and isotropy. Therefore, more promising roads for dark energy arise in the generalised Proca theories with broken gauge symmetry.


\subsection{New avenues in cosmology with massive vector fields}\label{sec_cosmo_GP}

As we mentioned above, vector fields have been also considered in cosmological applications. In contrast to scalars, the vector fields in cosmology have the additional difficulty of leading to the presence of large scale anisotropic expansion. This on the other hand would be in conflict with the high isotropy observed in the CMB. However, there is still some room for the vector fields if the anisotropies are not too large and a small amount of it could be actually used to explain the reported anomalies by WMAP and Planck in cosmological observations at large scales \cite{Bonvin:2017req,Tansella:2018hdm}. If one considers only an abelian gauge field, the only way of obtaining an isotropic expansion would consist of $N$ vector fields with a randomly distributed orientation. Another way is to consider a non-abelian vector field with interactions that exhibit a global $SO(3)$ symmetry. 

Within the context of massive vector fields, one can achieve isotropic expansion with another configuration, namely by the temporal component of the vector field, which is just an auxiliary field and does not propagate. The auxiliary mode gives rise to a modified Friedman equation. Let us consider the generalized Proca interactions that we discussed in section \ref{secProcacurved} in equation (\ref{vecGalcurv}). For the background metric we shall assume as usual a FLRW metric with the line element $ds^2=-dt^2+a^2(t)d\vec{x}^2$. Compatible with these symmetries of homogeneity and isotropy, the Ansatz for the background vector field is
\begin{equation}
A^{\mu}=(\phi(t),0,0,0)\,.
\end{equation}
From the action in equation (\ref{vecGalcurv}) we obtain the following background equations of motion
\begin{eqnarray}
& &G_2-G_{2,X}\phi^2-3G_{3,X}H \phi^3
-6(2G_{4,X}+G_{4,XX}\phi^2)H^2\phi^2 \nonumber\\
& &+6G_4H^2+G_{5,XX} H^3\phi^5+ 5G_{5,X} H^3\phi^3
=\rho_M\,,
\label{be1}\\
& &
G_2-\dot{\phi}\phi^2G_{3,X}+2G_4\,(3H^2+2\dot{H})
-2G_{4,X}\phi \, ( 3H^2\phi +2H\dot{\phi}
+2\dot{H} \phi )\nonumber\\
&& -4G_{4,XX}H\dot{\phi}\phi^3+G_{5,XX}H^2\dot{\phi} \phi^4+G_{5,X}
H \phi^2(2\dot{H}\phi +2H^2\phi+3H\dot{\phi})
=-P_M\,,
\label{be2}
\end{eqnarray}
for the background metric and similarly for the vector field 
\begin{equation}
\phi \left( G_{2,X}+3G_{3,X}H\phi +6G_{4,X}H^2
+6G_{4,XX}H^2\phi^2
-3G_{5,X}H^3\phi-G_{5,XX}H^3 \phi^3 \right)=0\,,
\label{be3}
\end{equation}
where $G_{i,X}$ stands for the derivative with respect to $X$ and $\rho_M$ and $P_M$ are the energy density and pressure of the matter fields. In the equation of motion for the vector field, one sees immediately that one consistent solution would be $\phi=0$. Of course, we will not be interested in these solutions. On the other hand the  $\phi\ne0$ solutions will result in the for us interesting de Sitter solutions. The de Sitter fixed point corresponds to $\dot{H}=0$ with $\rho_M=P_M=0$ and $\dot{\phi}=0$. 

On top of these background solutions, one has to guarantee that the tensor, vector and scalar perturbations do not have any pathologies. 
For the tensor perturbations we assume the transverse and traceless Ansatz $\delta g_{ij}=a^2 h_{ij}$ with $ \partial^i h_{ij}=0$ and $ {h_i}^i=0$. The action to second order in tensor perturbations becomes
\begin{equation}
S_T=\sum_{p={+},{\times}}\int dt\,d^3x\,
a^3\,\frac{q_T}{8}  \left[\dot{h}_p^2
-\frac{c_T^2}{a^2}(\partial h_p)^2\right]\,,
\label{ST}
\end{equation}
where $p$ denotes the two polarization modes and the coefficient $q_T$ is given by
\begin{equation}
q_T=2G_4-2\phi^{2}G_{{4,X}}+
H\phi^{3}G_{{5,X}}\,,
\label{qT}
\end{equation}
together with the sound speed
\begin{equation}
c_{T}^{2}=
\frac {2G_{4}+\phi^{2}\dot\phi\,G_{{5,X}}}{q_T}\,.
\label{cT}
\end{equation}
Accordingly, the tensor perturbation equation of motion in Fourier space reads
\begin{equation}
\ddot{h}_{\lambda}+\left( 3H+
\frac{\dot{q}_T}{q_T} \right) \dot{h}_{\lambda}
+c_T^2 \frac{k^2}{a^2}h_{\lambda}=0\,.
\end{equation}
In order to guarantee the absence of tensor ghost and small-scale Laplacian instabilities, one has to impose the conditions
 $q_T>0$ and $c_T^2>0$ respectively.
 
 In a similar way, we can work out the vector perturbations. In this case, not only the metric vector perturbations  $\delta g_{0i}=2V_i$ with $\partial^i V_i=0$ will contribute but also the vector perturbations of the vector field $\delta A^i=\frac{\delta^{ij}}{a^2}E_j$ with $E_j$ again satisfying the transverse condition. Similarly, there will be also the velocity perturbations of the matter fields. The action two second order in vector perturbations is of the form
 \begin{equation}
S_T=\sum_{i=1,2}\int dt\,d^3x\,
a^3\,\frac{q_V}{8}  \left[\dot{V}_i^2
-\frac{c_V^2}{a^2}(\partial V_i)^2-m_V^2V_i^2\right]\,,
\label{SV}
\end{equation}
with the corresponding coefficients this time
\begin{eqnarray}
q_V
&=& G_{2,F}+2G_{2,Y}\phi^2-4g_5H \phi
+2G_6H^2+2G_{6,X} H^2 \phi^2\label{qvProca}\,,\\
c_V^{2}&=&1+\frac{\phi^2(2G_{4,X}-G_{5,X}H \phi)^2}
{2q_Tq_V}+\frac{2[G_6 \dot{H}-G_{2,Y}\phi^2
-(H\phi-\dot{\phi})(H \phi G_{6,X}-g_5)]}
{q_V}\,,
\label{cv}\\
m_V^2
&=& \frac{1}{q_V}\Big(2( 2\,G_{{4,X}} -H\phi\,G_{{5,X}}) \dot{H} + ( G_{{3,X}}
+4\,\phi\,HG_{{4,XX}}-G_{{5,X}}{H}^{2}-{\phi}^{2}G_{{5,XX}}{H}^{2} )\dot\phi \Big)  \nonumber\\
&&+2 H^2+\frac{1}{q_V}\frac{d}{dt}(q_V H)\,.
\end{eqnarray}
The analytic estimation as well as the numerical solutions of the equations of motion of the vector perturbations reveals that the vector perturbations are nearly constants far outside the vector sound horizon and after horizon entry $c_V^2k^2/a^2>H^2$ they decay with oscillations \cite{DeFelice:2016uil}.

Last but not least, there are also the scalar perturbations. For the absence of ghost and gradient instabilities one has to put also the right constraints on these perturbations. From the vector field we have the scalar perturbations $\delta A^0=\delta\phi$ and $\delta A^i=\frac{\delta^{ij}}{a^2}\partial_j\chi_V$ and from the metric $\delta g_{00}=-2\alpha$ and $\delta g_{0i}=2\partial_i\chi$ and similarly from the matter fields for the energy density $\delta \rho_M$ and velocity potential $v$. The equations of motion of the scalar perturbations are given by
\begin{eqnarray}
& &
\delta \rho_M-2w_4 \alpha+\left( 3Hw_1-2w_4 \right)\frac{\delta \phi}{\phi}
+\frac{k^2}{a^2} \left( {\cal Y}
+w_1 \chi-w_6 \psi \right)=0\,,\label{per1} \\
& &
\left( \rho_M+P_M \right) v+
w_1 \alpha+\frac{w_2}{\phi} \delta \phi=0\,,\label{per2}\\
& &
\left( 3Hw_1-2w_4 \right)\alpha-2w_5 \frac{\delta \phi}{\phi}
+\frac{k^2}{a^2} \left[ \frac12 {\cal Y}
+w_2 \chi-\frac12 \left( \frac{w_2}{\phi}+w_6 \right) \psi
\right]=0\,,\label{per3} \\
& &
\dot{\delta \rho}_M+3H\left(1+c_M^2 \right) \delta \rho_M
+\frac{k^2}{a^2} \left( \rho_M+P_M \right)
\left( \chi+v \right)=0\,,
\label{per4}\\
& &
\dot{\cal Y}+\left( H -\frac{\dot{\phi}}{\phi} \right){\cal Y}
+2\phi \left( w_6 \alpha+w_7 \psi \right)
+\left( \frac{w_2}{\phi}+w_6 \right) \delta \phi
=0\,,\label{per5}\\
& &
\dot{v}-3Hc_M^2 v-c_M^2 \frac{\delta \rho_M}
{\rho_M+P_M}-\alpha=0\,,\label{per6}
\end{eqnarray}
with the introduced short-cut notations
\ba
w_{1} & = & {H}^{2} {\phi}^{3} (G_{{5,X}}+{\phi}^{2}G_{{5,{\it XX}}})
-4\,H(G_{{4}}+{\phi}^{4}G_{{4,{\it XX}}})-{\phi}^{3}G_{{3,X}}\,,
\label{w1}\\
w_{2} & = & w_1+2Hq_T\,,\label{w2}\\
w_{3} & = & -2{\phi}^{2}q_V\,,\label{w3} \\
w_{4} & = & \frac{1}{2}{H}^{3}\phi^{3}(9G_{{5,X}}-\phi^{4}G_{{5,{\it XXX}}})
-3\,H^{2} (2G_{{4}}+2\phi^{2}G_{{4,X}}+\phi^{4}G_{{4,{\it XX}}}-\phi^{6}G_{{4,{\it XXX}}}) \nonumber \\
 & &-\frac{3}{2}\,H\phi^{3}(G_{{3,X}}-\phi^{2}G_{{3,{\it XX}}})
 +\frac{1}{2}\,\phi^{4}G_{{2,{\it XX}}}\,,\label{w4} \\
w_{5} & = & w_{4}-\frac{3}{2}\,H(w_{1}+w_{2})\,, \label{w5} \\
w_{6} & = & -\phi\,\left[{H}^{2}\phi(G_{{5,X}}-{\phi}^{2}G_{{5,{\it XX}}})
-4\,H(G_{{4,X}}-{\phi}^{2}G_{{4,{\it XX}}})+\phi G_{{3,X}}\right]\,, \label{w6} \\
w_{7} & = & 2(H\phi G_{{5,X}}-2G_{{4,X}}) \dot{H}
+\left[H^{2}(G_{{5,X}}+{\phi}^{2}G_{{5,{\it XX}}})-4\,H\phi\,G_{{4,{\it XX}}}-G_{{3,X}}\right] \dot{\phi}\,, 
\label{w7}\nonumber \\
{\cal Y} &=&
 \frac{w_3}{\phi}
\left( \dot{\psi}+\delta \phi+2\alpha \phi \right)\,.
\label{Ydef}
\ea
and $c_M^2$ representing the matter propagation speed squared. For the purposes of observational applications we can define the gauge invariant Bardeen gravitational potentials $\Psi\equiv \alpha+\dot\chi$ and $\Phi\equiv H\chi$. Similarly, it is convenient to define the gauge invariant matter density contrast $\delta\equiv\delta \rho_M/\rho_M+3H(1+w_M)v$ with the matter equation of state $w_M \equiv P_M/\rho_M$. In terms of these gauge invariant quantities, the modified Poisson equation is given by
\be
\frac{k^2}{a^2}\Psi=-4\pi G_{\rm eff} \rho_M \delta\,,
\label{Geff}
\ee
with $G_{\rm eff}$ representing the effective gravitational coupling. Other important quantities are the slip parameter defined as $\eta \equiv -\frac{\Phi}{\Psi}$ and the growth rate of the matter density contrast $f \equiv \frac{\dot{\delta}}{H \delta}$. 

\begin{center}
\begin{figure}[h!]
\begin{center}
\includegraphics[width=9.2cm]{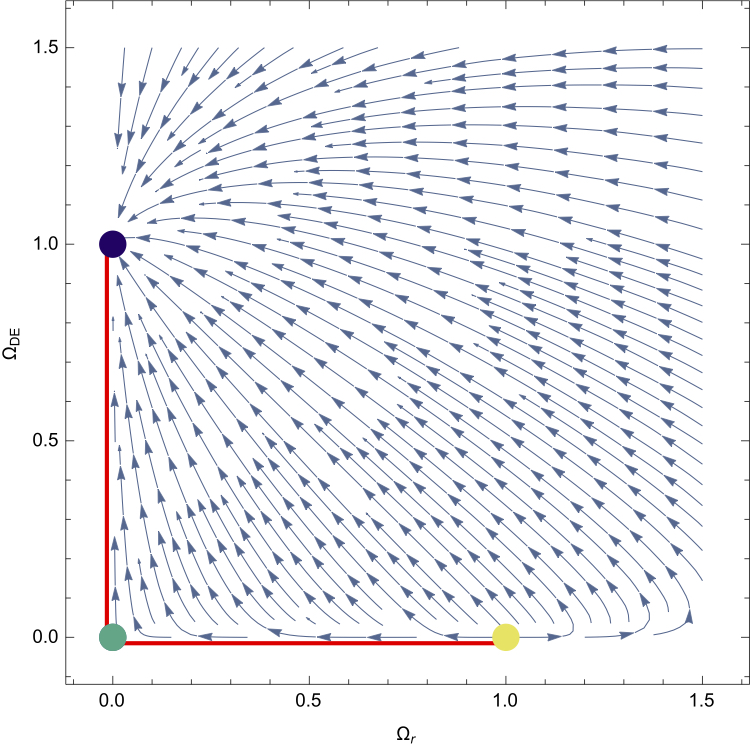}
\end{center}
\caption{\label{figBGproca}
In this figure the phase map portrait of
the dynamical autonomous system for $s=1$ is shown \cite{DeFelice:2016yws}. The yellow dot is the radiation fixed point, the green is the matter fixed point and
the blue is the de Sitter fixed point.}
\end{figure}
\end{center}

In order to gain a better intuition and for a concrete comparison to observations, let us assume specific forms for the general functions in the form
\ba
& &
G_2(X,Y,F)=b_2 X^{p_2}
+\left[ 1+g_2(X) \right]F\,,\qquad
G_3(X)=b_3X^{p_3}\,,\qquad
G_4(X)=\frac{1}{16\pi G}+b_4X^{p_4}\,,\nonumber \\
& &
G_5(X)=b_5X^{p_5}\,,\qquad
g_5(X)=\tilde{b}_5 X^{q_5}\,,\qquad
G_6(X)=b_6X^{p_6}\,.
\label{G23456}
\ea
For the specific relations of the powers $p_3=\frac12 \left( p+2p_2-1 \right)$, $p_4=p+p_2$ and $
p_5=\frac12 \left( 3p+2p_2-1 \right)$ one can construct interesting background solutions of the form $\phi^{p} \propto H^{-1}$ with a positive constant $p$. These solutions are interesting since the temporal vector component $\phi$ is negligible in the early cosmological epoch, but
with the decrease of $H$ it grows giving rise to the late-time cosmic acceleration. The background equations can be recast into
\be
\frac{\Omega_{\rm DE}\,\Omega_r^{3(1+s)}}
{(1-\Omega_{\rm DE}-\Omega_r)^{4(1+s)}}
=\left( \frac{\Omega_{\rm DE}}{\Omega_r^{1+s}}
\right)_0
\left( \frac{\Omega_{r}}{1-\Omega_{\rm DE}-\Omega_r}
\right)_0^{4(1+s)}\,,
\ee
where $\Omega_{\rm DE}$ is defined in terms of $p$, $p_2$ and $\phi$, $H$ and the lower subscript ``0'' representing the values today. The introduced new parameter $s$ stands for $s \equiv \frac{p_2}{p}$. For the exact definitions see \cite{DeFelice:2016yws}. The cosmological trajectories are illustrated in Figure \ref{figBGproca}. It can be clearly seen that de Sitter is an attractor. Independently of the initial conditions all the cosmological trajectories end there. 

\begin{center}
\begin{figure}[h!]
\begin{center}
\includegraphics[width=10cm]{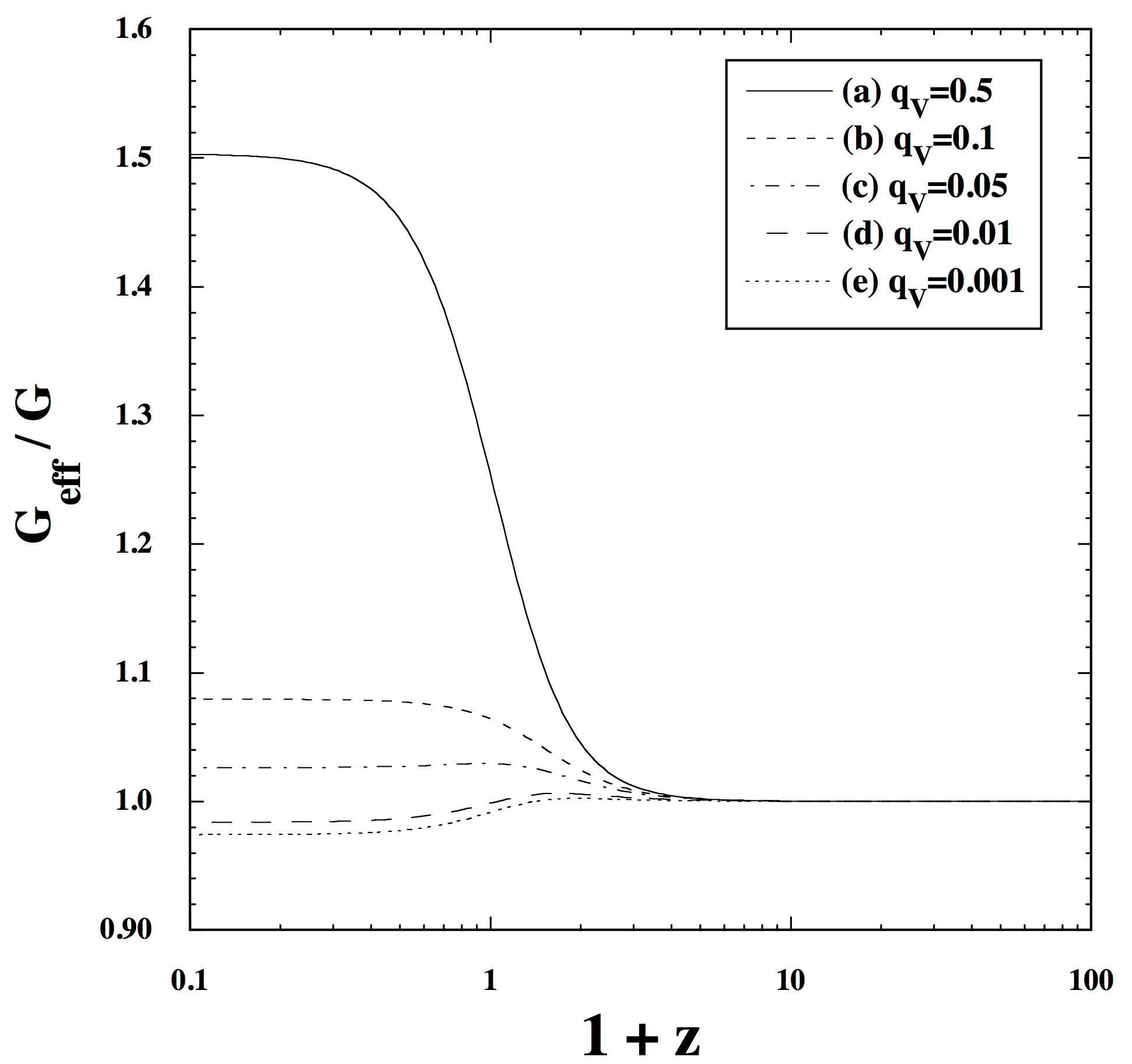}
\end{center}
\caption{\label{figGeffGP}
Evolution of $G_{\rm eff}/G$ for the model parameters
$p_2=1/2$, $p=p_6=5/2$, $g_2=0$,
$\tilde{b}_5=0$, $\beta_4=10^{-4}$, $\beta_5=0.052$ (related to $b_4$ and $b_5$),
$\lambda=1$ with $q_V=0.5,0.1,0.05,0.01,0.001$
(from top to bottom). More detail can be obtained from \cite{DeFelice:2016uil}, specially about the notation of the coefficients and their numerical values.}
\end{figure}
\end{center}

For modes of interest for the observations of large-scale structures, we can assume that they have physical momenta $k/a$ higher than the Hubble expansion rate, which is known as the sub-horizon approximation. Furthermore, for an adiabatic evolution we can neglect the fast modes, hence apply the quasi-static approximation. With these assumptions the effective gravitational coupling constant and the slip parameter take the form
\ba
G_{\rm eff}
&= & \frac{H(\mu_2 \mu_3-\mu_1 \mu_4)}
{4\pi \phi \mu_5}\,,
\label{Geffq}\\
\eta
&=& \frac{\phi^2[2w_2 \mu_2-w_3 \mu_4(w_1-2w_2)]}
{\mu_2 \mu_3-\mu_1 \mu_4}\,,
\label{etaq}
\ea
with the introduced variables $\mu_i$ for convenience
\ba
\mu_1 &\equiv&
\frac{\phi^2}{H} \left[ \left( \dot{w}_1-2\dot{w}_2+Hw_1
-\rho_M \right)w_3-2w_2(w_2+Hw_3) \right]\,,
\label{mu1}\\
\mu_2 &\equiv&
\phi\left( w_2^2+Hw_2w_3+\dot{w}_2 w_3 \right)
+w_2 ( w_6\phi^2-w_3 \dot{\phi})\,,
\label{mu2}\\
\mu_3 &\equiv& \frac{2\phi}{Hw_3} \mu_2\,,
\label{mu3}\\
\hspace{-0.8cm}
\mu_4 &\equiv&
-\frac{1}{w_3} \Big[ \phi^3 (w_6^2+2w_3w_7)
+\phi^2 (2w_2w_6+Hw_3w_6+w_3\dot{w}_6) \nonumber\\
&&+\phi \left\{ w_2^2+Hw_2w_3+w_3(\dot{w}_2-\dot{\phi}w_6)
\right\}-2\dot{\phi}w_2w_3 \Big] \,,
\label{mu4} \\
\mu_5 &\equiv&
(w_1-2w_2) \left[ \phi (w_1-2w_2)w_3 \mu_4
-2\phi w_2 \mu_2 \right]
+Hw_2 \left[ 2w_2(\mu_1+w_3 \mu_3)
-w_1 w_3 \mu_3 \right]\,.
\label{mu5}
\ea
An essential property of these generalized Proca theories is illustrated for the model parameters
$p_2=1/2$, $p=p_6=5/2$, $g_2=0$, $\tilde{b}_5=0$, $\beta_4=10^{-4}$, $\beta_5=0.052$,
$\lambda=1$ in figure \ref{figGeffGP}. 

As it can be seen in the figure \ref{figGeffGP}, one can realize weak gravity with $G_{\rm eff}$ smaller than the Newton gravitational constant $G$. Hereby, the contribution of the intrinsic vector modes in the quantity $q_V$ (kinetic term of the vector perturbations given by equation \eqref{qvProca}) plays a very crucial role.
For small values for $q_V$ the effective gravitational coupling
$G_{\rm eff}$ decreases. This illustrates that vector modes can significantly modify the gravitational interactions on cosmological scales with a large impact on the observations of large-scale structures and weak lensing. The realization of weak gravity could be used to explain the tension between the redshift space distortion measurements and the Planck data for $f \sigma_8$. Hence, it is possible to distinguish these models from the $\Lambda$CDM model according to both expansion history and cosmic growth.


\subsection{Cosmic structure formation with Kinetic Field Theory (KFT)}\label{sectionKFT}
\textbf{The concept of KFT}:\\
The formation of cosmic structures provides one of the most important and potentially most powerful testing grounds for theories of gravity. The earliest signature of cosmic structures that we can observe are the temperature and polarisation fluctuations in the Cosmic Microwave Background (CMB). We see them as they were when photons were first released from the cosmic plasma, approximately 400,000 years after the Big Bang. Their amplitude is characterised by the variance of the temperature fluctuations relative to the mean temperature, which is of order $10^{-5}$.

The angular power spectrum of the CMB temperature fluctuations indicates that these fluctuations were at least predominantly adiabatic and thus accompanied by corresponding fluctuations in the matter density. In our present universe, we see ourselves surrounded by pronounced cosmic structures marked by galaxies, galaxy clusters, large regions devoid of galaxies and huge, extended filaments. The quite obvious question whether, and if so, how the primordial density fluctuations traced by the CMB could have developed into the structures we observe today leads to the immediate conclusion that, at least in the standard cosmological model based on the General theory of Relativity, the cosmic structures must be dominated by a form of matter which cannot partake in the electromagnetic interaction \cite{1982ApJ...263L...1P}. This dark matter thus appears as an inevitable consequence of the standard cosmological model.

As long as the relative density fluctuations in cosmic structures are small compared to the mean cosmic matter density, the evolution of cosmic structures can safely be described as a linear process, i.e.\ as a process governed by a linearised set of equations. In the conventional analytic approach to cosmic structure formation, the density and velocity fields of the predominantly dark matter are both assumed to be smooth and differentiable fields whose behaviour is captured by the hydrodynamical equations, i.e.\ the continuity and the Euler equation, plus the Poisson equation as the Newtonian field equation for the fluctuations of the gravitational potential sourced by the density fluctuations. For structures with length scales small compared to the Hubble radius, moving with velocities small compared to the speed of light, this Euler-Poisson system appears adequate \cite[see][for a review]{2002PhR...367....1B}.

At late cosmic times and on small spatial scales, the evolution of cosmic structures becomes highly non-linear, i.e.\ the amplitude of the density fluctuations grows way beyond the mean density. Then, the approach based on the linearised hydrodynamical equations breaks down for two reasons. First, since the equations are linearised in the density, velocity, and gravitational-potential fluctuations, they cannot capture the formation of large density fluctuations. Second, and conceptually more importantly, a limitation in the essential assumptions of hydrodynamics becomes fundamentally important. Hydrodynamics assumes that the mean free path of the fluid particles is negligibly short compared to all other length scales relevant for the system under consideration. Specifically for dark matter, however, this cannot hold true. At the latest when particle streams converge and meet, discontinuities or shocks would form in a fluid, preventing the velocity field from becoming multi-valued. In cosmic structures, however, streams of the at most weakly interacting dark-matter particles can cross, leading to the formation of multi-valued velocity fields. At the latest when this happens, the hydrodynamical treatment must break down.

Conventionally, the most trustable way of studying non-linear cosmic structure formation are numerical simulations. They decompose the cosmic density field into pseudo-particles with a mass set by the spatial resolution of the simulation. Beginning with suitable initial conditions, the gravitational interaction of the dark matter particles is directly simulated. As highly resolved simulations show, the non-linearly evolved density field develops characteristic patterns and, in particular, gravitationally bound structures with universal properties, such as the radial density profiles with similar functional forms for dark matter haloes with masses between dwarf galaxies and massive galaxy clusters.

Much of what we can learn about the universe is due to observations of signals provided by non-linearly evolved structures. Moreover, as observations of the CMB temperature fluctuations show, cosmic structures begin as random fields which either are Gaussian random fields or have non-Gaussian contributions which are tightly constrained and compatible with zero. In the course of cosmic structure evolution, however, the Gaussian character of the structures gets lost by mode coupling due to the gravitational interaction. Non-Gaussian, non-linear structures thus carry much of the empirical evidence available in our observable universe.

Aiming at testing alternative theories of gravity on cosmological scales, the late-time, non-linear and non-Gaussian structures  supply a large and most valuable part of the potentially available signal. A thorough understanding of these aspects of structure formation is thus mandatory for sensitive tests of the cosmological consequences of gravity theories. This remark has two aspects. First, cosmic structure formation becomes most interesting in particular where density fluctuations become large and conventional analytic approaches break down. Second, the signal contained in these structures will not obey Gaussian statistics any more. For solid statistical inference from observations, therefore, higher than second-order density and velocity correlations are needed, allowing to assess the degree of mixing and of phase correlations between Fourier modes of the cosmic matter field. Simulating cosmic structures sufficiently deeply into the non-linear, non-Gaussian regimes for a possibly large variety of alternative theories of gravity, with an angular resolution required at least for galaxy clusters or groups and with simulation volumes large enough to reliably estimate non-Gaussian effects, would be forbiddingly time and resource consuming.

Effective field theories have been developed, based on the Euler-Poisson system, to study cosmic structure formation analytically \cite[e.g.][]{2014JCAP...07..057C, 2012JHEP...09..082C, 2014JCAP...05..022P}. They encounter two problems. First, by construction, they cannot bypass the conceptual problem that hydrodynamics does not allow the formation of multiple-valued velocity fields, i.e.\ of stream or shell crossing. Effective field theories of cosmic structure formation thus typically begin failing at wave numbers $k\gtrsim 0.3\,h\,\mathrm{Mpc}^{-1}$. Second, being effective field theories, they contain parameters which again require sufficiently large numerical simulations for their reliable calibration. Other analytic approaches based on Lagrangian perturbation theory may go further, but still assume the existence of a smooth spatial distribution functions e.g.\ for the cosmic matter \cite{1992MNRAS.254..729B, 1994MNRAS.267..811B, 1993MNRAS.264..375B, 1997GReGr..29..733E, 2012JCAP...06..021R}. Once multiple streams form, these functions are being folded and develop cusps where functional determinants relating the initial density distribution to its later states become singular.

For these and other reasons, a conceptually entirely different analytic approach to cosmic structure formation has recently been developed \cite{2016NJPh...18d3020B}, based on the approach by \cite{1973PhRvA...8..423M} and on the pioneering work by \cite{2013JSP...152..159D, 2012JSP...149..643D, 2011PhRvE..83d1125M, 2010PhRvE..81f1102M}. This approach, termed kinetic field theory or KFT for short, decomposes the cosmic matter field into classical pseudo-particles of a certain mass whose value is irrelevant. The dynamics of these particles in phase space is controlled by the Hamiltonian equations of motion. Being linear, the Hamiltonian equations allow the construction of a retarded Green's function for solving these microscopic equations of motion. Since Hamiltonian trajectories cannot cross in phase space due to their uniqueness, the problems related to shell crossing or multiple particle streams is absent by construction. In addition, the Hamiltonian flow on the phase space is diffeomorphic and thus free of singularities, and it is symplectic and thus, by Liouville's theorem, even preserves its volume in phase space.

KFT begins by covering phase space at a sufficiently early initial time, for example corresponding to the time of CMB decoupling, with a suitable distribution function quantifying the probability for a phase-space point to be occupied by a particle. This initial probability distribution must respect the relevant spatial density correlations corresponding to an assumed initial density-fluctuation power spectrum. In accordance with the Gaussian appearance of the CMB temperature fluctuations, modelling the initial density field as a Gaussian random field seems to be a safe procedure. By continuity then, the initial velocity field must also be correlated. The initial state of KFT is thus characterised by a set of $N$ classical point particles whose initial positions and momenta are correlated according to an initial Gaussian random potential field whose gradient determines the velocity and whose (negative) Laplacian determines the number density of the particles.

The Hamiltonian flow of these particles, expressed by the retarded Green's function of Hamilton's equations, then provides a volume-conserving, diffeomorphic map of the initial to any final phase space. It thus also maps the initial probability distribution to any later time. Very much in analogy to statistical quantum field theory, the statistical properties of the particle ensemble are encapsulated into a generating functional. Similar to a partition function in standard, equilibrium thermodynamics, this generating functional integrates over the probability for the particles to occupy the phase-space positions at any time. This probability for any phase-space position to be occupied at time $t$ is factorised into the probability an initial phase-space position to be occupied at the initial time $t = 0$, times the conditional or transition probability for a particle to move from the initial to the final position within the time $t$. Being a classical theory, this transition probability is deterministic, which is the main difference between KFT and a statistical quantum field theory.

Having augmented the generating functional with suitable generator or source fields, particle correlations of any order can then be derived by applying as many functional derivatives with respect to these source fields to the generating functional as correspond to the order of the correlations. Particle interactions are contained in an exponential interaction operator, again similar to quantum field theory. As there, expanding the interaction operator into a Taylor series leads to the Feynman diagrams of the theory.

Compared to most quantum field theories, applications of KFT to cosmology have two further important advantages. One concerns the interaction potential between the particles. Since the particles are diluted by cosmic expansion, the effective amplitude of their interaction potential decreases over time, corresponding to a particle mass shrinking with time. This prevents secular instabilities from occurring. The second concerns the properties of the initial density fluctuation power spectrum, which decreases quite steeply for $k\to0$ as well as for $k\to\infty$. Infrared or ultraviolet divergences thus do not occur either.

To a very large degree, it is arbitrary how a given Hamiltonian is split up into a free and an interaction part. This implies that the Green's function of the Hamiltonian can be varied to an equally large degree. In particular, it is possible to split the Hamiltonian such that already the Green's function of the free motion encapsulates part of the large scale gravitational interaction. One widely known example for this approach is the Zel'dovich approximation \cite{1970A&A.....5...84Z}, which could be further improved \cite{2015PhRvD..91h3524B}. On the basis of this improved Zel'dovich approximation, it could be shown that even first-order perturbation theory with KFT yields a non-linearly evolved density fluctuation power spectrum which, up to wave numbers $k\lesssim 10\,h\,\mathrm{Mpc}^{-1}$, very closely follows the power spectrum derived directly from numerical simulations \cite{2016NJPh...18d3020B}.

Further improvements are possible. First, beginning with a statistically homogeneous and isotropic, Gaussian random field for the initial conditions, the generating functional of KFT can be fully factorized into generic factors which can be evaluated once and for all \cite{2017NJPh...19h3001B}. This simplifies and speeds up even elaborate calculations with KFT, but also provides a conceptual understanding of the essential characteristics of cosmic structure formation. Second, the interaction operator can be evaluated in the Born approximation, i.e.\ along the inertial (free) particle trajectories described by the retarded Green's function of the Hamiltonian equations. This allows the derivation of a non-perturbative, closed, analytic and parameter free expression for the non-linearly evolved power spectrum of cosmic density fluctuations up to $k\lesssim 20,h\,\mathrm{Mpc}^{-1}$, thus reaching already quite deeply into the non-linear regime \cite{2017arXiv171007522B}. The evaluation of this expression is a matter of seconds on conventional laptops.

Even though KFT has been developed in the framework of the standard cosmological model, it can quite easily be generalised to alternative theories of gravity and different cosmological models. Such adaptations typically require only (1) a suitable choice of the time coordinate, which is conveniently (but not necessarily) chosen as the growth factor of linear structure formation; (2) the construction of a retarded Green's function for the Hamiltonian equations on an expanding background; and (3) the choice of the interaction potential between the particles. The correlated distribution function for the initial phase-space positions of the particles is defined by a suitable density-fluctuation power spectrum, typically set by the dark matter model. KFT then provides a general framework for developing the initial phase-space distribution forward in time.\\

\textbf{The formalism of KFT and its application to modified theories of gravity}:\\
The concepts of KFT are quite straightforwardly formalised. Its central object is a generating functional $Z$, analogous to a partition sum in equilibrium thermodynamics, which is defined as usual as an integral of the probability $P(\varphi)$ for a state $\varphi$ to be occupied over all possible states,
\begin{equation}
  Z = \int\mathcal{D}[\varphi]\,P(\varphi)\;,
\label{eq:1}
\end{equation} 
where $\mathcal{D}[\varphi]$ denotes the path integration over field configurations. We split the probability $P(\varphi)$ into the probability $P(\varphi_i)$ for the initial state and the conditional or transition probability $P(\varphi|\varphi_i)$ from the initial state to a state $\varphi$, hence
\begin{equation}
  Z = \int\mathcal{D}[\varphi]\int\mathcal{D}[\varphi_i]\,
  P(\varphi|\varphi_i)P(\varphi_i)\;.
\label{eq:2}
\end{equation}

For classical point particles subject to Hamiltonian dynamics, the fields are Dirac delta distributions at the particle positions $x = (q,p)$ in phase space, and thus the path integrations are reduced to ordinary integrations over phase-space points. Since the particle trajectories are deterministic, the transition probability also is a functional delta distribution of classical trajectories which is non-zero only along the particular trajectory which solves the Hamiltonian equation of motion. Let $G(t,t')$ be the retarded Green's function of the Hamiltonian equations, then the trajectories beginning at $x_i$ are
\begin{equation}
  \bar x(t) = G(t,0)x_i-\int_0^td t'\,G(t,t')\nabla V(t')\;,
\label{eq:3}
\end{equation}
if the initial time is chosen to be zero and $\nabla V(t')$ represents a force term to be specified via the potential $V$ in the Hamiltonian equations. Replacing the transition probability $P(x|x_i)$ by the Fourier representation of the functional delta distribution and introducing a generator field $J$ turns the generating functional into the form
\begin{equation}
  Z[J] = \int d\Gamma\,\e^{i\int_0^t\langle J,\bar x\rangle\,d t'}\;,
\label{eq:4}
\end{equation}
where $d\Gamma = P(x_i)\,d x$ is the initial phase-space measure and $\langle\cdot,\cdot\rangle$ denotes a suitably defined scalar product over all phase-space coordinates of all particles \cite{2016NJPh...18d3020B}. During the derivation of (\ref{eq:4}), it is being used that the Hamiltonian flow on the phase space is symplectic and thus has a unit functional determinant by Liouville's theorem.

Functional derivatives of $Z$ with respect to $J(t)$ pull the particle positions $\bar x(t)$ down from the exponent. Particle densities in configuration space are represented by delta distributions. Writing these in their Fourier representation, we can define an operator for density modes with wave number $\vec k$ in configuration space,
\begin{equation}
  \hat\rho\left(\vec k,t\right) =
  \exp\left(-i\vec k\cdot\frac{\delta}{\delta J_q(t)}\right)\;,
\label{eq:5}
\end{equation}
where $J_q(t)$ is the $q$ component of the generator field $J$ at time $t$. Since derivatives generate translations, exponentials of derivatives are finite translations. Applying a density operator $\hat\rho(1)$ with wave number $\vec k_1$ at time $t_1$ to the generating functional thus creates a shift $L(1)$ which, once the generator field $J$ has been set to zero, simply replaces $J$,
\begin{equation}
  \left.\hat\rho(1)Z[J]\right|_{J=0} = \langle\rho(1)\rangle =
  \int d\Gamma\,\E^{i\langle L(1),\bar x\rangle}\;.
\label{eq:6}
\end{equation}
The time integral over the phase in (\ref{eq:4}) disappears now because the functional derivative with respect to $J(t_1)$ returns a delta distribution in $t-t_1$. Density spectra of order $n$ can now be calculated by applying $n$ density operators to $Z$,
\begin{equation}
  G_n = \langle\rho(1)\ldots\rho(n)\rangle =
  \left.\prod_{i=1}^n\hat\rho(i)Z[J]\right|_{J = 0} = Z[L] =
  \int d\Gamma\,\E^{i\langle L,\bar x\rangle}\;,
\label{eq:7}
\end{equation}
where the shift $L = \sum L(i)$ is now a sum of contributions from density modes with wave numbers $\vec k_1, \ldots, \vec k_n$ taken at times $t_1,\ldots,t_n$. Up to this point, the theory is completely generic for classical Hamiltonian particles.

For cosmology, the probability distribution $P(x_i)$ of the initial phase-space positions $x_i$ must be specified. Since observations of the cosmic microwave background do not reveal any deviations from a Gaussian random field, it seems appropriate to assume that this $P(x_i)$ is a Gaussian, characterised only by the correlation matrices of particle positions and momenta, and between them. Assuming that the initial momenta are drawn from a vorticity-free field, $p=\nabla\psi$, and continuity then requires that $\delta=-\nabla^2\psi$. Specifying the correlation function of $\psi$ or, alternatively, the density power spectrum is thus sufficient to fix $P(x_i)$ \cite{2016NJPh...18d3020B}. Despite the correlations between all particle momenta, this initial probability distribution can be fully factorized, thus greatly simplifying the calculation of the generating functional $Z[L]$ and accordingly of $n$-th order power spectra \cite{2017NJPh...19h3001B}.

For cosmological applications, it is convenient to account for the expansion of spacetime by replacing spatial coordinates by comoving coordinates, and to adapt the Hamiltonian and the momenta accordingly. Effectively, this causes the interaction potential between the particles to decrease with time in response to the increasing mean particle separation. It is furthermore convenient to replace the cosmological time $t$ by a function $\tau$ growing monotonically with time and reflecting the linear growth of density fluctuations. A suitable choice for this function is the linear growth factor factor $D_+(t)$, shifted such that $\tau = 0$ at the initial time,
\begin{equation}
  \tau := D_+(t)-D_+(t_i)\;.
\label{eq:8}
\end{equation}

Simple as it looks, the expression (\ref{eq:7}) for the $n$-point spectra disguises one major difficulty, namely the interaction between the particles. The interaction term between two particles at positions $1$ and $2$ is described by the potential
\begin{equation}
  v(12) = \int_k\,\tilde v(k)\,\E^{i k\cdot (q_1(\tau_1)-q_2(\tau_1))}
\label{eq:9}
\end{equation} 
in terms of its Fourier transform $\tilde v(k)$. Note that $\tau_1 = \tau_2$ due to the instantaneous action of the Newtonian gravitational force. In (\ref{eq:9}), the particle positions can be replaced by functional derivatives with respect to the generator field component $J_q$. This allows to represent the particle interactions by an interaction operator appearing in an exponential function. Expanding this exponential operator into a Taylor series gives rise to the conventional perturbative approach graphically expressed by Feynman diagrams \cite{2016NJPh...18d3020B}. Another approach is enabled by Born's approximation, replacing an actual particle trajectory by its unperturbed trajectory
\begin{equation}
  q(\tau) = q+g_{qp}(\tau,0)p
\label{eq:10}
\end{equation} 
determined by the initial position $(q,p)$ and the component $g_{qp}(\tau,\tau')$ of the Green's function $G(\tau,\tau')$ \cite{2017arXiv171007522B}.

An important observation for both approaches is that the Green's function is determined by the Hamiltonian of a free particle. Splitting the Hamiltonian $H$ into a free part $H_0$ and an interaction part $H_I$ is, however, to a large degree arbitrary. This allows the construction of entire classes of Green's functions such that the interaction potential between the particles is minimized. One example for the benefit of this arbitrariness is the Zel'dovich approximation \cite{1970A&A.....5...84Z}, which describes free particle trajectories simply as
\begin{equation}
  q(\tau) = q+\tau p\;.
\label{eq:11}
\end{equation}
An improvement of the Zel'dovich approximation is given by
\begin{equation}
  g_{qp}(\tau,\tau') = \int_{\tau'}^\tau d\bar\tau\,\E^{h(\tau)-h(\bar\tau)}
\label{eq:12}
\end{equation}
with
\begin{equation}
  h(\tau) := \frac{1}{g(\tau)}-1\;,\quad
  g(\tau) := a^2D_+fH\;,
\label{eq:13}
\end{equation}
with the usual definition
\begin{equation}
  f = \frac{d\ln D_+}{d\ln a}
\label{eq:14}
\end{equation}
of the growth function $f$ \cite{2015PhRvD..91h3524B}. The function $g$ is to be normalized such that $g(0) = 1$. For $h = 0$, the Green's function (\ref{eq:12}) returns to the Zel'dovich approximation. The potential acting with respect to the trajectories (\ref{eq:10}) with the Green's function (\ref{eq:12}) can then easily be found. The Green's function (\ref{eq:12}) has the advantage of avoiding part of the re-expansion of cosmic structures after matter streams cross. The interaction term in Born's approximation has been derived in \cite{2017arXiv171007522B}, where it was shown that it reproduces the non-linear evolution of the cosmic matter power spectrum very well.

\begin{center}
\begin{figure}[h!]
\begin{center}
\includegraphics[width=14.0cm]{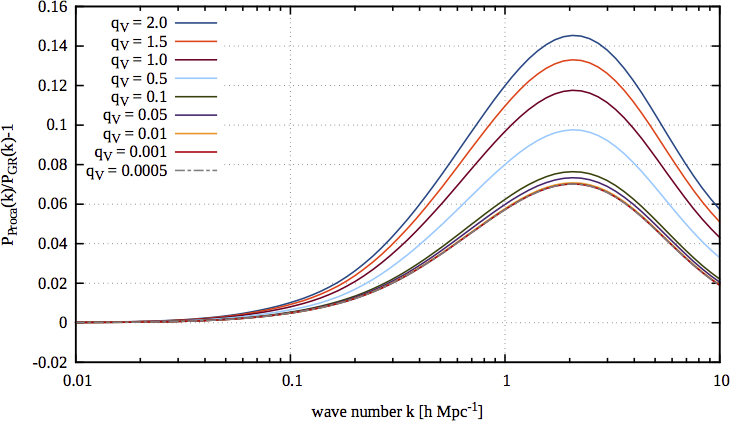}
\end{center}
\caption{\label{figKFT_GP}
The non-linear power spectrum of a dark energy model arising in generalised Proca theories with respect to the standard $\Lambda$CDM as a function of the wave number for different $q_v$ values (the quantity appearing in the kinetic term of vector perturbations of generalised Proca theories). The power spectra are normalised in the same way at the initial time. The model parameters are chosen as
$p_2=1/2$, $p=p_6=5/2$, $g_2=0$,
$\tilde{b}_5=0$, $\beta_4=0$, $\beta_5=0.0$,
$\lambda=1$.}
\end{figure}
\end{center}

Adapting KFT to alternative theories of gravity and using it to study non-linear structure formation as described by these theories is now quite straightforward. We need to specify
\begin{itemize}
  \item the background evolution in a cosmological model derived from the modified theory; this background evolution is characterised by the Hubble function $H$;
  \item a possible time dependence of the effective gravitational constant;
  \item the growth factor $D_+$ for linear perturbations in the modified theory; and
  \item possible changes to the gravitational potential.
\end{itemize}
The KFT formalism itself remains untouched by modifications of the theory of gravity, and the Green's function can be chosen as in General Relativity. In figure \ref{figKFT_GP} the application of KFT to generalised Proca theories is shown as a proof of concept for its applicability. We took the same dark energy model as illustrated in \ref{sec_cosmo_GP}. 

We first studied the background evolution of this specific dark energy model and determined the Hubble function (solved equations \eqref{be1},\eqref{be2} and \eqref{be3}). 
We used the Hamiltonian of the unperturbed background $\mathcal{H}=\Pi^\mu\dot{\mathcal{O}}-\mathcal{L}$ (with $\mathcal{O}$ representing $\mathcal{O}(N,a,\pi)$)
\begin{equation}
\mathcal{H}=-Na^3(G_2+6H2G_4-6G_{4,X}H^2\pi^2+2G_{5,X}H^3\pi^3
\end{equation}
and chosen the specific dark energy model given by the functions in equation \eqref{G23456}.
We have then used the linear growth factor as a suitable time component. For this, we solved the perturbation equation
\begin{equation}
\ddot{\delta}+2H\dot{\delta}-4\pi G_{\rm eff}\rho\delta=0,
\end{equation}
where the effective gravitational coupling $G_{\rm eff}$ is given by equation \eqref{Geffq}. The outcome of the KFT approach shows clearly significant deviations in the non-linear power spectrum with respect to General Relativity and could be tested against observations. The plot should be taken with a grain of salt, since we merely provide it just
to illustrate the applicability of KFT to modified gravity theories and represents preliminary results \cite{ProcaKFT}. it would be very interesting to introduce the new physics related to the generalised Proca field into a N-body simulation and compare the outcomes with the analytical KFT results.

\section{Gravitational Waves}

The breathtaking discovery of gravitational waves (GWs) \cite{Abbott:2016blz}, that received the well deserved Nobel prize, has triggered the field of GW astronomy. We are now able to listen to the mergers of binary black holes and neutron stars. This tremendous discovery has opened a new and unique opportunity to test gravitational theories beyond the regimes where it has been tested so far. With this new observational window we will enormously enhance our knowledge about the universe. The science of GWs will be a key field not only to astrophysics but also to cosmology. The LIGO/VIRGO team makes use of very advanced interferometers, which are able to detect strains with an accuracy corresponding to one hydrogen-atom radius compared to the astronomical unit. The ESA project eLISA will push this technology even further. These observational advances and technologies have to be accompanied tightly by the reinforced deeper theoretical understanding of the whole process of the merging of compact objects.
 
 Even before the direct detection of GWs we had tight constraints from indirect probes. For instance the period change of binary pulsars and energy loss were indirect evidence for GWS. The observations of the orbital decay rate of binary pulsars allowed to deduce that the inferred gravitational waves are predominantly of quadrupolar nature and the deviation of their propagation speed from the speed of light has to be of order $10^{-2}$ - $10^{-3}$ \cite{Yagi:2013qpa,Jimenez:2015bwa}. As it was shown in \cite{Jimenez:2015bwa} the anomalous propagation speed of GWs in Horndeski type of theories survives conventional screening due to the persistence of the gradient of the additional degree of freedom deep inside the virialized overdensities. Another tight constraint was already present from the absence of Cherenkov radiation, which puts a lower bound on the deviation of the order of $10^{-15}$, forbidding sub-luminal propagation in this way. 

These already tight constraints improved significantly with the direct detection of GWs. With the by now several events LIGO marked a milestone in GWs astronomy. With the two LIGO detectors the sensitivity to the polarisation of GWs and the source position was rather low, which was then increased significantly when the VIRGO collaboration joined the LIGO network. The $90\%$ credible region of the two detectors were then improved from $1160{\rm deg}^2$ down to $60{\rm deg}^2$. The observations indicate so far that the GW signals are predominantly of spin-2 nature, where vector and scalar polarisations, if present at all, have to be suppressed. 

Soon after VIRGO's participation GWs astronomy witnessed another major event with tremendous consequences. The GW170817 event \cite{TheLIGOScientific:2017qsa} of two merging neutron stars came with its electromagnetic counterpart as gamma-ray burst signal GRB170817A. The electromagnetic signal was arrived after $1.7$ seconds after the instant of merger of the neutron stars defined by the GW signal. This first multimessenger detection has outstanding repercussions of both astrophysical and cosmological nature. It not only allowed to measure the propagation speed with high precision, but also delivered essential evidence for heavy elements production. In the previous events of black hole binaries, the detectors were mainly sensitive to the merger phase and only a few cycles were observed before coalescence. However, in the case of neutron star binaries the detectors are sensitive to the inspiral phase prior to the merger, which allows the observation of tens of thousands cycles. In this way we can extract more information about the physical properties of compact objects.

One tremendous consequence of the GW170817 event is that the propagation speed of GWs is highly constrained to be $\vert c_{T}/c-1\vert\lesssim10^{-15}$, where $c$ represents the speed of light. This on the other hand tightly constraints many modified gravity theories for dark energy. In the following we shall discuss shortly the corresponding implications for different classes of modified gravity theories.

\subsection{Scalar-Tensor theories}
As we have discussed in a previous section, one can construct successfully the most general scalar-tensor theories with second order equations of motion, the Horndeski Lagrangians. The presence of derivative non-minimal couplings to gravity modifies among other things the propagation speed of GWs. These interactions come in at the forth and fifth order of Horndeski Lagrangians
\begin{eqnarray}
\mathcal{L}_4&=&G_4(\pi,X)R+G_{4,X}\left([\Pi]^2-[\Pi^2]\right)\nonumber\\
\mathcal{L}_5&=&G_5(\pi,X)G_{\mu\nu}\Pi^{\mu\nu}-\frac16G_{5,X}\left([\Pi]^3-3[\Pi][\Pi^2]+2[\Pi^3]\right)\,.
\end{eqnarray}
If one considers small perturbations on top of a homogeneous background $\bar{\pi}(t)$ and $\bar{g}_{\mu\nu}=(-N(t)^2,a(t)^2\delta_{ij})$, then the tensor perturbations on top of this background have the following generic form
\begin{equation}
\mathcal{S}^{(2)}_T=\sum_{\lambda}\int d^4x a^3 q_T(\dot{h}_\lambda^2-\frac{c_T^2}{a^2}(\partial h_\lambda)^2)
\end{equation}
where we note two relevant contributions in terms of modified gravitational coupling and propagation speed taking the explicit expressions
\begin{eqnarray}
q_T&=&\frac14\left( 2(G_4-2XG_{4,X})-2X(G_{5,X}\dot{\pi}H-G_{5,\pi}) \right) \nonumber\\
c_T^2&=&\frac{2G_4-2XG_{5,\pi}-2XG_{5,X}\ddot\pi}{4q_T}\,.
\end{eqnarray}
As we can see, the sound speed of tensor perturbations depends in a non-trivial way on the background functions $G_4$, $G_5$ and the background dynamics. The LIGO/VIRGO constraint on the speed of gravitational waves is restricted to frequencies $10-100$Hz, which corresponds to a scale that is at the edge of the strong coupling scale of the Horndeski theories, where the regime of validity of the effective field theory breaks down and new physics enter. As was pointed out in \cite{deRham:2018red}\footnote{This was also intensively discussed at the ITS-ETH workshop "Gravitational waves in modified gravity" on the 28-30.05.2018 and it is good that there is some awareness of this in the community.}, the new operators arising at the cutoff scale can significantly affect the speed of propagation of gravitational waves and make it even unity at those high energy scales. Note, that there is no need of finely-tuned cancellations and the new physics entering at that scale does not know anything about the very particular specific choice of the Lorentz-breaking background field configuration.

If we decide to be too conservative and ignore the presence of the higher order operators, that would enter naturally from the effective field theory point of view, we can then put very restrictive bounds on the modified gravity theories \cite{Lombriser:2016yzn,Ezquiaga:2017ekz,Creminelli:2017sry,Sakstein:2017xjx,Baker:2017hug,Langlois:2017dyl,Heisenberg:2017qka,Akrami:2018yjz}. Even if it is too early to say anything decisive with the LIGO measurement, future missions like LISA will tightly constrain non-luminal propagation in alternative theories on scales within the regime of validity of the constructed effective field theories. For the Horndeski theories, luminal propagation would require 
\begin{equation}
G_{4,X}=0 \qquad \text{and} \qquad G_5=0\,,
\end{equation}
if we do not allow any fine-tuning between these two functions and their specific dependence on $X$ and $\pi$.
One could be also attempted to fine-tune the functions $G_4$ and $G_5$ against each other in order to erase their contributions in the speed of GWs. However, such a fine-tuning would be valid only for a particular background evolution with a perfect homogeneous and isotropic symmetries. Any small deviation in the perturbations would break this fine-tuning. In order to guarantee a promising cancelation for arbitrary backgrounds, the contributions of these functions to the effective metric that determines the causal structure of the GWs should have the same tensor structure at the covariant level. For an arbitrary background the effective metric has the covariant form $\mathcal{G}_{\mu\nu}=Cg_{\mu\nu}+D\Pi_\mu\Pi_\nu+E\Pi_{\mu\nu}$ \cite{Bettoni:2016mij}. Whereas the function $G_4$ contributes to the $D$ term, the function $G_5$ contributes to the $E$ term and hence their cancelation for an arbitrary background is restrictively not possible \cite{Ezquiaga:2017ekz}. In other words, the luminal Horndeski interactions could be summarised in the simple Lagrangian as
\begin{equation}
\mathcal{L}_{\pi}^{\rm luminal}=G_2(\pi,X)+G_3(\pi,X)\Box\pi+G_4(\pi)R\,.
\end{equation}
As we mentioned in section \ref{subsec_GalQuantum}, a subclass of these theories with shift and Galileon symmetry on flat space-time has the nice property of technical naturalness. Even if this symmetry is slightly broken on curved space-time, their covariant version still allows a significant suppression of the quantum corrections within the regime of validity \cite{Pirtskhalava:2015nla}. Imposing the radiative stability would further narrow down the allowed Lagrangian to 
\begin{equation}
\mathcal{L}_{\pi}^{\rm luminal, natural}=G_2(X)+G_3(X)\Box\pi+\mpl^2R. 
\end{equation}
So far we only considered the second order Horndeski Lagrangians. Another possibility arises when one allows for higher order equations of motions, as it is the case in beyond Horndeski theories. Still within the restriction $G_5=0$, the following GLPV Lagrangian can be considered
\begin{equation}
\mathcal{L}^{\rm BH}_4=F_4(\pi, X)\left( \Pi_\mu \Pi^{\mu\nu}\Pi_{\nu\rho}\Pi^\rho-\Pi_\mu \Pi^{\mu\nu}\Pi_\nu [\Pi]-X([\Pi]^2-[\Pi^2]) \right)\,,
\end{equation}
which yields an additional contribution to the GW speed on top of the $G_4$ contribution
\begin{equation}
c_T^2=\frac{G_4}{G_4-2X(G_{4,X}-XF_4)}\,.
\end{equation}
From this expression it becomes clear that the contribution to the anomalous GW speed coming from $\mathcal{L}_4$ can be tuned away by the beyond Horndeski interaction in $\mathcal{L}^{\rm BH}$ by imposing the relation $F_4=G_{4,X}/X$. It is worth to emphasise that such a tuning among the functions $F_4$ and $G_4$ does not share the same problem as a possible cancelation among the $G_4$ and $G_5$ functions, since they contribute with the same tensor structure to the effective metric on top of which the gravitational waves propagate \cite{Ezquiaga:2017ekz}. 

In the language of the effective field theory of dark energy, luminal propagation will require quadratic terms with $a_1=0$ in $\mathcal{L}\supset (f-a_1X)K_{ij}K^{ij}$ and no cubic terms $K_ijK^{jl}K_l^i$, etc. Thus, among the DHOST theories, only those of the first category $\mathcal{C}_1$ with $a_1=0$ provide luminal propagation \cite{Langlois:2017dyl}. The luminal Lagrangian is given by
\begin{eqnarray}
\mathcal{L}_{\pi}^{\rm luminal}=f_2(\pi,X)R+P(\pi,X)+G_3(\pi,X)\Box\pi+a_3(\pi,X)\Box\pi \partial_\mu\pi \Pi^{\mu\nu}\partial_\nu\pi\nonumber\\
+a_4(\pi,X) \partial_\mu\pi \Pi^{\mu\nu}\Pi^{\mu\rho}\partial_\rho\pi+a_5(\pi,X)(\partial_\mu\pi \Pi^{\mu\nu}\partial_\nu\pi)^2\,.
\end{eqnarray}
As one can see, after imposing the luminal propagation on the DHOST theories, the remaining Lagrangian has four free functions, as many as Horndeski had before the condition $\vert c_{T}/c-1\vert\lesssim10^{-15}$.

Within these luminal theories it would be interesting to study their implications in the scalar perturbations for the large-scale structure formation \cite{Amendola:2017orw}. On top of the homogeneous and isotropic background we can consider small scalar perturbations $ds^2=-(1+2\Psi)dt^2+a^2(1-2\Phi)dx_i^2$ in Newtonian gauge. The interesting quantities are the effective gravitational coupling $G_{\rm eff}$ and the slip parameter $\eta=\Phi/\Psi$ in the small-scale limit within the quasi-static approximation. In this limit, the modified Poisson equation simplifies to $k^2\Psi\approx -4\pi a^2 G_{\rm eff} \rho_m \delta$ with the comoving non-relativistic matter density contrast $\delta$. With the remaining functions $G_2(\pi,X)$, $G_3(\pi,X)$ and $G_4(\pi)$ the resulting effective gravitational constant is always equal or larger than the Newton's constant $G_{\rm eff}\geq G_{\rm N}$. In other words, weak gravity can not be realised and hence matter cannot cluster slower than in the standard $\Lambda$CDM model. The gravitational slip parameter can be larger or smaller than unity but without scaling within the sound horizon or it is very close to unity above the Compton wavelength. If one enriches the allowed interactions by the $F_4$ term in the beyond Horndeski Lagrangian, then it is possible to have weak gravity with $G_{\rm eff}<G_{\rm N}$ and $\eta\ne1$.

\subsection{Vector-Tensor theories}
Similar implications of future missions like LISA apply also to a general class of vector-tensor theories. Within the generalised Proca interactions the exact type of scalar interactions are incorporated for the longitudinal mode of the vector field. In that limit $A_\mu\to \partial_\mu\pi$ the interactions reduce to the Horndeski scalar-tensor theories and hence one encounters exactly the same solutions with the same implications of the GW speed on them. Other independent solutions of the scalar counterpart arise if one considers field configurations with only the temporal component of the vector field being present $A_\mu=(A_0(t),0,0,0)$. In the same way as in the scalar-tensor theories the non-minimal derivative couplings in the forth and fifth Lagrangians
\begin{eqnarray}
\mathcal L_4 & = & G_{4}(X)R+G_{4,X} \left[(\nabla_\mu A^\mu)^2-\nabla_\rho A_\sigma \nabla^\sigma A^\rho\right] \nonumber\\
\mathcal L_5 & = & G_5(X)G_{\mu\nu}\nabla^\mu A^\nu-\frac{1}{6}G_{5,X} \Big[
(\nabla\cdot A)^3 \nonumber\\
&&+2\nabla_\rho A_\sigma \nabla^\gamma A^\rho \nabla^\sigma A_\gamma -3(\nabla\cdot A)\nabla_\rho A_\sigma \nabla^\sigma A^\rho \Big] 
\end{eqnarray}
contribute to the anomalous propagation speed of GWs. Note that the $g_5(X)$ term in the fifth order Lagrangian and the $G_6$ terms in the sixth order Lagrangian \eqref{vecGalcurv} are not sensitive to the background due to involved symmetries of the background and the interactions themselves. Therefore $g_5$ and $G_6$ do not modify the propagation speed. This crucial property will be relevant for us when we discuss interesting dark energy survivals in multi-Proca theories in the following subsection. The gravitational coupling and the propagation speed take this time the form \cite{DeFelice:2016uil}
\begin{eqnarray}
q_T&=&2G_4-2A_0^2G_{4,X}+HA_0^3G_{5,X}\nonumber\\
c_T^2&=&\frac{2G_4+A_0^2\dot{A}_0G_{5,X}}{q_T}\,.
\end{eqnarray}
The attempt of a cancelation between $G_4$ and $G_5$ fails in the same way as in the scalar case, since these functions contribute to the different tensor structure of the effective metric $\mathcal{G}_{\mu\nu}=Cg_{\mu\nu}+DA_\mu A_\nu+E\nabla_\mu A_{\nu}$ and hence for an arbitrary background a fine-tuning will not be possible. One is forced to demand $G_5=0$ and $G_{4,X}=0$, so that the luminal Lagrangian is of the following form
\begin{eqnarray}
\mathcal{L}_{A_\mu}^{\rm luminal}&=&G_2(X,Y,F)+G_3(X)\nabla_\mu A^\mu+\frac{\mpl^2}{2}R+g_5(X) \tilde{F}^{\alpha\mu}\tilde{F}^\beta_{\;\;\mu}\nabla_\alpha A_\beta \nonumber\\
& +& G_6(X)\mathcal{L}^{\mu\nu\alpha\beta}\nabla_\mu A_\nu \nabla_\alpha A_\beta 
+\frac{G_{6,X}}{2} \tilde{F}^{\alpha\beta}\tilde{F}^{\mu\nu}\nabla_\alpha A_\mu \nabla_\beta A_\nu \,.
\end{eqnarray}
In analogy with the scalar-tensor theories, another possibility arises from the presence of a specific beyond generalized Proca interaction, namely
\begin{equation}
\mathcal{L}^N_4=f_4 (X)\hat{\delta}_{\alpha_1 \alpha_2 \alpha_3 \gamma_4}^{\beta_1 \beta_2\beta_3\gamma_4}
A^{\alpha_1}A_{\beta_1}
\nabla^{\alpha_2}A_{\beta_2} 
\nabla^{\alpha_3}A_{\beta_3}
\end{equation}
where $G_5=0$ and the fifth order beyond generalized Proca interaction as well $f_5=0$ (for this background the beyond genealized Proca interactions $\tilde{f}_5=0$ and $f_6$ do not contribute). In the presence of this beyond generalized Proca interaction the propagation speed modifies into \cite{Heisenberg:2016eld}
\begin{equation}
c_T^2=\frac{2G_4}{2G_4-2A_0^2(G_{4,X}+f_4A_0^2)}\,.
\end{equation}
Hence, the tuning $f_4=G_{4,X}/A_0^2$ would achieve the cancellation of any contribution to the anomalous speed of GWs. This is again possible since $G_4$ and $f_4$ contribute to the effective metric of GWs with the same tensorial structure.

Concerning the implications of these restrictions for the phenomenology of large-scale structure formation, the same conclusions of the scalar case for the slip parameter and the effective gravitational coupling can be drawn also for the vector-tensor theories since one recovers the results of the scalar-tensor theories for the longitudinal part of the vector field $A_\mu=\partial_\mu\pi$. For this field configuration of the longitudinal mode of the vector field the limited Lagrangians satisfying $c_T^2=1$ will again allow only $G_{\rm eff}\geq G_{\rm N}$ and $\eta\ne1$. As regards the temporal field configuration $A_\mu=(A_0(t),0,0,0)$, there is no modification in the gravitational slip parameter at all, hence $\eta=1$ and one cannot realise weak gravity either, since $G_{\rm eff}\geq G_{\rm N}$ \cite{Amendola:2017orw}. For the beyond generalized Proca theories there are possibilities to achieve $\eta\ne1$ and $G_{\rm eff}< G_{\rm N}$ due to the presence of the additional function $f_4$.

\subsection{Tensor-Tensor theories}
In respect of the implications for the propagation speed of GWs massive gravity is very special since it does not give rise to any anomalous speed since it does not modify the derivative part of the Lagrangian but only contributes in form of potential term. If derivative interactions were possible, this would yield an anomalous propagation speed but there is a clear No-go theorem for this type of derivative interactions \cite{deRham:2015rxa}. Massive gravity will only introduce a mass term in the tensor perturbations
\begin{equation}
\mathcal{S}^{(2)}_T=\sum_{\lambda}\int d^4x a^3 \mpl^2\left\{\dot{h}_\lambda^2-\left(\frac{k^2}{a^2}+m_T^2 \right) h_\lambda^2\right\}\,.
\end{equation}
As one can see from the above expression, the gravitational coupling and the propagation speed are exactly the same as in General Relativity. For this reason massive gravity automatically satisfies the constraints on the anomalous propagation speed $c_T^2=1$. It naturally introduces an effective mass term $m_T^2 h_\lambda^2$, which incorporates the mass parameter $m^2$ and the background dynamics together with the dependence on the remaining parameters. The bounds on the mass of the graviton from GWs observations are rather moderate $m_T\leq7.7\times 10^{-23}$ eV compared to the tight cosmological constraints $m_T\lesssim 10^{-33}$ eV. The same conclusion can be made for the massive gravity scenario with doubly coupled matter fields. If there is a matter field that couples to the unique specific effective metric $g_{\rm eff}$, then the second order Lagrangian of the tensor perturbations maintain the property of $c_T=1$ and $q_T=\mpl^2$. However, one has to be careful with the propagation speed of the scalar sector. The matter field that couples to $g_{\rm eff}$ will obtain non-trivial contributions in the expression of the propagation speed $c_s^2$. This means that the parameters $\alpha$ and $\beta$ in $g_{\rm eff}$ will be exposed to tight constraints from Cherenkov radiation of order $10^{-15}$ at high frequencies. See also \cite{Akrami:2018yjz} for a recent work on this.

\subsection{Astrophysical applications}
The implications of modified gravity theories for astrophysical objects are tremendous and rich in phenomenology. We will not discuss this here in detail but just give a quick overview of the main features of the different large classes of alternative theories. For an exhaustive review on this topic we refer to \cite{Barack:2018yly}.

In standard General Relativity in the presence of a canonical scalar field and a Maxwell field, both minimally coupled to the gravity sector, there is a No-hair theorem for black hole solutions. This states the fact that the asymptotically flat and stationary black hole solutions are described by only three parameters, namely the mass, the charge and the angular momentum \cite{Chase1970,Bekenstein:1972ny}. This No-hair theorem can be avoided by going to alternative theories with non-minimal coupling to gravity as in the Hornsdeski theories. Restricting these interactions to the shift symmetric case allows to classify the conditions under which the No-hair properties of black holes remain. In \cite{Hui:2012qt}, this was done using the properties of the conserved current associated with the shift symmetry. Hairy black hole solutions can be constructed within this class of theories, if some of the assumptions are set aside, for instance solutions involving $\pi'\ne0$ \cite{Rinaldi:2012vy,Minamitsuji:2013ura,Sotiriou:2013qea,Babichev:2013cya}.

As we have seen, derivative self-interactions and non-minimal couplings also naturally arise in vector-tensor theories. Apart from their scalar analogon, there are also new purely genuine intrinsic vector interactions with no scalar counterpart. The vector nature of the involved interactions gives rise to astrophysical implications beyond Horndeski scalar. There have been already a multitude of exact and numerical black hole solutions with vector hair discussed in the literature \cite{Chagoya:2016aar,Minamitsuji:2016ydr,Heisenberg:2017xda,Heisenberg:2017hwb,Fan:2016jnz,Cisterna:2016nwq,Babichev:2017rti}.
In the presence of the cubic and quartic derivative interactions (including the vector Galileon as a specific case), one obtains black hole solutions with a primary Proca hair. Even though the quintic powerlaw couplings do not
accommodate black hole configurations regular throughout the horizon exterior, the sixth order as well as the purely intrinsic vector mode couplings allow hairy solutions with secondary Proca hair. In contrast to the scalar Horndeski theories, the presence of the temporal vector component on top of the longitudinal mode notably augments the possibility for the existence of hairy black hole solutions without the need of tuning the explicit models \cite{Heisenberg:2017xda,Heisenberg:2017hwb}. 

In the larger class of scalar-vector-tensor theories, the astrophysical phenomenology of both these classes is unified and also new effects due to genuine scalar-vector couplings arise. If one imposes the $U(1)$ gauge invariance, we saw that there are, among other things, two relevant scalar-vector couplings in $\mathcal{L}^3_{\rm SVT}$ and $\mathcal{L}^4_{\rm SVT}$ \eqref{genLagrangianSVT}. In the shift symmetric field space, cubic scalar-vector-tensor interactions in $\mathcal{L}^3_{\rm SVT}$ are essential for constructing black hole solutions with a scalar hair, that has significant effects around the event horizon. Moreover, the inclusion of the quartic order scalar-vector-tensor interactions in $\mathcal{L}^4_{\rm SVT}$ enriches the previous black hole solutions with an additional vector hair \cite{Heisenberg:2018vti}. Hence, the scalar-vector-tensor theories naturally give rise to black hole solutions endowed with scalar and vector hairs and are also stable in the presence of odd-parity perturbations \cite{Heisenberg:2018mgr}.

In the framework of massive (bi-gravity) the properties of the black hole solutions are diverse and depend strongly on the restrictions put on the two metrics \cite{Volkov:2012wp,Volkov:2013roa,Brito:2013xaa,Volkov:2014ooa,Babichev:2015xha}. For not simultaneously diagonal metrics, the theory admits exact Schwarzschild or Schwarzschild-dS solutions with no Yukawa suppression on large scales, which suffer from strong coupling issues. The other possibility is to have simultaneously diagonal metrics in the same coordinate system. However, the constructed solutions contain singularities at the horizon \cite{Deffayet:2011rh}. Furthermore, the study of linear perturbations admits an unstable mode with a Gregory-Laflamme instability \cite{Babichev:2013una,Brito:2013wya}, which seems to be persistent and surfaces also in bi-Kerr geometry. On the other hand, the analysis of perturbations in the non-bidiagonal geometry gives more promising results \cite{Babichev:2014oua}. Interestingly,  time dependent and non-singular numerical solutions were recently proposed in \cite{Rosen:2017dvn} and it would be worth investigating their stabilities further.

\subsection{Luminal dark energy in multi-Proca theories}
In the previous subsections we have seen the implications of the restriction of luminal propagation of GWs in Horndeski, generalized Proca and massive gravity theories. Dark energy models based on Horndeski type of Lagrangian with non-minimal derivative coupling are severely constrained by the requirement $\vert c_{T}/c-1\vert\lesssim10^{-15}$, even though these measurements so far were made on a scale, which is at the edge of the regime of validity of these effective field theories. If LISA measures luminal propagation of GWs also on significantly lower energy scales, some theories with non-minimal couplings will be strongly restricted. In \cite{BeltranJimenez:2018ymu} a class of dark energy models within multi-Proca theories were discussed, which incorporate non-minimal derivative couplings but nonetheless do not contribute to the GWs sector. Nevertheless, due to the presence of a second tensor sector they can give rise to new interesting features that can be probed by the GWs observations while giving rise to $c_T=1$ for the GWs sector. Specially, these models can accommodate interesting oscillation effects of GWs into additional tensor sectors. These dark energy models can be successfully realised in the frameworks of multi-Proca and Yang-Mills theories. Within these frameworks different suitable field configurations can be considered in order to obtain a homogeneous and isotropic background. In the following we shall go through these field configurations and highlight their common and different features.

\subsubsection{Pure temporal configuration}
The simplest field configuration is the pure temporal configuration where the only non-vanishing components are the temporal ones $A^a_\mu=\phi^a(t)\delta^0_\mu$. This field configuration can be realised in the framework of multi-Proca theories, which is nothing else but several copy of the single Proca field. For this field configuration the sixth order Proca Lagrangian with non-minimal derivative coupling behaves quite special
\begin{equation}
\mathcal{L}_6\supset G_6(X)L^{\mu\nu\alpha\beta} F^a_{\mu\nu}F_{a\alpha\beta}
+\frac{G'_{6}(X)}{2} \tilde{F}_a^{\alpha\beta}\tilde{F}_a^{\mu\nu}S^b_{\alpha\mu} S_{b\beta\nu}.
\label{L6_multiProca}
\end{equation}
As we have mentioned in section \ref{sec_VectorTensorTheories}, this sixth order Lagrangian is attributed only to the vector-tensor theories since there is no correspondence in the scalar Horndeski theories (note that the fifth order Proca interaction $\tilde{F}_{\mu\alpha}\tilde{F}^\alpha_\nu S^{\mu\nu}$ which also does not have its scalar correspondence, does not have the extension into multi-Proca interactions if one imposes an internal global $SO(3)$ symmetry as we saw in \ref{sec_multiProca}). 
For this field configuration the background field strength vanishes identically $\bar{F}_{\mu\nu}=0$, hence the contribution to the perturbations is only via
\be
\delta^{(2)}\mathcal{L}_6\supset\Big[G_6(Y)R^{\mu\nu\alpha\beta} 
+\frac{G'_{6}(Y)}{2}S^{b\alpha\mu} S_b^{\beta\nu}\Big]_0\delta\tilde{F}_{a\mu\nu}\delta\tilde{F}^a_{\alpha\beta}\,.
\label{L6_multiProcapert}
\ee
From this expression we immediately observe that it does not yield any effect on the GWs sector but contributes only through the vector field perturbations $\delta\tilde{F}_{\mu\nu}$ with non-trivial effects. It is also important to note that its contribution highly depends on the presence of the cubic interaction. If $G_3$ is absent, then we will not see any effects in the other sector of perturbations either \cite{BeltranJimenez:2018ymu}.

\subsubsection{Triad configuration}
For the existence of this field configuration the presence of an internal $SO(3)$ symmetry of an ensemble of fields is crucial. In this scenario the spatial components obey $A^a{}_\mu=A(t)\delta^a_i$. Note, that this field configuration is completely compatible with the symmetries of the FLRW background. It is true that this field configuration breaks both the spatial rotations and the internal symmetry, but it does it in a way that leaves a linear combination of the two unbroken. The symmetry breaking pattern follows $SO(3)\times ISO(3,1) \to ISO(3)_{\rm diagonal}$. The variation of the set of the vector fields under the spatial and internal rotations yields
\begin{eqnarray}
\delta A^a{}_i&=&w_i{}^j A^a{}_j+J^a_b A^b A^b{}_i \nonumber\\
&=& (w_i{}^j \delta^a_j+J^a_b\delta^b_i )A(t)\,,
\end{eqnarray}
where the the combination inside the brackets $(w_i{}^j \delta^a_j+J^a_b\delta^b_i )$ is a diagonal unbroken $SO(3)$ and this is the reason why this field configuration is compatible with a FLRW background. In difference to the pure temporal configuration we can not consider the sixth Lagrangian (\ref{L6_multiProca}) with the non-minimal coupling to the double dual Riemann tensor for this field configuration, since it would modify the GWs propagation speed. Thus, one has to enforce $\mathcal{L}_6=0$ for this field configuration. Even if the non-minimal couplings are absent, there is still non-trivial phenomenology of the GWs due to the presence of a second tensor mode associated to the multi-Proca fields. On top of the standard tensor modes of the metric perturbations $h_{ij}=\delta^{T}g_{ij}/a^2$, there are the tensor modes of the vector fields $t_{ij}=\delta^a_i \delta^{T}A^a_i$. These second tensor modes give rise to interesting feature since they can mix with the GWs sector in a non-trivial way yielding an oscillation of GWs into them. This is a very distinctive feature of these theories as dark energy models, since they can be interestingly probed by GWs astronomy without being in conflict with the tight constraints on the GWs speed. As we mentioned above, the constraints of the GWs speed does not allow the presence of non-minimal couplings for this field configuration. The allowed Lagrangian in this case would be of the form $\mathcal{L}=\mathcal{L}_{\rm EH}+\mathcal{K}(X,Z,\cdots)$. The dots denote the $SO(3)$ and Lorentz invariant contractions of $A_\mu$'s and $F_{\mu\nu}$'s, which do not contribute any new additional terms to the FLRW background and can be disregarded. 

\subsubsection{Temporally extended triad configuration}
The pure temporal configuration respected the spatial rotational invariance whereas the triad configuration a linear combination of internal and spatial rotations. There is another configuration that combines the two $A^a_\mu=\phi^a\delta^0_\mu+A(t)\delta^a_\mu$, which was first proposed in \cite{Jimenez:2016upj} in the framework of generalized multi-Proca theories. This combination does not respect any rotational symmetry anymore and one might worry that the underlying symmetries for a FLRW solution is lost. Nevertheless, if one restricts the interactions such that $A_\mu^a$'s and $F^a_{\mu\nu}$'s are never contracted with each other, in other words $Y=0$, then this field configuration for the remaining interactions gives rise to contributions to the energy momentum tensor that are isotropic on-shell. This constitutes a dramatically new mechanism to obtain homogeneous and isotropic cosmological solutions that had not been considered in the literature before \cite{Jimenez:2016upj}. In order to illustrate the underlying mechanism behind this field configuration let us consider the Lagrangian $\mathcal{L}=\mathcal{K}(X,Z)$ (note that $Y$ can not be present). Due to antisymmetric nature of the field strength $\phi^a$ does not contribute to $Z$ and hence the $Z$ sector will continue being isotropic. The $Y$ sector needs a little bit more care. The variation of the action is of the form 
\begin{equation}
\delta\mathcal{S}=\int d^4x\sqrt{-g}\mathcal{K}_X\delta X+ \text{isotropic terms}
\end{equation}
Due to the presence of $\phi^a$, the energy-momentum tensor receives an anisotropic contribution $T_{0i}\propto \mathcal{K}_XA(t)\phi^a(t) \delta^a{}_i$. The equations of motion for the $\phi^a$ components impose $\mathcal{K}_X\phi^a=0$. Therefore, satisfying the equations of motion of the temporal components makes the energy-momentum tensor isotropic on-shell. Since the branch $\phi^a=0$ corresponds to the triad configuration of the previous section, the genuine extended triad configuration only arises for $\mathcal{K}_X=0$. For the GWs phenomenology one does not gain any new effects for the tensor perturbations in this field configuration, since the $Z$ sector that contributes to the tensor modes share the same properties as in the triad configuration. 
On the other hand, the off-shell violation of isotropy in the extended triad configuration yields preferred direction effects in the vector and scalar perturbations, that makes them not decouple from each other. See \cite{BeltranJimenez:2018ymu} for more detail on this.

\section{Concluding remarks} \label{Sec:Conclusions}

This review was intended to give a comprehensive overview of modified gravity theories in the framework of effective field theories. Imposing Lorentz symmetry we have systematically constructed gravity theories based on different protagonists of particles. The Standard Model of Particle Physics makes an extensive use of massless and massive particles of spin-$0$, $1/2$, 1 and cosmology strongly relies on the consistent interactions of spin-2 particles. Since observations indicate that the gravitational force is a long-range attractive force and couples to electromagnetism, this uniquely filters out a massless (or very light) spin-2 particle as an exchange particle for the gravitational force.

We started our journey by recalling the properties of the Lorentz group and discussed its fundamental representations since invariance under rotations and boosts in space plays a crucial role in particle physics and cosmology as well as for the underlying laws of physics. We have described these rotation and boost operations by the generators of the Lorentz group together with their commutation relations. As a one dimensional representation we have introduced the Lorentz scalar, that remains unchanged under Lorentz transformations. Non-trivial representations came into operation after establishing the tensor and vector representations. Particles in Nature are classified by their spin and mass and carried by the associated fields. 

We have started with the simplest spin-0 particle and asked ourselves how we could construct consistent interactions for a Lorentz scalar field with second order equations of motion. We discussed the free scalar field with a kinetic and a mass term together with its Hamiltonian density and stress energy tensor. The sign of the kinetic term was crucial in order to have a Hamiltonian bounded from below. After writing down the quadratic action of perturbations on top of a general background, we obtained the corresponding propagator of the spin-0 particle and saw that the scalar exchange amplitude would give rise to an attractive force between two conserved sources. 

As possible extensions of the free theory, we discussed the K-essence scalar field in form of a general function of the kinetic term and the field itself $P(\pi,X=-\frac12\partial_\mu\pi\partial_\mu\pi)$. This immediately guided us to the question whether one could consider derivative self-interactions with two derivatives per field and still give rise to second order equations of motion. The answer to this question naturally led to the construction of the Galileon theories, that generalises the decoupling limit interactions of the DGP model. The resulting interactions were invariant under Galileon and shift transformations of the scalar field. In terms of the Levi-Civita tensors they correspond to the elementary polynomials of the symmetric fundamental matrix $\Pi_{\mu\nu}=\partial_\mu\partial_\nu\pi$. In this context, we have seen that the Hamiltonian of the Galileon becomes unbounded from below if the kinetic term is negative. The antisymmetric structure of the interactions in terms of the Levi-Civita tensors helped us to understand why the Galileon interactions do not obtain quantum corrections and therefore are technically natural. In other words, they are protected from quantum corrections not only by the underlying symmetries but also by the non-renormalisation theorem as a result of the antisymmetric nature of the Levi-Civita tensors. We also obtained important information on the low energy effective field theory based on the unitarity and analyticity requirements of the scattering amplitudes for a Lorentz invariant UV completion of the massive Galileon.

We then moved on to the case of massless spin-2 particle. We convinced ourselves why gravity should be described by a massless spin-2 field (rather than a spin-0 or spin-1 particle). We started with the consistent linear theory for a massless spin-2 particle. Writing the most general Lorentz invariant Lagrangian at quadratic order in the spin-2 field $h_{\mu\nu}$, we imposed restrictive conditions on the parameters in order to make the $00-$ and $0i-$ components of the spin-2 field non-dynamical. This resulted in a one parameter family of Lagrangians that had either the full diffeomorphism invariance yielding the Fierz-Pauli or a Weyl transverse diffeomorphism invariant theory. We saw that the difference between these two linear theories was just an integration constant. 

With this linear Fierz-Pauli theory at hand, we discussed the natural matter coupling and quickly encountered an inconsistency of the resulting equations of motion. Since the stress energy tensor is only conserved on-shell, the new additional matter coupling immediately spoiled this property. The attempt to fix this by including the next-to leading operator resulted in an infinite series of interactions. Imposing the full diffeomorphism invariance and the consistency of the equations of motion yielded a Taylor expansion of the matter action where the initial Minkowski metric had to be replaced by $h_{\mu\nu}$ with a corresponding prefactor and scaling. However, we very soon realised that we are still missing the stress energy tensor associated to the self-interactions of the spin-2 field itself. If one takes the linear theory in the weak field limit and compares it with observations, the predictions are off by a factor of $1/4$. In other words, in order to correctly account for the observations we had to find the non-linear completion of the theory. 

Starting from the linear theory as the zeroth order Lagrangian, we computed the self energy in terms of the stress energy tensor and added an additional coupling to itself. This again gives rise to an infinite series in the same manner as the matter coupling due to the inconsistency of the equations of motion. One has to resum an infinite series of self-interactions in order to realise the full diffeomorphism invariance at the non-linear level. We saw that using Deser's approach of the bootstrapping procedure allowed to perform the resummation in one single step using the Palatini language. The resulting theory is General Relativity up to a boundary term, that can be added by hand to recover the Einstein-Hilbert action. The underlying property of this Lagrangian is that it is second order in derivatives at the level of the equations of motion and propagates only two physical degrees of freedom. A crucial property of this action is that the associated Hamiltonian is linear in the lapse and the shifts, guaranteeing the presence of first class constraints that remove the non-physical degrees of freedom. The fact that the lapse and the shifts are Lagrange multipliers is a virtue of the full diffeomorphism symmetry. 

In fact, imposing the full diffeomorphism symmetry and second order equations of motion naturally filters out the Lovelock invariants as the unique consistent covariant Lagrangians. In four dimensions this corresponds exactly to the Einstein-Hilbert action with a cosmological constant. We saw that in higher dimensions one can construct further interesting Lovelock invariants like the Gauss-Bonnet term and beyond. Associated to the Lovelock invariant terms we constructed divergenceless tensors. In four dimensions these were the metric, the Einstein tensor and the double dual Riemann tensor. We also discussed how the well-posedness of the variational principle is jeopardised by the Einstein-Hilbert action since it contains second order derivatives for the metric. This made the introduction of a boundary term (the Gibbons-Hawking-York boundary term) unavoidable and introduced a disturbing ambiguity.

We continued our journey by abandoning the masslessness condition of the spin-2 particle. Adding a mass term explicitly breaks the diffeomorphism invariance and therefore the constraints with respect to the lapse and shift cease to be first class. Consequently one would naively have six propagating degrees of freedom, among which one is predestined to be a ghost degree of freedom since the massive spin-2 representation of the Lorentz group carries only five propagating degrees of freedom. At the linear order we have seen that the Fierz-Pauli tuning makes the mass of this sixth degree of freedom to become infinite and returns a unique linear theory. Comparing this linear theory with the observations in the massless limit reveals the vDVZ discontinuity with the consequence that General Relativity can not be recovered. This is due to the fact that non-linear interactions of the helicity-0 mode become very relevant in that limit. Therefore, the theory has to be non-linearly completed. 

Usually, the inclusion of non-linear interactions brings back the sixth ghostly degree of freedom and we saw that the potential interactions have to have a very specific square root structure in order to avoid the ghost. They correspond to the symmetric elementary polynomials of the fundamental square root matrix $\mathcal{K}_{\mu\nu}$. The construction of such terms relies on the existence of a second metric $f_{\mu\nu}$, that plays the role of a fiducial metric that can be used to contract indices. We also saw that the key properties of the theory become manifest already in the decoupling limit, where the helicity modes of the massive graviton decouple from each other and the helicity-0 mode resembles a scalar Galileon. The interactions are such that the pure helicity-0 sector comes in form of total derivatives and the leading order operators mix the helicity-0 and the helicity-2 modes. In this limit the helicity-1 mode usually decouples from the matter sources and does not yield extra fifth forces in contrast to the helicity-0 mode. The fifth force arising from the helicity-0 mode gets frozen on small scales via the Vainshtein mechanism, which guarantees a successful recovery of General Relativity on Solar System scales. 

Within the decoupling limit the leading order interactions are protected from quantum corrections by a similar non-renormalisation theorem as in the Galileon case. Beyond the decoupling limit, the contributions coming from graviton loops unfortunately detune the specific structure of the potential interactions. However, this detuning remains irrelevant below the Planck scale even for large background values after properly taking into account the Vainshtein mechanism. Furthermore, the graviton mass receives small quantum corrections, and hence, is technically natural. Another important question we discussed concerned the consistent matter couplings in the framework of massive gravity. We showed that there are two viable options (both at the classical and quantum level): the matter fields either couple to $g_{\mu\nu}$ or to $f_{\mu\nu}$ but never to both simultaneously or they couple to a composite effective metric $g_{\mu\nu}^{\rm eff}$. We then commented on the consequences of the requirement to have Lorentz invariant UV completion for massive gravity in the same way as for the Galileon interactions. The positivity bounds on the tree level scattering amplitudes drastically reduce the allowed parameter space of the low energy effective field theory.

After the successful construction of field theories for a spin-0 and a spin-2 field, we aimed at unifying both in form of scalar-tensor theories, in other words, we wanted to promote the scalar theories to the curved space-time. Concerning the k-essence type of interaction, this step is trivial in the sense that adding a volume element in form of a minimal coupling does not alter the physical degrees of freedom. However, if we want to promote Galileon-type of interactions to curved space-time, more caution is needed. If we naively covariantise the quartic and quintic Galileon interactions, we obtain higher order equations of motion. To counterbalance this we had to introduce non-minimal couplings between the spin-0 and the spin-2 field. The cubic Galileon is an exception since it is linear in the connection and therefore no counterterm in form of a non-minimal coupling is needed. This gave rise to the rediscovery of the Horndeski interactions. They constitute the most general scalar-tensor theories with second order equations of motion. We also saw that starting from the decoupling limit of massive gravity and covariantising the interactions after integration by parts, one obtains a specific subclass of Horndeski theories. In this way we could relate massive gravity to Horndeski interactions in the same way as we recovered in the decoupling limit the Galileon interactions for the helicity-0 mode of the massive graviton. Next, we addressed the question as to whether the requirement of second-order equations of motion could be abandoned. Horndeski interactions strongly rely on the assumption of having second order in derivatives at the level of the equations of motion. Renouncing this restriction allowed us to construct beyond Horndeski interactions (and DHOST theories) with higher order equations of motion but still avoiding the Ostrogradski instability.

Having overviewed the general landscape of scalar-tensor theories, we moved on to the different territories where the protagonist now was a vector field. We focused our attention on the spin-1 representation of the Lorentz group and asked ourselves the question how we could construct consistent field theories for a vector field. Starting with the most general quadratic Lagrangian for the massless spin-1 field, we worked out the conditions on the parameters that give rise to a first class constraint removing two degrees of freedom. This uniquely leads to Maxwell theory with $U(1)$ gauge invariance, where the latter is reflected in the vanishing of the Poisson bracket of the primary and the secondary constraint. We also saw that even if we do not impose directly the $U(1)$ symmetry on the matter coupling, the consistency of the equations of motion inevitably introduces the gauge symmetry into the matter sector as well. We illustrated that on the basis of a complex scalar field. We then continued our journey towards massive vector fields and pointed out the main differences between the massive and the massless spin-1 representations. We saw that adding a mass term drastically changes the fundamental degrees of freedom and the nature of the system of constraints. The mass term in the Proca theory explicitly breaks the gauge symmetry and therefore three degrees of freedom propagate. Now, one has a second class constraint with the Poission bracket of the primary and the secondary constraint being proportional to the mass. The temporal component of the vector field ceases playing the role of a Lagrange multiplier and becomes an auxiliary field, that can be integrated out. Finally, we confirmed that the exchange amplitude of a spin-1 particle between conserved currents gives rise to a repulsive force.

A natural immediate question in the framework of vector field theories is the possibility of constructing vector Galileon-type interactions. The attempt to construct derivative self-interactions for a massless vector field ends at a No-go result and we soon had to abandon this inhospitable place. However, this did not prevent us from conquering the landscape of Proca theories. We started building the allowed interactions by imposing two conditions: second order equations of motion and a non-dynamical temporal component of the vector field. In this way we were able to systematically construct order by order the allowed interactions in terms of the Levi-Civita tensors. In contrast to the scalar Galileon theories, the generalised Proca theories contain two additional purely intrinsic genuine vector interactions, that vanish in the limit of the longitudinal mode. The same interactions also arise by requiring that the determinant of the Hessian matrix vanishes at each order, which guarantees the presence of a primary constraint that removes the temporal component. We showed that the series stops after the sixth order Lagrangian by the virtue of the Cayley-Hamilton theorem. As an additional support we also constructed the allowed interactions from a bottom-up approach, where we first built the allowed interactions in the decoupling limit with the leading order terms and then promoted them to the corresponding interactions beyond that limit. 

Furthermore, we discussed the stability of the classical interactions under quantum corrections. For those interactions, which do have their scalar Galileon limit as the longitudinal mode of the vector field, there exists a similar non-renormalisation theorem. For the genuine intrinsic vector interactions the non-renormalisation theorem is lost but the quantum corrections might still be small. Concerning a Lorentz invariant UV completion, the presence of these purely intrinsic vector interactions seems to be crucial. We then tried to promote these theories based on vector fields to the curved space-time for both the $U(1)$ invariant and broken cases in order to construct consistent vector-tensor theories. In the case of the gauge invariant interactions we realised quickly that there is only a unique non-minimal coupling to the double dual Riemann tensor that gives rise to second order equations of motion. On the other hand, for the generalised Proca interactions when we plunged into the curved space-time, we came across similar findings as in the scalar Horndeski case. Namely, the covariantisation of the generalised Proca interactions in the quartic, quintic and sixth-order Lagrangians requires the introduction of non-minimal couplings in order to maintain second-order equations. Finally, following the spirit of the beyond Horndeski construction for scalar fields, we also established similar interactions beyond generalized Proca with higher order equations of motion but still five propagating physical degrees of freedom.

Having spent some time with these two important classes of modified gravity theories, that either contain an additional  scalar or a vector field, we aimed at unifying them in a single scalar-vector-tensor theory. The unification and the resulting underlying interactions strongly depend on the presence or absence of the gauge symmetry of the vector field. For the interactions with explicit gauge invariance we saw that one can successfully construct three different Lagrangians with non-trivial mixing of the scalar and vector fields. In the case of broken gauge invariance there are six propagating degrees of freedom and we were able to formulate six independent Lagrangians. These scalar-vector-tensor theories will have very rich applications to cosmology and astrophysics.

Still within the framework of vector theories, the case of single vector field interactions can be extended to a set of vector fields. In fact, the Standard Model of Particle Physics relies on the existence of non-abelian gauge fields, as described by the Yang-Mills theories. Unfortunately, for a set of vector fields with a non-abelian gauge symmetry one can not construct derivative self-interactions in four dimensions that would go beyond the standard terms of the field strength and its dual. However, if one is willing to explicitly break the non-abelian gauge symmetry, one can systematically build consistent derivative interactions for a set of massive vector fields, the multi-Proca fields. We have discussed the straightforward extension of the single Proca interactions but also emphasised the main differences coming from the genuine new multi-Proca interactions. 

After finishing the construction of this colourful landscape of gravitational theories, we moved on to the different scenery where all these theories find cosmological applications. We first established the standard perturbation theory in General Relativity and pointed out the role played by the gauge invariance. We convinced ourselves that out of the naively counted six scalar perturbations only one single scalar mode propagates, which is associated to the presence of a matter field. We discussed two prominent choices of gauge fixing, namely the synchronous and the conformal Newtonian gauges. Along an independent path, we also analysed the perturbations and obtained the right physical properties without fixing any gauge. After integrating out the non-dynamical modes by means of algebraic equations, the final differential equations contained the right combination of gauge invariant quantities. We then embraced the rich cosmological applications of massive gravity, scalar-tensor and vector-tensor theories and discussed their fundamental features and defining differences. Most of the applications in the literature concern the dark energy phenomenology, but these theories can also be naturally applied to early universe cosmology and to dark matter phenomenology. 

As we have seen, all these field theories of gravity result in alterations of the Hubble function at the background level. The first step would be to test these modifications against observations. Experimental tests of the expansion dynamics of the universe consist of the distance-redshift relation of supernovae and measurements of the angular diameter distance as a function of redshift (for instance as in the measurements of the Cosmic Microwave Background and the Baryon Acoustic Oscillations). These geometrical tests of the background evolution will not be powerful enough to disentangle between the degeneracies among the different field theories. 
The time dependence of evolving cosmic structures and the influence of the gravitational theory on the geodesics of relativistic and non-relativistic test particles will play a crucial role in distinguishing between field theories. The latter includes the growth of cosmic structures and the formation of galaxies and clusters of galaxies by gravitational collapse whereas the former encompasses the Sachs-Wolfe effects and gravitational lensing. 

The different gravitational theories will have modified time sequence of gravitational clustering, number density of collapsed objects and evolution of peculiar velocities and can be tested against each other. In this context, we discussed quickly a conceptually entirely different and new analytic approach to cosmic structure formation based on kinetic field theory and we believe that adapting this approach to alternative theories of gravity and using it to study non-linear structure formation will be very relevant and complementary to N-body simulations in order to test alternative theories. Furthermore, the tremendous discovery of gravitational waves has opened a new and unique opportunity to test gravitational theories and we already saw its constraining power for most of these alternative theories, that give rise to a different propagation speed of gravitational waves than the light speed. The adventure undertaken throughout this review granted us to gain an overview among existing consistent gravitational theories with fascinating novel effects and distinctive features and we believe that they offer unique opportunities for exploring the fundamental gravitational interactions and addressing pertinent open questions.





\section*{Acknowledgments}
\addcontentsline{toc}{section}{Acknowledgments}

We are grateful to Matthias Bartelmann, Jose Beltran, Luc, Blanchet, Claudia de Rham, Ryotaro Kase, Tomi Koivisto, Shinji Tsujikawa. L.H. is funded by Dr. Max R\"ossler, the Walter Haefner Foundation and the ETH Zurich Foundation. 

\appendix
\section{Instabilities}\label{AppendixInstabilities}
In most of modifications of gravity, additional extra degrees of freedom are present. One might be adding an additional tensor
field, or a vector field or simply an additional scalar field to the gravity sector. When constructing a consistent theory, one
has to enforce the absence of instabilities for these new degrees of freedom. There are ghosts, Laplacian and tachyons instabilities. 

A ghost is a field that carries the wrong sign for its kinetic term. Consider the following action with two fields
\be
{\cal S}=\int d^4x \left( -\frac{1}2 (\partial_\mu\pi)^2+\frac12 (\partial_\mu\pi)^2+\mathcal{L}^{\rm int}(\pi,\pi) \right)
\ee
where we immediately see that the $\pi$ field has the wrong sign for the kinetic term, which makes the Hamiltonian unbounded from below
independently of the interaction term $\mathcal{L}^{\rm int}(\pi,\pi)$
\be
H=\int d^3x \frac12 \left( \Pi_\pi^2- \Pi_\pi^2+\partial_i\pi^2-\partial_i\pi^2 \right)\,,
\ee
 with the conjugate momentum $\Pi_\pi$ and $\Pi_\pi$. The scale associated with the ghost instability is the momentum, which 
 has an arbitrarily fast instability as a consequence already at the classical level. Sometimes a ghostly degree of freedom might be
 hidden behind a higher derivative interactions of another field where the wrong sign of the kinetic term might not be as immediate as
 in the previous example. For instance consider the following Lagrangian
 \be
{\cal S}=\int d^4x \left( -\frac{1}2 (\partial_\mu\pi)^2+\frac{(\Box\pi)^3}{6\Lambda^5} \right)\,,
\ee
with a non-linear interaction of higher order in derivatives terms. Assume the background field configuration $\bar{\pi}=\Lambda^3c_1x_\mu x^\mu /8$ and perturbation on top of that background $\delta\pi$.
The above action at quadratic order in the perturbations takes the form
 \be
{\cal S}=\int d^4x \frac12 \left( \delta\pi \left[\Box+\frac{c_1\Box^2}{\Lambda^2} \right] \delta\pi  \right)\,.
\ee
The associated propagator can be read off and correspond to the expression in the square brackets
 \be
D=\frac{1}{\left(1+\frac{c_1\Box}{\Lambda^2} \right) \Box}=\frac{1}{\Box}-\frac{1}{\Box+\Lambda^2}
\ee
As it can be taken from the above expression of the graviton, there is an additional degree of freedom behind the interaction and it comes with
a wrong sign in the kinetic term.

Another severe instability that one has to avoid in any consistent theory is the Laplacian instability. Let us consider an arbitrary background configuration $\bar{\pi}$ and perturbations $\delta\pi$ on top of this background. The quadratic action will take the general form
 \be
{\cal S}=\int d^4x \frac12 \mathcal{Z} \left( \delta\dot\pi^2 -c_s^2\nabla\delta\pi^2 \right)-\frac12m^2\delta\pi^2
\ee
with the functions $\mathcal{Z}$, $c_s^2$ and $m^2$ depending on the background configuration. As we have seen above, in order to
guarantee the absence of ghost instabilities, we have to impose the right sign for the kinetic term, i.e. we have to require that $\mathcal{Z}>0$.
On the other hand, for the absence of any Laplacian instability one has to impose $c_s^2\ge0$.

Last but not least, one also has to impose conditions for the absence of tachyonic instabilities. This is related to the presence of 
imaginary mass. It means that one has the wrong sign for the mass term 
 \be
{\cal S}=\int d^4x \left( -\frac{1}2 (\partial_\mu\pi)^2+\frac{1}2m^2\pi^2\right).
\ee
In difference to the ghost instability, one has now an instability in the potential with the scale given by $m$. In the presence of a general potential interactions, tachyonic instabilities are present if second derivatives of the potential is negative. For a better illustration of the consequences of a tachyonic instability, let us consider the following action with the specific potential $V$
 \be
{\cal S}=\int d^4x \left( -\frac{1}2 (\partial_\mu\pi)^2-\frac{\Lambda}{4}(\pi^2-v^2)^2\right)\,.
\ee
The corresponding Klein-Gordon equation $\ddot\pi-\nabla^2\pi+\partial{V}/\partial\pi=0$ has critical points at $\pi=0$ and $\pi=\pm v$.
Consider perturbations $\delta\pi$ on top of the critical points as background. For instance, for the background $\bar\pi=0$ with $V''<0$ the perturbations equation in Fourier space $\delta \ddot{\pi}+k^2 \delta \pi-\Lambda v^2 \delta \pi=0$ can be solved easily and one obtains $\delta \pi\sim e^{\sqrt{-k^2t\Lambda v^2}}/\sqrt{2k}$. This means that modes with $k<v\sqrt{\Lambda}$ have an exponential growing. The perturbative analysis breaks down when the perturbations quickly become large. 

Another worrying phenomenon is the possibility of having superluminal propagation in the regime of interest. Special relativity has the important restriction that fluctuations should not propagate faster then the speed of light. If a theory allows for superluminal fluctuations, then it is considered to be a sick theory in some of the literature. This is still an ongoing debate. However, undoubtedly the real concern is about the possibility of creating closed timelike curves. In General Relativity, and also in some modifications of gravity like massive gravity and Galileon theories, the theory allows for the construction of timelike curves. However, once one tries to send information along these curves, the effective field theory breaks down. This is known as Hawking's chronology protection conjecture. 

A crucial point in General Relativity is the Null Energy Condition. It states that the energy momentum tensor satisfies $T_{\mu\nu}n^\mu n^\nu>0$ for 
any null vector obeying $g_{\mu\nu}n^\mu n^\nu>0$. In other words, it means that any observer travelling along a light-ray should measure a non-negative matter density. However, in some modifications of gravity like the scalar-tensor theories, this condition might be violated

\section{Perturbations of the Scalar-Tensor Theories}\label{AppendixHorndeski}
The coefficients of the Einstein field equations in the scalar-tensor theories in section \ref{sec_cosmology_horndeski} are given by
\begin{eqnarray}
 &  & A_{1}=6\Theta,\qquad A_{2}=-2(\Sigma+3H\Theta)/\dot{\pi},\qquad A_{3}=2{\cal G}_T,\qquad A_{4}=2\Sigma+\rho_{m},\nonumber \\
 &  & A_{5}=-2\Theta,\qquad A_{6}=2(\Theta-H{\cal G}_T)/\dot{\pi},\qquad\mu={\cal E}_{,\pi}\,,\label{Ai}
\end{eqnarray}
 
\begin{eqnarray}
 &  & B_{1}=6{\cal G}_T,\qquad B_{2}=6(\Theta-H{\cal G}_T)/\dot{\pi},\qquad B_{3}=6(\dot{{\cal G}_T}+3H{\cal G}_T),\nonumber \\
 &  & B_{4}=3\left[\left(4H\ddot{\pi}-4\dot{H}\dot{\pi}-6H^{2}\dot{\pi}\right){\cal G}_T-2H\dot{\pi}\,\dot{{\cal G}_T}-\left(4\ddot{\pi}-6H\dot{\pi}\right)\Theta+2\dot{\pi}\dot{\Theta}-\rho_{m}\dot{\pi}\right]/\dot{\pi}^{2}, \nonumber \\
 &  & B_{5}=-6\Theta, \qquad B_{6}=2{\cal F}_T,\qquad B_{7}=2\left[\dot{{\cal G}_T}+H\left({\cal G}_T-{\cal F}_T\right)\right]/\dot{\pi},\nonumber \\
 & &B_{8}=2{\cal G}_T,\qquad B_{9}=-6(\dot{\Theta}+3H\Theta),\nonumber \\
 &  & B_{10}=-2{\cal G}_T,\qquad B_{11}=-2(\dot{{\cal G}_T}+H{\cal G}_T),\qquad\nu={\cal P}_{,\pi}\,,\label{Bi}
\end{eqnarray}
 
\begin{eqnarray}
&&C_{1}=2{\cal G}_T,\qquad C_{2}=2(\Theta-H{\cal G}_T)/\dot{\pi},\qquad C_{3}=-2\Theta,\nonumber \\
&&C_{4}=\left[2(H\ddot{\pi}-\dot{H}\dot{\pi}){\cal G}_T-2\ddot{\pi}\,\Theta-\rho_{m}\dot{\pi}\right]/\dot{\pi}^{2},\label{Ci}
\end{eqnarray}
\begin{eqnarray}
 &  & D_{1}=6(\Theta-H{\cal G}_T)/\dot{\pi},\qquad D_{2}=2(3H^{2}{\cal G}_T-6H\Theta-\Sigma)/\dot{\pi}^{2},\nonumber \\
 &  & D_{3}=-3\left[2H(\dot{{\cal G}_T}+3H{\cal G}_T)-2(\dot{\Theta}+3H\Theta)-\rho_{m}\right]/\dot{\pi},\nonumber \\
 &  & D_{4}=2\Big[3H\{(3H^{2}+2\dot{H})\dot{\pi}-2H\ddot{\pi}\}{\cal G}_T+3H^{2}\dot{\pi}\dot{{\cal G}_T}+6\{2H\ddot{\pi}-(3H^{2}+\dot{H})\dot{\pi}\}\Theta\nonumber \\
 & &-6H\dot{\pi}\dot{\Theta}+(2\ddot{\pi}-3H\dot{\pi})\Sigma-\dot{\pi}\dot{\Sigma}\Big]/\dot{\pi}^{3},\nonumber \\
 &  & D_{5}=2(\Sigma+3H\Theta)/\dot{\pi},\qquad D_{6}=-2(\Theta-H{\cal G}_T)/\dot{\pi},\qquad D_{7}=2\left[\dot{{\cal G}_T}+H\left({\cal G}_T-{\cal F}_T\right)\right]/\dot{\pi},\nonumber \\
 &  & D_{8}=3\left[6(\dot{H}\dot{\pi}-H\ddot{\pi})\Theta-2\ddot{\pi}\Sigma+3H\rho_{m}\dot{\pi}-\mu\dot{\pi}^{2}\right]/\dot{\pi}^{2},\nonumber \\
 &  & D_{9}=\left[2H^{2}{\cal F}_T-4H(\dot{{\cal G}_T}+H{\cal G}_T)+2(\dot{\Theta}+H\Theta)+\rho_{m}\right]/\dot{\pi}^{2},\nonumber \\
 &  & D_{10}=2(\Theta-H{\cal G}_T)/\dot{\pi},\nonumber \\
 & & D_{11}=\left[6\{(3H^{2}+\dot{H})\dot{\pi}-H\ddot{\pi}\}\Theta+6H\dot{\pi}\dot{\Theta}+2(3H\dot{\pi}-\ddot{\pi})\Sigma+2\dot{\pi}\dot{\Sigma}-\mu\dot{\pi}^{2}\right]/\dot{\pi}^{2},\nonumber \\
 &  & D_{12}=\left[2H(\dot{{\cal G}_T}+H{\cal G}_T)-2(\dot{\Theta}+H\Theta)-\rho_{m}\right]/\dot{\pi}\,,\nonumber \\
 &  & M^{2}=\left[\dot{\mu}+3H(\mu+\nu)\right]/\dot{\pi}\nonumber \\
 &  & =-K_{,{\pi\pi}}+(\ddot{\pi}+3H\dot{\pi})K_{,\pi X}+2XK_{,\pi\pi X}+2X\ddot{\pi}K_{,\pi XX}\nonumber \\
 &  & +[6\, H(G_{{3,\pi{\it XX}}}X+G_{{3,\pi X}})\dot{\pi}-2\, G_{{3,\pi\pi X}}X-2\, G_{{3,\pi\pi}}]\ddot{\pi}+6\, H\left(G_{{3,\pi\pi X}}X-G_{{3,\pi\pi}}\right)\dot{\pi}\nonumber \\
 &  & +6\, G_{{3,\pi X}}X\dot{H}+2(9\,{H}^{2}G_{{3,\pi X}}-G_{{3,\pi\pi\pi}})X\nonumber \\
 &  &+[6\,{H}^{2}(4\, G_{{4,\pi{\it XXX}}}{X}^{2}+8\, G_{{4,\pi{\it XX}}}X+G_{{4,\pi X}})-6\, H(2\, G_{{4,\pi\pi{\it XX}}}X+3\, G_{{4,\pi\pi X}})\dot{\pi}]\ddot{\pi}\nonumber \\
 &  & +[12\, H(G_{{4,\pi X}}+2\, G_{{4,\pi{\it XX}}}X)\dot{H}+6\, H(6\,{H}^{2}G_{{4,\pi{\it XX}}}X-2\, G_{{4,\pi\pi\pi X}}X+3\,{H}^{2}G_{{4,\pi X}})]\dot{\pi}\nonumber \\
 &  &+12\,{H}^{2}\left(2\, G_{{4,\pi\pi{\it XX}}}{X}^{2}-3\, G_{{4,\pi\pi X}}X-G_{{4,\pi\pi}}\right)-6\left(2\, G_{{4,\pi\pi X}}X+G_{{4,\pi\pi}}\right)\dot{H}\nonumber \\
 &  &+[2\,{H}^{3}(2\, G_{{5,\pi{\it XXX}}}{X}^{2}+7\, G_{{5,\pi{\it XX}}}X+3\, G_{{5,\pi X}})\dot{\pi}-6\,{H}^{2}(5\, G_{{5,\pi\pi X}}X+G_{{5,\pi\pi}}+2\, G_{{5,\pi\pi{\it XX}}}{X}^{2})]\ddot{\pi}\nonumber \\
 &  & +[2H^{3}(2\, G_{{5,\pi\pi{\it XX}}}{X}^{2}-9\, G_{{5,\pi\pi}}-7\, G_{{5,\pi\pi X}}X)-12\, H(G_{{5,\pi\pi X}}X+G_{{5,\pi\pi}})\dot{H}]\dot{\pi}\nonumber \\
 &  & +6\,{H}^{2}X\left(3\, G_{{5,\pi X}}+2\, G_{{5,\pi{\it XX}}}X\right)\dot{H}+6\,{H}^{2}X\left(3\,{H}^{2}G_{{5,\pi X}}-G_{{5,\pi\pi\pi}}+2\,{H}^{2}G_{{5,\pi{\it XX}}}X-2\, G_{{5,\pi\pi\pi X}}X\right).\nonumber\\
 \label{masss}
\end{eqnarray}

\section*{References}
\addcontentsline{toc}{section}{References}

\bibliographystyle{elsarticle-num}
\bibliography{fundamentalfields}

\end{document}